\documentclass[12pt]{article}
\usepackage[cmtip,arrow]{xy}\usepackage{pb-diagram,pb-xy}%

\input xy
\xyoption{all}
\input xy
\xyoption{all}
\usepackage{heck}
\usepackage{cite}
\usepackage{graphicx}
\usepackage{makeidx}
\usepackage{multicol}
\usepackage{amsfonts}
\usepackage{mathrsfs}
\usepackage{amssymb}
\usepackage{amsmath}%

\providecommand{\U}[1]{\protect\rule{.1in}{.1in}}
\numberwithin{equation}{section}

\newcommand{\bea}{\begin{eqnarray}}
\newcommand{\eea}{\end{eqnarray}}
\newcommand{\be}{\begin{equation}}
\newcommand{\ee}{\end{equation}}

\newcommand{\bem}{\begin{pmatrix}}
\newcommand{\eem}{\end{pmatrix}}

\def\U{\Upsilon}

\def\ca{{\cal A}}

\def\ch{{\cal H}}

\def\ck{{\cal K}}
\def\cl{{\cal L}}

\def\cn{{\cal N}}
\def\co{{\cal O}}
\def\cp{{\cal P}}
\def\cq{{\cal Q}}

\def\ct{{\cal T}}

\def\cw{{\cal W}}

\def \Z {{\mathbb Z}}
\def \C {{\mathbb C}}
\def \R {{\mathbb R}}

\xyoption{arc}

\begin{document}

\bibliographystyle{utphys}

%\date{July, 2011}

%\preprint{arXiv:????.????}

\institution{SISSA}{Scuola Internazionale Superiore di Studi Avanzati, via Bonomea 265, I-34100 Trieste, ITALY}

\title{Categorical Tinkertoys \\ for $\cn=2$ Gauge Theories }
%EndExpansion
%

\authors{Sergio Cecotti\footnote{e-mail: {\tt cecotti@sissa.it}}}%
%EndExpansion

\abstract{In view of classification of the quiver 4d $\cn=2$ supersymmetric gauge theories, we discuss the characterization of the quivers with superpotential $(Q,\cw)$ associated to a $\cn=2$ QFT which, in some corner of its parameter space,
looks like a gauge theory with gauge group $G$. 
The basic idea is that the Abelian category $\mathsf{rep}(Q,\cw)$ of (finite--dimensional) representations of the Jacobian algebra $\C Q/(\partial\cw)$ should enjoy what we call the \textit{Ringel property of type $G$;} in particular, $\mathsf{rep}(Q,\cw)$ should contain a universal `generic' subcategory, which depends only on the gauge group $G$, capturing the universality of the gauge sector.  More precisely, there is a family of `light' subcategories $\mathscr{L}_\lambda\subset \mathsf{rep}(Q,\cw)$, indexed by points $\lambda\in N$, where $N$ is a projective variety whose irreducible components are copies of $\mathbb{P}^1$ in one--to--one correspondence with the simple factors of $G$. If $\lambda$ is the generic point of the $i$--th irreducible component, $\mathscr{L}_\lambda$ is the universal subcategory corresponding to the $i$--th simple factor of $G$. Matter, on the contrary, is encoded in the subcategories $\mathscr{L}_{\lambda_a}$ where $\{\lambda_a\}$ is a finite set of \textit{closed} points in $N$.

In particular, for a Gaiotto theory there is one such family  of subcategories, $\mathscr{L}_{\lambda\in N}$, for each maximal degeneration of the corresponding surface $\Sigma$, and the index variety $N$ may be identified with the \emph{degenerate} Gaiotto surface itself: generic light subcategories correspond to cylinders, while closed--point subcategories to `fixtures' (spheres with three punctures of various kinds) and higher--order generalizations. The rules for `gluing' categories are more general that the geometric gluing of surfaces, allowing for a few additional exceptional $\cn=2$ theories which are not of the Gaiotto class. 

 We include several examples and some amusing consequence, as the characterization in terms of quiver combinatorics of asymptotically free theories. 
}

%TCIMACRO{\TeXButton{Maketitle}{\maketitle}}%
%BeginExpansion
\maketitle
%EndExpansion

\tableofcontents

\newpage

\section{Introduction: the Ringel property}

There is a large and interesting class of 4d $\cn=2$ supersymmetric gauge theories \cite{SW1,SW2} whose BPS spectrum has a quiver description, \textit{i.e.}\!
their BPS states may be identified with the stable objects of the ($\C$--linear) Abelian category $\mathsf{rep}(Q,\cw)$ of representations of a (finite) quiver $Q$ bounded by the Jacobian relations $\partial\cw=0$; here $\cw$ is the superpotential of a quiver SQM, defined on $Q$, whose supersymmetric vacua correspond to the 4d BPS states (see the recent papers \cite{CV11,ACCERV1,ACCERV2}; for previous work see \cite{Denef00,DM07,DM,Dia99,DFR1,DFR2,FM00,Fiol,Denef,FHHI02,FHH00,Feng:2001xr,HK05,FHKVW05,Feng:2005gw}; for a nice review of the basic theory, see \cite{Aspinwall09}). More intrinsically, the BPS states are described by the corresponding bounded derived category $D^b(\mathsf{rep}(Q,\cw))$; indeed, for a given $\cn=2$ model the pair $(Q,\cw)$ is determined only up to derived equivalence or, more concretely, up to quiver mutations \cite{derksen1,kellerA} (\textit{a.k.a.}\! Seiberg dualities \cite{Seiberg}\!\!\cite{FHHI02}).

For such a \emph{quiver} 4d $\cn=2$ theory the computation of the BPS spectrum reduces to a standard problem in representation theory, and hence it is rather easy when compared with the difficulty of computing the non--perturbative spectrum of other strongly--interacting QFTs.

Given this state of affairs, it is natural to ask for an intrinsic characterization of the Abelian categories $\mathscr{A}$ (or, rather, their derived counterparts $D^b(\mathscr{A})$) which \emph{do} correspond to physically consistent 4d QFTs. The next step would be to fully classify such categories. 

In this direction, two results are known:

\noindent$\bullet\ $ \textbf{The $2d/4d$ correspondence \cite{CNV}\!\!\!\cite{CV11}.} According to this principle, the classification of the quiver four--dimensional $\cn=2$ theories is contained in the classification \cite{CV92}  of the {two--dimensional} $(2,2)$ theories with\footnote{\ $\hat c_\mathrm{uv}$ is one--third the Virasoro central charge at the UV fixed point. } $\hat c_\mathrm{uv}<2$. In other words, all  4d quiver theories {arise} from a 2d one. Comparing with ref.\!\cite{CV92} one gets, in particular, a \emph{necessary condition} the pair $(Q,\cw)$ must satisfy.
Mathematically, it may be rephrased as follows: the pair $(Q,\cw)$ is the $3$--CY completion\footnote{\ With some abuse of language, by the \textit{$3$--CY completion} of an algebra of the form $\C Q^\prime/I$ in this paper we simply mean the pair $(Q,\cw)$ of quiver with superpotential constructed in \S.\,6.9 of ref.\!\cite{kellerB}; see ref.\cite{kellerB} for the full story.} of an algebra $\mathsf{A}$ of global dimension $\leq 2$ whose Coxeter element $\Phi$ has spectral radius $1$.

Being just a necessary condition, this criterion is powerful only in the negative, that is, to prove that a given quiver $Q$ cannot correspond to any QFT. \textit{E.g.}\!\! by a theorem of Ringel \cite{Ring1} the only acyclic quivers with spectral radius $1$ are the ordinary and affine $ADE$ Dynkin quivers. Thus, all other acyclic quivers cannot correspond to any QFT. The class of QFTs associated to ordinary and affine $ADE$ quivers (respectively Argyres--Douglas \cite{AD} \& asymptotically free $SU(2)$ theories \cite{CV11}) is also characterized as the set of $\cn=2$ theories having a {very small} BPS chamber in which only the simple representations correspond to stable particles \cite{ACCERV1,ACCERV2}.

\noindent$\bullet\ $  \textbf{The classification of complete theories \cite{CV11}.} The classification of the $\cn=2$ 4d theories corresponding to the 2d ones with  $\hat c_\mathrm{uv}\leq 1$  is equivalent to the classification of the mutation--finite quivers (but the $k$--Kronecker ones for $k>2$ which are ruled out by Ringel's theorem). Luckily, this last classification is known \cite{felikson}: using it we recover all quiver $A_1$ Gaiotto theories \cite{Gaiotto,GMN09} plus 11 exceptional models \cite{CV11}. Roughly speaking, the complete class contains all the $\cn=2$ theories whose gauge group is a product of $SU(2)$'s in \emph{all} weak--coupling corners of their parameter space.   
\medskip

The purpose of the present paper is to lay some groundwork for a future extension of the classification to the general quiver case. Our immediate goal is to characterize pairs $(Q,\cw)$ corresponding to gauge theories with gauge group $G$.

The general idea is that  --- in order to correspond to a 4d $\cn=2$ QFT with gauge group $G$ ---  the Abelian category
$\mathsf{rep}(Q,\cw)$ must satisfy a necessary condition
 which we call  the \emph{Ringel property of type $G$}. However, having the Ringel property is certainly not sufficient for a category to  represent a sensible quantum field theory.

\subsection*{The physics of the Ringel property}
Let me first state the Ringel property in the physical language.
Suppose the pair $(Q,\cw)$ corresponds, in some limit, to a $\cn=2$ QFT with gauge group $G$. The central charge $Z$ gives a map between the physical parameters of the QFT and the stability conditions on $\mathsf{rep}(Q,\cw)$ under which  the BPS spectrum (at the given  point in parameter space) maps to the stable representations (for details see, \textit{e.g.}, \cite{ACCERV1,ACCERV2}). In particular, by varying the Yang--Mills coupling constants $g_i$ we get a $s$--dimensional family of such stability conditions, $Z_g$,  where $s$ is the number of simple factors in $G$.
Now take
the zero gauge--coupling limit $g_i\rightarrow 0$. The exact non--perturbative BPS spectrum should asymptotically match the perturbative spectrum of the QFT. Hence, the objects $X$ of $\mathsf{rep}(Q,\cw)$ which are asymptotically stable under $Z_{g\rightarrow 0}$ and have finite mass, $|Z_{g\rightarrow 0}(X)|=O(1)$, should consist of vector supermultiplets, making precisely \emph{one copy} of the adjoint representation of the gauge group $G$, plus matter transforming in definite representations of $G$. Moreover, the matter should be mutually--local (\textit{i.e.}\! have zero Dirac electric/magnetic pairing) with respect to the gauge vectors, and must also be consistent with the Weinberg--Witten theorem \cite{WW}, that is, the light matter should consist only of spin $\leq 1/2$ states.  All other BPS states which are stable in a weakly--coupled chamber, $g_i\ll 1$, must be solitonic objects, carrying non--Noetherian conserved charges, and having masses $O(1/g_i^2)$ as $g_i\rightarrow 0$ in order to asymptotically decouple from the perturbative spectrum. 

In summary, from the QFT side, the Ringel property is just the tautological statement
\begin{center}
\fbox{\ $\phantom{\Big|}$perturbation theory is asymptotically correct as $g\rightarrow 0$\ } 
\end{center} 
Of course, only the gauge couplings $g_i$ are treated perturbatively; the model may be strongly--coupled in other respects.

\subsection*{The mathematics of the Ringel property}

The statement `perturbation theory is asymptotically correct as $g\rightarrow 0$'  has a simple but deep translation in the categorical language. It says that the category $\mathsf{rep}(Q,\cw)$ contains a full exact Abelian subcategory $\mathscr{L}$, closed under direct sums, summands and extensions, whose stable objects are the perturbative states. The statement that the perturbative states give rise to a nice Abelian category of their own, $\mathscr{L}$, just reflects the fact that perturbation theory defines a \emph{formal} $\cn=2$ supersymmetric QFT, that is, that extended \textsc{susy} is a symmetry visible in perturbation theory. The statement that $\mathscr{L}$ is just a \emph{sub}category reflects the fact that perturbation theory is an incomplete description of the quantum dynamics which asks for a non--perturbative completion, that is for a larger category of BPS states $\mathsf{rep}(Q,\cw)$.  

The Abelian (sub)category $\mathscr{L}$ should be very simple: since it does not describe higher--spin particles, its bricks\footnote{\ We recall that an object $X$ of a $\C$--linear Abelian category is a \emph{brick} iff $\mathrm{End}_\C(X)=\C$. A brick is, in particular, indecomposable. All stable objects are bricks \cite{king}. The mathematical terminology we use is summarized in appendix \ref{TTTTTTT}. } belong to families of (non--isomorphic) representations having dimension at most $1$. Mutual locality of matter with the gauge vectors implies a stronger constraint on $\mathscr{L}$, namely the symmetry\footnote{\ If the theory has a Lagrangian description, we get the strong symmetry condition $\langle X_i, X_j\rangle_E=\langle X_j, X_i\rangle_E$ for all (stable) bricks $X_i,X_j$.}
\begin{equation*}
\langle  M,X\rangle_E=\langle X,M\rangle_E,\quad \text{where }\left[\:\begin{array}{l}X\ \text{any (stable) brick}\\ \begin{aligned}M\ &\text{a generic brick in}\\ &\text{a 1--dim. family,}\end{aligned}\end{array}\:\right.
\end{equation*}
and $\langle X,Y\rangle_E$ is the Euler form.

In facts, much more should be true. $\mathscr{L}$ should contain a sub--category,  \textit{universal} for a given gauge group $G$, describing the decoupled SYM gauge vectors in the adjoint of $G$. Let us focus on a simple Lie subgroup $G_k\subset G$. Since vectors have spin $1$, the $G_k$--SYM sector should correspond to a one--parameter family of sub--categories of $\mathscr{L}$ having the form\footnote{\ The notation is as in ref.\!\cite{RI} \S.\,2.2, and is reviewed in appendix \ref{TTTTTTT}. {Note} that the notation already implies that the $W$--bosons are mutually--local between themselves.} 
$$\bigvee_{\zeta\in U}\mathscr{L}_\zeta\subset \mathscr{L}\subset \mathsf{rep}(Q,\cw),$$ 
where $U$ is Zariski--open in $\mathbb{P}^1$. In addition, $\mathscr{L}$ should contain a finite collection of `matter' subcategories in one--to--one correspondence with the matter sectors which get decoupled as $g\rightarrow 0$. For instance, in SQCD with $N_f$ flavors, we should get $N_f$ such elementary subcategories, each one associated with a free elementary quark in the given representation of $G$. Classification of the possible matter subcategories is equivalent to the classification of possible matter subsectors (most of them not having any Lagrangian description) and hence of all four--dimensional $\cn=2$ gauge theories with a given group $G$ modulo the condition of UV--completeness, which requires that the sum of the contributions to the gauge--coupling beta function from the matter subcategories is not larger than twice the dual Coxeter number of $G$. The Gaiotto $\ct_N$ theories \cite{Gaiotto} obtained by wrapping $N$ M5--branes on a sphere with three maximal regular punctures, and, more generally, any tinkertoy `fixture'  \cite{CD1,CD2} may be seen as special instances of such `matter' categories.

For $G=SU(2)$ the complete matter subcategories with isospin $\tfrac{1}{2}$ are the standard stable tubes of period $p\geq 2$, while the SYM part is described by a one--dimensional family of homogeneous tubes \cite{RI,ASS}. Physically, a stable tube of period $p$ corresponds to an Argyres--Douglas system \cite{AD} of type $D_p$ with its $SU(2)$ symmetry gauged. In particular, the $p=2$ tube describes a gauged quark iso--doublet. Absence of Landau poles then restricts the number and types of tubes with $p\geq 2$; imposing this last condition we recover the full classification of ref.\!\cite{CV11}.

More precise statements of the Ringel property (of type $G$) are given in the next section and trough the paper.
\bigskip

We assume the gauge group $G$ to be a product $G=G_1\times G_2\times \cdots\times G_s$ of simple simply--laced Lie groups. The extension to \emph{non}--simply--laced gauge groups remains an open problem.

\subsection*{Organization of the paper}

After an overview of the main ideas in section 2, we start with two (rather long) review sections. In section 3 we review the quiver theories with gauge group $SU(2)$, which correspond to Ringel's canonical algebras (up to derived equivalence, of course).
In section 4 we review general complete theories with gauge group a \emph{power} of $SU(2)$, which correspond to ($3$--CY completions of) \textit{string} algebras\footnote{\ I thank William Crawley--Boevey for pointing out to me the relevance of the string algebras for 4d gauge theories.} \cite{butring} or to natural generalizations of such algebras.

In section 5 we collect some preliminary facts in view of the generalization to arbitrary (simply--laced) gauge groups. Section 6 discusses pure SYM, which is the universal sector of any gauge theory. In section 7 we introduce a class of algebras which are the straightforward generalization to $G=ADE$ of the usual canonical algebras for the $A_1$ case. In section 8 we discuss the matter subcategory in the simple `Lagrangian' case (\textit{i.e.}\! the light BPS states are required to be pairwise mutually--local). In section 9 we present some (non conclusive) comments on the general canonical case. In section 10 we show that all the algebras we consider are consistent with the necessary condition following from the $2d/4d$ correspondence. In section 11 we present conclusions and speculations.
Technicalities and boring computations are confined in the appendices.

\section{Overview}\label{intt}

In this section we give a discursive overview of the main ideas, motivating why the structures we describe are general features of quiver $\cn=2$ QFTs, without entering in the details. The statements may seem rather abstract at first, so the reader may prefer to give first a look to the several explicit examples in the following sections.\smallskip

To fix the notation, we first recall some basic facts (see \textit{e.g.}\! \cite{ACCERV1,ACCERV2}). Given a quiver $Q$ with superpotential $\cw$ we write $\mathsf{rep}(Q,\cw)$ for the Abelian category of the representations of $Q$ satisfying the relations $\partial\cw=0$. Equivalently, $\mathsf{rep}(Q,\cw)$ is the category of finite--dimensional modules of the Jacobian algebra $\C Q/(\partial\cw)$. A BPS chamber\footnote{\ To be precise, this is a \emph{formal} BPS chamber which may or may not be physically realizable, see ref.\cite{CV11}. } is specified by
a central charge (\textit{a.k.a.}\! stability function) $Z(\cdot)$ which is a homomorphism of Abelian groups from 
the charge/dimension lattice $K_0(\mathsf{rep}(Q,\cw))$ to $\C$ such that the positive cone in $K_0(\mathsf{rep}(Q,\cw))$ of dimensions of objects (\textit{i.e.}\! charge vectors of \emph{particles} as contrasted to \emph{anti}particles) is mapped in the upper--half plane $\ch\subset \C$.
An object $X\in \mathsf{rep}(Q,\cw)$ is stable --- that is, a physical BPS particle present in the BPS chamber $Z(\cdot)$ --- iff for all non--zero proper  subobjects $Y$ of $X$ one has
\begin{equation}
0< \arg Z(Y)< \arg Z(X)<\pi.
\end{equation}
The mass of the BPS particle corresponding to the stable object $X$ is $|Z(X)|$; by abuse of language, we will say that an object $X\in \mathsf{rep}(Q,\cw)$ has mass $|Z(X)|$.

It is well--known \cite{CNV}\!\!\cite{ACCERV2} that $\cn=2$ pure SYM with gauge group $G=ADE$ corresponds to the mutation--class of the quiver with superpotential $\cw$ 
\begin{equation}\label{sqaurepro}
\overleftrightarrow{G}\,\square\,\widehat{A}(1,1)\equiv \mathbb{G},
\end{equation}
where $\overleftrightarrow{G}$ is the Dynkin graph\footnote{\ I use the same letter $G$ to denote three different things: the gauge group, its Lie algebra, and its Dynkin diagram. I hope this will not cause confusion.} of the gauge group $G$ with the alternating orientation, $\widehat{A}(1,1)$ is the Kronecker quiver $1\rightrightarrows 2$, and $\square$ stands for the the square--tensor--product of two alternating quivers defined in ref.\!\cite{kellerP}. 
\textit{E.g.}, for $G=SU(5)$  the SYM quiver has the form
\begin{equation}
\mathbb{A}_4\colon\qquad
\begin{aligned}
\xymatrix{\bullet \ar[r] & \bullet \ar@<0.4ex>[d] \ar@<-0.4ex>[d] & \bullet \ar[l]\ar[r] &  \bullet\ar@<0.4ex>[d] \ar@<-0.4ex>[d]\\
\bullet \ar@<0.4ex>[u] \ar@<-0.4ex>[u] &\bullet\ar[r]\ar[l] & \bullet  \ar@<0.4ex>[u] \ar@<-0.4ex>[u] &\bullet \ar[l]}
\end{aligned}
\end{equation}
The quiver $\mathbb{G}$ has one Kronecker subquiver $\bullet\rightrightarrows \bullet$ per simple root of $G$, and the sink of the $i$--th Kronecker subquiver $\mathbf{Kr}_i$ is connected to the source of the $j$--th one $\mathbf{Kr}_j$ by $-C_{ij}$ arrows, where $C_{ij}$ is the Cartan matrix of $G$. The superpotential $\cw$ of $\mathbb{G}$ is also fixed by the square--tensor--product prescription and will be given in \S.\,\ref{pureSYMN}.
We write $\mathsf{rep}(\mathbb{G})$ for the Abelian category of the representations of $\mathbb{G}$ satisfying the relations $\partial\cw=0$.
The spectrum of pure SYM in the chamber $Z(\cdot)$ is then given by the stable objects of the  Abelian category $\mathsf{rep}(\mathbb{G})$.

\subsection{Light and heavy subcategories}\label{sec:light}

\subsubsection{$\cn=2$ SYM}

For pure SYM our first result is the decomposition
\begin{equation}
\mathsf{rep}(\mathbb{G})= \mathscr{L}^\mathrm{YM}(G)\vee \mathscr{H}(G), 
\end{equation}
where the SYM `light' subcategory $\mathscr{L}^\mathrm{YM}(G)$ is an exact Abelian subcategory of $\mathsf{rep}(\mathbb{G})$ closed under direct sums/summands, kernels, cokernel, and extensions, with the nice property that it contains all stable representations which have bounded mass in the limit $g_i\rightarrow 0$. The stable objects in $\mathscr{H}(G)$, instead, have masses of order $O(1/g_i^2)$. In \emph{all} weakly--coupled chambers the stable objects of the light subcategory $\mathscr{L}^\mathrm{YM}(G)$ correspond to the perturbative degrees of freedom of SYM, namely mutually--local vector--multiplets 
making one copy of the adjoint representation of $G$. The stable objects of $\mathscr{H}(G)$ are, at weak coupling, massive solitonic dyons.
The light subcategory $\mathscr{L}^\mathrm{YM}(G)$ may be further decomposed into one--dimensional families of Abelian subcategories
\begin{equation}\label{uuuer8}
\mathscr{L}^\mathrm{YM}(G)= \bigvee_{\text{simple factors}\atop G_k\ \text{of }G} \left(\,\bigvee_{\lambda\in\mathbb{P}^1} \mathscr{L}_\lambda^\mathrm{YM}(G_k)\right),
\end{equation} 
 the subcategories $\mathscr{L}_\lambda^\mathrm{YM}(G_k)$ with different $\lambda$ being pairwise orthogonal and  equivalent. Eqn.\eqref{uuuer8} just reflects the fact that the gauge interactions are mediated by (mutually--local) vectors.
 We call the category $\mathscr{L}_\lambda^\mathrm{YM}(G_k)$, which depends only on the simple Lie group $G_k$, the \emph{homogeneous $G_k$--tube}. 
 
 Explicitly,  $\mathscr{L}_\lambda^\mathrm{YM}(G_k)$ is the subcategory of the representations $Y$ of the SYM quiver $\overleftrightarrow{G_k}\,\square\,\widehat{A}(1,1)$  which restrict in each Kronecker subquiver $\mathbf{Kr}_{i,k}$ to the form 
\begin{equation}
Y\big|_{\mathbf{Kr}_{i,k}}\colon\quad\xymatrix{\C^\ell \ar@<0.6ex>[rr]^A\ar@<-0.6ex>[rr]_{\mathbf{1}} && \C^\ell},\quad\text{with } \det[x-A]=(x-\lambda)^\ell,
\end{equation}
that is, to a representation in the stable tube $\mathcal{T}_\lambda$ of the AR quiver of the  algebra $\C\widehat{A}(1,1)$ labelled by the point $\lambda\in \mathbb{P}^1$ \cite{RI,ASS,CB} (the \underline{same} $\lambda$ for all Kronecker subquivers $\mathbf{Kr}_{i,k}$ of $\overleftrightarrow{G_k}\,\square\,\widehat{A}(1,1)$).
Homogenous $SU(2)$--tubes are just the ordinary homogeneous stable tubes $\ct_\lambda$ \cite{RI,ASS,CB}.

\subsubsection{General $\cn=2$ gauge theories}

Universality of the gauge interactions at weak coupling suggests that the perturbative subcategory $\mathscr{L}^\mathrm{YM}(G)$ should be a subcategory of $\mathsf{rep}(Q,\cw)$ for all pairs $(Q,\cw)$ which correspond to a 4d $\cn=2$ QFT with gauge group $G$. In facts one must have
\begin{equation}
\mathscr{L}^\mathrm{YM}(G)\subset \mathscr{L}\subset\mathsf{rep}(Q,\cw)
\end{equation}
where $\mathscr{L}$ is the `light' exact subcategory containing all perturbative states of our $G$--gauge theory whose existence  is predicted by the Ringel property\footnote{\ Properly speaking, this is true provided we pick up a representative $(Q,\cw)$ in the mutation--class which `covers' the weak--coupling regime of interest. Otherwise the statement holds only up to derived equivalence.}.

This idea can be made more precise. First of all \cite{Gaiotto}, most 4d $\cn=2$ models have several distinct limits in which some gauge couplings get weak; these different weakly--coupled descriptions are related by generalized $S$--dualities \cite{Gaiotto}.  Therefore, most of the physically relevant Abelian categories $\mathsf{rep}(Q,\cw)$ should contain \emph{infinitely many} light subcategories of the form $\mathscr{L}$, while the $S$--duality group acts by permuting such light Abelian subcategories. More precisely, the statement applies to the corresponding \emph{derived} categories: \textit{e.g.}\! the derived category for $SU(2)$ SCQD with $N_f=4$ \cite{SW2} should decompose in a $\mathbb{P}^1\mathbb{Q}$--family of light subcategories $\{\mathscr{L}_{q}\}_{q\in \mathbb{P}^1\mathbb{Q}}$ on which the $S$--duality group $PSL(2,\Z)$ acts projectively; each $\mathscr{L}_q$ is a copy of the `perturbative' category whose only stable states are a $W$--boson and four fundamental quarks. That such a $\mathbb{P}^1\mathbb{Q}$--family of `perturbative' categories is the correct non--perturbative  category, $D^b(\mathsf{rep}(Q,\cw)$, for $N_f=4$ $SU(2)$ SQCD is in facts a known mathematical theorem (see \S.\,\ref{ttubular} for details and references).

Second: in an interacting theory there are pairs of matter objects $X,Y\in \mathscr{L}\setminus\mathscr{L}^\mathrm{YM}(G)$ whose extension $Z$,
\begin{equation}0\rightarrow X \rightarrow Z \rightarrow Y\rightarrow 0,\end{equation}
is an indecomposable object of the YM subcategory $\mathscr{L}^\mathrm{YM}(G)\subset \mathscr{L}$. Extensions of this kind correspond to physical processes and may be roughly identified with the perturbative vertices. For instance, $X,Y$ may correspond to a quark--antiquark pair which annihilates into a gauge vector $Z$. Charge conservation in such processes then naturally leads to the identification of the conserved charges with the Grothendieck group $K_0(\mathsf{rep}(Q,\cw))$.

Given that the matter consists of spin $\leq 1/2$ states, the stable matter objects $X,Y\in \mathscr{L}\setminus\mathscr{L}^\mathrm{YM}(G)$ are necessarily rigid. The closure under extensions in $\mathscr{L}$ of the class of matter objects  then contains at most countably many objects (up to isomorphism).  Consequently, the representations in $\mathscr{L}^\mathrm{YM}_\lambda(G)$ with $\lambda$ \emph{generic} are not the extension of any matter representations. Therefore, the light category of a general (quiver) gauge theory (in any given $S$--duality frame) must have the structure 
\begin{equation}
\mathscr{L}= \left(\,\bigvee_{r\in F} \mathscr{L}_r\right)\bigvee \left(\,\bigvee_{\text{simple factors}\atop G_k\ \text{of }G} \bigvee_{\lambda\in U_k} \mathscr{L}^\mathrm{YM}_\lambda(G_k)\right)\supset\bigvee_{\text{simple factors}\atop G_k\ \text{of }G} \bigvee_{\lambda\in \mathbb{P}^1} \mathscr{L}^\mathrm{YM}_\lambda(G_k),
\end{equation}
where $U_k\subseteq \mathbb{P}^1$ is Zariski--open and $F$ is a finite set.
Thus universality of the gauge interactions as $g_i\rightarrow 0$, together with the limits on the charges and spin of the light states \cite{WW}, produce the following description of the light subcategory $\mathscr{L}$ corresponding to a limit with a weakly coupled gauge group $G_1\times G_2\times \cdots \times G_s$: one has a orthogonal decomposition 
\begin{equation}\label{orthdec}
\mathscr{L}= \bigvee_{\lambda \in N} \mathscr{L}_\lambda,
\end{equation}
where $N$ is a closed projective variety with $s$ irreducible $\mathbb{P}^1$ components  in one--to--one correspondence with the simple factor groups $G_k$ of $G$, such that
\begin{align*}
&\bullet\ \lambda\ \text{is a generic point in the $k$--th irreducible component of }N\quad\Rightarrow\quad \mathscr{L}_\lambda= \mathscr{L}_\lambda^\mathrm{YM}(G_k)\\
&\bullet\ \lambda\ \text{is in the closure of the $k$--th irreducible component of }N\hskip 0.66cm \Rightarrow\quad \mathscr{L}_\lambda\supset \mathscr{L}_\lambda^\mathrm{YM}(G_k). 
\end{align*} 
The special subset $F\subset N$ of points $\lambda$ such that the light subcategory $\mathscr{L}_\lambda$ is \emph{not} a homogenous $G$--tube (that is, it is strictly larger than the one for pure SYM) consists of finitely many (closed) points.
The subcategories $\mathscr{L}_{\lambda\in F}$, called \emph{non--homogeneous} $G$--tubes, correspond to the various matter sectors which decouple from SYM in the limit $g\rightarrow 0$. To be more precise, for each $\lambda \in F$ there is a subcategory $\mathsf{matter}_{\lambda}\subset \mathscr{L}_\lambda$ whose stable objects are the BPS states of the corresponding decoupled matter subsystem.  In many instances, the $\cn=2$ matter sector associated to the point $\lambda_0\in F$ has a quiver description of its own, say $(Q_{\lambda_0},\cw_{\lambda_0})$. In this case one has
\begin{equation}
D^b(\mathsf{matter}_{\lambda_0})= D^b(\mathsf{rep}(Q_{\lambda_0},\cw_{\lambda_0})).
\end{equation}
In general $\mathsf{matter}_{\lambda_0}$ corresponds to an interacting $\cn=2$ theory, except when the full theory has a weakly--coupled Lagrangian description; in that case  $\mathsf{matter}_{\lambda_0}\subset \mathscr{L}_{\lambda_0}$ just describe a bunch of free hypermultiplets.

\begin{figure}
%\begin{center}
\begin{equation*}
\xymatrix{\ar@{-}[ddrr] && \ar@{-}[ddll]\\
\ar@{-}[rr] && \\
&& }
\end{equation*}
\caption{$\mathbb{P}^1$ components in Kodaira's singular fiber of type IV.}
\label{ssswwrtm}
%\end{center}
\end{figure}
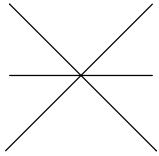

The matter sector encoded in the Abelian subcategory
$\mathscr{L}_{\lambda_0}$ is charged with respect to all gauge group factors $G_k$ such that the point $\lambda_0$ is in the closure of the generic point of the $\mathbb{P}^1$ component of $N$ associated to that group factor. Thus the matter in $\mathscr{L}_{\lambda_0}$ may be charged under two or more factor groups $G_{k_1},\dots, G_{k_r}$ only if $\lambda_0$ lays at the intersection of the associated irreducible components of $N$.
\textit{E.g.}, if we consider the gauge theory with $G=SU(2)^3$ coupled to a half--hypermultiplet in the representation $(\mathbf{2},\mathbf{2},\mathbf{2})$, $N$ is equal to the Kodaira singular fiber of type IV, and the $\tfrac{1}{2}(\mathbf{2},\mathbf{2},\mathbf{2})$ matter subcategory is contained in $\mathscr{L}_0$, where $0$ is the crossing point of the three lines (figure \ref{ssswwrtm}).
The same $N$ applies (in particular $S$--duality frames) to the complete $\cn=2$ theory described by the Derksen--Owen  quiver $X_7$ \cite{CV11}. 

\subsection{Relation with the Gaiotto construction \cite{Gaiotto,CD1}}

\emph{A priori} the above discussion is not related to the Gaiotto construction of a large class of $\cn=2$ gauge theories by compactifying the 6d $(2,0)$ theory on a Riemann surface $\Sigma$ with various insertions of defect operators. However, \emph{a posteriori}  
the two constructions turn out to be strictly related.

In the Gaiotto framework each weakly coupled regime corresponds to a maximal degeneration of the Riemann surface with punctures $\Sigma$. The degenerate surface consists of a number of infinitely long cylinders, \textit{i.e.} copies of $\C^*=\mathbb{P}^1\setminus\{0,\infty\}$, whose ends are attached to the punctures of spheres having 3 punctures of various kinds called `fixtures' \cite{Gaiotto,CD1}. We may collapse each fixture down to a point. These points close the cylinders, making them $\mathbb{P}^1$'s, crossing each other at the fixture points and having a finite number of marked points associated to the insertion of a fixture with two free punctures (\textit{i.e.}\! punctures to which no cylinder is attached). Since the long cylinders are in one--to--one correspondence with the simple factors of the gauge group \cite{Gaiotto}, namely with the $\mathbb{P}^1$ irreducible components of $N$, by collapsing the fixtures the degenerate Gaiotto curve gets identified with the index variety $N$ according to the dictionary\footnote{\ `Fixtures' here means generalized fixtures in the sense of \cite{CV11} for the $A_1$ case, that is, the resulting 4d theory is not required to be superconformal. }:
\vglue 6pt
\begin{center}
\begin{tabular}{|c|c|}\hline
\textbf{Ringel property} & \textbf{Gaiotto construction}\\\hline
index reducible variety $N$ & $\begin{matrix}\text{maximal degeneration of the curve $\Sigma$}\\ \text{with fixtures collapsed}\end{matrix}$\\\hline
1--parameter family of homogeneous $G_k$--tubes & cylinders with simple gauge group $G_k$\\\hline
non--homogenous $G_{k_1}\times G_{k_2}\times G_{k_3}$--tubes & fixtures with  $G_\text{flavor}\supset G_{k_1}\times G_{k_2}\times G_{k_3}$\\\hline
\end{tabular}
\end{center}
\vglue 6pt

Then we may think of the non--homogenous $G$--tubes as a categorical version of the Gaiotto tinkertoys \cite{Gaiotto,CD1,CD2}. The usual way of thinking of the tinkertoys identifies them with the subcategory (object class) $\mathsf{matter}_{\lambda_0}\subset \mathscr{L}_{\lambda_0}$, \textit{e.g.}\! the $A_1$ $3$--punctured sphere is seen as a half--hyper in the $(\mathbf{2},\mathbf{2},\mathbf{2})$ rep.\! of $SU(2)^3$, that is, four free hypers. However this conventional viewpoint is \emph{not} convenient in either approaches: from the categorical side the subcategory $\mathsf{matter}_{\lambda_0}$ is neither exact nor closed under extensions, while from the Gaiotto point of view, we do have fixtures which are matter--empty --- they represent zero free hypers --- yet non trivial and whose presence is required for the consistency of the construction, see \cite{CD1,CD2} for examples. As emphasized in ref.\!\cite{CD2} (see their remarks at the end of page 25) the empty fixtures carry, in some sense, a non--trivial flavor symmetry.  Categorically speaking, these empty fixtures correspond to situations where the subcategory $\mathsf{matter}_{\lambda_0}=0$, while $\mathscr{L}_{\lambda_0}$ is non--trivial, and in fact is a homogenous $H$--tube for some $H\supseteq G$.  

For these reasons, when referring to \textit{the} category of a fixture we always mean $\mathscr{L}_{\lambda_0}$ with $\lambda_0\in N$ the point at which the corresponding three $\mathbb{P}^1$'s cross.
Then, for the sake of comparison, we check that the matter subcategory $\mathsf{matter}_{\lambda_0}\subset  \mathscr{L}_{\lambda_0}$ is the expected one (we are able to make explicit checks only for $A_1$ theories). 

There is, however, an important difference between the two approaches. Even limiting ourselves to the categorical version of the $A_1$ tinkertoys, their allowed gluings are more general than the pant decompositions of punctured Riemann surfaces; roughly speaking, in the pant decompositions, the SYM sectors are represented by cylinders $\C^*$, that is, $\mathbb{P}^1$ less \textit{two} points, whereas categorically we can use any open set in $\mathbb{P}^1$, \textit{i.e.}\! we can delete \emph{any} finite number of points.  This allows us to construct additional theories, which are not of Gaiotto type, using the \emph{same} building blocks. However this flexibility comes in a package: whereas the Gaiotto construction produces models which are automatically quantum consistent, a general gluing produces a formal field theory which may have Landau poles or other pathologies. There are only finitely many non--Gaiotto gluings of the $A_1$ categorical tinkertoys which are free of Landau poles, and they correspond to the 11 exceptional complete models \cite{CV11}.  

\subsection{The classification problem}

\subsubsection{The three steps}

From the above results it follows that the classification of all (quiver) $\cn=2$ theories with gauge group $G$ (in some corner of its parameter space) consists of three main steps:
\begin{enumerate}
\item the classification of the non--homogenous $G$--tubes $\mathscr{L}_{\lambda_0}$,  which gives the list of all possible matter subsystems we may couple to $G$--SYM;
\item the conditions expressing the quantum consistency of the coupling of SYM with a given collection of such matter subsystems, that is, the rules specifying which combinations $\bigvee_{\lambda\in N}\mathscr{L}_\lambda$ of $G$--tubes, almost all homogeneous, are physically allowed (the `gluing rules');
\item determine which pairs $(Q,\cw)$ are such that 
$$\bigvee_{\lambda\in N}\mathscr{L}_\lambda \equiv \mathscr{L} \subset \mathsf{rep}(Q,\cw). $$ 
The category $D^b(\mathsf{rep}(Q,\cw))$ is the non--perturbative completion to the `perturbative' category $D^b(\mathscr{L})$. Two such  `perturbative' categories with the same completion 
\begin{equation*}D^b(\mathscr{L}_1),D^b(\mathscr{L}_2)\subset D^b(\mathsf{rep}(Q,\cw))\end{equation*} are then related by a generalized $S$--duality.
\end{enumerate}

For $G=SU(2)^k$ the list of the non--homogenous $G$--tubes with $\mathsf{matter}_{\lambda_0}$ containing only spin $\leq 1/2$ BPS states (in all chambers) is explicitly known. 
For $k=1$ the non--homogenous $SU(2)$--tubes representing matter in the fundamental representation are the well--known non--homogeneous stable tubes \cite{RI,ASS,CB} which are labelled by a number $p\geq 2$ (the period). Each such matter subsystem gives its contribution 
to the $\beta$--function of the gauge coupling. The consistency conditions of step 2.\! above is then simply the requirement that the $\beta$--function is non--positive for all $SU(2)$ gauge groups, as required by UV--completeness. 
Step 3.\! is solved, up to derived equivalence, by taking as $(Q,\cw)$ the $3$--CY completion \cite{kellerB} of the canonical algebra
\cite{RI} of type $\{p_1,p_2,\cdots, p_\ell\}$, where $p_i$ is the period of the $i$--th non--homogeneous tube. Equivalently, the non--perturbative category is $D^b(\mathrm{Coh}\,X)$, where $X$ is the weighted projective line of weights $\{p_1,p_2,\cdots, p_\ell\}$  \cite{LE1,LE2}.  

For $G=SU(2)$
(and more generally for $G=SU(2)^k$) one recovers in this way the classification obtained in \cite{CV11} using totally different ideas. The present approach, however, may be extended (at least in principle) to arbitrary gauge groups.
\smallskip

Therefore we pose the
\smallskip

\textbf{Problem.} \textit{Classify the non--homogeneous $G$--tubes $\mathscr{L}_{\lambda_0}$.}
 \smallskip
 
 Although the complete answer to this problem is not known for higher rank gauge groups, we do have a large supply of examples of $G$--tubes coming from two sources:
 \begin{itemize}
\item generalized Gaiotto tinkertoys;
\item canonical constructions (see subsection \ref{cangaugefunct}).
\end{itemize}

The first class is a bit implicit, while the second one is totally explicit.
For this reason, most of the paper is dedicated to the canonical examples. 
 
 \subsubsection{Non--perturbative completion of $\mathscr{L}$. {Non}--quiver $\cn=2$ theories}
 
Step 3.\! in the classification program asks for a non--perturbative completion of the `perturbative' category $\mathscr{L}$. One may wonder how such a completion may work in practice, and also to which extend the non--perturbative completion would be \emph{unique} (up to derived equivalence, of course).

A procedure for constructing the non--perturbative completion of $\mathscr{L}$ is equivalent to a method for \emph{solving} the QFT. Indeed, one starts with the basic degrees of freedom of the theory --- say, gluons and quarks --- which are encoded in $\mathscr{L}$, then constructs its non--perturbative completion and reads the full BPS spectrum of the QFT.

Let us discuss these matters in a simple example, namely $SU(2)$ SQCD with $N_f\leq 4$. The perturbative category $\mathscr{L}$ (see \S.\ref{CanRingel} for details and references) is the category of \emph{rank zero} coherent sheaves on the weighted projective line $X_{N_f}$, which is $\mathbb{P}^1$ (associated to the gauge vectors) with $N_f$ points of weight 2 (each such point representing a quark doublet). The simple objects of $\mathscr{L}$ are the skyscraper sheaves on $X_{N_f}$ (suitably defined on the points of weight 2).  The set of simple objects of $\mathscr{L}$ may  be identified with a geometric space, \textit{i.e.}\! $X_{N_f}$ itself.

From this point of view, the non--perturbative completion of $\mathscr{L}$ is quite obvious, say, from string theory. It is the full category
$D^b(\mathrm{Coh}\,X_{N_f})$, where we drop the perturbative restriction to sheaves of rank zero. This is indeed the right answer and the stable  objects in $D^b(\mathrm{Coh}\,X_{N_f})$ with positive rank are precisely the BPS dyons of $SU(2)$ SQCD. So, at least in this baby example, the completion appears to be uniquely dictated. We find the full BPS spectrum just by completing the category encoding the perturbative degrees of freedom (quarks and gauge supermultiplets).

If our $\cn=2$ model is not a quiver gauge theory, it still has a light subcategory $\mathscr{L}$. A typical example of a non--quiver model is an $A_1$ Gaiotto theory defined on a curve of higher genus, $g\geq 3$, without punctures \cite{CV11}. For such a theory, one constructs the light category $\mathscr{L}$ by going to a degenerate limit and using the correspondence of the previous subsection. The problem is its non--perturbative completion. It should be a category which is more general than (the derived category of) the representations of a quiver, but otherwise the structure described in this paper applies to these theories too.

 \subsection{The canonical gauge functor}\label{cangaugefunct}
 
 For a  $\cn=2$  $SU(2)$ gauge theory, there is a canonical construction of the `non--perturbative' category $D^b(\mathsf{rep}(Q,\cw))$ as the derived category of the corresponding \emph{completed}
 canonical algebra (see next section for a review). Indeed, this is precisely why these algebras are called `canonical'.
 
Similar canonical constructions work for many other $\cn=2$ models, \textit{e.g.}\! for SQCD with any number of colors $N_c$ and flavors $N_f$, or $SU(2)^3$ SYM coupled to a half three--fundamental, or else \emph{any} $\cn=2$ gauge theory which is `nice' in the sense of ref.\!\cite{Tack}. Whenever such a canonical construction exists (and is known) the computation of the BPS spectrum dramatically simplifies and the Ringel property becomes quite transparent.

In these situations there is a preferred pair $(Q,\cw)$ in the mutation class; its Jacobian algebra $\C Q/(\partial\cw)$ is the $3$--CY completion of an algebra $\mathsf{A}$ with the property that \emph{all} representations of $\C Q/(\partial\cw)$ which are stable in any weakly--coupled\footnote{\ With respect to a given $S$--duality frame.} chamber --- those representing light particles as well as the heavy dyon ones --- are in fact modules of $\mathsf{A}$ extended by zero to modules of the full Jacobian algebra $\C Q/(\partial\cw)$. Thus, at weak coupling we may replace, without loss of physical information, the category $\mathsf{rep}(Q,\cw)$ with the subcategory $\mathsf{mod}\,\mathsf{A}$ of (finite--dimensional) modules of $\mathsf{A}$. 

The crucial property of the canonical setting is the existence of a functor   
 \begin{equation}
 \mathscr{G}\colon \mathsf{mod}\,\mathsf{A} \rightarrow \mathsf{rep}(\mathbb{G}),
 \end{equation}
called the \emph{canonical gauge functor}, mapping a module $X$ into a representation $\mathscr{G}(X)$ of the pure SYM quiver $\mathbb{G}$. Roughly speaking, the functor $\mathscr{G}$ has the effect of projecting on the weakly--coupled gauge sector of the theory.
If the gauge group is a product $G=G_1\times\cdots \times G_s$,
the gauge functor $\mathscr{G}$ maps a representation $X\in \mathsf{mod}\,\mathsf{A}$ into a $s$--tuple of SYM representations $\mathscr{G}(X)_a$, $a=1,2,\dots,s$, with $$\mathscr{G}(X)_a\in \mathrm{rep}(\mathbb{G}_a).$$
 
The pair $(\mathsf{A},\mathscr{G})$, with the functor $\mathscr{G}$ satisfying the properties below, is called a canonical pair, and the algebra $\mathsf{A}$ a \emph{$G$--canonical algebra}; a $\cn=2$ theory admitting (in some weakly coupled limit) a canonical pair, will be called a \emph{canonical} $\cn=2$ QFT. As we already mentioned,
most of the `usual' $\cn=2$ gauge theories are indeed canonical in this sense.

The gauge functor $\mathscr{G}$ is required to enjoy the following properties:
\begin{enumerate}
\item the object $X$ is \textit{stable at weak coupling}  $\Rightarrow$ $\mathscr{G}(X)_a$ is either stable at weak coupling or zero;
\item the magnetic charges\footnote{\ ${}^LG$ stands for the Langlands dual of the group $G$.}
\begin{equation}
m\colon  \mathsf{rep}(Q,\cw)\rightarrow \Gamma_\mathrm{weight}({}^LG), 
\end{equation} 
are preserved\footnote{\ Instead the electric charges of the $G$--canonical theory and of pure SYM differ by a linear combination of the charges of the matter sector. Physically, Noether charges mix between themselves but not with topological ones.} by $\mathscr{G}$
\begin{equation}
m(\mathscr{G}(X))=m(X).
\end{equation}
So, a representation $X\in \mathsf{mod}\,\mathsf{A}$ corresponds to a dyon stable at weak coupling (with respect to the given $S$--duality frame) iff the representation $\mathscr{G}(X)$ corresponds to a stable dyon in pure SYM;
\item Let $\mathscr{L}\subset \mathsf{mod}\,\mathsf{A}$ be the exact Abelian subcategory (closed under extensions and direct summands) consisting of all objects $X$ such that $m(X)=0$ while\footnote{\ By $m(Y)\leq 0$ we mean that $-m(Y)$ is either zero or a positive weight of ${}^LG$.} $m(Y)\leq 0$ for all its subobjects $Y$. $\mathscr{L}$ is the `light' subcategory of \S.\ref{sec:light}, containing the states with zero magnetic charge. From 1.\! we get
\begin{equation}
X\in\mathscr{L}\ \text{indecomposable},\  \text{and}\ \mathscr{G}(X)_a\neq 0\quad\Rightarrow\quad\mathscr{G}(X)_a\in \mathscr{L}^\mathrm{YM}_{\lambda_a}(G_a)\end{equation}
for a \emph{unique} $\lambda_a\in \mathbb{P}^1$. Define
\begin{equation}
N=\Big\{(\lambda_1,\dots,\lambda_s)\in (\mathbb{P}^1)^s\:\Big|\: \exists\, X\in \mathscr{L}\ \text{such that } 0\neq \mathscr{G}(X)_a\in \mathscr{L}^\mathrm{YM}_{\lambda_a}(G_a)\ \,\forall\:a\Big\}.
\end{equation}
Then,
\begin{enumerate}
\item $\mathscr{L}= \bigvee_{\lambda\in N}\mathscr{L}_\lambda$;
\item each indecomposable $X\in\mathscr{L}$ belongs to a unique $\mathscr{L}_\lambda$.
\item $\mathscr{G}(\mathscr{L}_\lambda)_a\subseteq\mathscr{L}_{\lambda_a}^\mathrm{YM}(G_a)$.
\end{enumerate}
The category $\mathscr{L}_\lambda$ is precisely a \textit{$G$--tube} indexed by $\lambda\in N$ in the sense of \S.\,\ref{sec:light}.
Thus the $G$--tubes arising in the canonical context are easy to construct explicitly using the gauge functor. There is no reason, however, to expect that all $G$--tubes are canonical. Canonical $SU(2)$--tubes are just ordinary stable tubes \cite{RI,ASS,CB} or obvious generalizations.
\item
If, as $g\rightarrow 0$,  the matter sector consists of $n$ decoupled subsystems (\textit{e.g.}\! for a Lagrangian theory $n$ is the number of flavors of quarks) there are $n$ special points
$\lambda^{(\alpha)}\in N$ such that
$N\setminus \{\lambda^{(1)},\cdots, \lambda^{(n)}\}$ is the disjoint union of $s$ Zariski--open sets $U_a\subset \mathbb{P}^1$. 
The restriction of the gauge functor to $U_a$ has universal properties
\begin{align}\label{fffirsttt}
&\mathrm{(a)}\qquad \lambda\in U_a\hskip 1.67cm \Rightarrow\quad \mathscr{G}(\cdot)_a\Big|_{\mathscr{L}_\lambda}\colon\mathscr{L}_\lambda \rightarrow \mathscr{L}_\lambda^\mathrm{YM}(G_a)\quad \text{is an equivalence;}\\
&\mathrm{(b)}\qquad \lambda\in U_a,\; b\neq a\quad\Rightarrow\quad\mathscr{G}(\cdot)_b\Big|_{\mathscr{L}_\lambda}\colon\mathscr{L}_\lambda \rightarrow \mathscr{L}_\lambda^\mathrm{YM}(G_b) \quad\text{is locally constant}\label{fffirsttt2}\end{align}
in the sense that there is a \underline{fixed} $\mu_{a,b}\in\mathbb{P}^1$ such that $\mathscr{G}(\mathscr{L}_\lambda)_b\subseteq \mathscr{L}_{\mu_{a,b}}^\mathrm{YM}(G_b)$ for  $\lambda\in U_a$. In simple instances, in which the quiver representations associated to the $W$--bosons of distinct group factors have non--overlapping supports in $Q$, the restriction to $U_a$ of $\mathscr{G}(\cdot)_b$ 
($b\neq a$) is just zero. However, there are situations, \textit{e.g.}\! in presence of a \emph{half}--hypermultiplet charged with respect to several gauge group factors, where overlapping supports are  unavoidable \cite{CV11}. Physically, equations \eqref{fffirsttt}\eqref{fffirsttt2} say that the theory at hand has the same gauge vectors as pure SYM, guaranteeing universality of the gauge interactions. 

The $G$--tubes $\mathscr{L}_\lambda$ with $\lambda\in N\setminus \{\lambda^{(1)},\cdots, \lambda^{(n)}\}$ are {homogeneous}. For $G=SU(2)$ they are homogeneous in the usual sense \cite{RI,ASS,CB};
\item the $G$--tube over a special point $\lambda^{(\alpha)}$  contains the states of an irreducible decoupled (in the limit $g\rightarrow 0$) matter sector. For a Lagrangian model these are just quarks (free hypermultiplets). For a non--Lagrangian model they are more sophisticated forms of matter: \textit{e.g.}\! for $G=SU(2)$ the $SU(2)$--tubes are classified by a single integer $p\geq 1$, namely their \emph{period} ($p=1$ meaning a homogeneous tube). The matter corresponding to a tube of period $p$ is an Argyres--Douglas system of type $D_{p}$ \cite{CV11}
(where $D_1=\emptyset$, $D_2=A_1\oplus A_1$, $D_3=A_3$, of course);
\end{enumerate}

The canonical $G$--tubes themselves may be studied with the help of other natural functors. \textit{E.g.}\! the analysis of the homogeneous $G$--tubes gets reduced to the study of the representations of the preprojective algebra of the corresponding Dynkin graph, $\cp(G)$
\cite{pre1,pre2}. This is a nice finite--dimensional Frobenius algebra. Similar functors relate the `quark' non--homogeneous $G$--tubes to the representations of a preprojective algebra $\mathsf{mod}\,\cp(Q)$, where now the quiver $Q$ depends on $G$ and the gauge representation $R$ of the quark.
%\smallskip

\subsection{Graphical rules for asymptotic freedom} 

The classification of complete $\cn=2$ theories of ref.\!\cite{CV11} produced, in particular, graphical rules to decide whether a given quiver $Q$ corresponds to an  asymptotically free, a UV--superconformal, or UV--incomplete QFT.  Similar graphical criterions emerge for general gauge groups $G$. A simple and elegant example of such a rule is the following. Consider the class of $\cn=2$ with a simple--laced gauge group $G=ADE$ coupled to a single quark in a representation $R$ of $G$ specified by a Dynkin label of the form \begin{equation}[d_1,d_2,\cdots,d_r]=[0,\cdots, 0,1,0\cdots, 0].\end{equation}

We ask: \textit{i)} which pairs $(G,R)$ correspond to an asymptotically free theory? \textit{ii)} which pairs $(G,R)$ give $\cn=2$ theories which may be engineered by Type IIB on a (non--compact) Calabi--Yau hypersurface embedded in $\C^4$?

Of course, the answer to these questions is well--known, see \textit{e.g.}\! ref.\!\cite{Tack} for the list of the relevant pairs $(G,R)$. However, the gauge functor analysis gives a purely graphical interpretation of these lists.
Let us associate to a pair $(G,R)$ the augmented Dynkin graph obtained by adding a new vertex to the Dynkin diagram of $G$ connected to the $i$--th node of $G$ by $d_i$ links.
The $(G,R)$ theory is asymptotically free if and only if the augmented Dynkin graph is also a (finite--type) Dynkin graph.
See figure \ref{graphicalrule2} and the discussion around it.

\section{The $SU(2)$ case (Ringel's canonical algebras)}\label{CanRingel}

\subsection{Review of the complete case}

Let us start by considering the simplest situation, namely the $\cn=2$ quiver gauge theories having gauge group $SU(2)$ and matter in the fundamental representation. As will be clear below, their quivers $(Q,\cw)$ coincide, up to mutation, with the $3$--CY completion of Ringel's canonical algebras \cite{RI,LE1}. For this reason we shall refer to these $\cn=2$ models as the $SU(2)$--canonical theories. 

Limiting ourselves to models without Landau poles\footnote{\ This restriction will be eliminated in \S.\ref{makingsense} below.}, these theories are obviously \emph{complete}\footnote{\ We recall that a quiver $\cn=2$ theory is complete iff its quiver has mutation--finite type.}, and hence should be present in the classification of ref.\!\cite{CV11}.  In the notations of that paper, the full list of UV--complete $SU(2)$--canonical theories is
\begin{align}
&\widehat{A}(p,q),\ p\geq q\geq 1, &&\widehat{D}_r,\ r\geq 4 && \widehat{E}_6,\ \widehat{E}_7,\ \widehat{E}_8\\
&\widehat{\widehat{D}}_4,\ \widehat{\widehat{E}}_6,\ \widehat{\widehat{E}}_7,\ \widehat{\widehat{E}}_8,
\label{Tututubu}\end{align}
where the first line corresponds to the \textit{affine} (\textit{domestic canonical}) models, which are asymptotically free, and the second one to the \textit{elliptic} (\textit{tubular canonical}) models which are UV superconformal. The five $SU(2)$ SQCD models which are UV complete, namely those with $N_f=0,1,2,3,4$, correspond, respectively, to the following $SU(2)$--canonical models
\begin{equation}
\widehat{A}(1,1),\ \widehat{A}(2,1),\ \widehat{A}(2,2),\ \widehat{D}_4,\ \widehat{\widehat{D}}_4,
\end{equation}
while all the other $SU(2)$--canonical theories do not have a weakly--coupled Lagrangian description. Their physics is described in detail in \cite{CV11} and it is most easily understood by considering the full family rather than each model by itself. As we shall review below, they all correspond to $SU(2)$ SYM coupled to a number of $D_p$--type Argyres--Douglas models.

From the above list, we notice that the family of the $SU(2)$--canonical models contains all the complete theories with gauge group \textit{strictly} $SU(2)$ with the single exception of $\cn=2^*$. Indeed, $\cn=2^*$ is the only UV--complete $SU(2)$ gauge theory with matter not in the fundamental representation. However, $SU(2)$ $\cn=2^*$ \textit{is} canonical in the more broad sense of \S.\,\ref{cangaugefunct}, as we shall see in \S.\ref{makingsense} below; there we discuss also some wilder generalization to higher $SU(2)$ representations.

\subsubsection{Definitions and conventions}

To be specific, for $k\geq 0$, let 
\begin{equation}\kappa=\{n_1,n_2,\cdots, n_k\}
 \end{equation}
 be a set of $k$ integers with $n_i\geq 2$. The canonical algebra $\mathscr{C}(\kappa)$ of type  $\kappa\equiv \{n_1,n_2,\cdots, n_k\}$ may be described from many different viewpoints \cite{RI,ASS,LE1,LE2}, some of which are reviewed in appendix \ref{appcanonical}. 
Here we describe the $3$--CY completion of a canonical algebra of type $\kappa=\{n_1,n_2,\cdots, n_k\}$. Its \emph{canonical quiver} $Q(\kappa)$ consists of a `source' node\footnote{\ Sink and source nodes are so distinguished that we adopt a special symbol for them, $\diamondsuit$ and $\spadesuit$, respectively. All other nodes are represented by a plain $\bullet$ symbol.} $\spadesuit$ and a `sink' node $\diamondsuit$ connected by $k$ distinct paths --- one for each integer $n_a$ in the set $\kappa$ --- of the form
\begin{equation}\label{patthh}
\ell_a\colon\quad \overbrace{\xymatrix{\spadesuit \ar[r] & \bullet\ar[r] &\bullet \ar[r] & \cdots \ar[r] &\bullet \ar[r] &\diamondsuit}}^{n_a\ \text{arrows}}
\end{equation}
(each node $\bullet$ belonging to a unique such path), as well as by $k-2$ `inverse' arrows going directly from $\diamondsuit$ to $\spadesuit$ (with the usual convention that a negative number of arrows means arrows in the opposite direction). The total number of nodes of $Q(\kappa)$ is then \begin{equation}\label{numnodes}\#Q(\kappa)_0=2+\sum_a(n_a-1).
\end{equation} 

From the above description, it is obvious that the superpotential $\cw$ vanishes for $k\leq 2$ while for $k\geq 3$ $\cw$ is
\begin{equation}
\cw(\kappa)=\sum_{a=3}^k \mathrm{Tr}\Big[\eta_a \Big(\prod_{\ell_a}\psi-\lambda_a\,\prod_{\ell_1}\psi-\mu_a\prod_{\ell_2}\psi\Big)\Big]
\end{equation}
where $\eta_a$, $a=3,4,\dots, k$ are the Higgs fields associated to the $k-2$ `inverse' arrows $\xymatrix{\diamondsuit\ar[r]^{\eta_a} & \spadesuit}$, while $\prod_{\ell_a}\psi$ is a short--hand symbol for the ordered product of the Higgs fields associated to the arrows along the path $\ell_a$, cfr.\! eqn.\eqref{patthh}, which we shall write simply as $\prod_a\psi$; finally, 
\begin{equation}
 (\lambda_a:\mu_a)\in \mathbb{P}^1
\end{equation}
is a set of $k-2$ \emph{distinct} points in $\mathbb{P}^1\setminus\{0,\infty\}$.\smallskip

\textsc{Definition.} For $k>2$, the canonical algebra\footnote{\ For simplicity, we omit writing the dependence of the algebra form the points $(\lambda_a,\mu_a)\in \mathbb{P}^1$; of course the algebra depends on these parameter modulo projective transformations. } $\mathscr{C}(\kappa)$ is the subalgebra of $\C Q(\kappa)/(\partial\cw)$
\begin{equation}
 \C Q^{0}(\kappa)/(\partial_{\eta_a}\cw)
\end{equation}
where $Q^{0}(\kappa)$ is the subquiver of $Q(\kappa)$ obtained by eliminating the `inverse' arrows $\eta_a$. For $k\leq 2$,
$\mathscr{C}(\kappa)\equiv \C Q(\kappa)$. Then the $3$--CY completion of $\mathscr{C}(\kappa)$ is the full Jacobian algebra $\C Q(\kappa)/(\partial\cw)$.
\medskip

The BPS states correspond to the representations (modules) $X$ of the $3$--CY completed algebra which are stable with respect to the given stability function ($\equiv$ central charge)
\begin{equation}
 Z\colon K_0\big(\mathsf{rep}(Q,\cw)\big)\rightarrow \C\qquad \begin{pmatrix} \text{the positive--cone being mapped}\\ 
 \text{to the upper half--plane }\mathcal{H}\end{pmatrix}.
\end{equation}
We write $\mathsf{mod}\,\mathscr{C}(\kappa)$ for the Abelian category of modules over the canonical algebra $\mathscr{C}(\kappa)$ which we identify with the subcategory of $\mathsf{rep}(Q,\cw)$ of representations $X$ with vanishing inverse arrows, $X_{\eta_a}=0$.

The \emph{Euler characteristic}\footnote{\ $\chi(\kappa)$ is called the `Euler characteristic' because it is the orbifold Euler characteristic of the weighted projective line $X(\kappa)$  \cite{LE1,LE2} such that $D^b(\mathrm{Coh}\,X(\kappa))\simeq D^b(\mathsf{mod}\, \mathscr{C}(\kappa))$. For $\kappa=\emptyset$, $\mathscr{C}(\emptyset)$ is the Kronecker algebra (corresponding to pure $SU(2)$ SYM), and $X(\emptyset)\equiv \mathbb{P}^1$ which has $\chi=2$.} of a canonical algebra $\mathscr{C}(\kappa)$
is \cite{LE1,LE2}
\begin{equation}
\chi(\kappa)=2-\sum_i \left(1-\frac{1}{n_i}\right)
\end{equation}
which has the physical interpretation of the $\beta$--function of the $SU(2)$ gauge coupling $g$ \cite{CV11}
\begin{equation}\label{BETA}
\mu\frac{\partial}{\partial\mu}\frac{1}{g^2}=\frac{\chi(\kappa)}{\pi}.
\end{equation}
Hence the $\cn=2$ theory is \cite{RI,LE1,LE2}
\begin{align}
 \text{asymptotically free}\quad &\Longleftrightarrow\quad \chi>0\ \Leftrightarrow\ \mathscr{C}(\kappa)\ \text{is concealed Euclidean}\\
\text{UV superconformal}\quad &\Longleftrightarrow\quad \chi=0\ \Leftrightarrow\ \mathscr{C}(\kappa)\ \text{is tubular}\label{chizerozero}\\
\text{with Landau poles}\quad &\Longleftrightarrow\quad \chi<0\ \Leftrightarrow\ \mathscr{C}(\kappa)\ \text{is wild}.
\end{align}

In particular, for $\kappa=\emptyset$ we recover pure $SU(2)$ SYM, while $\kappa=\{2,2,\cdots,2\}$ with $N_f$ $2$'s gives $SU(2)$ SQCD with $N_f$ flavors. If the set $\kappa$ contains some $n_i>2$, the model has no Lagrangian formulation, but it may still be constructed \emph{\'a la} Gaiotto--Moore--Neitzke \cite{Gaiotto,GMN09} (but for six exceptional models \cite{CV11}). Note that the formula \eqref{BETA} for the $\beta$--function agrees with the usual one for models having a Lagrangian formulation. %\smallskip

It is remarkable that the representation theory of a canonical algebra is nice (that is, not strictly wild) if and only if the corresponding $\cn=2$ theory is UV--complete. 

\subsubsection{Magnetic charge and weak coupling}\label{maggg}

Suppose a given pair $(Q,\cw)$ corresponds to a $SU(2)$ gauge theory, and let $\delta$ be the charge (\textit{i.e.}\! dimension) vector  of the representations corresponding to the $W$--boson. 
Since the $W$ boson has electric charge $2$, the magnetic charge $m(X)$ of the BPS state described by a representation $X\in \mathsf{rep}(Q,\cw)$ is --- by definition --- given by
\begin{equation}\label{magmag}
m(X)\equiv \frac{1}{2} \langle \dim X,\delta\rangle_\text{Dirac}=
\frac{1}{2}
\Big(\langle \dim X, \delta\rangle_E- \langle \delta, \dim X\rangle_E\Big),
\end{equation} 
where $\langle \cdot, \cdot \rangle_E$ is the Euler form\footnote{\ See appendix \ref{appeuler} for the definitions and main properties.} of any algebra $\mathscr{A}$ whose $3$--CY completion is equal to $\C Q/(\partial\cw)$. 

If $(Q,\cw)$ is the $3$--CY completion of a canonical algebra, the above formula simplifies. A canonical algebra, has a natural candidate for the $W$--boson family of representations with charge/dimension vector
\begin{equation}\label{oooqq}\delta=(1,1,\cdots,1,1).\end{equation} In suitable chambers \eqref{oooqq} is a stable BPS vector multiplet. Note that $\langle \delta, \delta\rangle_E=0$.

 With respect to this canonical\footnote{\ As we shall see in \S.\,\ref{ttubular}, a canonical theory may have other vector multiplets besides the canonical one; the present argument may be repeated for each such vector. In these cases we get several distinct weakly coupled limits which are related by (generalized) $S$--dualities.} $W$--boson, eqn.\eqref{magmag} gives for the magnetic charge function $m\colon K_0\big(\mathsf{rep}(Q,\cw)\big)\rightarrow \Z$ the expression
\begin{equation}
 m(X)= (\dim X)_\spadesuit-(\dim X)_\diamondsuit.
\end{equation}
which restricts to $\mathsf{mod}\,\mathscr{C}(\kappa)$ to the Ringel linear form $\iota(\cdot)=\langle \cdot, \delta\rangle_E$ \cite{RI}.\medskip

The central charge $Z\colon K_0\big(\mathsf{rep}(Q,\cw)\big)\rightarrow \C$ is a function of the parameters of the physical theory and, in particular, of the Yang--Mills coupling constant $g$. Let us focus on the stability function $Z(\cdot)$ in the weak--coupling limit $g\rightarrow 0$. In this limit the classical expression for $Z(\cdot)$,
\begin{equation}\label{classZ}
Z(X)\approx  \langle a\rangle\,e(X)+ \left(\frac{\theta}{2\pi} + \frac{4\pi\,i}{g^2}\right)\langle a\rangle\, m(X) +\text{(mass terms),}
\end{equation}
is asymptotically correct. 
In eqn.\eqref{classZ} $\langle a\rangle$ is the v.e.v.\! of the the complex adjoint field in the vector supermultiplet, and $e(X), m(X)$ stand, respectively, for the electric and magnetic charges of the BPS state associated to the representation $X$. Since, in the region of parameter space covered by the given quiver, the magnetic charge $m(X)$ takes both positive and negative values, in order for $Z(X)\in \ch$ as $g\rightarrow 0$ for all $X$'s in $\mathsf{rep}(Q,\cw)$, we must have $\langle a\rangle= C\,i+O(g^2)$ 
with $C$ real positive.
Then,  at weak--coupling (with respect to the canonical $W$--boson) we have a stability function $Z(\cdot)$ of the form
\begin{equation}\label{stabfunctwc}
 Z(X)=Z_0(X)-\frac{C}{g^2}\, m(X),\qquad g\rightarrow 0,\  \text{and }C\; \text{real positive},
\end{equation}
where $Z_0(\cdot)$ is some admissible stability function, and $g$ is asymptotically small. Depending on the particular $Z_0(\cdot)$, there may be \textit{several} such weak--coupling chambers, with different BPS spectra (\textit{i.e.}\! different stable representations). Our statements will apply to all such weak--coupling chambers.
\smallskip

\textbf{Proposition.}\label{prop} \textit{Let $X\in \mathsf{rep}(Q(\kappa),\cw)$ be a representation which is stable in the canonical weak coupling ($g\rightarrow 0$). Then $X_{\eta_i}=0$ for all arrows $\eta_i$, that is, $X$ is a representation in $\mathsf{mod}\,\mathscr{C}(\kappa)$ extended by zero.
Hence, \emph{at weak coupling}, we may replace the $3$--CY completed category $\mathsf{rep}(Q(\kappa),\cw)$ by the simpler (sub)category $\mathsf{mod}\,\mathscr{C}(\kappa)$.}
\smallskip

\textsc{Proof.} There is nothing to show for $k\leq 2$. For $k>2$, consider the following map from 
the representations of our completed algebra $\C Q(\kappa)/(\partial\cw)$ to representations of the Kronecker one
\begin{equation}\label{toKr}
\mathscr{G}\colon\quad X\longmapsto\begin{aligned}
          \xymatrix{ X_\spadesuit\ar@<0.6ex>[rr]^{\prod_1X_\psi}\ar@<-0.6ex>[rr]_{\prod_2X_\psi} && X_\diamondsuit}
         \end{aligned}
\end{equation}
which preserves the magnetic charge (and hence the definition of weak coupling).
$\mathscr{G}(X)$ can be written as a direct sum of indecomposable, $\oplus_i Y_i$. By the relations
$\partial_{\eta_a}\cw=0$, one has
\begin{equation}
 \forall\: a\quad \prod_a \psi\big(X_\spadesuit\big|_{Y_i}\big)\subseteq X_\diamondsuit\big|_{Y_i}.
\end{equation}
The $Y_i$'s are either preprojective (the arrows are injective non--iso), or preinjective (the arrows are surjective non--iso), or regular (in which case we may assume $\prod_1\psi$ to be the identity).
The relations $\partial_\psi\cw=0$ imply
\begin{gather}
 \eta_a \prod_a \psi = \prod_a \psi\: \eta_a=0,\ \text{for }a\geq 3\\
 \intertext{and } \big(\sum_a\lambda_a\,\eta_a\big)\prod_1\psi=\prod_1\psi\big(\sum_a\lambda_a\,\eta_a\big)=\big(\sum_a\mu_a\,\eta_a\big)\prod_2\psi=\prod_2\psi\big(\sum_a\mu_a\,\eta_a\big)=0.
\end{gather} 
Thus the only non--zero part of $\eta_a$ is
\begin{equation}\label{wwwr}
 \eta_a \colon X_\diamondsuit\Big|_{Y_p}\rightarrow X_\spadesuit\Big|_{Y_q}.
\end{equation}
where $Y_p$ (resp.\! $Y_q$) is the direct sum of the preprojective (resp.\! preinjective) $Y_i$'s.
If $Y_p=0$, there is nothing to show. Assume $Y_p$ is not zero. If $\mathscr{G}(X)=Y_p$ the statement is also obvious, so we assume
$\mathscr{G}(X)\neq Y_p$.
Set to zero the vectors in the direct summand $X_\diamondsuit\Big|_{Y_p}$, $X_\spadesuit\Big|_{Y_p}$
(and their images in the $X_\bullet$'s trough the $\psi$'s). By the above relations and assumptions, we get a proper subrepresentation $U$ of $X$. Its central charge is
\begin{gather}
Z(U)=Z(X)+\frac{C}{g^2}\:\big((\dim Y_p)_\spadesuit-(\dim Y_p)_\diamondsuit\big)+O(1)
\end{gather}
and, for $g\ll 1$, $\arg Z(U)>\arg Z(X)$ contradicting weak--coupling stability. Then either $Y_p=0$ or $\mathscr{G}(X)=Y_p$ and (cfr.\! \eqref{wwwr}) $\eta_a=0$ for all $a$'s. \hfill $\square$ 
\bigskip

Therefore, if we are only interested in weak coupling, $g\approx 0$, we may limit ourselves to the category of (finite--dimensional) modules of the canonical algebra $\mathscr{C}(\kappa)$ which is well--understood. The basic result is
\vglue 12pt

\textbf{Theorem} (Ringel \cite{RI}).\label{ringeltheorem} \textit{One has
\begin{equation}
\mathsf{mod}\,\mathscr{C}(\kappa)= \mathcal{P}\vee \mathcal{T}\vee \mathcal{Q},
\end{equation}
and for an indecomposable module $X$
\begin{gather}
m(X)<0 \Leftrightarrow X\in \mathcal{P}\\
m(X)=0 \Leftrightarrow X\in \mathcal{T}\\
m(X)>0 \Leftrightarrow X\in \mathcal{Q}
\end{gather}}
\textit{Moreover
\begin{equation}\mathcal{T}\equiv \bigvee_{\lambda\in \mathbb{P}^1}\mathcal{T}_\lambda\end{equation} is a sincere stable tubular $\mathbb{P}^1$--family of type $\kappa=\{n_1,\dots, n_k\}$ separating $\mathcal{P}$ from $\mathcal{Q}$. $\mathcal{T}$ is controlled by the restriction of $\chi_C$ to $\ker m$.}\vglue 12pt

\textsc{Translation for physicists:} the category $\mathcal{P}$ (resp.\! $\mathcal{Q}$) is generated by the dyons of negative (resp.\! positive) magnetic charge. States with only electric charges (that is, \emph{mutually local} with respect to the gauge vectors) generate the Abelian category $\mathcal{T}$. The states in $\mathcal{T}$ are the only ones with bounded mass in the weak coupling limit $g\rightarrow 0$; states in $\mathcal{P}$, $\mathcal{Q}$ have masses $O(1/g^2)$, cfr.\! eqn.\eqref{stabfunctwc}. Hence the Abelian subcategory $\ct$ is precisely our `light' subcategory $\mathscr{L}$, and 
$\mathscr{L}_\lambda\equiv \ct_\lambda$. The purely electric stable states are precisely \emph{one} $W$--boson (associated to the generic brick of the $\mathbb{P}^1$--family) and $k$ matter subsystems each equal to a $D_{n_i}$ Argyres--Douglas theory associated to a tube of period $n_i$. In particular, there are no higher spin states in $\mathcal{T}$.
\medskip

Note that in this theorem there is no assumption on the type $\kappa$, that is, it is true even if $\mathscr{C}(\kappa)$ is wild (\textit{i.e.}\! the QFT has Landau poles). On the other hand, the properties of the dyonic subcategories $\cp, \cq$ dramatically depend on the type: in the domestic case ($\equiv$ AF) they contain only hypermultiplets, in the
tubular case ($\equiv$ UV superconformal) they contain both hypermultiplets and vector multiplets (which form $SL(2,\Z)$ orbits under $S$--duality, see \S.\,\ref{ttubular}), while in the wild case ($\equiv$ Landau pole) states of arbitrary high spin will be present. Representation theory and physical intuition match perfectly in all cases.    
\smallskip

Controlled by $\chi_C|_{\ker m}$ means that an electric state is a hypermultiplet (resp.\! a vector multiplet) iff its charge vector $\dim X$ is a real (resp.\! imaginary) root in Kac's sense \cite{kac1,kac2}.

\subsubsection{The gauge functor $\mathscr{G}$}

The operation in eqn.\eqref{toKr} defines a functor
\begin{equation}
\mathscr{G}\colon \mathsf{mod}\,\mathscr{C}(\kappa)\rightarrow \mathsf{mod}\,\mathbf{Kr}\equiv \mathsf{mod}\, \mathscr{C}(\emptyset),
\end{equation}
where the category $\mathsf{mod}\,\mathbf{Kr}$ of representations of the Kronecker quiver $Q(\emptyset)\colon\ \spadesuit \rightrightarrows \diamondsuit$ corresponds in physics to pure $SU(2)$ SYM. 

The functor $\mathscr{G}$ has all the properties stated in section
\ref{intt} 
for the {canonical gauge functor}. The Ringel theorem quoted on page \pageref{ringeltheorem}  is essentially the existence of $\mathscr{G}$ plus a very explicit description of the light subcategory  $\mathscr{L}\equiv\mathcal{T}$. We write $\mathscr{L}^\mathrm{YM}(SU(2))$ for the light category $\mathcal{T}_\mathbf{Kr}$ of the canonical algebra $\mathbf{Kr}\equiv \mathscr{C}(\emptyset)$, \textit{i.e.}\! for the light category of pure $SU(2)$ SYM.

Note that $X\in \mathcal{T}$ is indecomposable implies $\mathscr{G}(X)$ is a (possibly zero) indecomposable in $\mathscr{L}^\mathrm{YM}(SU(2))$.
If $\mathscr{G}(X)$ is a non--zero indecomposable in the tube $\mathscr{L}_\zeta^\mathrm{YM}(SU(2))$, from the relations of $\mathscr{C}(\kappa)$ we get
\begin{equation}
\forall a\colon\qquad \prod_a\psi = \lambda_a\, \mathbf{1}+\mu_a \,J(\zeta)
\end{equation}
where $(\lambda_1:\mu_1)=(1:0)$, $(\lambda_2:\mu_2)=(0:1)$, and $J(\zeta)$ is the Jordan block of eigenvalue $\zeta$. Hence, \textit{for generic} $\zeta\in\mathbb{P}^1$, all $\prod_a\psi$ are isomorphisms, and, the representation $X$ being irreducible, each arrow $\psi$ must be an isomorphism as well. Then there is precisely one simple object in the category $\ct_\zeta$ which is mapped by $\mathscr{G}$ into the brick representation of the $W$--family parameterized by the generic point $\zeta\in \mathbb{P}^1$, namely
\begin{equation}
 \begin{gathered}
  \xymatrix{\C_\spadesuit \ar@<0.9ex>[rr]^{\zeta} \ar@<-0.3ex>[rr]_{1} && \C_\diamondsuit.}
 \end{gathered}
\end{equation}

Precisely at the $k$ special points $\zeta^{(a)}\equiv (-\lambda_a:\mu_a)\in \mathbb{P}^1$, the map $\prod_a\psi$ fails to be an isomorphism, and we get additional brick representations, which correspond to $k$ matter sectors which we shall discuss momentarily. 

In conclusion, we see that, for $\zeta\in \mathbb{P}^1\setminus\{\zeta^{(1)},\cdots, \zeta^{(k)}\}$, the subcategory
$\mathscr{L}_\zeta\equiv \ct_\zeta$ coincides with the pure SYM one, $\mathscr{L}^\mathrm{YM}_\zeta(SU(2))$, \textit{i.e.}\! is a homogeneous stable tube.

\subsubsection{The matter category $\mathsf{add}\text{(rigid bricks)}$}\label{matterinsidetube}

At the special point $(\zeta^{(a)}:1)=(-\lambda_a:\mu_a)$ in $\mathbb{P}^1$, $\prod_a\psi$ becomes nilpotent (while $\prod_b\psi$ remains an isomorphism for $b\neq a$). 

In $\mathscr{L}_{\zeta^{(a)}}$ there is obviously an indecomposable representation of the form (we draw the case $a=1$)
\begin{equation}\label{rtps}
S_1^{(1)}=\begin{gathered}
\xymatrix{ & 0 \ar[r] & 0 \ar[r] & 0\ar[r] & \cdots\ar[r] & 0 \ar[ddr] &\\ & \C \ar[r] & \C \ar[r] & \C \ar[r] &\cdots\ar[r] & \C \ar[dr]\\ 
\C\ar[uur]\ar[ur]\ar[r]\ar[dr]\ar[ddr] 
& \vdots & \vdots & \vdots & & \vdots \ar[r] & \C\\
 & \C \ar[r] & \C \ar[r] & \C \ar[r] &\cdots\ar[r] & \C \ar[ur]\\
& \C \ar[r] & \C \ar[r] & \C \ar[r] &\cdots\ar[r] & \C \ar[uur]} 
\end{gathered}
\end{equation}
$S^{(a)}_1$ is a brick, $\mathrm{End}(S_1^{(a)})=\C$, and also rigid $\mathrm{Ext}^1(S^{(a)}_1,S^{(a)}_1)=0$. $S^{(a)}_1$ is a simple object of the  (sub)category $\ct_{\zeta^{(a)}}\equiv \mathscr{L}_{\zeta^{(a)}}$, and hence is automatically stable in all weakly--coupled chambers. Being rigid, $S^{(a)}_1$ represents a matter BPS hypermultiplet, belonging to the non--homogeneous tube $\ct_{\zeta^{(a)}}$.

As discussed in \cite{CB}\!\cite{ASS} the Auslander--Reiten translation $\tau$ is an autoequivalence of the category $\ct_{\zeta^{(a)}}$; then if $X$ is a rigid simple so is $\tau X$. Moreover, one has $\dim(\tau X)=\Phi\dim X$, with $\Phi$ the Coxeter element. Hence there are rigid simple objects in $\ct_{\zeta^{(a)}}$ having dimension vectors
\begin{equation}
\Phi^\ell \dim S^{(a)}_1
\end{equation}
for all $\ell\in \Z$, $a=1,\dots, n_a$.

One easily checks that $\Phi^\ell \dim S^{(a)}_1$ is periodic in $\ell$ with period $n_a$. Therefore there are $n_a$ non--isomorphic simples in $\ct_{\zeta^{(a)}}$, namely $\tau^{j-1} S^{(a)}_1$, $j=1,\dots, n_a$. Besides the one in eqn.\eqref{rtps}, the other $n_a-1$  simple representations have the form
\begin{equation}\label{kakaak}S^{(1)}_{i+1}=\begin{gathered}
\xymatrix{ & 0 \ar[r] & \cdots \ar[r] & \C_i\ar[r] & \cdots\ar[r] & 0 \ar[ddr] &\\ & 0 \ar[r] & 0 \ar[r] & 0 \ar[r] &\cdots\ar[r] & 0 \ar[dr]\\ 
0\ar[uur]\ar[ur]\ar[r]\ar[dr]\ar[ddr] 
& \vdots & \vdots & \vdots & & \vdots \ar[r] & 0\\
 & 0 \ar[r] & 0 \ar[r] & 0 \ar[r] &\cdots\ar[r] & 0 \ar[ur]\\
& 0 \ar[r] & 0 \ar[r] & 0 \ar[r] &\cdots\ar[r] & 0 \ar[uur]} 
\end{gathered}
\end{equation}
with a single $\C$ at the $i$--th position along the $a$--th path $\ell_a$. It is clear that these are the only other simple objects in $\ct_
{\zeta^{(a)}}$ besides $S^{(a)}_1$ (they are mapped to zero by $\mathscr{G}$ and belong to the tube $\ct_
{\zeta^{(a)}}$ since $S^{(a)}_1$ does).

Let $S^{(a)}_i=\tau^{i-1}S^{(a)}_1$ be the simples in a tube $\ct_{\zeta^{(a)}}$ of period $n_a$. The tube is precisely the minimal subcategory containing all the $S^{(a)}_i$ which is closed under direct sums and extensions.
However, 
we are presently interested in an even smaller subcategory --- not closed under extensions --- namely the \textit{matter subcategory} $\mathsf{matter}_{\zeta^{(a)}}$ of the non--homogeneous tube. The tube $\ct_{\zeta^{(a)}}$ is larger than $\mathsf{matter}_{\zeta^{(a)}}$; in particular, it contains also a representation $Y$ in the closure of the $\mathbb{P}^1$--family associated to the $W$--boson, which is a \emph{non--rigid} brick. 
Matter --- consisting of spin $\leq 1/2$ states --- corresponds to the
representations in $\ct_{\zeta^{(a)}}$ which are \emph{rigid} bricks, and the matter subcategory we are looking for is then\footnote{\ Again, see appendix \ref{TTTTTTT} for notation. } $\mathsf{add}\text{(rigid bricks)}$, that is, the full subcategory of $\ct_{\zeta^{(a)}}$ whose objects are direct sums of the rigid bricks of $\ct_{\zeta^{(a)}}$. 
We have
\smallskip 

\textbf{Proposition} \cite{CB}. \textit{An indecomposable representation $X$  in a periodic  tube is a rigid brick iff $\dim X<\delta$, where $\delta=(1,1,\cdots, 1)$. }\smallskip

Let us consider then the matter category $\mathsf{add}\text{(rigid bricks)}$. The Gabriel quiver of this matter category has one node associated to each simple $S^{(a)}_i$, $i=1,2,\dots, n_a$, while nodes $i,j$ are connected by $\dim\mathrm{Ext}^1(S^{(a)}_i,S^{(a)}_j)$ arrows. Since $\tau$ is an equivalence, one has
\begin{equation}
\dim \mathrm{Ext}^1(S^{(a)}_{i+r},S^{(a)}_{j+r})=\dim\mathrm{Ext}^1(\tau^rS^{(a)}_i,\tau^rS^{(a)}_j)=\dim\mathrm{Ext}^1(S^{(a)}_i,S^{(a)}_j) 
\end{equation}
so $\dim\mathrm{Ext}^1(S^{(a)}_i,S^{(a)}_j)$ depends only on $i-j \mod n_a$.
From the figure \eqref{kakaak} it is obvious that
\begin{equation}
\dim\mathrm{Ext}^1(S^{(a)}_i,S^{(a)}_j)= \delta_{i,j+1}.
\end{equation}
Hence the matter quiver is given by a single oriented cycle of  length $n_a$, that is, by the affine quiver $\widehat{A}(n_a,0)$. \textit{E.g.}\! for $n_a=8$ one gets
\begin{equation}\label{cycleor}
\begin{gathered}
\xymatrix{ & \bullet \ar[r]^{\phi_1} & \bullet\ar[dr]^{\phi_2} &\\
\bullet \ar[ur]^{\phi_8} & & & \bullet\ar[d]^{\phi_3} \\
\bullet\ar[u]^{\phi_7} & & & \bullet\ar[dl]^{\phi_4}\\
& \bullet\ar[ul]^{\phi_6} & \bullet\ar[l]^{\phi_5} &}
\end{gathered}
\end{equation} 
However this is not the end of the story. The category $\mathsf{nil}(\widehat{A}(n_a,0))$ of the \emph{nilpotent} representations of the quiver $\widehat{A}(n_a,0)$ corresponds to the full periodic tube of period  $n_a$. We are interested in its subcategory $\mathsf{add}\text{(rigid bricks)}$. The indecomposables in $\mathsf{add}\text{(rigid bricks)}$ satisfy
$\dim X<\delta=\sum_i \dim S^{(a)}_i$. Then, in all indecomposable representation $X\in\mathsf{add}(\text{rigid bricks)}$ --- seen as representations of \eqref{cycleor} ---  at least  one of the vector spaces $X_i$ , $i=1,\dots, n_a$, should vanish. Say, $X_{i_0}=0$. The two arrows ending and starting at the $i_0$--th node also vanish. Hence: \textit{the category $\mathsf{add}\mathrm{(rigid\ bricks)}$ is given by direct sums of indecomposable representations of the quiver  $\widehat{A}(n_a,0)$ with at least two arrows equal to zero.} This last requirement is equivalent to the relations $\partial\cw=0$ for $\cw$ the product of all arrows in the cycle, \begin{equation}\label{wwwwdn}\cw=\mathrm{Tr}[\phi_{n_a}\phi_{n_a-1}\cdots \phi_2\phi_1].\end{equation}

The algebra $\C\widehat{A}(n_a,0)/\partial\cw$ is a self--injective Nakayama algebra \cite{ASS1} which is derived equivalent to the 
hereditary algebra $\C D_{n_a}$, which corresponds to the Argyres--Douglas model of type $D_{n_a}$. Indeed, it is well--known that the quiver $\widehat{A}(n_a,0)$ with the superpotential \eqref{wwwwdn} is mutation--equivalent to a $D_{n_a}$ Dynkin quiver.

Thus we have shown that the insertion of a non--homogeneous tube of period $n_a\geq 2$ at a point $\zeta^{(a)}\in \mathbb{P}^1$ means that the $SU(2)$ SYM is coupled to a matter subsector which is a $D_{n_a}$ Argyres--Douglas system. This picture is physically consistent, since these $\cn=2$ systems have a $SU(2)$ global symmetry which may be gauged.
\medskip

A particular case occurs when $n_a=2$. As we have already anticipated, this is equivalent to a doublet of free hypermultiplets.
Indeed, in this case we get the Gabriel quiver $\xymatrix{\bullet\ar@<0.4ex>[r]^\phi& \bullet\ar@<0.4ex>[l]^\psi }$ with superpotential $\mathrm{Tr}[\psi\phi]$. The relations $\partial\cw=0$ set both arrows to zero, leaving the disconnected quiver $\xymatrix{\bullet & \bullet}$, corresponding to two copies of the free hypermultiplet. By the same token, for $n_a=1$ the matter subcategory is trivial, but not the full light subcategory $\mathscr{L}_{\zeta^{(a)}}$, which is a homogeneous tube.

Note that the $D_{n_a}$, $n_a>2$ Argyres--Douglas systems have several BPS chambers with different spectra. Correspondingly, the canonical $SU(2)$ gauge theory with $n_a>2$, \textit{i.e.}\! with non--Lagrangian matter, has several inequivalent weakly coupled BPS chambers.

\subsection{The tubular case and $S$--duality}\label{ttubular}

The charge vector $\delta$ of the canonical $W$--boson of a $\chi\neq 0$ canonical algebra,  has the following characterization. Consider the Tits quadratic form $q(\alpha)\equiv\langle \alpha, \alpha\rangle_E$ on the 
charge (dimension) lattice $\Gamma$. The isotropic elements
\begin{equation}
\Big\{\gamma\in \Gamma\ \Big|\ q(\gamma)=0\Big\}\equiv \mathrm{rad}\, q
\end{equation} 
form a sublattice $\mathrm{rad}\,q\subset \Gamma$ called the \emph{the radical} of $q$. If $\chi\neq 0$, the radical has rank one, and all its elements are of the form $n\, \delta$, the canonical $W$--boson charge vector $\delta$ being the primitive element of the radical sublattice \cite{RI}.

On the contrary, when the canonical algebra has vanishing Euler character, $\chi=0$, the radical lattice has rank $2$. 
Physically, the $\chi=0$ canonical algebras, called \emph{tubular} algebras, correspond to UV superconformal theories whose most spectacular property is $S$--duality. The basic example is $SU(2)$ SQCD with $4$ flavors  \cite{SW2}, or $\widehat{\widehat{D}}_4$ in the Lie theoretical notation.
In fact, as we are going to explain, $S$--duality is a \emph{consequence} of the enhanced rank of $\mathrm{rad}\,q$.

The representation theory of tubular algebras is well--understood thanks again to Ringel \cite{RI}.
There are four such canonical algebras, corresponding to four of the five $SU(2)$ superconformal theories, see eqn.\eqref{Tututubu}.
They are the canonical algebras of types $\kappa$ equal to
\begin{equation}\label{rrrttt}
 \{2,2,2,2\}, \qquad \{3,3,3\},\qquad \{2,4,4\}, \qquad \{2,3,6\}.
\end{equation}
Their completed quiver \cite{CV11} is the Dynkin graph of the elliptic root systems \cite{saito} of the types listed in eqn.\eqref{Tututubu}.  These root systems have a rank--two lattice of isotropic imaginary roots, and they are the root systems of the corresponding toroidal Lie algebras \cite{moody}. For the relation between the representation theory of a tubular algebra and that of the corresponding toroidal Lie algebra, see ref.\!\cite{CBBB}.
\smallskip

Explicitly,
for each quiver $Q(\kappa)$, with the set $\kappa$ as in eqn.\eqref{rrrttt}, one considers the two full subquivers, $Q(0)$ and $Q(\infty)$, obtained by deliting the node $\spadesuit$ and, respectively, $\diamondsuit$. The subquivers $Q(0)$ and $Q(\infty)$ are both affine Dynkin stars of type
$\widehat{D}_4$, $\widehat{E}_6$, $\widehat{E}_7$, and $\widehat{E}_8$, respectively. Let $\delta_0$ and $\delta_\infty$
be the minimal positive imaginary roots of the affine subquivers $Q(0)$ and $Q(\infty)$. $\delta_0$ and $\delta_\infty$ belong to $\mathrm{rad}\,q$, and are obviously the dimension vectors of two one--parameter families of bricks which, in suitable chambers, are stable BPS \emph{vector} multiplets.
One has
\begin{equation}
 \langle \delta_A,\delta_B\rangle_E= - \langle \delta_B,\delta_A\rangle_E= K\, \epsilon_{A,B},\qquad A,B=0,\infty,\ \ K\neq 0.
\label{yyyyywww}\end{equation}
Therefore any vector $v$ of the form $p\,\delta_0+q\,\delta_\infty$ ($p,q\in \mathbb{Q}$) has the property $q(v)\equiv\langle v,v\rangle_E=0$, and the integral such vectors form a rank $2$ lattice $\mathrm{rad}\,q$ equipped with the symplectic form \eqref{yyyyywww} on which the group $SL(2,\Z)$ naturally acts. 

It is natural to identify this $SL(2,\Z)$ with the $S$--duality group of the superconformal SCFTs. Choosing a duality--frame is then equivalent to choosing a point in the orbit of $SL(2,\Z)$, namely an element $p/q\in \mathbb{Q}^+$ which we use to define the magnetic charge $m_{p/q}(\cdot)$ in the chosen duality--frame (consequently introducing a duality--frame--dependent notion of weak--coupling). 
Up to overall normalization, we have
\begin{equation}
 m_{p/q}(X)=\langle p\,\delta_0+q\,\delta_\infty, \dim X\rangle_E.
\end{equation}
Then the basic result on the representation of tubular algebras \cite{RI} may be rephrased as follows: \textit{for $p/q\in \mathbb{Q}^+$ let $\ct_{p/q}$ be the light category in the weak--coupling limit defined by the above duality--frame, \textit{i.e.}\! the additive category generated by the indecomposable $X$'s with $m_{p/q}(X)=0$. Then
\begin{equation}
 \ct_{p/q}= \bigvee_{\zeta\in \mathbb{P}^1} \mathscr{L}_\zeta,
\end{equation}
where for generic $\zeta$ the category $\mathscr{L}_\zeta$ is a stable homogeneous tube, while at 4 (or 3) special points $\zeta^{(a)}\in\mathbb{P}^1$ the stable tube $\mathscr{L}_{\zeta^{(a)}}$ has period $n_a$, where $n_a$ are the 4 (or 3) integers in  eqn.\eqref{rrrttt}. The tube structure defines for each $p/q\in \mathbb{Q}^+$ a gauge functor with the desired properties.
} 

Stated differently, for each given $S$--duality frame we find exactly the expected `perturbative' spectrum, but the full theory contains \textit{infinitely many} copies of this `perturbative' spectrum, one for each $S$--duality frame (actually, the category of representations of the canonical algebra does not contain the full $SL(2,\Z)$ orbit: indeed, it is only a `piece' of the physical theory. To get the full $SL(2,\Z)$ story one has to go to the full derived category which is known\footnote{\ In particular, in ref.\!\cite{ringelderived} it is shown that the derived category of a tubular algebra is equivalent to the stable module category $\underline{\mathrm{mod}}\, \widehat{C}$ of a certain algebra $\widehat{C}$. This category has the same objects as $\mathrm{mod}\, \widehat{C}$ whose explicit form is given on page 165 of that reference: it is equal to $\bigvee_{\gamma\in \mathbb{Q}}\ct_\gamma$, where $\ct_\gamma$ is a $\mathbb{P}^1$--family of tube of type $\kappa=\{2,2,2,2\}$, $\{3,3,3\}$, $\{2,4,4\}$, or $\{2,3,6\}$, respectively. } \cite{ringelderived}\cite{tuderived}, or even better to the cluster category
(which is also known \cite{tubcluster}). 

Alternatively, one can use the derived equivalent picture that the non--perturbative category is $\mathrm{Coh}\,X(\kappa)$, where $X(\kappa)$ is the weighted projective line of type as in eqn.\eqref{rrrttt}. Then $S$--duality is equivalent to \textbf{Theorem 6.16} of the nice review \cite{llll}.

Of course, the resulting structure is exactly what is expected on physical grounds by the combined effect of $S$--duality and asymptotic weak--coupling analysis.
\medskip

The important lesson for our present purposes is that, in general, a quiver $\cn=2$ theory may have many different limits in which 
it asymptotes a SYM weakly coupled to some matter system. In the favorable cases, for each such limit there is 
a nice gauge functor. If the theory has a large $S$--duality group, the gauge functors organize themselves into orbits of this duality group, producing a very elegant structure.

\subsection{Making physical sense of wild canonical algebras}\label{makingsense}
 
Wild canonical algebras have Euler character $\chi<0$, which means a Landau pole for the corresponding QFT. Hence wild canonical algebras represent formal field theories which do not exist at the quantum level because they are not UV complete. Of course, we may still use them as effective field theories at energies less than a certain cut--off scale $\Lambda$. On the other hand, the (stable) representations of the wild canonical algebras make perfect sense (even if they are hard to study). We can think of this representation theory as a particular `UV completion' of the corresponding effective theory. We want to understand the properties of such a completion.

\subsubsection{Ringel's property again}\label{yyyeooe}
At first sight, one is tempted to say that the wild $SU(2)$--canonical theories do not exist at the quantum level, and just forget them. However this is too naive.
Consider the canonical algebra of type $\{2,2,2,2,2\}$, which is $SU(2)$ SQCD with $N_f=5$, whose quiver (after $3$--CY completion)
is \begin{equation}\label{nF5}
\begin{gathered}
\xymatrix{\diamondsuit\ar@<0.6ex>[dd]\ar[dd]\ar@<-0.6ex>[dd]\\
& \bullet\ar[ul] & \bullet\ar[ull] & \bullet\ar[ulll] &\bullet \ar[ullll] &\bullet\ar[ulllll]\\
\spadesuit \ar[ur]\ar[urr]\ar[urrr]\ar[urrrr]\ar[urrrrr]}
\end{gathered}
\end{equation}
This model has a $\beta$--coefficient $b\equiv 2N_c-N_f=-1$, and hence a Landau pole. Let us compare \eqref{nF5} with the quiver
  \begin{equation}\label{nF52}
\begin{gathered}
\xymatrix{\circ\ar@<0.6ex>[dd]\ar@<-0.6ex>[dd] &&\diamondsuit\ar[ll]\ar@<0.6ex>[dd]\ar[dd]\ar@<-0.6ex>[dd]\\
&&& \bullet\ar[ul] & \bullet\ar[ull] & \bullet\ar[ulll] &\bullet \ar[ullll] &\bullet\ar[ulllll]\\
\circ\ar[rr] &&\spadesuit \ar[ur]\ar[urr]\ar[urrr]\ar[urrrr]\ar[urrrrr]}
\end{gathered}
\end{equation}
 which corresponds \cite{ACCERV2} to $SU(3)$ SQCD with $N_f=5$ (up to mutation at the nodes $\bullet$). Now
 $b\equiv 2N_c-N_f=+1$ and the QFT is asymptotically free, hence physically well defined. It is pretty obvious that the representations of the quiver \eqref{nF5} are, in particular, representations of \eqref{nF52}. Thus the representations of the first quiver give the UV completion of the $SU(2)$ $N_f=5$ SQCD corresponding to embedding this theory in $SU(3)$ SQCD with an adjoint Higgs background $\langle\Phi\rangle$ breaking $SU(3)\rightarrow SU(2)\times U(1)$ in the regime of  $\langle\Phi\rangle$ parametrically large. Note that $\langle\Phi\rangle\rightarrow\infty$ corresponds to the decoupling limit $|Z_\circ|\rightarrow +\infty$. The stable representation of \eqref{nF52} which have bounded mass in this limit are precisely the stable objects of the category of representations of \eqref{nF5}.
 
 Clearly, this argument applies to all $N_f$ by embedding the canonical algebra of type $\overbrace{\{2,2,\cdots, 2,2\}}^{N_f\ \text{terms}}$, that is $SU(2)$ SQCD with $N_f$ flavors, into the quiver algebra associated to some $SU(N_c)$ SQCD with $N_c>N_f/2$.
 
 In order for this interpretation to be coherent, we must have that the BPS states of the quiver \eqref{nF5} (and its generalization for higher $N_f$) which in the decoupling limit $g\rightarrow 0$ have \emph{bounded masses} are well--behaved, while all \emph{wild} representations of \eqref{nF5} should correspond to states of masses $O(1/g^2)$ as $g\rightarrow 0$ (\textit{i.e.}\! they must correspond to solitons of the QFT).
 
The BPS states with masses $O(1)$ in the $g\rightarrow 0$ limit are precisely those with magnetic charge zero, $m(X)=0$, that is --- in Ringel's notation --- those belonging to the module class
 $\mathcal{T}$. Then the content of the above physical consistency requirement is precisely the statement in \textbf{Remark 2.} on page 166 of Ringel's book \cite{RI}. Due to its importance we reproduce the remark here
 
 \begin{quotation}
 \textbf{\underline{Remark 2.}\,} It seem curious that for a canonical algebra $C$, the module class $\mathcal{T}$ is always very well behaved, whereas the two classes $\mathcal{P}$ and $\mathcal{Q}$ may be wild. The representation type of both $\mathcal{P}$ and $\mathcal{Q}$ very much depends on the type of $C$. [...]
 \end{quotation} 
 
In facts the module classes $\mathcal{P}$, $\mathcal{Q}$ --- which correspond to the dyons ($\equiv$ solitons) --- are well--behaved if and only if $C$ is either concealed Euclidean, that is the QFT is asymptotically free, or tubular, that is the QFT is UV superconformal. This corresponds precisely to the predictions of Quantum Field Theory. We see that Ringel's `curious' phenomenon  is a direct consequence of locality of the microscopic forces of nature. 
\smallskip

On the other hand, this argument shows that the solitonic (dyonic) sector of a quiver gauge theory with a gauge group which is not a product of copies of $SU(2)$ is typically \emph{strictly wild}, that is, it contains BPS particles of arbitrarily high spin. Indeed, the spectrum of $2N_c> N_f\geq 5$ SQCD will contain the strictly wild spectrum\footnote{\ Here a \emph{caveat} is in order. For $N_c>2$ the theory is not complete in the sense of \cite{CV11} and the above chambers may be \emph{formal} (not physically realizable \cite{CV11}).} of the $\{2,2,2,2,2\}$ canonical theory, and then will have stable BPS states (at weak coupling) of arbitrarily high spin. This was shown directly in \cite{ACCERV2}. An alternative characterization of the complete theories of ref.\!\cite{CV11} is the non--wildness of the BPS spectrum in all chambers.
\medskip

Further examples of the Ringel property in non--UV complete theories are discussed in \S.\,\ref{uuurujsnssxx} below.

\subsection{Digression: decoupling and controlled subcategories}

A basic method used to extract the physics of $\cn=2$ models is to take a decoupling limit to reduce to a simpler system. The typical examples are taking a mass parameter $m$ or a Higgs background $\langle\Phi\rangle$ to infinity, and sending a YM coupling constant $g$ to zero. In all these examples a subsector of the theory decouples from the rest, and we remain with a simpler QFT.

If both the original and the decoupled QFTs are quiver $\cn=2$ theories, it is natural to ask what is the relation between the Abelian categories of the representations of the two quivers (or, rather, their derived categories). We have already encontered examples of this decoupling at the level of categories in the form of the Ringel decoupling as $g\rightarrow 0$. Here we digress from our main theme to formalize the categorical version of the standard decoupling theorems of QFT.  

Let $\lambda\colon K_0(\mathsf{rep}(Q,\cw))\rightarrow \Z$ be a group homomorphism from the charge lattice to $\Z$. Let $\mathsf{C}(\lambda)\subset
\mathsf{rep}(Q,\cw))$ be the full subcategory consisting of the objects $X$ of $\mathsf{rep}(Q,\cw)$ satisfying the two conditions
\begin{align*}
 &(i)\hskip 0.96cm \lambda(X)=0,\\
&(ii) \qquad \lambda(Y)\leq \lambda(X)\ \text{for each subobject $Y$ of $X$.}
\end{align*}
$\mathsf{C}(\lambda)$ is called the \emph{subcategory controlled by} $\lambda$. It is easy to see \cite{LE7} that the full subcategory $\mathsf{C}(\lambda)$ is exact Abelian (\textit{i.e.}\! closed in $\mathsf{rep}(Q,\cw)$ under kernels, cokernels and directs sums) as well as closed under extensions.

In the present context, all QFT decoupling theorems correspond to replacing the category of the original theory
$\mathsf{rep}(Q,\cw)$ (or, more precisely, its derived category) with the (derived category of) the controlled subcategory for a suitable choice of $\lambda$. The decoupling limit is defined by considering the $1$--parameter family of central charges
\begin{equation}
 Z_t(X)= Z_0(X)-t\, \lambda(X),\qquad t\in \R,
\end{equation}
and taking the limit $t\rightarrow +\infty$.

The decoupling limit of small YM coupling we introduced in the previous section, see eqn.\eqref{stabfunctwc}, corresponds to taking $\lambda$ equal to the magnetic charge $m(\cdot)$ and the category $\mathsf{C}(m(\cdot))$ is equal to $\ct$. The limit of large mass $m\rightarrow\infty$ is typically given by taking $\lambda$ equal to the dimension at the node representing the quark, so that $\mathsf{C}(\lambda)$ is the category of the representations of the full subquiver with that node deleted; \textit{etc.}

What is important is that the decoupled subcategory is as nice as it may be. This corresponds to the obvious physical fact that the decoupling limit theory is a good QFT of its own.

 \section{The Ringel property for general complete theories}\label{XXXTT}

In this paper we mainly consider the case in which the gauge group $G$ is simple; nevertheless one wishes to understand also the case of a general semi--simple gauge group
$$G=G_1\times G_2\times\cdots\times G_s.$$
In this section we present a discussion of the new phenomena that appear in the general case trough the example of complete theories.
 %\smallskip

\subsection{\emph{Gentle} algebras and WKB methods}

The simplest instances of gauge theories with a non--simple gauge group are the complete theories  \cite{CV11}. Their 
 gauge group has the form
 $G=SU(2)^k$ with $k\in \mathbb{N}$.
 The quiver with superpotential $(Q,\cw)$ of a complete theory --- but for 11 exceptional models \cite{CV11} --- arises from an ideal triangulation of a surface $\Sigma$ having punctures and marked points on its boundary components \cite{triangulation1,traingulation2}. The surface $\Sigma$ is obtained from the Gaiotto curve \cite{Gaiotto} of the corresponding $A_1$--theory by replacing irregular poles of the quadratic differential $\phi_2$, having order $c+2\geq 3$,  by boundary components with $c$ marked points \cite{CV11}. $3$ out of the $11$ exceptional cases have no SYM subsector \cite{CV11}, and $6$ are  $SU(2)$ canonical models (3 domestic, 3 tubular) already described in section \ref{CanRingel}. The  last two exceptional complete models, corresponding to the Derksen--Owen quivers $X_7$ and $X_6$ \cite{derksen}, are easy to study directly.

It is convenient to start with a special class of such Gaiotto $A_1$--theories, namely those defined by a genus $g$ surface with only irregular punctures, and at least one such puncture. The corresponding $(Q,\cw)$ arise from the triangulation of a surface with at least one boundary component and no (regular) puncture.

 \subsubsection{The Assem--Br\"ustle--Charbonneau--Plamondon theorem}\label{sec:ABCP}

A finite--dimensional algebra $\mathscr{A}$ is called a \emph{string} algebra \cite{butring} iff
it has the form $\C Q /I$ where:
\begin{description}
\item[S1] each node of the quiver $Q$ is the starting point of at most two arrows;
\item[S2] each node of the quiver $Q$ is the ending point of at most two arrows;
\item[S3] for each arrow $\alpha$ in $Q$ there is at most one arrow $\beta$ with $\alpha\beta\not\in I$;
\item[S4] for each arrow $\alpha$ in $Q$ there is at most one arrow $\beta$ with $\beta\alpha\not\in I$;
\item[S5] the ideal $I$ is generated by zero--relations.
 \end{description}

Every string algebra is, in particular, special biserial and tame. Physically, tame means that the BPS spectrum consists only of hypermultiplets and vector--multiplets, higher spin states been forbidden. The meaning of biserial would be clear below.

A \emph{gentle} algebra is a string algebra where the ideal $I$ may be generated by zero--relations of lenght two (\textit{i.e.}\! involving just two arrows).
\vglue 12pt

\textbf{Theorem.} (Assem--Br\"ustle--Charbonneau--Plamondon \cite{assemgentle}) \textit{Let $(Q,\cw)$ be a quiver with superpotential corresponding to a Gaiotto $A_1$--theory with only irregular punctures and at least one such puncture. Then the Jacobian algebra $\mathscr{A}\equiv \C Q/(\partial \cw)$ is gentle.}
\vglue 12pt

Thus for such an `irregular' Gaiotto theory the BPS states are stable modules of a gentle algebra.
The good news is that the indecomposable modules of a gentle algebra are explicitly known \cite{butring}.
Moreover, they can be constructed in terms of WKB geodesics \cite{assemgentle,LF2} in agreement with the physical analysis of Gaiotto--Moore--Neitzke \cite{GMN09} (see also \cite{ACCERV1}).

\subsubsection{\emph{String} and \emph{band} modules}

A string algebra has
two kinds of indecomposable representations \cite{butring}:
\begin{enumerate}
 \item \emph{string} representations, which have no free parameters;
\item \emph{band} representations which come in one--parameter families.
\end{enumerate}

By a \textit{string} $C$  we mean a generalized path in the quiver $Q$ where we are allowed to pass from one node to another one either following an arrow connecting them or going backwards such an arrow. We write a string as a sequence of nodes connected by direct or inverse arrows, \textit{i.e.}\! in the form
\begin{equation}
C\colon\qquad  i_0\xleftarrow{\phi_{\alpha_1}} i_1\xrightarrow{\phi_{\alpha_2}} i_2 \xrightarrow{\phi_{\alpha_3}}\: \cdots\cdots\ i_{n-1}\xleftarrow{\phi_{\alpha_\ell}} i_\ell,
\end{equation}
where the nodes in the sequence are labelled as $\{i_0,i_1,i_2,\cdots, i_\ell\}$ and the direct/inverse arrows as $\{\phi_{\alpha_1},\phi_{\alpha_2},\cdots, \phi_{\alpha_\ell}\}$. 
A string is \underline{not} allowed to contain:
\begin{description}
          \item[C1] adiacent direct/inverse arrow pairs, as $\cdots i\xrightarrow{\phi_a} j\xleftarrow{\phi_a} i \cdots$ or
$\cdots i\xleftarrow{\phi_a} j\xrightarrow{\phi_a} i \cdots$;
\item[C2] adiacent pairs of (direct or inverse) arrows whose composition vanishes by the relations $\partial\cw=0$.
         \end{description}
  
A string $C$ is identified with its inverse $C^{-1}$ (\textit{i.e.}\! the string obtained by reading  $C$ starting from the right instead than from the left). 

Given a string $C$ of length $\ell+1$, the corresponding string module $M(C)$ is obtained in the following way. For each vertex $v$ of $Q$ let
\begin{equation}
 I_v=\{ a \;|\; \text{the }a\text{--th node in the string $C$ is }v \}\subset\{0,1,2,\cdots, \ell\}.
\end{equation}
As $M(C)_v$ we take the vector space of dimension $\#I_v$ with base vectors $z_a$, $a\in I_v$. We define
\begin{equation}\label{kklklkd}
 \begin{cases}\phi_{\alpha_a}(z_a)=z_{a+1} & \phi_{\alpha_a}\ \text{is direct}\\
  \phi_{\alpha_a}(z_{a+1})=z_a & \phi_{\alpha_a}\ \text{is inverse}
 \end{cases}
\end{equation}
Finally, for all arrows $\gamma$ and vectors $z_a$ such that $\gamma(z_a)$ is not of the form in \eqref{kklklkd}, we set $\gamma(z_a)=0$. 

Physically, a string module corresponds to a open WKB trajectory \cite{GMN09} on the Gaiotto surface of the theory, see \cite{assemgentle,LF2} for the explicit construction of the string module out of the WKB geodesic and \textit{viceversa}.\smallskip

A \emph{band} is a cyclic sequence $C$ of nodes of $Q$ and direct/inverse arrows, again forbidding sequences as in \textbf{C1} and \textbf{C2} above.
Moreover, $C$ is required to be such that all powers $C^n$ are well--defined (cyclic) strings, but $C$ itself is not the power of a   string of smaller length. $C$ is identified up to cyclic rearrangement and overall inversion. 

Given such a band $C$ of length $\ell$, we define a family $M(C,\lambda, n)$ of indecomposable band modules labelled by an integer $n\geq 1$ and a point $\lambda\in \C^*$.
Explicitly, we set $$M(C,\lambda,n)_v=\bigoplus_{a\in I_v} Z_a, \qquad\quad (a=1,2,\dots,\ell),$$ where the $Z_a$'s are copies of $\C^n$, and 
\begin{equation}\label{uuus}
 \begin{cases}
  \phi_{\alpha_a}\Big|_{Z_a}\colon Z_a\xrightarrow{1} Z_{a+1} & \phi_{\alpha_a}\ \text{direct } a\neq \ell\\
  \phi_{\alpha_a}\Big|_{Z_{a+1}}\colon Z_{a+1}\xrightarrow{1} Z_{a} & \phi_{\alpha_a}\ \text{inverse } a\neq \ell\\
  \phi_{\alpha_\ell}\Big|_{Z_\ell}\colon Z_\ell\xrightarrow{J(\lambda)} Z_{1} & \phi_{\alpha_\ell}\ \text{direct }\\ 
  \phi_{\alpha_\ell}\Big|_{Z_{1}}\colon Z_{1}\xrightarrow{J(\lambda)^{-1}} Z_{\ell} & \phi_{\alpha_\ell}\ \text{inverse }
\end{cases}
\end{equation}
where $J(\lambda)$ is the $n\times n$ Jordan block of eigenvalue $\lambda\neq 0$. All other linear maps which are not defined in 
\eqref{uuus} are set to zero.

The above string and band modules give a complete list of indecomposable (and pairwise non--isomorphic) $\mathscr{A}$--modules \cite{butring}.
Note also that a brick has necessarily $n=1$ since $J(\lambda)\in\mathrm{End}\, M(C,\lambda, n)$ is non trivial for $n\geq 2$.

A family $M(C,\lambda,1)$ of representations is associated to a primitive\footnote{\ By primitive homotopy class we mean a class which is not a non--trivial integral multiple of an integral class. Note also that the identification $C\leftrightarrow C^{-1}$ corresponds to the fact that the WKB geodesics have no orientation due to the $\Z_2$--monodromy of the differential $\sqrt{\phi_2}$.} homotopy class of  closed WKB gedesics on $\Sigma$. Hence the theorem of ref.\!\cite{assemgentle} just says that the WKB analysis is exact in the present context (a fact which is, of course, physically obvious, see \cite{GMN09}).

\subsection{Example of $SU(2)^3$ gauge functor}\label{su(2)3333}

As a first relevant example, let us consider
$SU(2)^3$ SYM coupled to a three--fundamental half--hypermultiplet $\tfrac{1}{2}(\mathbf{2},\mathbf{2},\mathbf{2})$.
This is the `irregular' Gaiotto theory on the sphere with three boundary components each with one marked point.
In the decoupling limit this models contains the sphere with three ordinary punctures, namely the Gaiotto $\ct_2$ theory \cite{Gaiotto}, which is the main building block of the complete theories. This motivates our detailed discussion of this theory.
The reader may prefer to skip the example and jump ahead to the general gentle case in \S.\,\ref{generalcomplete}.

This model is `canonical' in the broad sense of \S.\,\ref{cangaugefunct}.
The canonical form of its quiver is the $\Z_3$--symmetric `triangular prism' form $Q$ 
\begin{equation}\label{prisms}
 \begin{gathered}
\xymatrix{ & 1,2\ar[dr]^{H_1} &\\
1,1\ar[ur]^{H_3} & & 1,3\ar[ll]_{\hskip 1.14cm H_2}\\
\\
& 2,2\ar[uuu]^{V_2}\ar[dl]_{h_3}\\
2,1\ar[uuu]^{V_1}\ar[rr]_{h_2} && 2,3\ar[uuu]_{V_3}\ar[lu]_{h_1}}
 \end{gathered}
\end{equation}
with superpotential
\begin{equation}
 \cw= \mathrm{Tr}(H_1H_3H_2)+\mathrm{Tr}(h_1h_2h_3).
\end{equation}
Note that
all the vertical arrows $V_i$ point upward and that each (vertical) face of the prism is an $\widehat{A}(3,1)$ affine (full) subquiver.

\medskip

For each face of the prism $Q$ we have a restriction functor
\begin{equation}
\mathscr{F}_i\colon \mathsf{mod}\,\mathscr{A} \rightarrow \mathsf{mod}\,\C\widehat{A}(3,1),\qquad i=1,2,3,
\end{equation}
which composed with the gauge functor of the $\widehat{A}(3,1)
$ canonical theory, $\mathscr{G}\colon \mathsf{mod}\,\widehat{A}(3,1)\rightarrow \mathsf{mod}\,\mathbf{Kr}$, gives
a $SU(2)$ gauge functor. 
The full $SU(2)^3$  gauge functor $\mathfrak{G}$ is
\begin{align}
&\mathfrak{G}\colon \mathsf{mod}\,\mathscr{A} \rightarrow (\mathrm{mod}\,\mathbf{Kr})^3\\
& X\longmapsto \big(\mathscr{GF}_1(X), \mathscr{GF}_2(X),\mathscr{GF}_3(X)\big)
\end{align}
which preserves the three magnetic charges
\begin{equation}
m_i(X)= \dim X_{2,i}-\dim X_{1,i}, \qquad i=1,2,3.
\end{equation}
dual to the $W$--bosons associated to the imaginary roots of the three affine subquivers.  The corresponding  weak--coupling regime is given by
\begin{equation}\label{Pooiuuu}
\arg Z(1,i) =O (g_i^2), \qquad \arg Z(2,i)=\pi-O(g_i^2)
\end{equation}
where $g_1,g_2,g_3\rightarrow 0$ are the three YM couplings.
\smallskip

We write $\mathscr{S}_i\colon \mathsf{mod}\,\C\widehat{A}(3,1)\rightarrow \mathsf{mod}\,\mathscr{A}$ for the extension by zero of a representation of the $i$--th affine subquiver.
\smallskip

The light category of the $\tfrac{1}{2}(\mathbf{2},\mathbf{2},\mathbf{2})$ model in the weak coupling regime \eqref{Pooiuuu}, $\mathscr{L}$, is the controlled subcategory of $\mathsf{mod}\,\mathscr{A}$ consisting of objects $X$ such that $m_i(X)=0$ for $i=1,2,3$, while for all their subobjects $Y$ one has $m_i(Y)\leq 0$. All BPS states which are stable and light at weak coupling correspond to brick representations in $\mathscr{L}$.

 Let $X\in\mathscr{L}$ be an indecomposable;
for each $j=1,2,3$, $\mathscr{F}_j(X)$ is a representation of
$\widehat{A}(3,1)$ which, by the theorem quoted on page \pageref{ringeltheorem}, may be decomposed as follows
\begin{equation}\label{ppttqq}
\mathscr{F}_j(X)= X_p\oplus X_q\oplus X_t
\end{equation}
where $X_p\in \cp$, $X_q\in \cq$, and $X_t\in \ct\equiv \bigvee_{\lambda\in\mathbb{P}1}\ct_\lambda$. 
In fact, one has the more precise result \vglue 6pt

\textbf{Lemma.} \textit{$X\in\mathscr{L}$ and indecomposable. For each $j=1,2,3$ there is a point $\lambda_j\in \mathbb{P}^1$ such that
\begin{equation}
\mathscr{F}_j(X) \in  \ct_{\lambda_j}.
\end{equation}
Moreover, if $\mathscr{F}_j(X)\in \ct_{\lambda_j}$ with $\lambda_j\neq 0$, $X$ is the extension by zero of an indecomposable representation of the $j$--th $\widehat{A}(3,1)$ subquiver. }
\vglue 6pt

\textsc{Proof.} We have to show that in eqn.\eqref{ppttqq} $X_p=X_q=0$, while $X_t\in \ct_{\lambda}$ for a unique $\lambda\in\mathbb{P}^1$.
We show first $X_p=0$, the argument for $X_q=0$ being dual. 

Consider, say, the $\widehat{A}(3,1)$ subquiver over the vertices $(1,1)$, $(1,3)$, $(2,1)$ and $(2,3)$ of figure \eqref{prisms}. Recall that, for a canonical algebra like $\C\widehat{A}(3,1)$,
a module $Y\in \cp$ iff all arrows are injective and not all isomorphisms \cite{RI}. In particular, the two horizontal arrows in our $\widehat{A}(3,1)$ full subquiver are injective when restricted to $X_p$. Since the relations $\partial \cw=0$ imply that the composition of any two horizontal arrows, $H_l$, $h_l$, vanishes, we get 
\begin{equation}
X_{1,3}\Big|_{X_p} \subseteq \mathrm{coker}\,H_1,\qquad X_{2,1}\Big|_{X_p} \subseteq \mathrm{coker}\,h_3.
\end{equation}
Hence we have an exact sequence of the form
\begin{equation}
0\rightarrow \tilde X \rightarrow X \rightarrow {X_p}\rightarrow 0,
\end{equation}
where ${X_p}$ stands for the representation $X_p$ of the subquiver $\widehat{A}(3,1)$ extended by zero to a representation of full quiver \eqref{prisms}. Now
\begin{equation}
m_1(\tilde X)= \dim X_{1,1}\Big|_{X_p}-\dim X_{2,1}\Big|_{X_p},
\end{equation}
while
\begin{equation}
 0\neq X_p\in \cp\qquad\Rightarrow\qquad\dim X_{1,1}\Big|_{X_p} >\dim X_{2,1}\Big|_{X_p},\end{equation}
  and hence $X_p\neq 0$ implies
\begin{equation}
m_1(\tilde X)>0
\end{equation}
which is a contradiction since $X\in\mathscr{L}$.

Next let us show that $\mathscr{F}_j(X)\in\ct_\lambda$ for some $\lambda$. Let
\begin{equation}
\mathscr{G}_j(X)= X_{\lambda_1}\oplus X_{\lambda_2}\oplus\cdots \oplus X_{\lambda_s},\qquad X_{\lambda_r}\in \ct_{\lambda_r}
\end{equation}
with the $\lambda_r$ pairwise distinct points in $\mathbb{P}^1$. We have to show that $s=1$. Suppose on the contrary that $s>1$, then at least one point, say $\lambda_s$, is not the origin in $\mathbb{P}^1$.
Restricted to $X_{\lambda_s}$, all arrows in the $j$--th face of $Q$ are isomorphisms, in particular the two horizontal ones. Then, from the relations $\partial\cw=0$ it follows that the vector spaces
\begin{equation}
X_{i,k}\Big|_{X_{\lambda_s}} \subseteq \mathrm{ker}\,H \ \ (\mathrm{resp. }\subseteq\mathrm{coker}\,H^\prime)
\end{equation}
where $H$ (resp.\! $H^\prime$) stands for the horizontal
arrow starting (resp.\! ending) at the $i,k$ node and not belonging to the $j$--th face full subquiver.  Hence we have
\begin{equation}
X=X^\prime \oplus \mathscr{S}_j(X_{\lambda_s})
\end{equation}
with $X^\prime\neq 0$. Again this is a contradiction since $X$ is indecomposable.
\hfill $\square$

\subsubsection{$SU(2)^3$--tubes for the $\tfrac{1}{2}(\mathbf{2},\mathbf{2},\mathbf{2})$ theory}

From the lemma we have a well--defined map
\begin{equation}
\varpi\colon\ \big\{\,\text{indecomposables of }\mathscr{L}\,\big\} \longrightarrow \mathbb{P}^1\times \mathbb{P}^1\times \mathbb{P}^1
\end{equation}
which associates to $X$ the labels $(\lambda_1,\lambda_2,\lambda_3)$ of the stable tubes containing the regular indecomposable $\C\widehat{A}(3,1)$--modules $\mathscr{G}_1(X)$,
$\mathscr{G}_2(X)$, and $\mathscr{G}_3(X)$.
From the lemma, we also see that, if the representation $X$ has one label $\lambda_j\neq 0$ ($j=1,2$ or $3$), then it is necessarily the extension by zero of a light stable representation of the $j$--th $\C\widehat{A}(3,1)$ subalgebra, and hence the other two labels of $X$ must vanish. Thus the image of $\varpi$ is the reducible `star'
\begin{equation}
N\equiv \mathbb{P}^1\times \{0\}\times\{0\}\,\bigcup\,
\{0\}\times\mathbb{P}^1\times\{0\}\,\bigcup\,\{0\}\times\{0\}\times\mathbb{P}^1.
\end{equation}
As an algebraic variety $N$ has three irreducible components, namely the three $\mathbb{P}^1$'s, and we have three generic points of $N$, one for each component.

 The $SU(2)^3$--tube over the point $\lambda\in N$,
$\mathscr{L}_\lambda$, is the subcategory generated by the indecomposable object $X\in\mathscr{L}$ with
$\varpi(X)=\lambda\equiv (\lambda_1,\lambda_2,\lambda_3)$. Then the light category has the form
$\mathscr{L}=\bigvee_{\lambda\in N} \mathscr{L}_\lambda$, as expected on general grounds.

Moreover the pairs of functors $\mathscr{F}_j$, $\mathscr{S}_j$ give an equivalence
\begin{equation}
\lambda\neq 0\qquad\Rightarrow\qquad  \mathscr{L}_{\lambda,0,0} \simeq\mathscr{L}_{0,\lambda,0} \simeq\mathscr{L}_{0,0,\lambda} \simeq \ct_\lambda,
\end{equation}
that is \textit{the $SU(2)^3$--tube over each of the three generic points of $N$ is an ordinary stable homogeneous tube.} Since each irreducible component of $N$ corresponds to a $W$--boson --- hence to a $SU(2)$ subgroup of $SU(2)^3$ --- we get precisely the structured claimed in \S.\,\ref{sec:light}: the generic tube in a $\mathbb{P}^1$--family depends only on the simple gauge group $G_k$ associated to that generic point, and hence for $G_k=SU(2)$ it should be a standard homogenous tube.

From the above discussion we see that the charge (dimension) vector of the $j$--th $W$--boson is equal to the primitive imaginary root of the $j$--th $\widehat{A}(3,1)$ affine subquiver. We write $\delta_j$ for this dimension vector ($j=1,2,3$). 

\subsubsection{Relations with band modules and WKB geodesics}

\begin{figure}
 \centering
\includegraphics[width=0.34\textwidth, height=0.34\textwidth]{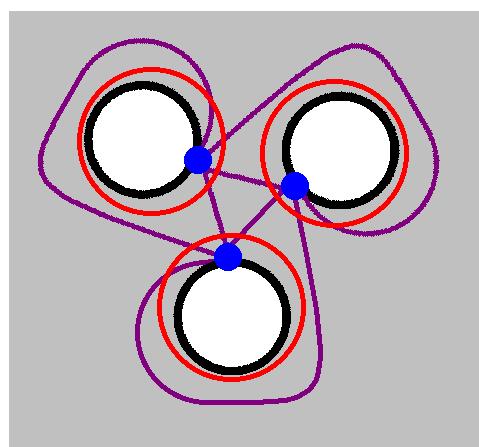}   
 \caption{The ideal triangulation of the sphere with three boundary components corresponding to the `prism' quiver \eqref{prisms}. The arcs of the triangulation are the purple curves.
The WKB geodesics corresponding to the three light stable $SU(2)$ $W$--bosons are draw in red. The corresponding bands $C$ are given by the cyclically ordered list of arcs crossed by each WKB geodesic. }
 \label{fig:xxxxx}
\end{figure}

By the Assem--Br\"ustle--Charbonneau--Plamondon theorem \cite{assemgentle}, the Jacobian algebra $\mathscr{A}\equiv \C Q/(\partial\cw)$ of \eqref{prisms} is gentle, and the stable BPS vectors should correspond to band modules. The band modules are in one--to--one correspondence with the primitive homotopy classes of WKB geodesics \cite{assemgentle}. Moreover, geometrically a weak coupling limit is a
degeneration limit for the associated triangulated surface \cite{Gaiotto}, and a module $M(C,\lambda,1)$ is `light' in the corresponding weak coupling limit precisely if the length of the corresponding geodesic goes to zero in that degeneration limit. In the present example $\Sigma$ is a sphere with three holes, and the degeneration limit $g_j\rightarrow 0$ corresponds to the limit in which the $j$--th hole contracts to a puncture. Then the three light vectors at weak coupling correspond to the three red curves in figure \ref{fig:xxxxx}.

We can read the three light bands directly from the figure \ref{fig:xxxxx}. They are
\begin{align}
 &\xymatrix{2,a\ar[r]\ar@/^1.1pc/[rrr]& 2,a+2\ar[r]& 1,a+2\ar[r] & 1,a} && a=1,2,3,\ \text{and }a+3\equiv a,
\end{align}
which coincide with the representations of the $\widehat{A}(3,1)$ affine subquivers of dimension $\delta_j$, as predicted by the canonical gauge functor $\mathfrak{G}$.

\subsubsection{The $\tfrac{1}{2}(\mathbf{2},\mathbf{2},\mathbf{2})$ light category}

With reference to the canonical quiver \eqref{prisms}, the only `new' $SU(2)^3$--tube, associated to the $\tfrac{1}{2}(\mathbf{2},\mathbf{2},\mathbf{2})$ matter, is $\mathscr{L}_{0,0,0}$, corresponding to the unique point in $N$ where the three rational curves meet. Indeed, in general, if we have
a gauge group of the form $G=G_1\times \cdots \times G_s$ with $G_k$ simple, the light subcategory will be of the form
\begin{equation}
 \mathscr{L}= \bigvee_{\lambda\in N} \mathscr{L}_\lambda
\end{equation}
with $N$ a reducible variety with $s$ irreducible components each equal to $\mathbb{P}^1$. Each generic point of a component is associated to a simple gauge factor $G_k$, and a subcategory $\mathscr{L}_\lambda$ may contain representations corresponding to BPS matter charged under a plurality of gauge factors, say $G_{k_1}, G_{k_2},\cdots, G_{k_r}$ only if the point $\lambda$ belongs to the intersection $\mathbb{P}^1_{k_1}\cap \mathbb{P}^1_{k_2}\cap\cdots\cap \mathbb{P}_{k_r}^1$ of the corresponding irreducible components. In particular, in the present example, the three--fundamental half--hyper, being charged with respect to the three $SU(2)$'s, should lay in the category associated to the unique intersection point of the three lines. 
\medskip

Let $0\in N$ be the crossing point. As in the canonical examples of \S.\,\ref{matterinsidetube}, the special light category $\mathscr{L}_0$ contains the matter subcategory $$\mathsf{matter}_0\equiv \mathsf{add}\text{(rigid bricks of $\mathscr{L}_0$)}.$$ 
Matter modules are, in particular, rigid bricks of $\mathscr{A}$.
If $X$ is a rigid indecomposable representation of the gentle algebra $\mathscr{A}$, it should be a \emph{string} module $M(C)$, and, in particular, a nilpotent module. Define $\mathfrak{d}$ to be the dimension vector $\mathfrak{d}=(1,1,1,1,1,1,1)$.
Using the string module construction, in appendix \ref{bricksofffB} it is shown the following
\vglue 12pt

\textbf{Lemma.} \textit{Let $X\in \mathscr{L}_{0}$ be indecomposable. Then \begin{equation}
\mathrm{End}(X)=\C\quad\Longleftrightarrow\quad \dim X\leq \mathfrak{d}.\end{equation}}
\noindent Note that for non--sincere modules this reduces to the usual characterization of bricks inside the stable tubes \cite{CB}.
\medskip

Thus a light indecomposable of $\mathscr{L}_0$ is a rigid brick iff its dimension vector is one of the following four
\begin{align}\label{chv1}
&\tfrac{1}{2}\Big(\delta_1-\delta_2+\delta_3\Big)\\
&\tfrac{1}{2}\Big(-\delta_1+\delta_2+\delta_3\Big)\\
&\tfrac{1}{2}\Big(\delta_1+\delta_2-\delta_3\Big)\label{chv3}\\
&\tfrac{1}{2}\Big(\delta_1+\delta_2+\delta_3\Big)\equiv \mathfrak{d}
\label{chv4}\end{align} 
Adding the PCT conjugates, we get eight possible states which have charge vectors equal to the $SU(2)^3$ weights of the $(\mathbf{2},\mathbf{2},\mathbf{2})$ representation, that is precisely the states of our three--fundamental half--hyper as predicted by physics.

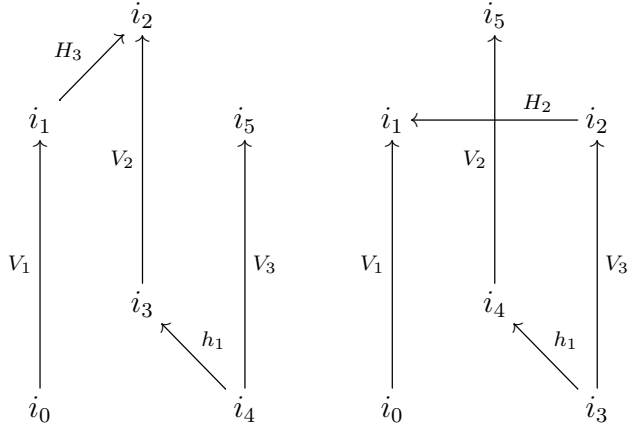
\begin{figure}
 \begin{equation*}
 \begin{gathered}
\xymatrix{ & i_2 &\\
i_1\ar[ur]^{H_3} & & i_5\\
\\
& i_3\ar[uuu]^{V_2}\\
i_0\ar[uuu]^{V_1} && i_4\ar[uuu]_{V_3}\ar[lu]_{h_1}}
 \end{gathered}\qquad 
 \begin{gathered}
\xymatrix{ & i_5 &\\
i_1 & & i_2\ar[ll]_{\hskip 1.14cm H_2}\\
\\
& i_4\ar[uuu]^{V_2}\\
i_0\ar[uuu]^{V_1} && i_3\ar[uuu]_{V_3}\ar[lu]_{h_1}}
 \end{gathered}
\end{equation*} 
\caption{\label{deltareppp}The strings of two bricks with dimension $\mathfrak{d}$. All others are obtained from these two by acting with the $\Z_3$ automorphism of the prism quiver $Q$.}
\end{figure}

There is a (slightly) subtle point however: while the first three dimension vectors \eqref{chv1}--\eqref{chv3} correspond to a unique representation, which is stable in any weakly coupled chamber, there are six bricks with the charge vector $\mathfrak{d}$. The strings of two of them are represented in figure \ref{deltareppp}, the other four may be obtained using the automorphisms of $Q$. It is easy to convince ourselves that, in each weakly coupled chamber, precisely one representation with dimension $\mathfrak{d}$ is stable, which one of the six depending on the particular chamber. (We don't prove this here, since we shall give a general argument in section \ref{chambindependence}). 

Hence the `naive' matter Krull--Schmidt linear subcategory
\begin{equation}
 \mathsf{matter}\equiv \mathsf{add}\text{(stable rigid)}
\end{equation}
 is independent on the BPS chamber only up to equivalence, but its image in $\mathsf{rep}(Q,\cw)$ does depend on the (weakly--coupled) chamber.
 In facts for the class of models having a weakly coupled Lagrangian description, the BPS spectrum is the same in all weakly--coupled chambers, see \S.\ref{chambindependence} for details. In  particular, one gets the same light spectrum starting from any quiver in the mutation--class. But the chamber--independence should be understood as an \emph{equivalence} not an equality.

\subsubsection{The naive category $\mathsf{matter}$ as Gaiotto's $\ct_2$ theory}\label{222matcategory}

Let $X$ be an indecomposable representation of the gentle algebra $\mathscr{A}$ with $X\in \mathscr{L}_0$. The fact that $\mathscr{F}_j(X)\in \ct_0$ implies that the vertical arrows $V_j$ in \eqref{prisms} are isomorphisms. Hence we may identify upper and lower nodes in pairs, so that $X$ may be seen as an indecomposable representation of the quiver with superpotential
\begin{equation}\label{uuuuwr}
\widetilde{Q}\colon\: \begin{gathered}
\xymatrix{& &\circ \ar@<-0.5ex>[ddll]_{h_2} \ar@<-0.5ex>[ddrr]_{H_3} & &
\\
\\
\circ\ar@<-0.5ex>[uurr]_{H_2}\ar@<-1ex>[rrrr]_{h_1} &&&& \circ\ar@<0ex>[llll]_{H_1}\ar@<-0.5ex>[uull]_{h_3}}
\end{gathered}\qquad \cw=H_3H_2H_1+h_1h_2h_3.
\end{equation}

The corresponding Jacobian algebra $\C\widetilde{Q}/\partial\cw$ is not a string algebra any longer, because it is not finite--dimensional. However, since $X\in \mathscr{L}_0$, one has
\begin{equation}
X_{H_1}X_{h_1},\ X_{H_2}X_{h_2},\ \text{and }X_{H_3}X_{h_3} \ \text{are nilpotent}
\end{equation}
which implies that $X\in \mathsf{nil}(Q,\cw)$, the category of \emph{nilpotent} modules of the Jacobian algebra $\C\widetilde{Q}/\partial\cw$. Indeed, the special light category $\mathscr{L}_0$ is given by the nilpotent modules
\begin{equation}
\mathscr{L}_0\equiv \mathsf{nil}(\widetilde{Q},\cw).
\end{equation}
The indecomposable objects of $\mathsf{nil}(\widetilde{Q},\cw)$ are given by string and band modules, and we may interpret 
$\mathsf{nil}(\widetilde{Q},\cw)$ in terms of WKB geodesics on the three--punctured sphere \cite{ACCERV2}. 
This identification holds, in particular, for the stable rigid representations generating $\mathsf{matter}$.

Precisely the quiver \eqref{uuuuwr} was shown in ref.\cite{ACCERV2} to correspond to the Gaiotto $A_1$ trinion theory $\ct_2$ \cite{Gaiotto}, namely  the model obtained by compactifying the 6d $A_1$ $(2,0)$ theory on a sphere with three \emph{ordinary} punctures. The theory $\ct_2$ represents physically a free half--hypermultiplet in the $(\mathbf{2},\mathbf{2},\mathbf{2})$ representation of a flavor $SU(2)^3$ group, so it is exactly the matter sector of our complete model with gauge group $SU(2)^3$.

So far so good. However, for the $\ct_2$ theory the relations on the quiver \eqref{uuuuwr} follow from the superpotential \cite{ACCERV2}
\begin{equation}
 \cw= H_1h_1+H_2h_2+H_3h_3+ H_1H_3H_2+h_1h_2h_3
\end{equation}
and don't look the same as the relations inherited by the quiver \eqref{uuuuwr} from the parent gentle algebra $\mathscr{A}$.

Is there a paradox? Not at all. The matter categories are just equivalent
\begin{equation}
 \mathsf{matter}\simeq \mathsf{matter}(\ct_2),
\end{equation}
 but, if we look at $\mathsf{matter}$ as an explicit subcategory of $\mathsf{mod}\,\mathscr{A}$, the form of the functors
$\mathsf{matter}\rightleftarrows \mathsf{matter}(\ct_2)$ will depend on the chosen BPS chambers, just because of the subtlety mentioned at the end of the last subsection (which motivates us to define the matter category $\mathsf{matter}_0$ as the object class of \textit{rigid} bricks rather than that of \emph{stable} rigid bricks). This subtlety is relevant only for representations of dimension $\mathfrak{d}$. From the $\ct_2$ viewpoint, there is a unique such representation described in eqn.(6.10) of ref.\!\cite{ACCERV2}. From the $SU(2)^3$--tube perspective there is one such representation per given (weak--coupled) chamber.  \textit{E.g.}, in the BPS chambers in which the state of charge $\mathfrak{d}$ is given by the string modules in figure \ref{deltareppp} the identifications between the tube horizontal maps $H_i, h_i$ and the $\ct_2$ maps $Y_i,X_i$ (notation of \cite{ACCERV2}) are, respectively, as follows
\begin{align}
&\left\{\begin{aligned}
  H_1&=X_3+Y_1Y_2\\
H_2&=X_1+Y_2Y_3\\
H_3&=X_2\\
h_1&=Y_3\\
h_2&=Y_1+X_3X_2\\
h_3&=Y_2+X_1X_3
 \end{aligned}\right.&&
\left\{\begin{aligned}
  H_1&=X_3+Y_1Y_2\\
H_2&=X_1\\
H_3&=X_2+Y_3Y_1\\
h_1&=Y_3\\
h_2&=Y_1+X_3X_2\\
h_3&=Y_2+X_1X_3
 \end{aligned}\right.
\end{align}

Since the $SU(2)^3$ symmetry and its gauging 
are manifest in the $SU(2)^3$--tube perspective, they \underline{cannot} be explicit in the $\ct_2$ framework; then gauging in that formalism is rather subtle (see \cite{ACCERV2}). However the two viewpoints are equivalent and using one or the other is a matter of convenience.

There is a third realization of the naive matter category: indeed a free half--hyper in the $\tfrac{1}{2}(\mathbf{2},\mathbf{2},\mathbf{2})$ is just four free hypers which corresponds to the disconnected quiver $\bullet\ \bullet\ \bullet\ \bullet$. The objects are just direct sums of four copies of representations of the trivial $A_1$ quiver $\bullet$ (equivalently, this is the tensor product $D_2\,\square\, D_2$).

The existence of many different quiver realizations of the same matter category, all with a transparent physical interpretation, may be seen as a novel class of $\cn=2$ dualities.

\subsubsection{Other quivers in the mutation class}

The mutation class of the quiver \eqref{prisms} contains four quivers. Of course, the physics of the light states is independent of the quiver we use by the argument in \S.\,\ref{chambindependence}. For the sake of comparison, we study the physics as seen from a \emph{non}--canonical quiver. For instance, a second quiver in the class is
\begin{gather}\label{firstquiver}
\begin{gathered}\xymatrix{1\ar@<0.4ex>[dd]\ar@<-0.4ex>[dd] & & & 5\ar@<0.4ex>[dd]\ar@<-0.4ex>[dd]\\
& 3\ar[ul]\ar[r] & 4\ar[ur] &\\
2 \ar[ur] & & & 6\ar[ul]}\end{gathered}\end{gather}
Drawing the corresponding triangulation of the three--holed sphere, one sees that the three bands which are light in the hole shrinking limit are the obvious two associated to the $W$--bosons of the Kronecker subquivers (with dimension vectors $\delta_1,\delta_2$) and \vglue3pt
\begin{equation}\label{99eerrr}
 \xymatrix{3\ar[r]\ar@/^1pc/[rrrrrrr]& 1\ar[r] & 2\ \ar[r] & 3 \ar[r] & 4\ar[r] & 5\ar[r] & 6\ar[r] & 4 }
\end{equation}
with dimension vector
\begin{equation}
\delta_3=(1,1,2,2,1,1),
\end{equation}
and dual magnetic charge\footnote{\ Here and trough the paper $B$ stands for the exchange matrix of the quiver, equal to the Dirac pairing in the basis of the simple representations $S_i$ of the quiver.}
$$m_3(X)\equiv -\tfrac{1}{2}\,\delta_3^t\, B\,\dim X = \dim X_3-\dim X_4.$$
Besides the three $W$--bosons the other light stable representations are string modules having dimension vectors
\begin{equation}\label{rrrrr}
(0,0,1,1,0,0),\quad (1,1,1,1,1,1),\quad (1,1,1,1,0,0),\quad (0,0,1,1,1,1)
\end{equation}
or 
\begin{equation}
\tfrac{1}{2}(\delta_3-\delta_1-\delta_2),\quad 
\tfrac{1}{2}(\delta_3+\delta_1+\delta_2),\quad
\tfrac{1}{2}(\delta_3+\delta_1-\delta_2),\quad
\tfrac{1}{2}(\delta_3-\delta_1+\delta_2),
\end{equation}
one representation per dimension vector being stable in each weakly coupled chamber (the actual representation which is stable depends on the chamber). The charge vectors represent a half--hyper in the $(\mathbf{2},\mathbf{2},\mathbf{2})$ representation, as they should.

Again we may write the Abelian light subcategory $\mathscr{L}$ in the form $$\mathscr{L}=\bigvee_{\lambda\in N}\mathscr{L}_\lambda$$  where $N$ is the configuration of three projective lines crossing at a single point. If $\lambda$ is a generic point on the $k$--th $\mathbb{P}^1$ component, $\mathscr{L}_\lambda$ is the subcategory generated by all indecomposable representations
of the form $M(C_k,\lambda,n)$, where $C_k$ is the corresponding band. This generic category is a stable homogeneous tube \cite{butring}, in agreement with the universality of the gauge sector. If $0$ is the crossing point, $\mathscr{L}_{0}$ is a non--homogeneous subcategory containing all the half--hyper states which are stable in some weak coupling chambers and all their extensions.

The main point is that $\mathscr{L}_{0}$ --- which is the closure with respect to kernel, cokernels, and extensions of the Krull--Schmidt subcategory generated by the indecomposable objects \eqref{rrrrr} --- contains indecomposable representations with dimension vectors $\delta_k$, $k=1,2,3$, which are in the projective closure of the family $M(C_k,\lambda,1)_{\lambda\in \C^*}$. In this sense, the point $\lambda_0$ does belong to each one of the three $\mathbb{P}^1$ lines.  As always, the notation
$\mathscr{L}=\bigvee_{\lambda\in N}\mathscr{L}_\lambda$ means that two objects
$X\in \mathscr{L}_\lambda$ and $Y\in \mathscr{L}_\mu$, with $\lambda\neq \mu$ are \emph{orthogonal} 
in the sense
\begin{equation}\label{ortttg}
\mathrm{Hom}(X,Y)=\mathrm{Hom}(Y,X)=\mathrm{Ext}^1(X,Y)=\mathrm{Ext}^1(Y,X)=0.
\end{equation}
Physical consistency requires that
\begin{equation}
D^b(\mathscr{L}_0)\simeq D^b\big(\mathscr{L}_0\big|_\text{canonical quiver}\big).
\end{equation}

\subsection{$\mathscr{G}$ for general \emph{gentle} $\cn=2$ models}\label{generalcomplete}

\begin{figure}
 \centering
\includegraphics[width=0.46\textwidth, height=0.40\textwidth]{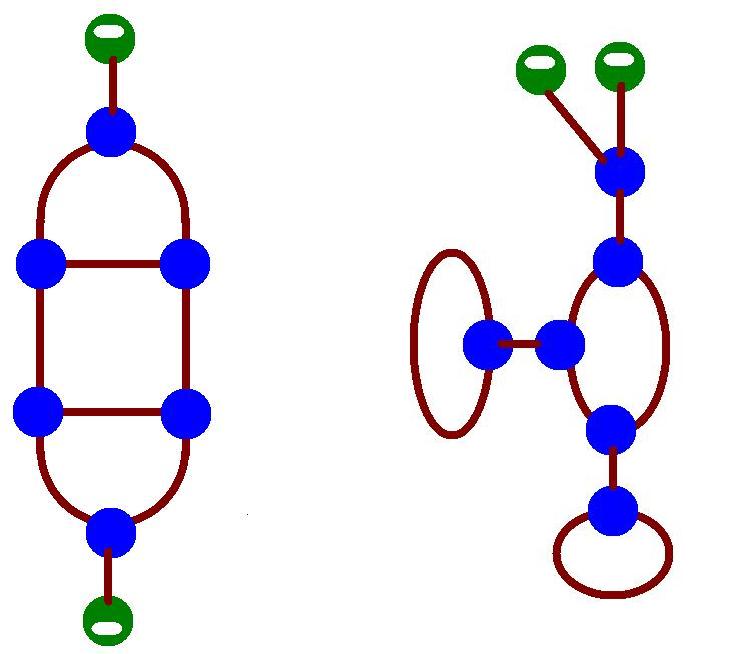}   
 \caption{Two inequivalent degeneration of the $g=3$, $b=2$ surface: the surface is decomposed into $3$--punctured spheres (blue), punctured disks (green) and infinitely thin cylinders (red). Each such maximal degeneration defines a weakly coupled regime.}
 \label{fig:g3b2}
\end{figure}

The analysis for a general gentle $\cn=2$ theory --- that is, a $A_1$ Gaiotto theory \cite{Gaiotto} with only irregular poles and at least one such pole --- is rather similar to the one in the previous subsection except for one important aspect: most of the examples considered so far have a canonical weak coupling limit for the YM sector. A general Gaiotto theory has, instead, many inequivalent weak coupling limits in one-to--one correspondence with the maximal degenerations of the surface $\Sigma_{g,b}$ having genus $g$ and $b$ boundary components.
Each such degeneration produces a different weakly coupled description which is related to the other ones by a generalized $\cn=2$ duality \cite{Gaiotto}. Each weakly coupled regime corresponds to a limit in which $\Sigma_{g,b}$ degenerates into a collection of $2(g-1)+b$ spheres with $3$ punctures and $b$ punctured disks with $c_i$ marked points on their $S^1$ boundary. The several punctures are connected in pairs by $3(g-1)+2b$ infinitely long and thin plumber cylinders, the plumbing parameter of the $j$--th cylinder $q_j$ being related to the (complexified) square--inverse gauge coupling of the $j$--th $SU(2)$ gauge factor, $-i\tau_j$ by the formula $q_j=\exp(2\pi i \tau_j)$. See figure \ref{fig:g3b2} for an example.

Therefore, for each such maximal degeneration of the surface $\Sigma_{g,b}$ we have a light (derived) subcategory $D^b(\mathscr{L})$. The various light subcategories are related to each other by the dualities of the $\cn=2$ theory. The simplest instance of these dualities being the $SL(2,\Z)$ action on the tubes of the (derived) module category of a tubular algebra, cfr.\! \S.\ref{ttubular}, or of the models discussed in \S\S.\,\ref{remmodel}
and \ref{cn2*} below.

Choosing an ideal triangulation of $\Sigma_{g,b}$ defines an incidence quiver $Q$ and a superpotential $\cw$ as in refs.\!\cite{triangulation1,traingulation2}; two triangulations give mutation--equivalent pairs $(Q,\cw)$.
By the Assem--Br\"ustle--Charbonneau--Pladmondon theorem \cite{assemgentle} the algebra $\C Q/(\partial\cw)$ is gentle.
The indecomposable modules are either string modules, which do not have free parameters, or band modules which come in one--dimensional families labelled by $\lambda\in \C^*$. Hence the BPS states are either hypermultiplets, which should correspond to string modules, or vector multiplets which correspond to the projective closure of a family of band modules $M(C,\lambda,1)$.

For a general string algebra one has\vglue 12pt

\textbf{Theorem.} (Butler --Ringel \cite{butring}) \textit{The band modules belong to a homogeneous tube.}\vglue 6pt
\vglue 12pt

This result is already enough to guarantee that each light category $\mathscr{L}$ has the form
\begin{equation}
\mathscr{L}=\bigvee_{\lambda\in N}\mathscr{L}_\lambda
\end{equation}
where $N$ is a space which is the union of $3(g-1)+2b$ copies of $\C^*$  (one copy per light vector supermultiplet) together with finitely many closed points. Each (light) indecomposable $X$ belongs to a unique $\mathscr{L}_\lambda$,
and two indecomposable $X\in \mathscr{L}_\lambda$, $Y\in\mathscr{L}_\mu$ are orthogonal in the sense \eqref{ortttg}
for $\lambda\neq \mu$.
By the above theorem, the generic $\mathscr{L}_\lambda$ is a standard homogenous tube, as required by the universality of the gauge interactions. Thus the Ringel's property holds true for all gentle $\cn=2$ theories.

We may be more precise.  From \cite{assemgentle} \textbf{Proposition 4.2}  we know that the bands $C$ are in one--to--one correspondence with the indivisible cycles in $H_1(\Sigma_{g,b},\Z)$. A cycle $\gamma$ defines a band $C(\gamma)$ which is light in the limit $g_j\rightarrow 0$ precisely if it can be represented by a WKB curve of vanishing length in the corresponding degeneration limit. Hence the light bands are the WKB geodesics wrapping once around each narrow plumber cylinder. The free parameter $\lambda\in \C^*$ in the band module $M(C(\gamma),\lambda,1)$ may then be identified with the freedom of moving the curve $\gamma$ along the infinitely long cylinder which, as a complex analytic variety, is a copy of $\C^*$.
In particular, the number of $\C^*$ components in $N$ is equal to the number of thin cylinders in $\Sigma$, \textit{i.e.}\! $3(g-1)+2b$.

Likewise, the string modules are in one--to--one correspondence with the homology class of open paths in $\Sigma_{g,b}$ (not homotopic to an arc in the triangulation) \cite{assemgentle}. Those of bounded mass in the limit $g_i\rightarrow 0$ have vanishing length. Therefore,
\begin{enumerate}
 \item if they correspond to matter charged under more than one $SU(2)$, they should be localized in a $3$--punctured sphere where the cylinders corresponding to those $SU(2)$ group factors meet. As a particular case, two punctures of the same sphere may be connected by a long cylinder (as in the second figure \ref{fig:g3b2}), producing a line with a node where a second line may cross it.  The light category $\mathscr{L}_{\lambda_\star}$  over the the nodal crossing $\lambda_\star\in N$ corresponds to matter in the representation  $\mathbf{3}\oplus\mathbf{1}$ of the $SU(2)$ associated to the nodal curve (tensor the fundamental rep.\! of the other line $SU(2)$, if present), see \S.\,\ref{remmodel}, \S.\,\ref{cn2*};
\item matter charged under a single $SU(2)$ corresponds to WKB geodesics lying in a small punctured disk. A punctured disk with $c_i$ marked points on the boundary corresponds to a a $D_{c_i}$ Argyres--Douglas system \cite{CV11}. In particular, for $c_i=1$ we get back a homogeneous tube whose only function is to ensure the projective closure to the $\mathbb{P}^1$--family.
\end{enumerate}

In conclusion: {in the gentle case the light subcategory with respect to a given maximal degeneration limit of $\Sigma_{g,b}$ has the form
$$\mathscr{L}=\bigvee_{\lambda\in N} \mathscr{L}_\lambda$$
where  $N$ is the reducible curve which is the degenerate limit of the Gaiotto curve $\Sigma_{g,b}$ itself, where the $3$--punctured spheres are contracted to the crossing points $\lambda_\alpha$ and the holes (\emph{i.e.}\! irregular punctures) are contracted to smooth closed points $\lambda_i$. The subcategories $\mathscr{L}_{\lambda_i}$ are stable tubes of period $p_i\geq 1$, where $p_i$ is the number of markings on the $S^1$ boundary of the $i$--th hole\footnote{\ Pictorially, the boundary looks like the mouth of the corresponding stable tube, the markings being associated to the regular simple reps. }. The normalization of each connected component of $N$ is a $\mathbb{P}^1$; the $\mathscr{L}_\lambda$ over the generic point of each irreducible component is a homogeneous stable tube. The category over the intersection points $\lambda_\alpha$ of three distinct $\mathbb{P}^1$ components is the closure with respect to kernel, cokernels and extensions of the $\tfrac{1}{2}(\mathbf{2},\mathbf{2},\mathbf{2})$ matter subcategory of \S.\,\ref{222matcategory}, while the category over a triple point which is a node of a $\mathbb{P}^1$ component corresponds to the analogous Abelian category associated to a half--hyper in the representation $\tfrac{1}{2}(\mathbf{3},\mathbf{2})\oplus \tfrac{1}{2}(\mathbf{1},\mathbf{2})$ of $SU(2)^2$, to be discussed in \S.\,\ref{cn2*} below.}

This result gives a new `categorical' interpretation of the Gaiotto curve and its degeneration limits. For instance, in the examples of figure \ref{fig:g3b2}, $N$ is the union of 10 $\mathbb{P}^1$'s crossing at six points and having two marked points over which we put the stable tubes associated to 
the green punctured disks.

Writing $\mathbb{P}_j^1$ for the irreducible component of $N$ associated to the $j$--th factor in the gauge group  $SU(2)\times SU(2)\times \cdots \times SU(2)$, we conclude
\begin{align*}
&\begin{aligned}&X\, \text{a string module corresponding to a light hyper}\\
&\text{charged under }SU(2)_{j_1}, SU(2)_{j_2},\cdots, SU(2)_{j_r}\quad\Longrightarrow\end{aligned}\\
&\hskip 3cm\Longrightarrow\quad X\in \mathscr{L}_\lambda,\ \text{with }\lambda\in \mathbb{P}^1_{j_1}\cap \mathbb{P}^1_{j_2}\cap\cdots\cap \mathbb{P}^1_{j_r}.
\end{align*}

Concretely \cite{ACCERV1}, the band representation $M(C,\lambda,1)$ (resp.\! the string representation $M(C)$) is given as follows: draw on the degenerate surface $\Sigma_{g,b}$ the ideal triangulation $T$ whose incidence quiver is $Q$. There is a one--to--one correspondence between nodes of $Q$ and arcs of $T$, as well as arrows $i \rightarrow j$ and triangles having sides the arcs $j$, $i$ in the \emph{same} order (with respect to the orientation of $\Sigma_{g,b}$). The string $C(\gamma)$ associated to a curve on $\Sigma_{g,b}$ is then just the sequence of nodes and (direct or inverse) arrows corresponding to the sequence of arcs and triangles of $T$ crossed by $\gamma$.

Using this rule, we easily construct the charge (dimension) vector of the representation corresponding to the $j$--th $W$--boson (namely the generic representation in the family $M(C(\gamma_j), \lambda, 1)$ with $\gamma_j$ a curve of vanishing length wrapping the $j$--th plumbing cylinder). Call $\delta_j$ this vector.

The magnetic charges $m_j(\cdot)$ (with respect to the duality frame associated to the particular degeneration limit of $\Sigma_{g,b}$ under consideration) are given by the Dirac pairing with the corresponding electric charge, that is
\begin{equation}\label{dirrrp}
m_j(X)= - \frac{1}{2}\delta_j^t\,B\, \dim X,\qquad j=1,2,\dots, 3(g-1)+2b,
\end{equation}
where  $B$ is the incidence matrix of the triangulation $T$, equal under the correspondences $$\text{indecomposable modules}\ \longleftrightarrow\ \text{strings/bands}\ \longleftrightarrow\ \text{homotopy classes of curves}$$ to the intersection form in homology. 
The factor $1/2$ in front of the \textsc{rhs} of \eqref{dirrrp} reflects the fact that the $W$--boson has electric charge $+2$, so the charge vector corresponding to the unit electric charge is $\delta_j/2$. Then, \emph{a priori}, $m_j(X)\in \tfrac{1}{2}\Z$, while, physically, we know that the correct charge quantization requires
the map $m_j\colon \mathsf{rep}(Q,\cw)\rightarrow \Z$ to have image precisely $\Z$. This integrality property of $m_j(\cdot)$ gives a useful check on the identification of the $W$--boson charges $\delta_j$.  

In particular
\begin{equation}
m_j(\delta_i)=0,
\end{equation}
since the corresponding geodesics have well--separated supports in $\Sigma_{g,b}$. Hence the above $W$--bosons are mutually local (as they should). More generally, a representation $X$ may belong to the light category $\mathscr{L}$ (for the given maximal degeneration) only if $m_j(X)=0$ for all $j$'s.
\smallskip

Let $C(\gamma_j)$ be the bands corresponding to the vanishing length geodesics \textit{i.e.}\! to the light $W$--bosons. The generic representation over $\mathbb{P}_j^1$ defines  a representation of an affine quiver $\widehat A_{n_j}$ (where $n_j+1$ is length of the band) with an orientation given by the sequence of direct/inverse arrows in $C(\gamma_j)$. Let $n_+$ (resp.\! $n_-$) be the number of positively (resp.\! negatively) oriented arrows in $C(\gamma_j)$; $n_+n_-\neq 0$ since the gentle algebra of a triangulation has oriented cycles only of length $3$ which have radical square zero \cite{assemgentle}. 
Then, in some sense, the $W$--bosons of a gentle $\cn=2$ model `look like' those of the $SU(2)$ model corresponding to the affine algebra associated to $C(\gamma_j)$, derived equivalent to the canonical algebra $\C \widehat{A}(n_+,n_-)$. In the next subsection we make this more precise.

\subsubsection{Bands \textit{vs.}\! magnetic charges}

Given a maximal degeneration limit of the Gaiotto curve of a gentle $\cn=2$ model, we have $3(g-1)+2b$ short WKB geodesics corresponding to $3(g-1)+2b$ light bands $C_a$, $a=1,2,\dots, 3(g-1)+2b$, associated to the $3(g-1)+2b$ $W$--bosons. Let $\delta_a$ be the corresponding charge (dimension) vectors. The dual magnetic charges are given by $m_a(X)=-\tfrac{1}{2}\,\delta^t_a\, B\,\dim X$. In the gentle case the magnetic charges have a simple and illuminating form.
  
We write the bands $C_a$ as a cyclic sequence of nodes $j(k)\in Q$ ($k=1,2,\dots, \ell$ with $j(k+\ell)\equiv j(k)$) separated by direct/inverse arrows; we number the arrows of $C_a$ so that the $k$--th arrow is the one between the nodes $j(k)$ and $j(k+1)$ in the sequence. We say that the node $j(k)\in Q$ a `sink in $C_a$' (resp.\! a `source in $C_a$') if it is a sink (resp.\! a source) of the resulting affine quiver, that is, if the $(k-1)$--th arrow is direct and the $k$--th one is inverse (resp.\! if the $(k-1)$--th arrow is inverse and the $k$--th one is direct).
Then 
\begin{equation}\label{uuuwuw}
m_a(X) =\sum_{k\colon j(k)\ \text{is a}\atop \text{source in }C_a} \dim X_{j(k)}-
\sum_{k\colon j(k)\ \text{is a}\atop \text{sink in }C_a} \dim X_{j(k)}.
\end{equation}
%\medskip

To see this, notice first of all that whenever $\mathrm{Supp}\,\dim X \cap \mathrm{Supp}\,\delta_a=\emptyset$, one has $m_a(X)=0$.
 Indeed, take $X$ to be the simple $S(i)$ with $i\not\in \mathrm{Supp}\,\delta_a$. 
Let $a_{i,j(k)}$ be the signed number of arrows from node $i$ to node $j(k)$ (a negative number meaning arrows in the opposite direction). By definition,
\begin{equation}
m_a(S(i))= \tfrac{1}{2}\,\sum_{k=1}^m a_{i,j(k)}.
\end{equation}
Fix $k$. The possible values of $a_{i,j(k)}$ are $a_{i,j(k)}=-2,-1,0,1,2$. If $a_{i,j(k)}=-2$ (resp.\! $2$), $j(k)$ must be a sink (resp.\! a source) in $C_a$. Using the fact that the Jacobian algebra $\C Q/(\partial\cw)$ is both $3$--CY complete and gentle, we see that locally we must have a configuration of the form
\begin{equation}
\xymatrix{ && \ar[ld] i \ar[rd]\\
\cdots & j(k-1) \ar[r] & j(k)\ar@<0.4ex>[u]\ar@<-0.4ex>[u] & j(k+1)\ar[l] &\cdots}
\end{equation} 
\textit{i.e.}\! $a_{i,j(k-1)}=a_{i,j(k+1)}=1$, $a_{i,j(k)}=-2$, which gives a total contribution zero to $m_a(S(i))$. Likewise, if $a_{i,j(k)}$ is $-1$ (resp.\! $+1$), $j(k)$ cannot be  a sink or a source in $C_a$, and the definition of a band together with the $3$--CY completeness implies
a configuration like
\begin{equation}
\xymatrix{ && \ar[ld] i \\
\cdots & j(k-1) \ar[r] & j(k)\ar[u]\ar[r] & j(k+1) &\cdots}
\end{equation}  
(resp.\! its dual) which again gives a vanishing total contribution to $m_a(S(i))$. \hfill $\square$

Thus $m_a(S(i))\neq 0$ requires $i\equiv j(k)$ for some $k$. If $Q$ is simply--laced, eqn.\eqref{uuuwuw} follows. If it is not simply-laced, we have to work a bit more.
Suppose, say, that $i\equiv j(k)$ is the source of a Kronecker subquiver of $Q$ while it is not a sink in $C_a$. Since a Kronecker subquiver should be attached to $j(k-1)$ by an oriented triangle \cite{CV11}, we must have a configuration of the form
\vglue 6pt
\begin{equation}
\xymatrix{\cdots & j(k-1) \ar[r] & i\equiv j(k) \ar@<0.5ex>[r]\ar@{-->}@<-0.5ex>[r] & j(k+1)\ar@{-->}@/_1.4pc/[ll]\ar@{-->}[r] & \bullet \ar@{-->}@/^1.4pc/[ll]\cdots }
\end{equation}
\vglue 6pt
\noindent where the solid arrows correspond to arrows that certainly are in $C_a$, while the other (possible) arrows of $Q$ are dashed, and we used $3$--CY completeness and gentleness to restrict the configuration in the vicinity of a Kronecker full subquiver. From the rules for forming bands, we infer that the arrow going out from the sink of the Kronecker subquiver which is in $C_a$ is the curved one, that is
\vglue 6pt
\begin{equation}
\xymatrix{\cdots & j(k-1)\equiv j(k+2) \ar[r] & i\equiv j(k) \ar@<0.5ex>[r]\ar@{-->}@<-0.5ex>[r] & j(k+1)\ar@/_1.4pc/[ll]\ar@{-->}[r] & \bullet \ar@{-->}@/^1.4pc/[ll]\cdots }
\end{equation}
\vglue 6pt
\noindent and this local configuration contributes zero to $m_a(S(i))$. The same argument work for $i$ the sink of a Kronecker subquiver which is not a source in $C_a$.
This completes the argument for eqn.\eqref{uuuwuw}. In particular, $m_a(X)$ is always an integral form, in agreement with Dirac quantization.
\medskip

Formally, eqn.\eqref{uuuwuw} suggests to associate to each light band $C_a$ and $X\in \mathsf{rep}(Q,\cw)$ a representation of the Kronecker quiver of the following form
\begin{equation}
 \xymatrix{\bigoplus\limits_{j(k)\ \text{is a}\atop \text{source in }C_a} X_{j(k)}\ar@<2ex>[rr]
\ar@<1ex>[rr] && \bigoplus\limits_{j(k)\ \text{is a}\atop \text{sink in }C_a}X_{j(k)}}
\end{equation}
which preserves the magnetic charges. However, this construction is not really uselful since for a typical indecomposable $X$ the resulting Kronecker representation is decomposable.

\subsubsection{Example: the $g=b=c=1$ model}\label{remmodel}

As a further illustration, we consider the `remarkable' $\cn=2$ model of ref.\!\cite{CV11} corresponding to the torus with one boundary component having a marked point.
Cutting open the torus, we have the ideal triangulation in the figure
\begin{gather}
\begin{gathered} \xymatrix{\bullet \ar@{=}@<-0.7ex>@/^2pc/[ddrr]
 \ar@{=}@<0.7ex>@/_2pc/[ddrr]\ar@{-}[rrrr]^1 \ar@{-}@/^3.5pc/[ddddrrrr]^3\ar@{-}@/_3.5pc/[ddddrrrr]_4 & & & &\bullet\ar@{-}[dddd]^2\\
 & & & &\\
 && &&\\
 &&&&\\
 \bullet\ar@{-}[uuuu]^2 &&&&\bullet\ar@{-}[llll]^1 }
\end{gathered}\end{gather}
where the double line stands for the boundary of the surface and the bullet $\bullet$ denotes the marked point. With the labelling of arcs in the figure, the adjacency matrix reads
\begin{align}
B=\begin{pmatrix} 0 & 2 & -1 & -1\\
   -2 & 0 & 1 & 1\\
1 & -1 & 0 & 1\\
1 & -1 & -1 & 0
  \end{pmatrix}
\end{align}
corresponding to the quiver
\begin{equation}\label{remquiver}
\begin{gathered}
\xymatrix{ 4 \ar[rr]^{B_2} &&1\ar@<-0.5ex>[dd]_{A_1}\ar@<0.5ex>[dd]^{B_1} && 3\ar[ll]_{A_2}\ar@/_2.5pc/[llll]_{\Phi}\\
&&& &\\
 && 2 \ar[uull]^{B_3}\ar[rruu]_{A_3}}
\end{gathered}
\end{equation}
with the relations
\begin{gather}
 A_1A_2=A_2A_3=A_3A_1=0\\
B_1B_2=B_2B_3=B_3B_1=0.
\end{gather}
Using Keller's mutation applet \cite{kellerapp}, one checks that this quiver is unique in its mutation class. Folding the quiver along the curved arrow $\Phi$, its relation with the $SU(2)$ $\cn=2^*$ quiver (\S.\ref{cn2*}) gets obvious.

In the (maximal) degeneration limit there are two short geodesics corresponding to the light $W$--bosons.
$\gamma_1$ corresponds to a curve homotopic to the shrinking  boundary, while $\gamma_2$ to the primitive homology class of the torus which vanishes in the degeneration limit. Writing $A$, $B$ for the canonical basis of $H_1(\text{torus}, \Z)$, we have
$[\gamma_2]=p A+ q B$, with $(p,q)$ coprime integers. The numbers $p$, $q$ depend on the particular $S$--duality frame in which we take the weak--coupling limit. Of course, the possible limits (and the corresponding gauge functors $\mathscr{G}$) form an orbit of $PSL(2,\Z)$. 

The dimension vector of $M(C(\gamma_1),\lambda,1)$ is independent of the chosen duality frame
\begin{equation}
 \delta_1\equiv \dim M(C(\gamma_1),\lambda,1)= (2,2,2,2),
\end{equation}
which corresponds to the band\vglue 12pt
\begin{equation}
C(\gamma_1)\colon\quad \begin{gathered} \xymatrix{3\ar@<1ex>@/^2pc/[rrrrrrr] \ar[r]& 1 \ar[r]^{B_1}& 2 \ar[r]& 3 \ar[r]& 4\ar[r]& 1 \ar[r]^{A_1} & 2 \ar[r]& 4}
                       \end{gathered}
\end{equation}
(notice that $C(\gamma_1)$ satisfies the constraints arising from $\partial\cw=0$ and it is not a power of any shorter band). The dual magnetic charge is
\begin{equation}
 m_1(X)\equiv -\frac{1}{2}\, \delta_1^t\, B\, \dim X= \dim X_3-\dim X_4
\end{equation}
in agreement with the general formula \eqref{uuuwuw}.

The second light vector depends on the duality frame, that is, on the coprime integers $p$, $q$. Here are a few choices of duality frames\vglue 12pt

\begin{center}
\begin{tabular}{|c|c|c|c|c|}\hline
 $p$, $q$ & $\delta_2$ & band $C$ & $m_2(X)$ & notes\\\hline
$1$, $0$ &$\phantom{\Bigg|}(0,1,1,1)$ & $\xymatrix{4 & 3\ar[l] & 2\ar[l] \ar@<-0.2ex>@/_0.8pc/[ll]}$ & $\dim X_2-\dim X_4$& rep.\! $\widehat{A}(2,1)$\\\hline
$0$, $1$ &$\phantom{\Bigg|}(1,0,1,1)$ & $\xymatrix{4\ar@<0.2ex>@/^0.8pc/[rr] & 3\ar[l]\ar[r] & 1}$ & $\dim X_3-\dim X_1$& rep.\! $\widehat{A}(2,1)$\\\hline
$1$, $-1$ &$\phantom{\Bigg|}(1,1,0,0)$ & $\xymatrix{1\ar@/^0.8pc/@<0.4ex>[r]^\alpha\ar@<-0.4ex>[r]_\beta &2}$ & $\dim X_1-\dim X_2$& rep.\! $\widehat{A}(1,1)$\\\hline
$1$, $1$ &$\phantom{\Bigg|}(1,1,2,2)$ & $\xymatrix{1 & 4\ar[l] & 3\ar[l] & 2\ar[l] \ar[r] & 4 & 3\ar[l] \ar@<-0.3ex>@/_0.8pc/[lllll]}$ & $\begin{matrix}\dim X_2+\dim X_3-\\ -\dim X_1-\dim X_4\end{matrix}$& \\\hline
\end{tabular}
\end{center}
\vglue 12pt

In the first three frames the second $W$--boson representation is actually a representation of a canonical affine subquiver, and we have a canonical gauge functor.
\medskip

One can easily compute the matter representations, says in the duality frame in which the vector associated to the Kronecker subquiver is the second light weakly coupled $W$--boson. Using the string modules, one sees that in each BPS chamber with both $W$--bosons light and weakly coupled
\begin{equation}
\begin{aligned}
 &\arg Z_3=\pi-O(g_1^2),&& &\arg Z_4=O(g_1^2),\\
&\arg Z_1=\pi-O(g_2^2),&& &\arg Z_2=O(g_1^2) 
\end{aligned}
\end{equation}
we have four BPS hypermultiplets with charge vectors\footnote{\ Which light representations are stable depend on the particular weakly coupled chamber, but not the set of charge vectors of light stable representation. See discussion in \S.\ref{chambindependence}. }
\begin{equation}
\frac{1}{2}\delta_1-\delta_2,\qquad \frac{1}{2}\delta_1,\qquad \frac{1}{2}\delta_1,\qquad \frac{1}{2}\delta_1+\delta_2 
\end{equation}
This corresponds to the \emph{half}--hyper $\Z_2$--anomaly--free $SU(2)\times SU(2)$ representation
\begin{equation}
 (\mathbf{2},\mathbf{3})\oplus (\mathbf{2},\mathbf{1}).
\end{equation}
\textit{E.g.}\! for $\arg(Z_1+Z_2)<\arg(Z_3+Z_4)$ the four stable light strings (which are not in the closure of the $\mathbb{P}^1$ families of the vectors) are
\begin{align}
&3\rightarrow 4 && 3\rightarrow 4\rightarrow 1\rightarrow 2\\
& 4\leftarrow 3\rightarrow 1\rightarrow 2 && 2\leftarrow  1 \leftarrow 3\rightarrow 4\rightarrow 1\rightarrow 2
\end{align}

\subsection{Complete non--gentle $\cn=2$ models}

Physically we expect that the discussion of the previous subsection applies, with mild modifications, to all complete $\cn=2$ theories.
The Assem--Br\"ustle--Charbonneau--Plamondon theorem has been (in part) generalized to the case of surfaces with ordinary punctures, provided there is at least one boundary component \cite{LF2}. For the more general case of no--boundary --- which, typically, lead to a non--finite--dimensional Jacobian algebra $\C Q/(\partial\cw)$ --- in most cases one may choose the free parameters in $\cw$ in such a way that there is a \emph{string} algebra $\mathscr{A}$ whose  $3$--Calabi--Yau completion is the Jacobian algebra
$\C Q/(\partial\cw)$ of an ideal triangulation of $\Sigma_g$\begin{equation}\C Q/(\partial\cw)=\Pi_3(\mathscr{A}).\end{equation}

The gentle algebra $\mathscr{A}$ (when present) is typically non unique. In favorable cases one may find a duality frame and a gentle $\mathscr{A}$ so that the light category $\mathscr{L}$ sits in $\mathsf{mod}\,\mathscr{A}$, and hence consists of string and band modules and the previous constructions work. 

\subsubsection{Categorical fixtures}

As in the previous section, we expect that the light subcategory $\mathscr{L}$ (with respect to a fixed maximal degeneration) has the form
\begin{equation}
\mathscr{L}=\bigvee_{\lambda\in N}\mathscr{L}_\lambda,
\end{equation}
where the reducible variety $N$ is again the maximal degeneration of the Gaiotto curve with the $3$--punctured spheres contracted to crossing points. With respect to the case of no (regular) puncture there are just three new special categories $\mathscr{L}_{\lambda_a}$ corresponding to the $3$--punctured spheres (`fixtures') with, respectively,
\begin{enumerate} \item one puncture connected to a long tube and two free punctures;\item two punctures connected to two different long tubes and a free puncture;
\item two punctures connected to the same long tube and a free puncture. 
\end{enumerate}
These three special subcategories may be extracted from the simplest $\cn=2$ theories having the corresponding type of matter. These theories are for the three fixtures above, respectively:
\begin{enumerate} 
\item $SU(2)$ SQCD with $N_f=2$, which is associated to a canonical algebra; \item the 
$SU(2)^2$ gauge theory with a bi--fundamental quark (see \S.\ref{bifund}); \item the $SU(2)$ $\cn=2^*$ model (see \S.\ref{cn2*}).
\end{enumerate}

In particular, a sphere with two free punctures and the third one connected to a thin cylinder just inserts two copies of the stable tube of period $2$ in the one--parameter family of the attached cylinder over two distinct points $\lambda_1,\lambda_2\in \mathbb{P}^1$. The result for the light category $\mathscr{L}$ is consistent even if the corresponding Jacobian algebra is not the $3$--CY completion of a gentle algebra; \textit{e.g.}\! if the surface $\Sigma$ is a twice punctured disk with $c\geq 1$ marked points on the boundary, the above rules gives a $\mathbb{P}^1$--family of stable tubes, almost all homogeneous, except for those over the points $0,1,\infty$ which have periods, respectively $2$, $2$, and $c$. This agrees with the fact that corresponding algebra is (derived equivalent to) the Euclidean algebra $\C \widehat{D}_c$ (the orientation of the affine quiver $\widehat{D}_c$ being irrelevant up to derived equivalence).

Even when there is no quiver, as in the case of the Gaiotto $A_1$ theories associated to $g\geq 3$ surfaces with no punctures nor boundaries, we can perform the above construction in each given maximal degeneration limit. We get a well defined light subcategory
$\mathscr{L}$ which has the standard form, even if there is no clear understanding of the `full' non--perturbative category, heavy objects included.

\subsubsection{$SU(2)\times SU(2)$ bifundamental}\label{bifund}

Here we check explicitly our claims for the bi--fundamental fixture. Again, readers not interested in (too much) details may prefer to jump ahead.

We start from the $SU(2)\times SU(2)$ gauge theory with a quark in the bi--fundamental.
The corresponding Gaiotto surface is the punctured cylinder with one marking on each boundary component. The most symmetric ideal triangulation leads to the $\Z_2$--symmetric quiver
\begin{equation}\label{8888}
\begin{gathered}
\xymatrix{\diamondsuit_1\ar[dr]^{\phi_1} & & \diamondsuit_2\ar[dl]_{\phi_2}\\
& \bullet\ar[dr]_{\psi_2}\ar[dl]^{\psi_1}\\
\spadesuit_1\ar@<0.4ex>[uu]^{A_1} \ar@<-0.4ex>[uu]_{B_1} && \spadesuit_2 \ar@<0.4ex>[uu]^{A_2} \ar@<-0.4ex>[uu]_{B_2}}
\end{gathered}
\end{equation}
The superpotential $\cw$ may be read directly from the triangulation \cite{traingulation2,ACCERV1}
\begin{equation}\label{uuuuqqq}
\cw=\mathrm{Tr}\big[A_1\psi_1\phi_1+A_2\psi_2\phi_2+\phi_2B_2\psi_2\phi_1B_1\psi_1\big].
\end{equation}

The quiver \eqref{8888} has an obvious candidate $SU(2)\times SU(2)$ gauge functor from $\mathsf{rep}(Q,\cw)$ to $(\mathsf{mod}\,\mathbf{Kr})^2$ given by restriction to the two Kronecker subquivers. It manifestly preserves the magnetic charges.
We claim that the restriction of an indecomposable is either zero or indecomposable. This follows from the fact that there are no non--zero paths in $\C Q/(\partial\cw)$ from $\diamondsuit_1$ to $\spadesuit_1$. Indeed, 
$\partial_{A_1}\cw=0$ gives $\psi_1\phi_1=0$, so any path connecting these two nodes should contain a subpath of the form
\begin{equation}
\psi_1\phi_2 B_2\psi_2\phi_1\quad \text{or}\quad \psi_1\phi_2 A_2\psi_2\phi_1. 
\end{equation}
The first one is $\partial_{B_1}\cw$ and vanishes. The relation $\partial_{\phi_2}\cw=0$ gives $A_2\psi_2+B_2\psi_2\phi_1B_1\psi_1=0$ and the second one also vanishes 
\begin{equation}
\psi_1\phi_2 A_2\psi_2\phi_1=-\psi_1\phi_2(B_2\psi_2\phi_1B_1\psi_1)\phi_1 =0.
\end{equation}

Therefore, the restriction to the Kronecker subquivers $\mathbf{Kr}_a$ ($a=1,2$) defines a canonical gauge functor in the sense of \S.\,\ref{cangaugefunct}. Note that the presence of the sextic term in the superpotential
\eqref{uuuuqqq} is absolutely necessary for the existence of a \emph{canonical} gauge functor\footnote{\ I thank Clay C\'ordova for this remark.}.

Consequently, for a light representation $X$ we have
$$\mathscr{G}_a(X)\equiv X\big|_{\mathbf{Kr}_a}\in\ct\equiv \bigvee_{\lambda\in \mathbb{P}^1}\ct_\lambda.$$
Without loss of generality, we may assume $B_1, B_2$ to be the identity. Then it is easy to check that the map
\begin{equation}\label{KKKKt}
\big(X_{\spadesuit_1},X_{\diamondsuit_1},X_\bullet,X_{\spadesuit_2},X_{\diamondsuit_2}\big)\mapsto \big(0,0, -\phi_1\psi_1X_\bullet, A_2X_{\spadesuit_2}, A_2X_{\diamondsuit_2}\big)
\end{equation}
is an element of $\mathrm{End}\,X$; a second element of $\mathrm{End}\,X$ is obtained by applying the $\Z_2$ automorphism to \eqref{KKKKt}. Thus, if $X$ is a light brick, $A_1,A_2$ should be elements of $\C$ (say, $A_i=\lambda_i\in\C$), and the restrictions $X\big|_{\mathbf{Kr}_i}$ (if non--zero) are regular bricks. Moreover, for a light brick we have
$\dim X_{\spadesuit_1}=\dim X_{\diamondsuit_1} \leq 1$. 
Among the relations, for such a light brick, we have
\begin{align}
&\psi_1(\lambda_1+\phi_2\psi_2)=0 && (\lambda_1+\phi_2\psi_2)\phi_1=0\\
&\psi_2(\lambda_2+\phi_1\psi_1)=0 && (\lambda_2+\phi_1\psi_1)\phi_2=0,
\end{align} 
so that if $\lambda_1$ (resp.\! $\lambda_2$) is not an eigenvalue of $\phi_2\psi_2$ (resp.\! $\phi_1\psi_1$) we have $\psi_1=\phi_1=0$ (resp.\! $\psi_2=\phi_2=0$) and our representation is, in fact, a module of the $N_f=1$ SQCD algebra (derived equivalent to the canonical algebra of type $\{2\}$).
But the relation $\psi_2\phi_2=0$ implies $\phi_2\psi_2$ nilpotent, and the only  eigenvalue is $\lambda_2=0$. Thus if $\lambda_2\neq 0$, the brick $X$ is a Kronecker indecomposable representation of dimension $\delta_a$ (the minimal imaginary root) extended by zero (and similarly for $\lambda_1$). The closure of the object class of the $\lambda_1\neq 0$ (resp.\! $\lambda_2\neq 0$) modules with respect to the extensions is then a standard homogenous tube.

In conclusion, setting $N$ to be the union of two projective lines meeting at the origin, we have the orthogonal decomposition (in the sense \eqref{ortttg})
\begin{equation}
\mathscr{L}=\bigvee_{\lambda\in N}\mathscr{L}_\lambda
\end{equation}
with $\mathscr{L}_\lambda$ a standard homogeneous tube for $\lambda\not=0$; $\mathscr{L}_0$ is the closure in $\mathsf{rep}(Q,\cw)$ under kernels, cokernels and extensions  of the Krull--Schmidt category generated by the indecomposable of dimension vector
\begin{equation}
\dim X_{\spadesuit_a}=\dim X_{\diamondsuit_a}\leq 1,\ a=1,2\qquad \dim X_\bullet\leq 1,
\end{equation}
with $A_a=0$ and $B_a=1$.
There are six such vectors (we write the dimension/charge of the simple $S(\bullet)$ in its phisical form $B-(\delta_1+\delta_2)/2$ where $B$ is the unit `baryon' charge)
\begin{equation}\label{listchar}
\delta_1,\, \delta_2,\ B\pm \frac{1}{2}\delta_1\pm \frac{1}{2}\delta_2,
\end{equation}
which correspond to the quantum numbers of the two $W$--bosons and a quark of unit baryon number in the $(\mathbf{2},\mathbf{2})$ rep.\! of $SU(2)^2$.
In general, there are several distinct brick representations having the \underline{same} dimension vector in the list \eqref{listchar}; again, precisely one representation per dimension vector is stable in any given weakly--coupled chamber. Hence, the matter category $\mathsf{matter}_0\subset \mathscr{L}_0$ is larger than the naive one, as in most examples.

The two bricks in $\mathscr{L}_0$ of dimension $\delta_1,\delta_2$ are not rigid: indeed they belong to the closure of the $\mathbb{P}^1$--family with the same dimension vector; this shows that the light category $\mathscr{L}_0$ lays `over' the intersection of the two projective lines.

The presence in $\mathscr{L}_0$ of non--rigid bricks with the quantum numbers of the two $W$--bosons signals the fact that the $SU(2)\times SU(2)$ bi--fundamental fixture has a  \emph{gaugeable} $SU(2)\times SU(2)$ symmetry. In Gaiotto language, it may be connected to infinite cylinders. The presence in a light non--homogeneous category of such non--rigid bricks with the quantum numbers of the $W$--bosons is the landmark that the light category represents matter with a gaugeable symmetry.

 \subsubsection{$SU(2)$ $\cn=2^*$}\label{cn2*}

 This is the unique complete theory with gauge group $SU(2)$ which is not canonical in the sense of the previous section. However most properties must still be valid, since they follow from physical locality and universality of the gauge sector.

It is often stated that the quiver of $SU(2)$ $\cn=2^*$ is the Markoff one
 \begin{equation}\label{mark}
 \begin{gathered}
 \xymatrix{\diamondsuit \ar@<0.4ex>[dd]^{Y_3}\ar@<-0.4ex>[dd]_{X_3} \\
 && \bullet\ar@<0.4ex>[ull]^{Y_1}\ar@<-0.4ex>[ull]_{X_1}\\
 \spadesuit\ar@<0.4ex>[urr]^{Y_2}\ar@<-0.4ex>[urr]_{X_2}}
 \end{gathered}\qquad \begin{aligned}\cw= &X_1X_2X_3+Y_1Y_2Y_3+\\
&+ \lambda\, X_1Y_2X_3Y_1X_2Y_3, \quad\lambda\in \C;
 \end{aligned}\end{equation}
 however, as explained in \cite{ACCERV2}, this is not strictly true: the Markoff quiver corresponds to $SU(2)$ SYM coupled to a quark in the representation
 \begin{equation}\label{[2][0]}
 [1]\otimes [1]= [2]\oplus [0],
 \end{equation}
 that is, to $SU(2)$ $\cn=2^*$ \emph{plus} a free decoupled hypermultiplet (carrying the same flavor charge as the triplet quark). This is obvious from the construction of the model as the 6d $A_1$ $(2,0)$ theory compactified on a torus with an ordinary puncture (\textit{i.e.}\! the surface whose triangulation has \eqref{mark} as incidence quiver): one starts with the sphere with three punctures \begin{large}$\ast$\end{large}  --- which corresponds to Gaiotto's $\ct_2$, \textit{i.e.}\! a half--hyper in the $(\mathbf{2},\mathbf{2},\mathbf{2})$ of  $SU(2)^3$ (cfr.\! \S.\,\ref{222matcategory}) --- and connects two of the punctures with a long plumbing cylinder. This gives a once--punctured torus:
\begin{figure}[here!]
 \centering
\includegraphics[width=0.38\textwidth, height=0.15\textwidth]{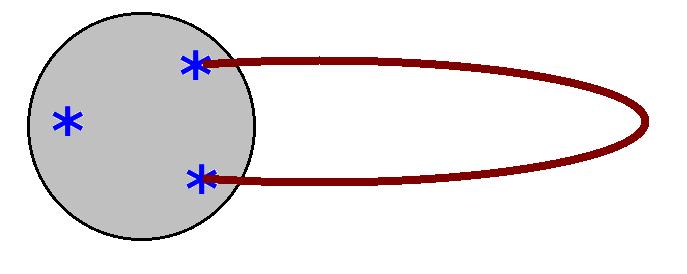}   
 % \caption{}
  %\label{fig:mspart}
\end{figure}

 The long thin cylinder corresponds to a $SU(2)$ SYM gauging the diagonal subgroup $SU(2)_\mathrm{diag}$ of the two $SU(2)$'s associated to the punctures at its ends. Under the subgroup
 $SU(2)_\mathrm{diag}\times SU(2)\subset SU(2)\times SU(2)\times SU(2)$ we have
 \begin{equation}
 (\mathbf{2},\mathbf{2},\mathbf{2})\longrightarrow (\mathbf{3},\mathbf{2})\oplus (\mathbf{1},\mathbf{2}).
 \end{equation}
 which corresponds to \eqref{[2][0]}. In \S.\,\ref{iirep} below we shall describe a remedy to the nuisance of having a spurious hypermultiplet in the spectrum. For the moment let us work in the full Gaiotto theory described by the Markoff quiver \eqref{mark}. For simplicity, we shall also set the free parameter $\lambda$ in eqn.\eqref{mark} to zero.
 \medskip

 The Gaiotto theory is, of course, UV--superconformal, and hence has an $S$--duality group $SL(2,\Z)$ acting on the vectors as in \S.\,\ref{ttubular}. Fixing (arbitrarily) a duality--frame, that is, declaring a certain BPS vector to be `electric', we may speak of `weak--coupling' with respect to the electric charge of this chosen vector. There are two choices which appear particulary natural:
 \begin{align}
 &(I) &&\begin{cases}\arg Z_\spadesuit =O(g^2)\\
 \arg Z_\diamondsuit=\pi - O(g^2)\end{cases}\qquad g\rightarrow 0\\
 &(II) &&\begin{cases}\arg Z_\diamondsuit =O(g^2)\\
 \arg Z_\spadesuit=\pi - O(g^2)\end{cases}\qquad g\rightarrow 0.\label{ttrupq}
 \end{align}

The first one corresponds to chosing the vertical Kronecker subquiver in eqn.\eqref{mark} as the support of the light $W$--boson representation. This is the conventional choice, and we leave to the reader the details of it. Instead we choose the less obvious duality frame corresponding to (II). 

\medskip

The Jacobian algebra of the Markoff quiver, $\C Q/(\partial \cw)$ is not gentle. The conditions \textbf{S1}--\textbf{S5} of \S.\,\ref{sec:ABCP} still hold true, but the algebra $\C Q/(\partial \cw)$ is not finite--dimensional.
However consider the (non--completed) algebra
$\mathscr{B}$ defined by the quiver
 \begin{equation}\label{biquiver}
 \begin{gathered}
 \xymatrix{\diamondsuit \ar@{..>}@<0.3ex>[dd]\ar@{..>}@<-0.3ex>[dd] \\
 && \bullet\ar@<0.3ex>[ull]^{Y_1}\ar@<-0.3ex>[ull]_{X_1}\\
 \spadesuit\ar@<0.3ex>[urr]^{Y_2}\ar@<-0.3ex>[urr]_{X_2}}
 \end{gathered}\qquad\text{with the relations }\quad Y_1Y_2=X_1X_2=0.
 \end{equation}
Clearly, the Jacobian algebra \eqref{mark} is a (possibly deformed) $3$--CY completion of $\mathscr{B}$.
Now $\mathscr{B}$ is finite--dimensional and hence gentle (and easy).

We claim that there is a duality frame, in facts the one corresponding to the choice (II) of eqn.\eqref{ttrupq}, in which all light stable BPS states correspond to representations $X$ with the dashed arrows of the quiver \eqref{biquiver} vanishing. In other words, as long as we are interested in the light category $\mathscr{L}$ in the duality frame (II) we may completely forget about the infinite--dimensional completed algebra $\C Q/(\partial \cw)$ and work with the nice gentle algebra $\mathscr{B}$.
In particular, $\mathscr{L}$ consists of string and band modules and WKB is (asymptotically) exact.

The claim may be shown in two ways. 
Either one goes trough a formal argument of the kind we used in
in \S.\ref{maggg} to establish a similar weak--coupling zero--arrow property, or one just looks to the  figure
\begin{figure}[here!]
 \centering
\includegraphics[width=0.45\textwidth, height=0.16\textwidth]{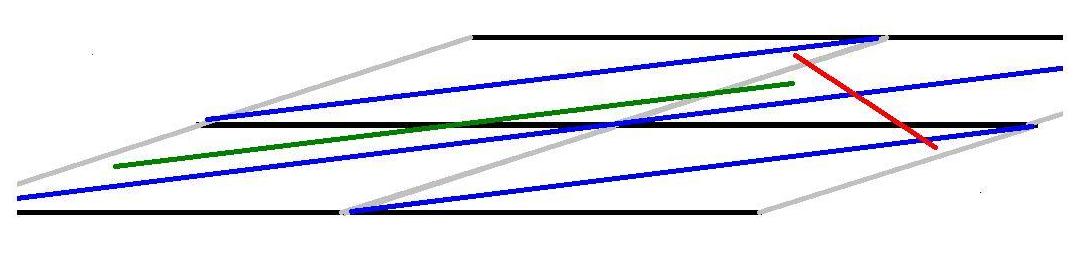}   
 % \caption{}
  %\label{fig:mspart}
\end{figure}
which represents a periodic ideal triangulation of the punctured torus where black, gray, and blue arcs correspond, respectively, to the $\spadesuit$, $\diamondsuit$, and $\bullet$ nodes of the quiver \eqref{mark}, while the red and green closed curves define, respectively, the bands
\begin{align}
 &C(green)\colon \quad \begin{gathered}
              \xymatrix{\diamondsuit \ar[rr]\ar@/^1.2pc/[rr] && \spadesuit } 
             \end{gathered},&&\dim M(C(green),\lambda,1)= \big(1_\spadesuit, 1_\diamondsuit, 0_\bullet\big)\\
&C(red)\colon \quad \begin{gathered}
              \xymatrix{& \bullet \ar@/^0.5pc/[rd]^{X_1} &\\
\spadesuit\ar@/^0.5pc/[ru]^{Y_2}\ar@/_0.5pc/[rd]_{X_2} & & \diamondsuit\\
& \bullet \ar@/_0.5pc/[ru]_{Y_1}} 
             \end{gathered},&&\dim M(C(red),\lambda,1)= \big(1_\spadesuit, 1_\diamondsuit, 2_\bullet\big).\label{cred}
\end{align}
It is clear from the picture that it is the red curve which has vanishing length in this particular degeneration limit.
Then, in our duality frame, the $W$--boson has a charge vector
\begin{equation}
 \delta=\big(1_\spadesuit, 1_\diamondsuit, 2_\bullet\big)
\end{equation}
and the magnetic charge form is
\begin{equation}
 m(X)= -\frac{1}{2}\, \delta^t\, B\, \dim X=\dim X_\spadesuit-\dim X_\diamondsuit,
\end{equation}
having the usual form for a $G=SU(2)$ theory.
\medskip

The gauge functor corresponding to our choice of duality frame
\begin{align}
 &\mathscr{G}\colon \mathsf{mod}\,\mathscr{B}\rightarrow \mathsf{mod}\,\mathbf{Kr},\\
\intertext{is explicitly}
&\mathscr{G}\!\!\left(  \begin{gathered}
 \xymatrix{\C^{n_\diamondsuit} \ar@{..>}@<0.3ex>[dd]^0\ar@{..>}@<-0.3ex>[dd]_0 \\
 && \C^{n_\bullet}\ar@<0.3ex>[ull]^{Y_1}\ar@<-0.3ex>[ull]_{X_1}\\
 \C^{n_\spadesuit}\ar@<0.3ex>[urr]^{Y_2}\ar@<-0.3ex>[urr]_{X_2}}
 \end{gathered}\right)= \begin{gathered}
 \xymatrix{\C^{n_\diamondsuit}\\ \\
 \C^{n_\spadesuit}\ar@<0.8ex>[uu]^{X_1Y_2}\ar@<-0.8ex>[uu]_{Y_1X_2}}
 \end{gathered}\ \in\ \mathsf{mod}\,\mathbf{Kr}.
\end{align}
and obviously preserves the magnetic charge.

 We are interested in the light category $\mathscr{L}\subset \mathsf{mod}\,\mathscr{B}$ containing all representations associated to BPS particle of bounded mass in the $g\rightarrow 0$ limit, \textit{i.e.}\! all stable representations with $m(X)=0$.
 
 Let $X$ be a module of $\mathscr{B}$ which is both stable and light in the weak coupling regime (II) (cfr.\ eqn.\eqref{ttrupq}). Then
\begin{equation}
 \mathscr{G}(X)\in \cp\vee \ct\vee \cq \equiv \cp\vee \left(\,\bigvee_{\lambda\in \mathbb{P}^1} \ct_\lambda\right)\vee \cq.
\end{equation}
In all canonical or gentle  models considered so far one has
$\mathscr{G}(X)\Big|_\cp=\mathscr{G}(X)\Big|_\cq=0$. As we shall show momentarily, this remains true in the present case too. However, in the canonical/gentle case we have that $\mathscr{G}(X)$ is either zero or belongs to a unique $\ct_\lambda$. This is no longer true, and there is a physical reason for this (Bose statistics) that will be explained in \S.\,\ref{iirep} below. For that reason we write
\begin{align}
&\ct= \bigvee_{(s:t)\in \mathbb{P}^1/\Z_2} \widehat{\ct}_{(s:t)}\\
&\text{where } \widehat{\ct}_{(s:t)}= \begin{cases}
\ct_{(s:t)}\vee \ct_{(t:s)} & (s:t)\neq (1:\pm 1)\\
\ct_{(s:t)} & (s:t)=(1,\pm1),
\end{cases}
\end{align}
where $\Z_2$ is the involution $(s:t)\leftrightarrow (s:t)$ of $\mathbb{P}^1$. Physics requires that if $X$ is a light stable object (in regime (II)) $\mathscr{G}(X)$, if not zero, belongs to a unique $\widehat{T}_{(s:t)}$ subcategory. 

By the adjoint--representation $SU(2)$ tube $\widehat{L}_\lambda$ we mean the Krull--Schimdt subcategory of $\mathsf{mod}\text{--}\mathscr{B}$ generated by the indecomposables $X$ which are mapped by $\mathscr{G}$ into $\widehat{\ct}_\lambda$. Universality of the gauge interactions requires all but finitely many adjoint $SU(2)$--tubes to be standard homogeneous tubes.  Again, this turns out to be true, as we show next.

\subsubsection{Band and string modules of $\mathscr{B}$}

$\mathscr{B}$ is a gentle algebra, hence all its indecomposable modules may be explicitly constructed as band or string modules.
We note that the algebra $\mathscr{B}$ has a $\Z_2$ automorphism $\sigma$ fixing all the nodes of the quiver and acting on the arrows as
\begin{equation}
\sigma\colon\ X_i \longleftrightarrow Y_i,\quad \forall\, i.
\end{equation} 
If $X$ is a stable module in some chamber, so is $\sigma(X)$. Notice that the $\Z_2$ \textit{orbit} of a string (resp.\! band) is uniquely fixed by the node sequence. Then it is convenient (and physically correct) to list the stable $\Z_2$--orbits rather than the single stable representations.

The complete list of the $\Z_2$--orbits of $\mathscr{B}$--modules which are stable and light in \underline{some} BPS chamber of the regime (II) is given in table \ref{tablestring}.
\begin{table}
\begin{tabular}{|c|c|c|c|c|}\hline
dimension & type & string/band & $\#(\Z_2$--orbit) & stable for\\\hline
& band & $C(red):\ \spadesuit\,\bullet\,\diamondsuit\,\bullet$ & 1$\phantom{\Big|}$ & all (II)\\ \cline{2-5}
$(1,1,2)$ & string & $\diamondsuit\,\bullet\,\spadesuit\,\bullet$ & 2$\phantom{\Big|}$ & $\arg Z(\bullet)< \arg[Z(\spadesuit)+Z(\diamondsuit)]$\\ \cline{2-5}
& string & $\spadesuit\,\bullet\,\diamondsuit\,\bullet$ & 2$\phantom{\Big|}$ & $\arg Z(\bullet)>\arg[Z(\spadesuit)+Z(\diamondsuit)]$\\\hline
$(0,0,1)$ & string & $\bullet$ & 1$\phantom{\Big|}$ & all (II)\\\hline
$(1,1,1)$ & string & $\spadesuit\,\bullet\,\diamondsuit$ & 2$\phantom{\Big|}$ & all (II)\\\hline
$(2,2,3)$ & string & $\diamondsuit\,\bullet\,\spadesuit\,\bullet\,\spadesuit\,\bullet\diamondsuit$ & 1$\phantom{\Big|}$ &
$\arg Z(\bullet)< \arg[Z(\spadesuit)+Z(\diamondsuit)]$\\
\cline{2-5}
& string & $\spadesuit\,\bullet\,\diamondsuit\,\bullet\,\diamondsuit\,\bullet\,\spadesuit$ & 1$\phantom{\Big|}$& $\arg Z(\bullet)> \arg[Z(\spadesuit)+Z(\diamondsuit)]$\\\hline
\end{tabular}
\caption{\label{tablestring} Table of string/band light stable modules for $\cn=2^*$ in regime (II).}
\end{table}
One has \begin{enumerate}
\item the string representations of dimension $(1,1,2)$ belong to the closure in $\mathbb{P}^1$ of the band representation $M(C(red),\lambda,1)$, $\lambda\in \C^*$;
\item in each chamber of the (II) weak coupling regime we have precisely four light hypermultiplets of charges
\begin{equation}
(0,0,1) \qquad 2\,\times\, (1,1,1) \qquad (2,2,3).
\end{equation}
or, in physical notations ($B$ unit baryon number, $\delta$ twice the unit electric charge),
\begin{equation}
\delta-B,\quad B, \quad B+\delta,\qquad \text{and } B
\end{equation}
\end{enumerate}

Thus the light spectrum is the expected perturbative spectrum: one $W$--boson, three hypers with the weights of the adjoint representation of $SU(2)$, and a free neutral hyper.
\medskip

Here we give the basics of the argument leading to table \ref{tablestring}, confining the details to appendix \ref{detailsforcn2*}.
We call  nodes $\spadesuit, \diamondsuit$ odd and $\bullet$ even. Along a string or band even/odd node alternates.
As already mentioned, given the sequence of nodes in the string/band $C$ there are only two allowed arrow assignments interchanged by the $\Z_2$ automorphism of $\mathscr{B}$; the two representations (if distinct) are either both stable or both unstable, and it is enough to check the stability of one of them.

Suppose $C$ (or $C^{-1}$) has the form
\begin{equation}\label{CCminuis}
  \cdots\cdots\ \spadesuit \rightarrow \bullet \leftarrow \spadesuit\ \cdots\cdots\ \diamondsuit \leftarrow \bullet\rightarrow \diamondsuit\ \cdots\cdots
\end{equation}
We claim that the corresponding string/band module $V=M(C)$ is unstable for all central charges $Z(\cdot)$ belonging to the (II) weak coupling regime.
Indeed, let $z_i,z_j$ be the basis vectors of $V_\bullet$ corresponding to the two $\bullet$'s visible in \eqref{CCminuis}.
One has $z_i\in \mathrm{ker}\, X_1\cap \mathrm{ker}\, Y_1$ and $z_j\in  \mathrm{coker}\, X_2\cap \mathrm{coker}\, Y_2$.
Since the vector subspaces $\mathrm{ker}\, X_1\cap \mathrm{ker}\, Y_1$ and $\mathrm{coker}\, X_2\cap \mathrm{coker}\, Y_2$ give, respectively, a subrepresentation and a quotient of $V$ with support on the node $\bullet$, if both subspaces are non zero the stability of $V$ requires
\begin{equation}
 \arg Z(\bullet) < \arg Z(V) < \arg Z(\bullet),
\end{equation}
which is impossible. Thus strings/bands of the form \eqref{CCminuis} are ruled out. 

The same argument shows that if --- in a given chamber --- a representation with a string/band containing
$\cdots\ \spadesuit\rightarrow \bullet \leftarrow \spadesuit\ \cdots$ is stable, all representations containing
$\cdots\ \diamondsuit \leftarrow \bullet\rightarrow \diamondsuit\ \cdots$ are unstable and \textit{viceversa}.

We focus on the first possibility, that is, $\arg Z(\bullet) <\arg Z(V)$, or equivalently
\begin{equation}\label{starstar}
\arg Z(\bullet) < \arg[Z(\spadesuit)+Z(\diamondsuit)].
\end{equation}
The situation in the chamber with the opposite inequality is dual. 
In the regime \eqref{starstar}, the string/band $C$ of a stable representation  is --- except for `boundary' terms at its ends --- a sequence of bullet configurations of the form
$\spadesuit\rightarrow \bullet\leftarrow\spadesuit$, or
$\spadesuit\rightarrow\bullet\rightarrow\diamondsuit$ and its inverse.

Bands are easier since they have no ends.
Let $C$ be a band such that $V\equiv M(C,\lambda,1)$ is both stable and light in our chamber. Then
\begin{gather}\label{09w}
\begin{split}
 0=m(V)&=\dim V_\spadesuit-\dim V_\diamondsuit =\\
 &=  \#\big\{\text{number of bullet configurations }\spadesuit\rightarrow \bullet \leftarrow \spadesuit\ \text{in } C\big\}.
\end{split}
\end{gather}
This leaves only one possibility
\begin{equation}
C=\big( \spadesuit \xrightarrow{X_2} \bullet \xrightarrow{Y_1} \diamondsuit \xleftarrow{X_2} \bullet \xleftarrow{Y_1} \big)^n,\qquad n\in \mathbb{N}.
\end{equation}
(Note that $C$ is $\Z_2$--invariant up to band equivalence). A band, by definition, cannot be a power of a shorter string, and hence we must have $n=1$, getting the band $C(red)$ we previously got from the WKB analysis. The same unique  band is obtained using the opposite inequality in \eqref{starstar}.

The case of a string module is slightly more complicated, since we have to discuss the several possible boundary configurations of $C$. The analysis is presented in appendix \ref{detailsforcn2*} with the result in table \ref{tablestring}.

\subsubsection{$ii$--representations of a quiver with automorphism}\label{iirep}

The $\Z_2$ symmetry $\sigma$ of the previous subsection has a clear physical meaning. Recall that the stable representations $X$ of a quiver $Q$ with superpotential $\cw$  with $\dim X_i=d_i$ are meant to correspond to the supersymmetric vacua of a SQM with quiver $Q$, gauge group $G=\prod_{i\in Q_0}U(d_i)$. For each arrow $i \xrightarrow{\alpha} j$ we have a bi--fundamental Higgs field
$h_\alpha$ transforming in the $(\overline{\mathbf{d}_i}, \mathbf{d}_j)$ representation of $U(d_i)\times U(d_j)$. 

Now, suppose our quiver has an automorphism $\sigma$, fixing all nodes, such that $\sigma(\cw)=\cw$. The simplest case is the Kronecker quiver
\begin{equation}
\begin{gathered}
\xymatrix{\bullet\ar@<0.5ex>[rrr]^A\ar@<-0.5ex>[rrr]_B &&& \bullet}\qquad \sigma\colon A\leftrightarrow B.
\end{gathered}
\end{equation}
Clearly, the SQM Hamiltonian is invariant under the arrow permutation $\sigma$. In Quantum Mechanics, when our system is invariant under a permutation symmetry, we have to decide whether the permuted subsystems are \emph{distinguishable} or not. If they aren't, we have to specify their statistics: Bose, Fermi, or maybe a fancier one. This amounts to projecting the Hilbert space to the subspace in which permutations act by the proper irreducible representation.

Considering the stable representations of $(Q,\cw)$ in the ordinary sense is equivalent to considering all arrows distinguishable, and keeping all states in the SQM Hilbert space. This clearly sounds odd, since we wish to think of SQM as a physical theory which still respects the Spin \& Statistics theorem. This would require to keep only the states which transform trivially under $\Z_2$.

In the case of the Kronecker quiver this would have no effect.
The bricks in $\cp$, $\cq$ (\textit{i.e.}\! the dyons) are $\Z_2$--invariant (up to gauge equivalence), while the brick in $\ct_{(s:t)}$ map in that in $\ct_{(t:s)}$. Since to get the quantum state we have to \emph{integrate} over the $\mathbb{P}^1$ family, the resulting $W$ is $\sigma$ invariant. Then all states are $\sigma$--invariant, the Bose--statistics projection is just the identity, and we may be safely ignore all subtleties. One could associate to pure $SU(2)$ SYM a $\mathbb{P}^1/\Z_2$--family of $\Z_2$--symmetrized tubes $\widehat{T}_{(s:t)}$; it would be both pedantic and unnatural.

However, for the Markoff quiver \eqref{mark} it makes a difference to consider the $X_i$ arrows to be or not to be distinguishable from the $Y_i$ ones. Projecting on the invariant subspace has precisely the effect of getting rid of the spurious free hypermultiplet, getting the $SU(2)$ $\cn=2^*$ theory by itself.

This can be understood from a different point of view. The Markoff quiver may be seen as arising from the quiver for the case of $N_f=2$ fundamentals by fusion of the two heavy quark nodes
\begin{equation}
\begin{gathered}
\begin{xy} 0;<1pt,0pt>:<0pt,-1pt>:: 
(0,0) *+{\diamondsuit} ="0",
(7,0) *+{\phantom{\diamondsuit}} ="1",
(0,71) *+{\spadesuit} ="2",
(7,71) *+{\phantom{\spadesuit}} ="3",
(41,37) *+{\bullet} ="4",
(63,37) *+{\bullet} ="5",
"0", {\ar"2"},
"1", {\ar"3"},
"4", {\ar"1"},
"5", {\ar"1"},
"3", {\ar"4"},
"3", {\ar"5"},
\end{xy}\end{gathered}
\quad\longrightarrow\quad
 \begin{gathered}
 \xymatrix{\diamondsuit \ar@<0.6ex>[dd]\ar@<-0.6ex>[dd]\\
&& \bullet\ar@<0.4ex>[ull]\ar@<-0.4ex>[ull]\\
 \spadesuit\ar@<0.4ex>[urr]\ar@<-0.4ex>[urr]}
 \end{gathered}
 \end{equation}
 from this perspective it is clear that each arrow trough the node $\bullet$ carries an $SU(2)$ fundamental index $a=1,2$, so that the pair of arrows correspond to $\mathbf{2}\otimes\mathbf{2}$ and projecting on the symmetric (resp.\! antisymmetric) part corresponds   to projecting on the $\mathbf{3}$ (resp.\! $\mathbf{1}$) irreducible $SU(2)$ representation. Then it is natural to identify the states of the $SU(2)$ $\cn=2^*$ theory (without extra spurious fields) with the $\sigma$--invariant representations of $(Q,\cw)$. As we have shown before, doing this one gets the right weak coupling spectrum.
 \medskip
 
 Representations invariants under an automorphism of a quiver have been studied before in the math literature. The corresponding indecomposable objects are called $ii$--representations in \cite{hubery}. 
For instance, looking to the $ii$--representations of the Kronecker quiver we are lead to replace the usual tube $\ct_\lambda$ by the $ii$--tubes $\widehat{\ct}_\lambda$ as we did in \S.\,\ref{cn2*}.

 \subsection{$SU(2)$ coupled to a quark of isospin $m/2$ }\label{uuurujsnssxx}
 
 Coupling a quark of isospin $m/2>1$ to $SU(2)$ SYM leads to a UV non--complete theory. Even if the theory makes no sense non--perturbatively, it has a good quiver consistent with $2d/4d$ \cite{CNV} and its perturbative spectrum makes sense as $g\rightarrow 0$, so the light category $\mathscr{L}$ should still be well defined and enjoy the Ringel property in the spirit of \S.\,\ref{makingsense}. Then the higher--isospin $SU(2)$--tubes still exist and should be `nice'. The full module category should, instead, be quite wild due to the UV incompleteness of the model.
 
The isospin $m/2$ quiver is obtained from the solutions to the Markoff diophantine equation (which is equivalent to the condition that  $\Phi$ has spectral radius $1$) \cite{CV92}\!\cite{CNV}. The convenient form of the quiver is obtained by a mutation at the $\bullet$ node of the one described in \cite{CNV}; one gets
 \begin{equation}\label{nJ}
 Q_m\colon\qquad\begin{gathered}
\begin{xy} 0;<1pt,0pt>:<0pt,-1pt>:: 
(0,0) *+{\diamondsuit} ="0",
(0,121) *+{\spadesuit} ="1",
(72,67) *+{\bullet} ="2",
"0", {\ar|*+{\boxed{m^2-2}}"1"},
"2", {\ar|*+{\boxed{m}}"0"},
"1", {\ar|*+{\boxed{m}}"2"},
\end{xy}
\end{gathered}
\end{equation}
where the boxed integer attached to each arrow denotes its multiplicity ($m\in\mathbb{N}$, and a negative number means arrows in the opposite direction).

\subsubsection{The $\mathscr{B}_m$ algebra}
We consider the weak coupling limit \eqref{ttrupq}. In this regime the algebra $\C Q/(\partial \cw)$ is best understood as the $3$--CY completion of the algebra of the quiver
\begin{equation}\label{rrrmms}
\xymatrix{\spadesuit \ar@<0.8ex>[rrrr]^{\phi_1,\phi_2,\cdots,\phi_m}\ar@<0.4ex>[rrrr]\ar[rrrr]\ar@<-0.8ex>[rrrr]\ar@<-0.4ex>[rrrr] &&&& \bullet\ar@<0.8ex>[rrrr]^{\psi_1,\psi_2,\cdots,\psi_m}\ar@<0.4ex>[rrrr]\ar[rrrr]\ar@<-0.8ex>[rrrr]\ar@<-0.4ex>[rrrr] &&&& \diamondsuit}
\end{equation} 
subjected to $m^2-2$ relations. 

The previous discussion of the $ii$--representations gives a nice interpretation of these relations.
Indeed, each arrow represents an $SU(2)$ fundamental index, and $m$ parallel arrows `propagate' the reducible $SU(2)$ representation
$\overbrace{\mathbf{2}\otimes\mathbf{2}\otimes\cdots\otimes\mathbf{2}}^{m\ \text{factors}}$. To project on the irreducible representation of dimension $(\mathbf{m+1})$ we have to project on the trivial representation of the symmetric group $S_m$.
Consider the $m^2$ paths of length $2$
\begin{equation}
\psi_a\,\phi_b,\qquad a,b=1,2,\dots,m.
\end{equation}
Under $S_m$ they transform according to the representation $(V_1\oplus V_{m-1})\otimes (V_1\oplus V_{m-1})$ where $V_1$ stands for the trivial representation and $V_{m-1}$ for the irreducible $(m-1)$--dimensional one. The relations are
\begin{equation}\label{ttmmaw}
\psi_a\,\phi_b \in V_1
\end{equation}
that is, all projections into irreducible $S_m$ representations vanish, but the trivial one. The number of linearly independent
paths of length $2$ is then equal to the multiplicity of the trivial representation in  $(V_1\oplus V_{m-1})\otimes (V_1\oplus V_{m-1})$ which is $2$.  Correspondingly, we get  $m^2-2$ independent relations, in agreement with the completed quiver of ref.\!\cite{CNV}.

The path algebra of the quiver in eqn.\eqref{rrrmms} subjected to the relations \eqref{ttmmaw} will be called the $\mathscr{B}_m$ algebra. The $\mathscr{B}$ algebra of \S.\,\ref{cn2*} is isomorphic to $\mathscr{B}_2$ trough the identifications
\begin{equation}
\begin{aligned}
 &  X_1= \psi_1-\psi_2, &&\quad X_2=\phi_1+\phi_2,\\
& Y_1=\psi_1+\psi_2, &&\quad Y_2=\phi_1-\phi_2.
\end{aligned}
\end{equation}
For $m>2$, $\mathscr{B}_m$ is not a string algebra; this is parallel to the fact that for $m>2$ the QFT is not UV complete.
However, we have still a good gauge functor given by the two non--zero paths from $\spadesuit$ to $\diamondsuit$
\begin{align}
 &\mathscr{G}\colon \mathsf{mod}\,\mathscr{B}_m\rightarrow \mathsf{mod}\,\mathbf{Kr}\\
 &\mathscr{G}(X)\longmapsto \xymatrix{X_\spadesuit \ar@<0.6ex>[rrr]^{\sum_a \psi_a\phi_a}\ar@<-0.6ex>[rrr]_{(\sum_a \psi_a)(\sum_b\phi_b)} &&& X_\diamondsuit}.
\end{align}
The magnetic charge is again $m(X)=\dim X_\spadesuit-\dim X_\diamondsuit$, and light objects are defined as before.

Again, we expect the Ringel property to hold, that is,
\begin{itemize}
 \item $X$ is a light stable object (at weak coupling \eqref{ttrupq}) then $\mathscr{G}(X)\in \ct$;
\item For all but finitely many light stable objects $X$
\begin{equation}\label{zmmmmv}
 \mathscr{G}(X)= \xymatrix{\C \ar@<0.5ex>[rr]^\lambda\ar@<-0.5ex>[rr]_1 &&\C},\qquad \lambda\in \mathbb{P}^1,
\end{equation}
and, moreover, for a generic $\lambda$ there is precisely one stable light object such that $0\neq \mathscr{G}(X)\in \ct_\lambda$ (up, possibly, to the action of the $S_m$ automorphism group of $\mathscr{B}_m$);
\item the matter objects $X\in \mathsf{matter}$ are (direct sums of) stable light representations with
$\mathscr{G}(X)\in \bigvee_{\lambda\in E}\ct_\lambda$ where $E\subset \mathbb{P}^1$ is the set of exceptional points.
\end{itemize}

We give preliminary evidence that these properties do hold. First of all,
we note that, as in \S.\,\ref{maggg}, if $X$ is stable and light at least \textit{one} of the following two conditions must be fullfilled in each chamber
\begin{gather}
 \bigcap_{a=1}^m \mathrm{coker}\,\phi_a=0\label{seccse}\\
\bigcap_{a=1}^m \mathrm{ker}\,\psi_a=0.
\end{gather}
As before, this implies that $\mathscr{G}(X)|_\cp=\mathscr{G}(X)|_\cq=0$ and hence $\mathscr{G}\in\ct$.
We assume, say, that \eqref{seccse} holds.

Let $X$ be a (finite--dimensional) light indecomposable $\mathscr{B}_m$--module with $X_\spadesuit\neq 0$. Since $\mathscr{G}(X)\in\ct$,
there are infinitely many $\xi\in \C$ such that $\sum_a \psi_a\phi_a+\xi(\sum_a\psi_a)(\sum_b\phi_b)$ is an isomorphism which may be taken to be the identity $\mathbf{1}$. We write $\kappa\equiv (\sum_a\psi_a)(\sum_b\phi_b)$ and we may decompose
\begin{equation}
 \kappa= J(\mu_1,n_1)\oplus J(\mu_2,n_2)\oplus\cdots\oplus J(\mu_r,n_r),
\end{equation}
where $J(\lambda,n)$ is the Jordan block of size $n$ and eigenvalue $\lambda$. 
The relations of $\mathscr{B}_m$ then give
\begin{equation}
\begin{aligned}
 M_{ab}&\equiv \psi_a\phi_b\colon X_\spadesuit\rightarrow X_\diamondsuit\\
&= \frac{1}{m}\delta_{ab} \big(\mathbf{1}-\xi\,\kappa\big)+\frac{1}{m(m-1)}\,\big(E_{ab}-\delta_{ab}\big)\,\Big((1+\xi)\,\kappa-\mathbf{1}\Big),
\end{aligned}
\end{equation}
  where $E$ is the matrix with all entries $1$. If $\dim X_\spadesuit=k$, taking bases, we may see $M_{ab}$ as a $(m\,k)\times (m\,k)$ matrix $M$. One has
\begin{equation}\label{ZZZXXU}
 \det M= (m-1)^{-(m-1)k}\, m^{-k}\, \prod_{\ell=1}^r \mu_\ell^{n_\ell} \Big(1-(\xi+1/m)\,\mu_\ell\Big)^{(m-1)n_\ell}.
\end{equation}
If all $\mu_\ell\neq 0,(\xi+1/m)^{-1}$, the matrix $M$ has maximal rank $m\,k$. Let $z_1, z_2,\dots, z_k$ be a basis of $X_\spadesuit$.
Saying that $M$ has rank $m\,k$ is equivalent to saying that the $m\,k$ vectors
$$\phi_a(z_i)\in X_\bullet,\qquad a=1,2,\dots,m,\ i=1,2,\dots,k,$$
are linearly independent. If, in addition, $X$ is stable light, from eqn.\eqref{seccse} we get that they form a basis of $X_\bullet$.
Let $\{z^{(i)}_\alpha\}$ and $\{z^{(j)}_\beta\}$ the basis vectors of $X_\spadesuit$ corresponding, respectively, to the $i$--th and $j$--th Jordan blocks of $\kappa$. Then for $i\neq j$
$$\mathrm{span}(\{\phi_a(z^{(i)}_\alpha)\})\bigcap \mathrm{span}(\{\phi_a(z^{(j)}_\beta)\})=0 $$
$$\mathrm{span}(\{\psi_b\phi_a(z^{(i)}_\alpha)\})\bigcap \mathrm{span}(\{\psi_b\phi_a(z^{(j)}_\beta)\})=0$$
so $X$ is decomposable unless $\kappa$ consists of a single Jordan block $J(\lambda,n)$. Requiring $\mathrm{End}(X)=\C$ then gives 
$n=1$. Therefore the \textit{generic} stable light module has dimension
\begin{equation}\label{ttwtwtw}
 \delta\equiv(1_\spadesuit, 1_\diamondsuit, m_\bullet),
\end{equation}
and this is the charge vector of the unique light $W$--boson (in our duality frame). As a check, let us compute the magnetic charge linear form
\begin{equation}
 m(X)\equiv -\tfrac{1}{2}\: \delta^t\, B\, \dim X= \dim X_\spadesuit-\dim X_\diamondsuit,
\end{equation}
as expected.

Let us focus on an indecomposable representation $X$ of dimension
\eqref{ttwtwtw} such that $\mathscr{G}(X)\in\ct_\lambda$. From the above argument we see that if $0\neq z\in X_\spadesuit$, the $m$ vectors $w_a=\phi_a(z)$ form a basis of $X_\bullet$.
Then the maps $\psi_b\colon X_\bullet\rightarrow X_\diamondsuit\simeq \C$ are totally specified by
\begin{equation}\label{zzzrrt}
 \psi_b(w_a)= M_{bc}\equiv \frac{1}{m}\delta_{ab}(1-\xi\,\lambda)+ \frac{1}{m(m-1)}(E_{ab}-\delta_{ab})\big((1+\xi)\lambda-1\big),
\end{equation}
 where for $\lambda\neq 1$ we may choose $\xi=0$. In particular the generic  brick of dimension \eqref{ttwtwtw} is $S_m$ invariant.
 
 It is then clear that, for $\lambda$ generic, the light category $\mathscr{L}_\lambda$ is a homogeneous tube associated to the $W$--boson of charge $\delta$. The matter lives on the light subcategories $\mathscr{L}_{\lambda_t}$ where $\lambda=\lambda_t$ is a zero of the determinant in the \textsc{rhs} of 
 \eqref{ZZZXXU}.

\section{General gauge group $G$: basic facts}

Our goal in the present paper is to generalize the structures discussed in the previous sections for the $\cn=2$ theories with gauge group $SU(2)^k$ to general (simply--laced) gauge groups.
Our immediate objective is to discuss the Ringel property and the light category $\mathscr{L}$ 
 in the general case.
 
In this section we collect some elementary considerations, mostly taken from ref.\!\cite{ACCERV2}, which we will need to simplify the analysis or as a check on the results.

\subsection{Magnetic charges}

For a quiver $\cn=2$ theory with gauge group $G=SU(2)^k$, the magnetic charges of a representation $X$ is given by eqn.\eqref{dirrrp}, that is by
$m_a(X)=- \frac{1}{2}\, \delta_a^t\, B\, \dim X,$
where $B$ is the exchange matrix of the quiver $Q$ (assumed to be $2$--acyclic) and $\delta_a$ is the charge vector of the $a$--th $W$--boson. This formula just expresses the facts that the integral skew--symmetric bilinear form $B$ is the pairing given by the Dirac electric/magnetic charge quantization, and that $W$ has electric charge $+2$, so $\delta_a$ is \emph{twice} the unit $a$--th electric charge. 

If we have a general (semi--simple) gauge group $G$ broken down to its Cartan torus, $\exp\mathfrak{h}\simeq U(1)^{r(G)}$,
the unbroken electric charges are identified with the simple coroots $\alpha_j^\vee\in \mathfrak{h}$, $j=1,2,\dots, r(G)$. So normalized, the $U(1)^{r(G)}$ electric charges are integral. The $U(1)^{r(G)}$ electric charges of the $W$--boson associated to the $i$--th positive root $\alpha_i$ is  given by the corresponding weights  
\begin{equation}\label{iiinnnp}
e_j(W_i)=\alpha_i(\alpha_j^\vee)\equiv C_{ij}\qquad \text{(Cartan matrix)}.
\end{equation}
Now
suppose that our quiver with superpotential $(Q,\cw)$ corresponds (in some duality frame) to a $\cn=2$ gauge theory with gauge group $G$. Let $$\delta_i\in \Gamma\equiv K_0(\mathsf{rep}(Q,\cw))$$ be the dimension vector of the $1$--parameter family of representations associated to the $i$--th simple--root $W$--boson. Comparing with eqn.\eqref{iiinnnp}, we see that the  following vector in $\Gamma\otimes \mathbb{Q}$
\begin{equation}
(C^{-1})_{ji}\,\delta_i
\end{equation}
corresponds to unit $e_j$ electric charge and zero $e_k$ charge for $k\neq j$.
Inserting this unit charge vector in the Dirac pairing (given by the exchange matrix $B$ of $Q$) we get the formula for the $j$--th magnetic charge of a representation $X$
\begin{equation}
m_j(X)= - C^{-1}_{ik}\: \big(\delta^t_k\,B\, \dim X\big).
\end{equation} 
\emph{A priori}, this expression gives a group homomorphism
\begin{equation}\label{genmagneticcharge}
m_j\colon \Gamma \rightarrow \mathbb{Q}^{r(G)},
\end{equation}
but physics requires its image to be contained in $\Z^{r(G)}$. In facts, depending on the representation of the matter, we know that the image should be a precise integral lattice, namely the weight lattice of the Langlands dual to the \emph{effective} gauge group
$G_\text{eff}=G/G_0$, where $G_0\subseteq Z(G)$ is the subgroup of the center of $G$ acting trivially on all degrees of freedom \cite{gukovrrr}. 
\smallskip

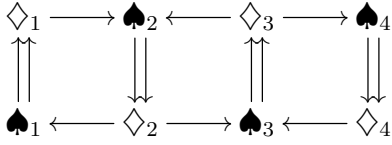
\begin{figure}
\begin{equation*}
\begin{aligned}
\xymatrix{\diamondsuit_1 \ar[r] &\spadesuit_2  \ar@<0.4ex>[d] \ar@<-0.4ex>[d] & \diamondsuit_3 \ar[l]\ar[r] &  \spadesuit_4\ar@<0.4ex>[d] \ar@<-0.4ex>[d]\\
\spadesuit_1 \ar@<0.4ex>[u] \ar@<-0.4ex>[u] &\diamondsuit_2\ar[r]\ar[l] & \spadesuit_3  \ar@<0.4ex>[u] \ar@<-0.4ex>[u] &\diamondsuit_4 \ar[l]}
\end{aligned}
\end{equation*}
\caption{\label{ZZZX135} The quiver $\mathbb{A}_4$.}
\end{figure}

In simple cases the above integrality property of $m_j(\cdot)$ may be used to guess the matrix $B$ and hence the quiver of a QFT. For instance, consider pure SYM with a simple simply--laced gauge group $G=ADE$. Its quiver has rank $2\,r(G)$, and the charge lattice $\Gamma$ should be isomorphic to
$\Gamma_\text{root}\oplus \Gamma_\text{coweight}$. Then $\det B=\pm \#Z(G)$, and $B$ must be such that the forms $m_j(\cdot)$ in \eqref{genmagneticcharge} are integral. The simplest solution to these conditions is
\begin{equation}
B= C\otimes S,\qquad S\equiv \begin{pmatrix} 0 & 1\\ -1 & 0\end{pmatrix}\in SL(2,\Z)
\end{equation}
where $C$ is the Cartan matrix of $G$. This solution reproduces the pure SYM quiver discussed in refs.\!\cite{CNV,ACCERV2}, namely the 
 quiver (and associated superpotential $\cw$) given by the square--tensor product \cite{kellerP}
\begin{equation}\label{sqaurepro}
\overleftrightarrow{G}\,\square\,\widehat{A}(1,1)\equiv \mathbb{G},
\end{equation}
where $\overleftrightarrow{G}$ is the Dynkin graph of the gauge group $G$ with the alternating orientation, and $\widehat{A}(1,1)$ is the Kronecker quiver $\spadesuit\rightrightarrows \diamondsuit$ whose exchange matrix is twice the modular matrix $S$. 
\textit{E.g.}\! for $G=SU(5)$  the quiver has the form in figure \ref{ZZZX135}.

The quiver $\mathbb{G}$ has one Kronecker subquiver \begin{equation}\mathbf{Kr}_i\colon \quad \xymatrix{\spadesuit_i \ar@<0.5ex>[rr]^{A_i}\ar@<-0.5ex>[rr]_{B_i} && \diamondsuit_i}\end{equation} per simple root $\alpha_i$ of $G$
whose source (resp.\! sink) node we denote as $\spadesuit_i$ (resp.\! $\diamondsuit_i$). We say that the Kronecker subquiver $\mathbf{Kr}_i$ is \emph{over} the $i$--node of the Dynkin graph\footnote{\ Recall that, by abuse of notation, $G$ stands --- depending on the context --- for the gauge group, its Lie algebra, or (as in the present case) its Dynkin graph. } $G$.

Then the electric/magnetic charges of the nodes of $\mathbb{G}$ are (with our sign conventions)
\begin{equation}\label{chargeeeer}
\big(e_j(\spadesuit_i),m_k(\spadesuit_i)\big)= \big(C_{ij}, \delta_{ik}\big),\qquad 
\big(e_j(\diamondsuit_i),m_k(\diamondsuit_i)\big)= \big(0, -\delta_{ik}\big).
\end{equation}
In particular, the magnetic charges of a representation $X$ of $\mathbb{G}$ are
\begin{equation}\label{Gmagneticharges}
m_i(X)=\dim X_{\spadesuit_i}-\dim X_{\diamondsuit_i}. 
\end{equation}

The superpotential $\cw$ is dictated by the square--tensor product form of $\mathbb{G}$ \cite{kellerP}. $\cw$ is a sum over the links of the Dynkin graph $G$. Since $G$ is bipartite, we distinguish between even and odd nodes, a link in $G$ connecting nodes of opposite parity. We write the links in the form $\xymatrix{i \ar@{-}[r] & j}$ with the first node, $i$, odd and the second one, $j$, even. With these conventions 
\begin{equation}\label{teseconventions}
 \cw=\sum_{ i \text{---} j \ \in G}\mathrm{Tr}\Big(A_i \,\phi_{ij}\, B_j\, \phi_{ji}- B_i\, \phi_{ij}\, A_j \,\phi_{ji}\Big),
\end{equation}
where $\phi_{ij}$ is the arrow $\xymatrix{\diamondsuit_j \ar[rr]^{\phi_{ij}} && \spadesuit_i}$.

The above quiver was obtained in \cite{CNV} using the geometrical engineering of pure SYM with simply--laced gauge group $G$ as Type IIB on the local Calabi--Yau hypersurface $\mathscr{H}_G\subset \C^4$
\begin{equation}\label{geoinSW}
 \mathscr{H}_G\colon\qquad e^z+e^{-z}+W_G(x_1,x_2,x_3)=0,
\end{equation}
together with the $2d/4d$ correspondence. In eqn.\eqref{geoinSW} $W_G(x_1,x_2,x_3)$ stands for (the universal unfolding of) the surface canonical singularity of type $G=ADE$, see \textit{e.g.}\! \cite{Tack} for explicit expressions.

\subsection{`Nice' and `reasonable' gauge representations}

\subsubsection{The `heavy quark' construction}

As shown in ref.\!\cite{ACCERV2}, the above analysis may be extended to $\cn=2$ SQCD with  (massive) quarks in the representation $R$ of the gauge group $G$.
Assuming that the resulting model is a \textit{quiver} theory in the sense of refs.\!\cite{CV11,ACCERV2}, we wish to guess the associated quiver and $\cw$. In refs.\!\cite{CV11,ACCERV2} one argued as follow. The  theory has a free mass parameter $m$ dual to the quark number (say, for $N_f=1$). Sending $m$ to infinity (along a suitable ray in $\C$), we end up with pure SYM and a decoupled heavy quark which contains states of electric/flavor charges $(\omega_\alpha,\pm 1)$ where $\omega_\alpha$ are the weights of $R$. 
Let $\varrho$ be the highest weight of the representation $R$. By the \emph{quiver} assumption, either the charge vector $(-\varrho,+1)$ or the charge vector $(-\varrho,-1)$ belongs to the the positive cone in $K_0(\mathsf{rep}(Q,\cw))$.  In facts one of the two vectors must be a generator of this cone. Then the corresponding representation $S\in\mathsf{rep}(Q,\cw)$ must have support on a single `massive' node  $\bullet$ of $Q$, while the representations associated to the SYM sector, which remain light in the $m\rightarrow\infty$ limit, must have support in the full subquiver obtained by removing the `massive' node $\bullet$. This last subquiver then belongs to the pure SYM mutation--class, and we may assume it to be in the canonical form $\mathbb{G}$.

The `massive' node  $\bullet$ is connected to the nodes $\spadesuit_j,\diamondsuit_j$ of $\mathbb{G}$ by as many arrows as the Dirac pairing of the electric/magnetic charge vector of the `massive' node $\bullet$, \textit{i.e.}\! $(-\varrho,0)$, with the charge vectors of the $\spadesuit_j,\diamondsuit_j$ nodes, see eqn.\eqref{chargeeeer}.
This gives
\begin{equation}
\forall\:j\in G\colon\qquad  \#\big\{\text{arrows } \xymatrix{\bullet \ar[r]&\spadesuit_j}\big\}= \#\big\{\text{arrows } \xymatrix{\diamondsuit_j \ar[r]&\bullet}\big\}=\alpha_j(\varrho)\geq 0.
\end{equation}
\textit{E.g.}\! $SU(5)$ with a quark in the $\mathbf{10}$ would correspond to the quiver  
\begin{equation}
\begin{aligned}
\xymatrix{\bullet\ar@/^1pc/[rr] &\diamondsuit_1 \ar[r] & \spadesuit_2 \ar@<0.4ex>[d] \ar@<-0.4ex>[d] & \diamondsuit_3 \ar[l]\ar[r] &  \spadesuit_4\ar@<0.4ex>[d] \ar@<-0.4ex>[d]\\
& \spadesuit_1 \ar@<0.4ex>[u] \ar@<-0.4ex>[u] &
\diamondsuit_2\ar@/^2.6pc/[llu]\ar[r]\ar[l] & \spadesuit_3  \ar@<0.4ex>[u] \ar@<-0.4ex>[u] &\diamondsuit_4 \ar[l]}
\end{aligned}
\end{equation}
If we have $N_f$ quarks, we just iterate the operation $N_f$ times, adding one extra node at each iteration.

However this `massive quark' procedure is too naive. The above argument guarantees that, for large $m$, the resulting module category $\mathrm{rep}(Q,\cw)$ contains stable representations corresponding to the various states of the massive quark in the given representation $R$,  but says nothing about the possibility that $\mathrm{rep}(Q,\cw)$ contains additional stable representations which remain light in the zero YM coupling constant limit, and hence should be interpreted as elementary fields (with respect to the given weakly coupled Lagrangian description).
If the extra states are just decoupled free fields, as was the case in $SU(2)$ $\cn=2^*$, this is not really a problem; when these states are charged under the gauge group, the quiver describes (if any) a totally different $\cn=2^*$ theory.

In facts in ref.\!\cite{ACCERV2} it is explained that one has to restrict the above quiver construction to quarks in a fundamental representation of $G$, that is, a representations whose Dynkin label has the form $[0,\cdots, 0,1,0,\cdots, 0]$. In this case the massive node $\bullet$ is attached to just one node pair, $\spadesuit_j,\diamondsuit_j$, by single arrows.

The restriction to fundamental representations is not enough to guarantee that no extra light BPS states are present. 
The point is that the above argument relies on the assumption that $G$--SYM coupled to a quark in the representation $R$ {is} a \textit{quiver} theory, and this may or may not be true for a given pair $(G,R)$.
 There are, however, models of this class which we  know \emph{a priori} to be quiver theories. Indeed, if a $\cn=2$ theory may be geometrically engineered by Type IIB superstring on a hypersurface, it has a quiver which --- in principle --- may be determined by geometric means thanks to the $2d/4d$ correspondence \cite{CNV}.
Therefore the `heavy quark' procedure must give the correct quiver at least for the pairs $(G,R)$ having a geometrical engineering. Such pairs $(G,R)$ are called `nice' representations in ref.\!\cite{Tack}; we list them in the first column of table \ref{niceres}. They are precisely the pairs with $b(R)\leq h(G)$ where $b(R)$ is the contribution of the representation $R$ to the $\beta$--function  coefficient \cite{Tack}. `Nice' theories are, in particular, asymptotically--free. The second column of table \ref{niceres} lists the `reasonable' pairs $(G,R)$: they correspond to models which are UV--complete but have no geometric engineering \cite{Tack}. Again we have divided asymptotically--free and UV--superconformal models.

We shall see that the light states of the `heavy quark' quiver of the pair $(G,R)$ will consist of vector supermultiplets, making one copy of the adjoint representations of $G$, plus hypermultiplets making one copy (or $N_f$ copies, in general) of the representation $R$ \textit{if and only if} the pair $(G,R)$ appears in the first column of table \ref{niceres}. If $(G,R)$ is in the list of AF representation of the second column, the light sector would still consists of vectors, forming one copy of the adjoint of $G$, together with finitely many hypermultiplets. We stress that this is a \emph{if and only if} statement
$$\begin{bmatrix}\text{the light spectrum of }\mathsf{rep}(Q_\text{heavy quark},\cw)\\
\text{consists of vectors in the adjoint of $G$}\\
 \text{plus finitely many hypers} \end{bmatrix}\quad \Longleftrightarrow \quad \begin{bmatrix}\text{the pair }(G,R)\ \text{is }\\
 \text{Asymptotically Free}\end{bmatrix}$$
The hypermultiplets of baryon number $1$ give the expected representation $R$, but for non--`nice' AF pairs $(G,R)$, the category $\mathsf{rep}(Q_\text{heavy quark},\cw)$ contains in addition light stable hypers of baryon number
$\geq 2$ in smaller representations of $G$.

\begin{table}
 \begin{tabular}{|c|c|c|c|}
\multicolumn{4}{c}{`nice' representations $b\leq h$}\\\hline 
  $G$ & $R$ & $b$ & $h$\\\hline
$SU(n)$ & $\mathbf{n}$ & $1$ & $n$\\\hline
$SU(n)$ & $\mathbf{n(n-1)/2}$ & $n-2$ & $n$\\\hline
$SU(6)$ & $\mathbf{20}$ & $6$ & $6$\\\hline
$SO(2n)$ & $\mathbf{2n}$ & $2$ & $2n-2$\\\hline
$SO(8)$ & spinor & $2$ & $6$\\\hline
 $SO(10)$ & spinor & $4$ & $8$\\\hline
$SO(12)$ & spinor & $8$ & $10$\\\hline
$E_6$ & $\mathbf{27}$ & $6$ & $12$\\\hline
$E_7$ & $\mathbf{56}$ & $12$ & $18$\\\hline
 \end{tabular} \qquad
\begin{tabular}{|c|c|c|c|}
\multicolumn{4}{c}{AF representations with $h<b<2h$}\\\hline 
  $G$ & $R$ & $b$ & $h$\\\hline
$SU(n)$, $n>2$ & $\mathbf{n(n+1)/2}$ & $n+2$ & $n$\\\hline
$SU(7)$ & $\mathbf{35}$ & $10$ & $7$\\\hline
$SU(8)$ & $\mathbf{56}$ & $15$ & $8$\\\hline
$SO(14)$ & spinor & $16$ & $12$\\\hline
\multicolumn{4}{c}{}\\
\multicolumn{4}{c}{Representations with $b=2h$}\\\hline
$G=ADE$ & adjoint & &\\\hline  
 \end{tabular}
\caption{\label{niceres}`Nice' and `reasonable' representations for $G=ADE$ gauge groups. `spinor' stands for both $c$ and $s$ spinorial representations of $SO(2n)$.}
\end{table}

\subsubsection{Representation--theoretical interpretation of the `heavy quark' quiver}

To simplify some calculations, it is convenient to give a representation--theoretical interpretation of the `heavy quark' $(G,R)$ quiver (with $N_f$ copies of the quark in a fundamental representation of $G$).
\vglue 12pt

\textbf{Claim.} \textit{The quiver with superpotential $(Q,\cw)$ describing $G=ADE$ SYM coupled to $k$ quarks in the fundamental representations $i_1$, $i_2$, $\cdots$ $i_k$ is given by the 3--CY completion of the algebra
\begin{equation}\label{rrr66u}
\mathscr{A}= \mathscr{A}_0[X_{i_1}][X_{i_2}]\cdots [X_{i_k}]
\end{equation} 
where $\mathscr{A}_0=\C \mathbb{G}/(\partial \cw)$ is the Jacobian algebra of the pure SYM quiver \eqref{sqaurepro},  and $\mathscr{C}[X]$ denotes the one point extension of the algebra $\mathscr{C}$ at the module $X$ (see appendix \ref{oneppoiiint} for definitions and details). In \eqref{rrr66u} $X_{i}$ stands for a module obtained by extending by zero a module $\xymatrix{\C \ar@<0.5ex>[r]^\lambda\ar@<-0.5ex>[r]_1 & \C}$ of the $i$--th Kronecker subquiver for a \emph{particular} $\lambda$.}   

Equivalently, we may describe $(Q,\cw)$ in terms of $3$--CY completions of one--point \emph{co}extensions.
See section \ref{Proofconjectr} for further details.

\section{Pure SYM and  the preprojective algebras $\cp(G)$}\label{pureSYMN}

For a quiver $\cn=2$ gauge theory with a simple simply--laced gauge group $G$ we expect that the light subcategory $\mathscr{L}$
has the form  $\bigvee_{\lambda\in \mathbb{P}^1}\mathscr{L}_\lambda$, with the generic subcategory $\mathscr{L}_\lambda$ equivalent to the corresponding subcategory for the pure SYM case (with the same gauge group). Physically this reflects the universality of the SYM sector. From physics we know that the light states of SYM should consist only of vector multiplets making precisely \emph{one copy} of the adjoint representation of $G$.

The purpose of the present section is to study in depth the properties of $\mathsf{rep}(Q,\cw)$ in the pure SYM case. 
We assume the gauge group $G$ to be simply--laced, and use the square--tensor product form of the quiver $\mathbb{G}$, eqn.\eqref{sqaurepro}. The magnetic charges $m_i(\cdot)$ are defined in eqn.\eqref{Gmagneticharges}, and the weak coupling limit is
\begin{equation}
\arg Z(\diamondsuit_i)=O(g^2),\qquad \arg Z(\spadesuit_i)=\pi-O(g^2),\qquad g\rightarrow 0.
\end{equation}
The light BPS particles correspond to the representations which are stable in this regime and have $m_i(X)=0$ for all $i\in G$.

\subsection{The controlled Abelian subcategory $\mathscr{L}^\mathrm{YM}$}

We consider the Abelian subcategory $\mathscr{L}^\mathrm{YM}$ of $\mathsf{rep}(\mathbb{G})$ controlled by the $m_i(\cdot)$'s, that is, the full subcategory of all objects $X\in \mathsf{rep}(\mathbb{G})$ such that $$\forall\, i\in G\colon\qquad \begin{array}{l}\text{\textit{i)} }\  m_i(X)=0\\ \text{\textit{ii)} } m_i(Y)\leq 0\ \text{for all subobjects }Y\  \text{of }X.\end{array}$$ 
$\mathscr{L}^\mathrm{YM}$ is an Abelian (full, exact) subcategory  closed under direct sums/summands and extensions.
Locally, around the Kronecker subquiver $\mathbf{Kr}_i$ over the node  $i\in G$, the pure SYM quiver $\mathbb{G}$ looks as in the figure
\begin{equation}
\begin{aligned}
\xymatrix{ \cdot\cdots && \diamondsuit_i\ar[ll]_{\phi_{ji}} \ar[rr]\ar@<-0.8ex>[rrd]&&\cdots\cdots \\
& &&&\ \ \ \cdots\cdots\\
&&&&\ \ \ \cdots\cdots \ar[lld]\\
\cdots\cdots \ar[rr]_{\phi_{ij}} && \spadesuit_i\ar@<0.6ex>[uuu]^{A_i}\ar@<-0.6ex>[uuu]_{B_i} && \cdots\cdots\ar[ll] }
\end{aligned}
\end{equation}
with two or three arrows $\phi_{ji}$ starting (resp.\! $\phi_{ij}$ ending) at $\diamondsuit_i$ (resp.\! $\spadesuit_i$).
Note that under the correspondence\footnote{\ As a matter of notation, given a node $a\in Q$ we write $s_a$ for the dimension vector of the simple representation $S(a)$ with support at the $a$--th node. } $\alpha_i \leftrightarrow \delta_i\equiv s_{\spadesuit_i}+s_{\diamondsuit_i}$ we have the identification
\begin{equation}
K_0(\mathscr{L}^\mathrm{YM})\simeq \Gamma_\mathrm{root}(G),
\end{equation}
with the \emph{root lattice} of $G$.

It is physically intuitive that $\mathscr{L}^\mathrm{YM}$ contains all representations which are stable and light at weak coupling. For a formal argument see appendix \ref{strongerstatement} which also describes the \textit{heavy} stable states (dyons).
\medskip

The crucial observation is the following analogue of Ringel's theorem (cfr.\! page \pageref{ringeltheorem})\vglue 9pt

\textbf{Proposition.}\label{genX1} \textit{Let $X$ be an indecomposable object of $\mathscr{L}^\mathrm{YM}$. The restriction of $X$ to each $\mathbf{Kr}_i$ belongs to the homogeneous tube $\mathcal{T}_\lambda$ for a \emph{fixed} $\lambda\in \mathbb{P}^1$ (that is, $\lambda$ is the \emph{same} for all $\mathbf{Kr}_i$'s in $\mathrm{supp}\,X$). Hence
\begin{equation}
\mathscr{L}^\mathrm{YM}=\bigvee_{\lambda\in\mathbb{P}^1} \mathscr{L}^\mathrm{YM}_\lambda,
\end{equation}
with
\begin{equation}
 X\in \mathscr{L}^\mathrm{YM}_\lambda\quad\Longleftrightarrow\quad X\big|_{\mathbf{Kr}_i}\in \ct_\lambda,\quad \forall\: i\in G.
\end{equation}
Moreover,
\begin{equation}
 X\in\mathscr{L}^\mathrm{YM}\ \text{is a brick}\quad\Longrightarrow\quad X\big|_{\mathbf{Kr}_i}= \xymatrix{\C^\ell \ar@<0.5ex>[rr]^{\lambda\cdot \text{Id}}\ar@<-0.5ex>[rr]_{\text{Id}} && \C^\ell}.
\end{equation}
In particular, the category $\mathscr{L}^\mathrm{YM}_\lambda$ is independent of $\lambda$ up to equivalence.
}
\smallskip

\textsc{Definition.} The category $\mathscr{L}^\mathrm{YM}_\lambda$  is called the \textit{homogeneous $G$--tube}. For $G=SU(2)$ we recover the usual homogeneous tube.
\medskip

\textsc{Proof.}
The proof is based on the following \textbf{Lemma} whose proof is presented in appendix \ref{applemma}.
\medskip

\textbf{Lemma.} \textit{$X\in \mathscr{L}^\mathrm{YM}$. Then $X\big|_{\mathbf{Kr}_i}\in\ct$ for all $i
\in G$.}
\medskip

Let $i$, $j$ be two nodes of $G$ with $C_{ij}=-1$. 
By the Lemma, we may write the restriction of $X$ to the subquiver over the $A_2$ subgraph of $G$,  $\xymatrix{i \ar@{-}[r] & j,}$ in the form
\begin{equation}
\begin{aligned}
\xymatrix{\C^k \ar[rr]^\phi && \C^\ell \ar@<0.5ex>[dd]^{B^\prime}\ar@<-0.5ex>[dd]_1\\
\\
\C^k\ar@<0.5ex>[uu]^1\ar@<-0.5ex>[uu]_B && \C^\ell \ar[ll]_{\tilde\phi}}
\end{aligned}
\end{equation} 
while the relations $\partial_\phi\cw=\partial_{\tilde\phi}\cw=0$ give
\begin{gather}\label{com1}
\phi\,B=B^\prime\,\phi\\
B\,\tilde\phi=\tilde\phi\, B^\prime.\label{com2}
\end{gather} 
Let $v\in X_{\diamondsuit_i}\equiv \C^k$ be a non--zero vector such that $(B-\lambda)^s\,v=0$ for some $\lambda\in \mathbb{P}^1$ and $s\in \mathbb{N}$. 
From eqn.\eqref{com1} one has
\begin{equation}\label{yyeyeyew}
0= \phi\,(B-\lambda)^s\,v= (B^\prime-\lambda)^s\,\phi\,v
\end{equation}
so, if $\phi\,v$ is not zero, it belongs to a Jordan block of $B^\prime$ associated to the same eigenvalue $\lambda$. The same holds for $\tilde\phi$. Hence the pair $\phi,\tilde\phi$ map Jordan blocks of $B$ to Jordan blocks of $B^\prime$ with the same eigenvalue and \textit{viceversa}. The representation $X$ is then the direct sum of representations with a fixed value of $\lambda$; since $X$ is indecomposable, there is only one direct summand, and there exists a $\lambda\in\mathbb{P}^1$ so that $X\in\mathscr{L}^\mathrm{YM}_\lambda$, where $\mathscr{L}^\mathrm{YM}_\lambda$ is the exact subcategory of $\mathscr{L}^\mathrm{YM}$ generated by the indecomposable $X$ such that $X|_{\mathbf{Kr}_i}\in\ct_\lambda$.

Having set $X_{A_i}= \boldsymbol{1}_{\dim X_{\spadesuit_i}}$ as a choice of basis, from eqns.\eqref{com1}\eqref{com2} we see that the map
\begin{equation}
(X_{\spadesuit_i}, X_{\diamondsuit_i}) \longmapsto (B_i\, X_{\spadesuit_i}, B_i\, X_{\diamondsuit_i})
\end{equation}
is an element of $\mathrm{End}\,X$. Hence if $X$ is a brick
\begin{align}\label{ccc-1}
X_{A_i}&= \boldsymbol{1}_{\dim X_{\spadesuit_i}}\\
X_{B_i}&= \lambda\,\boldsymbol{1}_{\dim X_{\spadesuit_i}},\qquad \lambda\in\mathbb{P}^1.
\label{ccc0}\end{align}
\hfill $\square$
 \smallskip

 \textbf{Corollary.} \textit{In pure SYM (at weak coupling) there are no \underline{light} BPS hypermultiplets.}
\smallskip

Indeed, the above arguments shows that all stable objects in $\mathscr{L}^\mathrm{YM}$ come in families of dimension at least one, since a brick $X$ will have Kronecker arrows as in eqns.\eqref{ccc-1}\eqref{ccc0}, and we are free to change $\lambda\in \mathbb{P}^1$ as we please since the relations $\partial\cw=0$ are homogenous in $\lambda$. 

In the next subsection we shall show that light higher spin states are also ruled out, and the only stable states are vector multiplets, as physically expected.
%\medskip

\subsection{The functor $\mathscr{E}$ and the preprojective algebras}

Again, let $G$ be a simply--laced Dynkin graph; it is a bi--partite graph, and we shall distinguish its even and odd nodes. Out of $G$ we construct a quiver by replacing each unoriented link by a pair of arrows going in opposite directions, $\leftrightarrows$, and adding a loop $\circlearrowleft$ at each node. 
Calling $H_{ij}$ (the bi--fundamental Higgs field of) the arrow $i\rightarrow j$ and $\Phi_i$ (the adjoint Higgs of) the loop at the $i$--th node, we associate to the resulting quiver the superpotential 
\begin{equation}\label{tty}
\widetilde{\cw}= \sum_{ij} \mathrm{Tr}\big[H_{ij}\,\Phi_j\, H_{ji}].
\end{equation}
The path algebra of this quiver subject to the relations $\partial\widetilde{\cw}=0$ is known as \emph{the (completed) preprojective algebra of the Dynkin quiver $G$}. We write $\Pi$ for the category of the representations (modules) of the preprojective algebra of the Dynkin graph $G$ at hand.
%\smallskip

\begin{figure}
 \begin{equation*}
\begin{gathered}
\xymatrix{X_{1} \ar@/^1.5pc/[rr]^{X_{\alpha_1}} && X_{2}
\ar@/^1.5pc/[ll]^{X_{\alpha_1^*}}\ar@/^1.5pc/[rr]^{X_{\alpha_2}}
&& X_{3}
\ar@/^1.5pc/[ll]^{X_{\alpha_2^*}}\ar@/^1.5pc/[rr]^{X_{\alpha_3}}
&& X_{4} 
\ar@/^1.5pc/[ll]^{X_{\alpha_3^*}}\ar@/^1.5pc/[rr]^{X_{\alpha_4}}
&& \ar@/^1.5pc/[ll]^{X_{\alpha^*_4}}
}\end{gathered}\ \cdots\ 
\begin{gathered}
\xymatrix{\ar@/^1.5pc/[rr]^{X_{\alpha_{n-1}}} && X_n \ar@/^1.5pc/[ll]^{X_{\alpha_{n-1}^*}}
}
\end{gathered}
\end{equation*}
\caption{\label{luian} Example: a representation $X$ of $\cp(A_n)$.}
\end{figure}
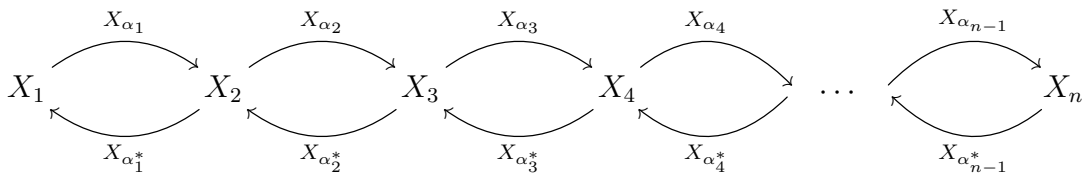

There is an obvious functor
\begin{equation}
\mathscr{E}\colon \mathscr{L}^\mathrm{YM}\rightarrow \Pi,
\end{equation}
defined by
\begin{align}
\mathscr{E}(X)_i&= X_{\spadesuit_i}\\
\mathscr{E}(X)_{H_{ij}}&= X_{\phi_{ij}}\\
\mathscr{E}(X)_{\Phi_i}&= \epsilon_i\, X_{A_i}.
\end{align}
where $\epsilon_i=\pm 1$ for even/odd nodes.

It is easy to check that $\mathscr{E}(X)$ satisfies the constraints coming from the superpotential \eqref{tty} since 
\begin{gather}0=\partial_{B_i}\cw=\epsilon_i\, \sum_j \phi_{ij}\, A_j\,\phi_{ji}\equiv -\sum_j H_{ij}\,H_{ji}=-\partial_{\Phi_i}\widetilde{\cw}\\
0=\partial_{\phi_{ij}}\cw\ \Rightarrow \ 0=\Phi_j\,H_{ji}+H_{ji}\Phi_i=\partial_{H_{ij}}\widetilde{\cw}.
\end{gather}

Usually in the math literature one defines the preprojective algebra $\cp(G)$ associated to a graph (here the Dynkin graph $G$) `forgetting' the arrows corresponding to the loops $\circlearrowleft$ (but keeping the relations $\partial_{\Phi_i}\cw=0$ they imply), see \textit{e.g.}\! Gelfand and Ponomarev \cite{pre1}.  We write  $\mathsf{mod}\,\cp(G)$ for the category of modules with the $\circlearrowleft$ maps so forgotten. Then it is clear that\smallskip

\textbf{Proposition.}\label{genX2} \textit{For fixed $\lambda\in\mathbb{P}^1$, let $\cl(\lambda)$ be the object class in $\mathscr{L}^\mathrm{YM}$ generated by the indecomposable satisfying  eqns.\eqref{ccc-1}\eqref{ccc0}. There is an equivalence}
\begin{equation}
\mathscr{E}\colon \cl(\lambda)\rightarrow \mathsf{mod}\,\cp(G).
\end{equation}
\textit{In particular, the functor $\mathscr{E}$ sets a one--to--one correspondence between the bricks of $\mathscr{L}^\mathrm{YM}_\lambda$ and the bricks of $\cp(G)$.}
\vglue 9pt

Thus, in order to find the BPS particles which are stable and of bounded mass in the limit $g\rightarrow 0$ we have to look for the brick representations of the preprojective algebra $\cp(G)$. It is convenient to transport the stability condition of representations from $\mathscr{L}_\lambda^\mathrm{YM}$ to $\mathsf{mod}\,\cp(G)$.
Thus, to a representation $X$ of $\cp(G)$ we associate the central charge
\begin{equation}
 Z(X)=\sum_i \dim X_i\, \Big(Z(\spadesuit_i)+Z(\diamondsuit_i)\Big).
\end{equation}
An object $X\in \mathsf{mod}\,\cp(G)$ will be said to be \emph{stable} iff for all proper non--zero subobject $Y$, $\arg Z(Y)<\arg Z(X)$. SYM light BPS states then correspond to stable objects in $\mathsf{mod}\,\cp(G)$. 
\smallskip

The above statement gives a complete characterization of the homogeneous $G$--tubes for $G$ simply--laced.

\subsubsection{The double $\cp(Q)$ of a quiver $Q$} The previous construction is a special instance of a general one called the \emph{double of a quiver}. One starts with an acyclic quiver $Q$; for  each arrow in the original quiver $Q$ one adds a new arrow $\alpha^*$ going in the opposite direction  
\begin{equation}
\xymatrix{i \ar[r]^\alpha & j}\quad\longrightarrow\quad\xymatrix{i \ar@<0.4ex>[r]^\alpha & j \ar@<0.4ex>[l]^{\alpha^*}} ,
\end{equation}
and imposes the relations corresponding to the adjoint Higgs, namely \begin{equation}\sum_\text{arrows}(\alpha\alpha^*-\alpha^*\alpha)=0.\end{equation}
 Physically, the passage from $Q$ to its double $\cp(Q)$ corresponds (in the language of quiver gauge theories) to replacing $\cn=1$ supersymmetry with $\cn=2$ \textsc{susy} while keeping the same gauge group and the same matter representation content. Note that for a tree quiver, $\cp(Q)$ is independent of the orientation of $Q$, and hence we may speak of the double of a Dynkin graph without ambiguity.  
\smallskip

One has the following deep result

\textbf{Theorem.} (Gelfand and Ponomariov \cite{pre1}, Ringel \cite{pre2}) \textit{The path algebra of $\cp(Q)$ (with the above relations) is finite dimensional iff $Q$ is an $ADE$ quiver.}

This result explains why in the family of the quivers of the form
$\overleftrightarrow{T}\,\square\,\widehat{A}(1,1)$, $T$ a three graph, the class of those with $T$ a Dynkin diagram is very special. The good ones are precisely the quivers associated with pure SYM with gauge group $G$, while there is no consistent
QFT associated to a general tree graph $T$.

From the math literature we know \cite{geissX} that the preprojective algebra of a Dynkin graph has finite--representation type iff $G=A_n$ with $n\leq 4$, it is tame for $A_5$ and $D_4$, and wild otherwise. Notice that $\cp(G)$ is representation--finite (resp.\! tame) if and only if the square--tensor--product $G\,\square\,A_2$ is representation--finite (resp.\! tame tubular).

However, this last result does not mean that for the other gauge groups $G$ we have infinitely many light BPS states. Not all indecomposable may correspond to stable particles, only the bricks. Bricks are in finite number, and have the right quantum numbers as the next subsection will show.

\subsection{The main claim}

\subsubsection{Functorial projection and sections}

Let $G_\Omega$ be the quiver obtained by giving the orientation $\Omega$ to the Dynkin graph $G$; there are $2^{\# G_1}$ distinct orientations, where $\#G_1$ is the number of links in $G$. 
Clearly, each $G_\Omega$ is a subquiver of $\cp(G)$; it is obtained from $\cp(G)$ by  eliminating, for each pair of arrows $\xymatrix{i \ar@<0.4ex>[r]^\alpha & j \ar@<0.4ex>[l]^{\alpha^*}}$, the unique arrow whose direction does not agree with the orientation of the arrow between nodes $i$, $j$ in the quiver $G_\Omega$. 

Then, for each choice of $\Omega$ we have two functors
\begin{align}
&\mathscr{P}_\Omega\colon \mathsf{mod}\,\cp(G) \rightarrow \mathsf{rep}(G_\Omega)\\ 
&\mathscr{S}_\Omega\colon \mathsf{rep}(G_\Omega)\rightarrow \mathsf{mod}\,\cp(G),
\end{align}
with $\mathscr{P}_\Omega\mathscr{S}_\Omega=\mathrm{Id}_{\mathsf{rep}(G_\Omega)}$, which we call, respectively, the `projection' and the `zero section'. 
The projection functor $\mathscr{P}_\Omega$ is obtained forgetting the linear maps associated to the arrows in $\cp(G)_1\setminus (G_\Omega)_1$. The zero--section functor is defined by seeing a representation of the subquiver $G_\Omega$ as a representation of the 
full quiver $\cp(G)$ with the extra arrows set  to zero (automatically satisfying the relations in $\cp(G)$).

\subsubsection{The claim}
Physics requires the following statement to be true
\vglue 6pt

\textbf{Main claim} (SYM case). \textit{In any chamber at weak coupling:
\begin{enumerate}
 \item the BPS states of bounded mass in the limit $g\rightarrow 0$ are vector multiplets with the quantum numbers of the positive roots of the gauge group $G$;
\item in each weakly coupled chamber there is precisely \emph{one} stable vector multiplet per positive root.
\end{enumerate}}
\smallskip

Of course, which representation $X$ of $\cp(G)$, with $\dim X$ equal to a given root $\alpha$ of $G$, is stable will depend on the specific chamber (\textit{i.e.}\! on the specific stability function $Z(\cdot)$). Changing $Z(\cdot)$ will change the set of stable representations, but the light particle spectrum
\begin{equation} \mathsf{Sp}_{Z(\cdot)}\equiv \Big\{\big(\dim X, \dim \mathrm{Ext}^1(X,X)\big)\:\Big|\: X\in\ \mathsf{mod}\,\cp(G)\ \text{is stable w.r.t. }Z(\cdot)\Big\}\end{equation}
 is independent of $Z(\cdot)$, as long as it corresponds to a small YM coupling.   
In facts, there is evidence for a stronger statement:
For each weakly--coupled chamber, there is an orientation $\Omega$ (depending on the particular chamber!) such that the \emph{stable light} representations (\textit{i.e.}\! BPS states) are given by the images trough $\mathscr{S}_\Omega$ of the bricks in $\mathsf{rep}(G_\Omega)$. Conversely, given a representation $\mathscr{S}_\Omega(Y)$, with $Y$ a brick of $G_\Omega$, there exists a weakly coupled BPS chamber in which it is stable.

\subsubsection{Chamber independence of the light particle spectrum $\mathsf{Sp}_{Z(\cdot)}$}\label{chambindependence}

Before going to the proofs, we note that it suffices to extablish the result in a particular chamber thanks to the Kontsevich--Solbelman wall crossing formula \cite{KS}\!\cite{DG,GMN09,CV09,DGS,GMN10,CNV,ADJM1,ADJM2,MPS,GMN11}.

We want to show that, in pure SYM, the spectrum of BPS states which are light at weak coupling, $g\rightarrow 0$, is independent of the BPS chamber (as long as it is a weak--coupling chamber!!).

Recall that the central charge $Z(\cdot)$ in a weak coupling chamber has the form
\begin{equation}\label{mmmdjjx}
Z(X)= \sum_{i\in Q_0} Z_i \, \dim X_i- \frac{1}{g^2}\, \sum_{a=1}^{r(G)}\, C_a\,m_a(X),\qquad g\rightarrow 0,
\end{equation}
where the $O(1)$ complex numbers $Z_i$ belong to the upper half--plane,  $m_a(X)\in \Z$ are the magnetic weights, and the $C_a$ are real positive.

What we have to show is that the spectrum of light BPS states, corresponding to the set of stable representations $X$ with $m_a(X)=0$ for all $a$, is independent of the $O(1)$ complex numbers $Z_i$ (with $\mathrm{Im}\, Z_i>0$) in \eqref{mmmdjjx}. We stress that this property is peculiar of SYM, or more generally, of theories which have a Lagrangian description (it is, in facts, a characterization of the class of $\cn=2$ Lagrangian theories).

The argument is straightforward\footnote{\ We use the results of appendix \ref{strongerstatement}.}. First we note that the stability conditions for a representation $X\in \mathscr{L}^\mathrm{YM}$ are insensitive to the parameters $C_a> 0$. Hence, as long as we are interested only in the stability of BPS states described by $\mathscr{L}^\mathrm{YM}$, we may fix them once and for all to any convenient value; we choose the  real numbers $C_a$ to be linear independent over $\mathbb{Q}$, \textit{i.e.}\! suitably generic. Then $\sum_aC_a m_a(X)=0$ implies $m_a(X)=0$ for all $a$'s.
No matter how we change the $Z_i$'s, the subset of states with $\sum_a C_a m_a(X)>0$ (resp.\! $<0$) will have central charges $Z(X)$ very near the negative (resp.\! positive) infinite real axis, and, by the Konsevitch--Soilbelman formula, any rearrangement of their relative phase order will produce states with phases very near the negative (resp.\! positive) real axis, which have again $\sum_a C_a m_a(X)>0$ (resp.\! $<0$), and hence are infinitely heavy in the $g\rightarrow 0$ limit. No heavy state gets aligned with a light one in the process. On the other hand, rearrangement of the relative phases of the light  BPS states, will not produce any new state, since these are mutually local, hence the associated quantum torus variables \emph{commute}, and the rearrangement of the order of the quantum dilogarithms in the KS wall--crossing formula is trivial.

The absence of wall-crossing in the light sector of the (weakly coupled) SYM theory may also be understood as a consequence of gauge--invariance. The central charges $Z(\delta_i)$ associated to the simple roots of $G$ may be identified, asymptotically as $g\rightarrow 0$, with the vacuum expectation valued of the adjoint Higgs field $\langle\Phi\rangle$ which is an element of $\mathrm{Lie}(G)/G$, that is, an element of the Lie algebra $\mathrm{Lie}(G)$ defined up to conjugacy. We see $\langle\Phi\rangle$ as an element of the Cartan algebra $\mathfrak{h}$, with the understanding that $\langle \Phi\rangle$ and $w(\langle \Phi\rangle)$ represent the same physical configuration for all $w\in \mathrm{Weyl}(G)$. Then, as $g\rightarrow 0$,
\begin{equation}
Z(\delta_i)\approx \alpha_i(\langle \Phi\rangle),\quad \text{where }\alpha_i\colon \mathfrak{h}\rightarrow \C,\ \text{is a simple root of }G.
\end{equation}
Then we see that two (weakly--coupled) stability functions $Z(\delta_i)$ and $Z(\delta_i)^\prime$ which are related by an element of $\mathrm{Weyl}(G)$ represent the same physical configuration and hence have the same BPS spectrum.

\subsubsection{Proof of part 1) of main claim}\label{zzzzzz1}

 We have to show that $X$ is stable in $\cp(G)$ for \emph{any} weak--coupling choice of $Z(\cdot)$ implies that $\dim X$ is a positive root of $G$ and that $X$ is rigid (that is, it does not belong to a continuous family of non--isomorphic representations of positive dimension).

We recall that by a theorem of King \cite{king} $X$ is stable for some choice of $Z(\cdot)$  $\Rightarrow $ $X$ is a brick, \textit{i.e.}\! $\mathrm{End}_\C(X)=\C$.  We need the following  
 \smallskip
 
\textbf{Lemma} (Crawley--Boevey \cite{CBlemma}, Geiss--Leclerc--Schr\"oer \cite{GLS}) \textit{Let $Q$ be a quiver, and let $X,Y\in \mathsf{mod}\,\cp(Q)$. Then
\begin{equation}\label{mainlemma}
 (\dim X,\dim Y)_C= \dim\mathrm{Hom}(X,Y)+\dim\mathrm{Hom}(Y,X)-\dim \mathrm{Ext}^1(Y,X)
\end{equation}
where the integral quadratic form $(v,w)_C= v^tCw$, with $C$ the Cartan matrix of $Q$ in the sense of Kac \cite{kac1,kac2}.}
\smallskip

In our case $Q$ is a Dynkin quiver of type $G=ADE$, so $C$ is the usual Cartan matrix of $G$.
Let $X$ be a non--zero brick. Using eqn.\eqref{mainlemma} with $X=Y$, we have
\begin{equation}
 0 <(\dim X, \dim X)_C\leq  2,
\end{equation}
where we used that the quadratic form $(\cdot,\cdot)_C$ is positive definite and $\dim \mathrm{Ext}^1(X,X)\geq 0$. The minimal non--zero value of the integral form $(\cdot, \cdot)_C$ is $2$, and hence we must have $\dim \mathrm{Ext}^1(X,X)= 0$ and
\begin{equation}
 (\dim X, \dim X)_C=2,
\end{equation}
so $\dim X$ is a positive root of $G$. On the other hand, we have the following result of Lusztig (\!\cite{LU} section 12), valid for all representations of $\cp(Q)$,
\begin{equation}
 \dim \mathrm{Ext}^1(X,X)=2\, \mathrm{codim}\,\co_X,
\end{equation}
where $\co_X$ is the orbit of $X$ in the (affine) representation space of $\cp(Q)$. In physical language, $\mathrm{codim}\,\co_X$ is the number of Higgs fields, living on the arrows of the quiver $\cp(Q)$, which are not `eaten up' by the gauge transformations or killed by the constraints $\partial\cw=0$, that is, they are the $1d$ gauge--invariant physical degrees of freedom. $\mathrm{Ext}^1(X,X)=0$ then implies no more physical global coordinates to quantize, that is, the corresponding particle is a vector (its spin coming from the quantization of the $\lambda$ labeling the $\mathbb{P}^1$--family of representations of the original quiver $\mathbb{G}$ associated to a given brick of $\cp(G)$). 

We conclude that all stable light states of $\mathbb{G}$ are vector supermultiplets with charge vector a positive root of $G$.

\subsubsection{Part 2)}

We have to show that, for any choice of $Z(\cdot)$, there is one and only one stable representation $X(\alpha)$ with
$\dim X(\alpha)=\alpha$ for each positive root $\alpha$ of $G$. By the Kontsevich--Soibelman formula it is enough to establish the result for a particular $Z(\cdot)$.

Existence of $X(\alpha)$ is easy (in particular for `nice' choices of $Z(\cdot)$).
Given a central charge function $Z(\cdot)$ we define orientation $\Omega$ of $G$ adapted to $Z(\cdot)$ in the following way.
Re--order the vertices of $G$ as $\{i_1, i_2, \dots, i_r\}$ in such a way that
\begin{equation}
 \arg Z(\delta_{i_1})>\arg Z(\delta_{i_2}) >\cdots > \arg Z(\delta_{i_r})
\end{equation}
and choose the orientation $\Omega$ such that the node sequence $\{i_1,i_2,\cdots, i_r\}$ is a source sequence.
We mention that, as a consequence of the mutation algorithm \cite{ACCERV1,ACCERV2}, this notion of adapted coincides with the one based on the possible minimal words for the longest element $w_0\in \mathrm{Weyl}(G)$ used by Geiss--Leclerc--Schr\"oer \cite{GLS}.

With this orientation, each brick $X$ of $\C G_\Omega$, is a stable object, and $\mathscr{S}_\Omega(X)$ is stable in $\cp(G)$ for the given stability function $Z(\cdot)$. Gabriel theorem \cite{GAB} guarantees that such bricks are in one--to--one correspondence with the positive roots of $G$ \emph{via} the dimension vector, and the result of Lusztig already cited \cite{LU} guarantees that $\mathscr{S}_\Omega(X)$ is isolated (no continuous moduli) in $\mathsf{mod}\,\cp(G)$. Hence, for $\Omega$--adapted, the rappresentations $$\Big\{\mathscr{S}_\Omega(X)\ \Big|\ X\ \text{a brick of }\C G_\Omega\Big\},$$
form a complete set of stable vector multiplets with the quantum number of the positive roots of $G$.
\medskip

For $G=A_n$ uniqueness is also easy. Since $\dim X$ is a root, each component of the dimension vector, $(\dim X)_i$, is either $0$ or $1$, and all maps are either isomorphisms or zero. Since the relations of $\cp(G)$ require $X_{\alpha}X_{\alpha^*}$ to be nilpotent for all $\alpha$'s, at least one map in each pair $X_\alpha, X_{\alpha^*}$ should vanish. Hence all bricks are of the form $\mathscr{S}_\Omega(X)$ for some $\Omega$. The support of $X$ is some $A_m\subset A_n$ and, replacing $A_n$ with $A_m$, we may assume $X$ to be sincere. Consider the $i$--th link of $G$. Orienting the link as $\rightarrow$ the representation $X_{(i)}$ obtained by setting all vector spaces over the nodes to the left  of the $i$--th arrow to zero is a subobject of $X$, while with the orientation $\leftarrow$ it is a quotient. Since, for generic $Z(\cdot)$ precisely one of the two conflicting inequalities
\begin{equation}
\arg Z(X_{(i)})< \arg Z(X),\qquad \arg Z(X_{(i)})> \arg Z(X)
\end{equation} 
is satisfied, there is exactly one orientation $\Omega$ of the (sub)quiver $\cp(A_m)$ consistent with stability. Thus uniqueness is true for $G=A_n$. 
\medskip

In the general $ADE$ case one chooses a special chamber.
$\cp(G)$ is a bi--partite quiver, and we distinguish between `even' and `odd' nodes. 
We choose the following special central charge
\begin{equation}\label{specialZ}
Z(\delta_j)= \begin{cases} 1 & j \ \text{is an odd node}\\
\exp(i\pi/2) & j \ \text{is an even node.}
\end{cases}
\end{equation} 
We set $e(X)=\sum_{i\ \text{even}} (\dim X)_i$, and 
$o(X)=\sum_{i\ \text{odd}} (\dim X)_i$, so that \begin{equation}Z(X)= o(X)+i\,e(X).\end{equation}

The \textbf{Statement} says that a representation $X$ of $\cp(G)$, which is stable with respect to the stability function \eqref{specialZ}, has the form $\mathscr{S}_{\overleftrightarrow{\Omega}}(X^\prime)$ where $\overleftrightarrow{\Omega}$ is the alternating orientation of the Dynkin graph $G$ with even nodes sources and the odd ones sinks. Note that $X=\mathscr{S}_{\overleftrightarrow{\Omega}}$ for a certain $X^\prime\in \mathsf{mod}\text{--}\C G$ iff there is an exact sequence
\begin{equation}\label{exevodd}
0\rightarrow \bigoplus_{i\ \text{odd}} S(i) \rightarrow X \rightarrow \bigoplus_{j\ \text{even}}S(j)\rightarrow 0.
\end{equation}

Let $X$ be such a stable representation. In particular $\dim X$ is a root $\alpha$ of $G$. If $\alpha$ is simple, there is nothing to show.
Otherwise  we may assume $e(X)\,o(X)\neq 0$. Let
\begin{equation}\label{filtration}
0=X_0\subset X_1 \subset X_2 \subset \cdots \subset X_\ell =X
\end{equation}
be a filtration such that 
\begin{equation}
X_{j+1}/X_j =\mathrm{soc}\big(X/X_j\big).
\end{equation}
By construction, each composition factor $X_{j+1}/X_j$ is semisimple. 

Let $X_1=\oplus_a S(i_a)$. We claim that all summands $S(i_a)$
have $i_a$ odd. Indeed, assume that $i_{a_0}$ is even. Then
the simple object $S(i_{a_0})$ is a subrepresentation of $X$. Stability of $X$ gives
\begin{equation}
+\infty =\tan \arg Z(S(i_{a_0})) < \tan \arg Z(X) = e(X)/o(X)
\end{equation}
which is a contradiction. Since $\cp(G)$ is bi--partite, we have
$X_2/X_1=\oplus_b S(i_b)$ with all $i_b$ even (because there is no non--trivial extension $0\rightarrow X_1 \rightarrow Y \rightarrow S(i)\rightarrow 0$ for $i$ odd). 
Recursively, we see that $X_k/X_{k-1}$ with $k$ even (resp.\! odd) is a direct sum of even simples  (resp.\! odd simples). Dually, we may start from the top of $X$ rather than from the socle. Stability requires $\mathrm{top}(X)$ to be a direct sum of even simples. Thus, the length $\ell$ of the filtration \eqref{filtration} should be an even number. 

Uniqueness in the general case boils down to show that $\ell=2$.
Let us check it in the simplest non--trivial case, namely $\dim X$ the highest root of $D_4$. In the convention in which the triple node is even, if $\ell>2$ we have only two possible composition series with the above properties (up to permutation of the three external nodes $1_a$ of the $D_4$ graph)
\begin{equation}
 \begin{array}{c}
S(2)\\\hline
S(1_2)\oplus S(1_3)\\\hline
S(2)\\\hline
S(1_1)
\end{array}\hskip 3cm
 \begin{array}{c}
S(2)\\\hline
S(1_3)\\\hline
S(2)\\\hline
S(1_1)\oplus S(1_2)
\end{array}
\end{equation}
The first one is obviously instable in our special chamber. The second one is in fact a representation of the subquiver
\begin{equation}
 \begin{gathered}
  \xymatrix{1_1\\
& 2 \ar[ul]_{\alpha_1}\ar[dl]^{\alpha_2} \ar@/^1.5pc/[rr]^{\alpha_3} && 1_3\ar@/^1.5pc/[ll]^{\alpha_3^*}\\
1_2}
 \end{gathered}
\end{equation}
and the relations give $\alpha_3\alpha^*_3=\alpha_3^*\alpha_3=0$. Hence, again, precisely one of two arrows $\alpha_3$, $\alpha_3^*$ vanishes. $\alpha_3=0$ would lead to an unstable representation, and we remain with a \emph{unique} possibility, $\alpha_3^*=0$ which is of the expected $\mathscr{S}_{\overleftrightarrow{\Omega}}(Y)$ form.

Similar gynmastic may be used to show the claim for all $D_r$. It should also be true for $E_6,E_7,E_8$ since the argument from the physical side is very robust.

\section{$G$--canonical algebras}

We wish to generalize the analysis of sections \ref{CanRingel} and \ref{XXXTT}, which was valid for $G=SU(2)^k$ to a general (simply--laced) gauge group. We know that a quiver $\cn=2$ model which --- in some duality frame --- may be seen as a gauge theory with gauge group $$G=G_1\times G_2\times\cdots\times G_s,\qquad (G_i\ \text{simple})$$
 must enjoy the Ringel property. The light category
 $$\mathscr{L}=\bigvee_{\lambda\in N} \mathscr{L}_\lambda,$$
 where $N$
is the union of $s$ $\mathbb{P}^1$'s in one--to--one correspondence with the simple gauge group factors $G_a$ in analogy with what we saw in sec.\,\ref{XXXTT}. Over the generic point of the $a$--th  irreducible component we must have the equivalence
$$\mathscr{L}_\lambda\simeq \mathscr{L}^\mathrm{YM}_\lambda(G_a),$$
where $\mathscr{L}^\mathrm{YM}_\lambda(G_a)$ is the homogeneous $G_a$--tube described in section \ref{pureSYMN}.
Each matter sector is contained in a subcategory $\mathscr{L}_{\lambda_\alpha}$ over a specific (closed) point $\lambda_\alpha\in N$. 
\smallskip

The simplest class of theories is obtained by extending to general $G$ the canonical construction which works for the $SU(2)$ models of \S.\,\ref{CanRingel}. These models have a canonical gauge functor $\mathscr{G}$ which makes the analysis simple and  nice.
In this section we propose a class of $G$--canonical algebras for all $G=ADE$. They satisfy the Ringel property in the desired form, and are also consistent with the $4d/2d$ correspondence as we shall show in sec.\,\ref{Proofconjectr}.
Thus these associative algebras satisfy the necessary conditions to correspond to good $\cn=2$ gauge theories with gauge group $G$. However not all $G$--canonical algebras are expected to be good QFTs: indeed, already in the $SU(2)$ case the QFT associated to a canonical algebra may fail to be UV--complete. 

For the classification program, one needs --- as we had in the $SU(2)$ case --- a criterion to decide which $G$--canonical algebra gives a consistent QFT. In this last direction we shall give only very partial results.

\subsection{The quiver}
We start with the Dynkin graph\footnote{\ Recall that the Dynkin \emph{graph} is unoriented.} $G$, and to each node $i\in {G}$ we attach a set $\kappa_i$ of $k_i\geq 0$ integers, $\kappa_i\equiv \{n_{i,1},\cdots,n_{i,k_i}\}$ with $n_{i,a}\geq 2$. \textit{E.g.}\! for $G=E_8$ we have
\begin{equation}\label{genege}\begin{gathered}
 \xymatrix{&&&& \kappa_8 & &\\
\kappa_1\ar@{-}[r] & \kappa_2 & \kappa_3\ar@{-}[l]\ar@{-}[r] & \kappa_4 & \kappa_5\ar@{-}[l]\ar@{-}[u]\ar@{-}[r] & \kappa_6 &\kappa_7\ar@{-}[l]}
\end{gathered}\end{equation}

To the data $\{\kappa_i\}_{i\in G}$ we associate the quiver $Q(G,\{\kappa_i\})$ so defined: we start with the disconnected quiver
\begin{equation}
 \coprod_{i\in G} Q(\kappa_i)
\end{equation}
which is the disjoint union of the (3--CY completed) canonical quivers $Q(\kappa_i)$ associated with the sets $\kappa_i$ (we use the same notations as in \S.\,\ref{CanRingel}). This quiver has
$2\,r(G)$ distinguished nodes, namely $\spadesuit_i$, $\diamondsuit_i$, for $i\in G$. For $i\neq j$, we draw $|C_{ij}|$ arrows from
$\diamondsuit_i$ to $\spadesuit_j$, where $C_{ij}$ is the Cartan matrix of $G$. The resulting connected quiver is our 
$Q(G,\{\kappa_i\})$.\smallskip

We shall say that the full subquiver $Q(\kappa_i)\subset Q(G,\{\kappa_j\})$ is the canonical subquiver \emph{over} the $i$--th node of the Dynkin graph, or that \textit{it covers} that node of $G$.

\subsection{Examples}

\begin{itemize}
\item $\kappa_i=\emptyset$ for all $i$. $Q(G,\{\emptyset,\cdots, \emptyset\})= \overleftrightarrow{G}\,\square\,\widehat{A}(1,1)$, \textit{i.e.}\! pure SYM with gauge group $G$. Over each node of $G$ there is a Kronecker subquiver $\widehat{A}(1,1)$;
\item $\kappa_1=\overbrace{\{2,2,2,\dots,2\}}^{N_f\ \text{terms}}$, and $\kappa_i=\emptyset$ for $i>1$. $Q(G,\{\kappa_i\})$ is the quiver of SQCD with gauge group $G$ and $N_f$ fundamental quarks \cite{ACCERV2};
\item $\kappa_i=\overbrace{\{2,2,2,\dots,2\}}^{N_i\ \text{terms}}$. $Q(G,\{\kappa_i\})$ is the `massive quark' quiver for SYM with gauge group $G$ coupled to $N_i$ (full) hypermultiplets in the $i$--th fundamental representation \cite{ACCERV2}. From \eqref{numnodes} we see that the total number of nodes in the quiver is
\begin{equation}
 2\,r(G)+\sum_i N_i,
\end{equation}
which is precisely equal to the number of electric, magnetic and flavor charges;
\item if some $n_{i,a}>2$, the model has no Lagrangian description (in terms of free fields with only gauge couplings).
One needs a criterion to distinguish the $\kappa_i$ which do arise in physical theories. In the $SU(2)$ case the criterion was just $\chi\geq 0$.
\end{itemize}

\subsection{Superpotential}

For each link $\xymatrix{i \ar@{-}[r] & j}$ in the Dynkin graph $G$ we get two arrows in our quiver, namely
$$\xymatrix{\diamondsuit_i\ar[r]^{\phi_{ji}}& \spadesuit_j}\qquad \text{and}\qquad \xymatrix{\spadesuit_i &\ar[l]_{\phi_{ij}} \diamondsuit_j}.$$
Moreover, in each (full) subquiver $Q(\kappa_i)$ over the $i$--th node we have the two paths $\ell_1(i)$ and $\ell_2(i)$ defined as in eqn.\eqref{patthh}. We write $\prod_1\psi_i$ and $\prod_2\psi_i$ for the product of (the bifundamental Higgs fields associated to) the arrows in the path $\ell_1(i)$, resp.\! $\ell_2(i)$.

With these notations the superpotential on the quiver $Q(G,\{\kappa_i\})$ is
\begin{equation}\label{superpotential}
 \cw= \sum_{\text{nodes}\atop \text{of }G}\cw(\kappa_i)+\sum_{\text{links}\atop \text{of }G}\mathrm{Tr}\Big(\prod_1\psi_i\:\phi_{ij}\prod_2\psi_j\:\phi_{ji}-\prod_2\psi_i\:\phi_{ij}\prod_1\psi_j\:\phi_{ji}\Big).
\end{equation}
where in  the sum over links we mean that $i$ (resp.\! $j$) is an odd (resp.\! even) node of the bi--partite graph $G$ and $\cw(\kappa_i)$ is as in \S.\,\ref{CanRingel}.

It is elementary to check that the above formula reproduces the known superpotential for the SQCD case (with any number of flavors), or any other model having a Lagrangian formulation.

\subsection{Magnetic charges and weak coupling}

Taking the charge vectors $\delta_i$ of the simple--root $W$--bosons to be the characteristic function of the subquiver over the $i$--th node of $G$,
the magnetic charges are given by the usual formula
\begin{equation}\label{mmmmagnetic}
m_i(X)= (\dim X)_{\spadesuit_i}-(\dim X)_{\diamondsuit_i},
\end{equation}
and the weak coupling is still defined by the limit $g\rightarrow 0$ with
\begin{align}
\arg Z(\spadesuit_i)&= \pi -O(g^2)\label{WE1}\\
\arg Z(\diamondsuit_i) &= O(g^2).\label{WE2}
\end{align}
%\smallskip

\textbf{Lemma.} \textit{Let $X$ a representation of $(Q(G,\{\kappa_i\}), \cw)$ which is stable at weak coupling. Then, for all $i\in G$, all arrows $\eta_{i,a}\subset Q(\kappa_i)_1$ vanish.}
\smallskip

\textsc{Proof.}  Same argument as in the \textbf{Proposition} on page \pageref{prop}. \hfill $\square$
\medskip

Hence all representations stable at weak coupling are in fact representations of the \textit{$G$--canonical algebra}, namely the path algebra of the above quiver with the $\eta$--arrows omitted and bounded by the relations $\partial\cw=0$ with the $\eta$'s set to zero. \smallskip

The category of the (finite--dimensional) representations of the $G$--canonical algebra will be denoted as  $\mathsf{C}(G,\{\kappa_i\})$.
Again, we are interested in the light category $\mathscr{L}\subset \mathsf{C}(G,\{\kappa_i\})$: by definition one has $X\in \mathscr{L}$ if $m_i(X)=0$ for all $i$'s while for all its subobjects $Y$ one has $m_i(Y)\leq 0$.

For $G=SU(2)$, $\mathscr{L}$ is the same as Ringel's category $\mathcal{T}$ \cite{RI}; moreover, if the $SU(2)$ model is asymptotically free, $\mathscr{L}$ coincides with the subcategory of regular modules. 

A light \textit{stable} representation (at weak coupling) is precisely a BPS state with bounded mass as $g\rightarrow 0$, that is a state which is a `local fundamental degree of freedom' of the corresponding QFT and not a soliton carrying the (topological) magnetic charge. 

\subsection{The Coxeter--Dynkin form of $(Q,\cw)$}

The quiver of a $\cn=2$ theory is defined only up to mutation (Seiberg duality in the SQM language), the module categories of mutation--equivalent quivers (with superpotentials) being derived equivalent. 

The canonical pair $(Q,\cw)$ given above is particular suitable for computations thanks to the gauge functor $\mathscr{G}$; there is another special form of $(Q,\cw)$ which may be convenient for specific purposes.
In the case of the $SU(2)$ models associated to the canonical algebras, \S.\,\ref{CanRingel}, this second form of $(Q,\cw)$ corresponds to (the $3$--CY completion of) a \emph{Coxeter--Dynkin algebra of canonical type} \cite{Len34}. We use the same terminology in the general $G$ case, and call the corresponding pair $(Q,\cw)$ the Coxeter--Dynkin form of the quiver.

The Coxeter--Dynkin version
of the $G$--canonical quiver --- specified by the set $\{\kappa_i\}_{i\in G}$, $\kappa_i=\{n_{1,i}, n_{2,i},\cdots, n_{k_i,i}\}$ --- is obtained by replacing  the subquiver $Q(\kappa_i)$ over the $i$--th node 
by a subquiver consisting of a Kronecker quiver $\spadesuit_i\rightrightarrows \diamondsuit_i$ with attached $k_i$ oriented triangles (of sides $\alpha_i,\beta_i$ and the Kronecker arrows); to the third node $\bullet_a$ ($a=1,2,\dots, k_i$) of each such triangle it is attached a $A_{n_{a,i}-2}$ subquiver with a linear orientation, as in the figure
\begin{equation}
 \begin{gathered}
  \xymatrix{\bullet\ar[r] & \bullet\ar@{..}[r] &\bullet \ar[r] &\bullet\ar[ddr]_{\alpha_1}&
\diamondsuit_i\ar[l]_{\beta_1} \ar[r]\ar[rdd] & \bullet\ar[ddl] &\bullet\ar[l] \ar@{..}[r] & \bullet &\bullet\ar[l]\\
&&&&&  & \vdots \ar@{..} & \vdots & \vdots\\
&&&&\spadesuit_i\ar@<0.4ex>@{-->}[uu]\ar@<-0.4ex>@{-->}[uu]& \bullet\ar[l] & \bullet\ar[l] \ar@{..}[r] & \bullet &\bullet\ar[l]}
 \end{gathered}
\end{equation}
(The Kronecker arrows are drawn dashed since they correspond to relations in the uncompleted Coxeter--Dynkin algebra).

The superpotential term involving only arrows of the $i$--th subquiver reads
\begin{equation}
 \eta_1\sum_{i=2}^k \alpha_i\beta_i+ \eta_2\big(\alpha_1\beta_2-\sum_{i=3}^k\lambda_i\, \alpha_i\beta_i\big),
\end{equation}
 where $\eta_1,\eta_2$ are the (dashed) arrows of the Kronecker subquiver.

The Coxeter--Dynkin form is the one most frequently used in ref.\!\cite{CV11} since it is the one with the most direct physical interpretation in terms of a generalized `heavy quark' argument.

\subsection{The gauge functor $\mathscr{G}$} \label{GGGG}

The gauge functor $\mathscr{G}$ maps a representation $X$ of the $G$--canonical quiver (with superpotential) $Q(G,\{\kappa_i\})$ into a representation of the quiver for pure SYM, $Q(G,\{\emptyset_i\})$.
Explicitly\footnote{\ Here and below we use the short--hand notations of section \ref{CanRingel}: then the composition of all arrows of $X$ belonging to the $a$--th path in the canonical subquiver over the node $i\in G$,  $\prod_{\ell_a(i)}X_{\psi_i}$ is written simply as $\prod_a\psi_i$.} ,
\begin{align}
&\mathscr{G}(X)_{\spadesuit_i} = X_{\spadesuit_i}\label{www1} 
&&\mathscr{G}(X)_{\diamondsuit_i} = X_{\diamondsuit_i}\\
&\mathscr{G}(X)_{\phi_{ij}} = X_{\phi_{ij}}\\
&\mathscr{G}(X)_{\psi_{1,i}}= X_{\prod_1\psi_i}
&&\mathscr{G}(X)_{\psi_{2,i}}= X_{\prod_2\psi_i}.
\end{align}
It is clear that $\mathscr{G}(X)$ is a representation of the pure SYM quiver $Q(G,\{\emptyset_i\})$ satisfying the correct relations $\partial\cw=0$.
\smallskip

At weak coupling we may limit ourselves to the subcategory
$\mathsf{C}(G,\{\kappa_i\})$. Then all the $\eta$'s vanish, and all the products along the paths $\ell_a(i)$
in the canonical subquivers $Q(\kappa_i)$ over each node $i\in G$ are uniquely determined by $\mathscr{G}(X)$, indeed
\begin{equation}\label{UUUBBB}
\prod_a\psi_i \equiv \lambda_{a,i}\,\prod_1 \psi_i+\mu_{a,i}\,\prod_2\psi_i= \lambda_{a,i}\,\mathscr{G}(X)_{\psi_{1,i}}+\mu_{a,i}\,\mathscr{G}(X)_{\psi_{1,i}}.
\end{equation}
From its canonical subcategories $\mathsf{C}(\kappa_i)$,  the category $\mathsf{C}(G,\{\kappa_i\})$ inherits the following useful property: assume $X\in\mathsf{C}(G,\{\kappa_i\})$ is indecomposable; then, if the map along the path $\ell_a(i)$ in the subquiver $Q(\kappa_i)$
$$\prod_a\psi_i\equiv \prod_{\ell_a(i)} X_{\psi_i}$$
 is, respectively, injective, an isomorphism, or surjective, the same is true for each arrow $X_{\psi_i}$ along the given path $\ell_a(i)$ in $Q(\kappa_i)$.
\medskip

From eqn.\eqref{www1} and the definition \eqref{mmmmagnetic} it is obvious that $\mathscr{G}$ preserves the magnetic charges
\begin{equation}
 m_i(\mathscr{G}(X))=m_i(X),
\end{equation}
 and hence maps the \emph{light} subcategory $\mathscr{L}\subset \mathsf{C}(G,\{\kappa_i\})$ of the given $G$--canonical theory to the 
corresponding subcategory $\mathscr{L}^\mathrm{YM}$ of pure SYM (with the same $G$)\footnote{\ One has to show that if $X\in \mathscr{L}$ then there is no subobject $Y$ of $\mathscr{G}(X)$ with $m_i(Y)>0$ for some $i\in G$. This is equivalent to the statement that the for all $i$ $\mathscr{G}(X)\big|_{\mathbf{Kr}_i}$ has no direct summand in $\cq$. In turn, this is also equivalent to $X\big|_{C(\kappa_i)}$ not having direct summand in $\cp$. This follows from the same argument as in eqns.\eqref{krqwz}--\eqref{trnnnkspg}}. Again, at weak coupling, $\mathscr{L}$ is the main object of interest, and we focus on it. Restricting to this subcategory at least one arrow in each subquiver $\mathbf{Kr}_i$, $\mathscr{G}(X)_{\psi_{1,i}},
\mathscr{G}(X)_{\psi_{2,i}}$, is an isomorphisms.

By the same argument as in the pure SYM case, for an indecomposable $X\in \mathscr{L}$ the matrix
\begin{equation}
 \prod_2\psi_i\left(\prod_1\psi_i\right)^{-1}
\end{equation}
decomposes in Jordan blocks of eigenvalue $\zeta\in \mathbb{P}^1$ which is the {same} for all Kronecker subquivers $\mathbf{Kr}_i$ over the nodes $i\in G$. Hence, again,
\begin{equation}
 \mathscr{L}\subseteq\bigvee_{\zeta\in \mathbb{P}^1} \mathscr{L}_\zeta,
\end{equation}
and \begin{equation}\mathscr{G}(\mathscr{L}_\zeta)\subseteq \mathscr{L}^\mathrm{YM}_\zeta.\end{equation}

In particular, for $$\zeta\in \mathbb{P}^1\,\setminus\, \bigcup_{a,i}\: (\lambda_{a,i}:\mu_{a,i})$$ all maps $\psi_i$ are isomorphisms, and
$\mathscr{G}\colon \mathscr{L}_\zeta \rightarrow \mathscr{L}^\mathrm{YM}_\zeta$ is an equivalence.
Then --- as in the classical $SU(2)$ case --- for all, but finitely many, points in $\mathbb{P}^1$ the category 
$\mathscr{L}_\zeta$ is the same as the corresponding category for the pure SYM case namely a {homogeneous $G$--tube}.
\medskip

If $\zeta$ is equal to one of the special points $(\lambda_{a,i}:\mu_{a,i})$ --- which we assume to be all distinct --- new phenomena appear. Restricted to the 
canonical subquiver $Q(\kappa_j)$ over a node $j\neq i$, the situation looks exactly the same as in pure SYM. But the restriction to the canonical subquiver
$Q(\kappa_i)$ over the $i$--th node is now a representation (not necessarily indecomposable) belonging to a stable tube of period $n_{a,i}$. 

Therefore, limiting ourselves to the special $G$--tubes arising from this canonical construction, the classification is as follows: besides the homogeneous $G$--tube, there is a distinct non--homogeneous $G$--tube for each pair $(i,n)$ where $i$ is a node in the Dynkin graph $G$ and $n$ is an integer $>2$.

\section{`Lagrangian' matter: $\cp(Q)$ algebras again}\label{iiiiiaaa}

Having understood the homogeneous $G$--tubes, it remains to study the non--homogeneous ones which, roughly speaking, correspond to the possible matter subsystems with a gaugeable global symmetry $G$. $G$--canonical algebras give a (small) special class of such $G$--tubes.

In this section we consider the case in which the restriction of the Dirac pairing to the light category $\mathscr{L}$ vanishes 
\begin{equation}\label{LLagrangianN}\langle\cdot,\cdot\rangle_\text{Dirac}\Big|_\mathscr{L}=0,\end{equation}
that is, the Euler form is \emph{symmetric} when restricted to $\mathscr{L}$.

Models with the property \eqref{LLagrangianN} will be called \emph{`Lagrangian'}, as all $\cn=2$ gauge theories with a weakly coupled Lagrangian description satisfy \eqref{LLagrangianN} since the fundamental fields of a Lagrangian description are mutually local. Under this assumption, the argument of \S.\,\ref{chambindependence} applies, and the set of dimension vectors of the light stable representations is independent of the particular weakly--coupled chamber. Hence a light BPS state which is present in one such chamber is present in all. That argument, based on the Kontsevich--Soibelman formula \cite{KS}, has also another useful consequence: if \eqref{LLagrangianN} holds, the physical quantum numbers of the light BPS states are the same for the quivers (with superpotential) $Q$ and $\mu(Q)$ provided the mutation $\mu$ preserves the magnetic charges (and hence the notion of weak--coupling). Of course, the quiver mutation $\mu$  \emph{tilts} the positive--cone in $K_0(\mathsf{rep}(Q,\cw))$, and the full BPS spectrum --- including the anti--particles with charge vector \emph{minus} the dimension vector of a stable representation --- should be taken into account.
This reflects the fact that the module categories of the two Jacobian algebras $\C Q/(\partial\cw)$ and $\C \mu(Q)/(\partial\mu(\cw))$ are derived equivalent \cite{kellerA} but different.

The $G$--canonical algebras which are `Lagrangian' in this sense are those associated to a family of sets $\{\kappa_i\}_{i\in G}$ of the form
\begin{equation}
 \kappa_i =\overbrace{\{2,2,\dots, 2\}}^{N_i\ \text{times}}.
\end{equation}
The $N_i$ points $\zeta_{a,i}\equiv (\lambda_{a,i}:\mu_{a,i})\in\mathbb{P}^1$ (cfr. eqn.\eqref{UUUBBB}) are assumed to be generic, \textit{i.e.}\! pairwise distinct. For $\zeta\neq \zeta_{a,i}$, the $G$--tube $\mathscr{L}_\lambda$ is equivalent to the homogeneous $G$--tube $\mathscr{L}^\mathrm{YM}_\lambda$, while 
$\mathscr{L}_{\zeta_{a,i}}$ is equivalent to the unique non--homogeneous $G$--tube of the canonical algebra with
\begin{equation}\label{specialcasse}
\kappa_j = \begin{cases} \{2\} & j=i\\
\emptyset & j\neq i,
\end{cases}
\end{equation} 
which heuristically corresponds to $G$ SYM coupled to a single massive quark in the $i$--th fundamental representation of $G$.
In other words, we are reduced to study the $N_i=1$ case, the $G$--tubes of the several `quarks' laying over distinct points of $\mathbb{P}^1$.

The quiver of the $N_i=1$ model is the pure SYM quiver $\mathbb{G}$ with the $i$--th Kronecker subquiver replaced by a $\widehat{A}(2,1)$ subquiver
\begin{equation}\label{yyybbbhhg}
 \begin{gathered}
  \xymatrix{\cdots&\diamondsuit_{i_0}\ar[l]&\\
\\
\cdots\ar[r]&\spadesuit_{i}\ar@<1ex>[uu]^{A_{i}}\ar\ar@<0.2ex>[uu]_{B_{i}}}
 \end{gathered}\quad\longrightarrow\quad
\begin{gathered}
  \xymatrix{\cdots&\diamondsuit_{i}\ar[l]\\
&& \bullet\ar[ul]_{\psi_1}\\
\cdots \ar[r] &\spadesuit_{i}\ar@<0.7ex>[uu]^{A_{i}}\ar[ur]_{\psi_2}}
 \end{gathered}\qquad \text{where }B_{i}\equiv \psi_1\psi_2.
\end{equation}
Mutating at the node $\bullet$ we get the Coxeter--Dynkin form of the quiver
\begin{equation}\label{cosseter}
\begin{gathered}
  \xymatrix{\cdots&\diamondsuit_{i}\ar[l]\\
&& \bullet\ar[ul]_{\psi_1}\\
\cdots \ar[r] &\spadesuit_{i}\ar@<0.7ex>[uu]^{A_{i}}\ar[ur]_{\psi_2}}
 \end{gathered}
 \quad\longrightarrow\quad
\begin{gathered}
  \xymatrix{\cdots&\diamondsuit_{i}\ar[l]\ar[dr]^{\phi_1}&\\
&& \bullet\ar[dl]^{\phi_2}\\
\cdots\ar[r]&\spadesuit_{i}\ar@<1ex>[uu]^{A_{i}}\ar\ar@<0.2ex>[uu]_{B_{i}}}
 \end{gathered}\end{equation}
which has the same magnetic charges $m_j(X)=\dim X_{\spadesuit_j}-\dim X_{\diamondsuit_j}$.
The quantum number identifications in the two quivers are related by mutation; writing $\delta_j$ for the charge vector of the $W$--boson associated to the $j$--th simple root of $G$ and $\mathfrak{q}$ for the charge vector of the basic quark state, we have
\begin{equation}\label{uuuwuwu}
\begin{aligned}
&\bullet\ \text{Coxeter--Dynkin} && \mathfrak{q}= s_\bullet, &&\delta_j=s_{\spadesuit_j}+s_{\diamondsuit_j}\\
&\bullet\ \text{canonical} &&  \mathfrak{q}= -s_\bullet, &&\delta_j=\begin{cases}s_{\spadesuit_j}+s_{\diamondsuit_j} &j\neq i\\
s_{\spadesuit_{i}}+s_{\diamondsuit_{i}}+s_\bullet &j=i.\end{cases}
\end{aligned}
\end{equation}
and the argument of \S.\,\ref{chambindependence} implies that the set of the quantum numbers of the light BPS states 
\begin{equation}\label{rwquiv}
 \left\{\pm(f,q_j)_{j\in G}\:\Bigg|\:\begin{aligned}& f\,\mathfrak{q}+\sum\nolimits_j q_j\,\delta_j=\dim X,\\
& X\ \text{a light stable representation of }(Q,\cw)\end{aligned}\right\}
\end{equation}
is the same for the two mutation--equivalent quivers.

\subsection{Relation to $\cp(Q)$}

In the present conventions, the non--homogeneous $G$--tube of the algebra \eqref{specialcasse} corresponds to $\zeta=0$. If $X\in \mathscr{L}_0$, focusing on the subquiver  \eqref{yyybbbhhg}, we have  
\begin{equation}X_{\diamondsuit_{i}}\simeq
X_{\spadesuit_{i}},\qquad 
X_{A_{i}}=\text{Id},\qquad X_{\psi_1}X_{\psi_2}\ \ \text{is nilpotent.}
\end{equation}
 If, in addition, $X$ is a brick $X_{\psi_1}X_{\psi_2}=0$. Moreover, for $X$ stable (in a weak coupling regime) either $X_{\psi_1}$ is injective or $X_{\psi_2}$ is surjective. Thus for $X$ light stable $X_{\psi_1}$ or $X_{\psi_2}$ vanishes.

For $X$ a brick in $\mathscr{L}_0$, the maps $A_j=\mathrm{Id}\colon \spadesuit_j \rightarrow \diamondsuit_j$, $j\in G$, identify
$X_{\diamondsuit_j}\simeq X_{\spadesuit_j}$, leaving effectively with a representation of the quiver $\overline{G[i]}$ obtained by
the following two--step procedure: 
\begin{enumerate}
 \item One constructs the \emph{augmented} Dynkin graph $G[i]$ by adding  to the Dynkin graph of $G$ an extra node, $\bullet$, connected by a link to the vertex $i\in G$. \textit{E.g.} if $G=A_7$ and $i=3$, corresponding to the $3$--index antisymmetric rep., we have
\begin{equation}\label{eeeexampleee}
A_7[3]\colon\qquad \begin{aligned}
\xymatrix{&& \bullet \ar@{-}[d]\\
\diamondsuit_1 \ar@{-}[r] &\diamondsuit_2 \ar@{-}[r] &\diamondsuit_3 \ar@{-}[r] &\diamondsuit_4 \ar@{-}[r] &\diamondsuit_5 \ar@{-}[r] &\diamondsuit_6 \ar@{-}[r] &\diamondsuit_7 }
\end{aligned}
\end{equation}
which, in this case, is the Dynkin graph $E_8$;
\item one replaces each link in $G[i]$ with a pair of opposite arrows $\leftrightarrows$, \textit{i.e.}\! forms the double quiver of $G[i]$. \textit{E.g.}, for the example \eqref{eeeexampleee} one gets
\begin{equation}
\overline{A_7[3]}\colon\begin{aligned}
\qquad\xymatrix{&& \bullet \ar@<0.4ex>[d]^{\psi_1}\\
\diamondsuit_1 \ar@<0.4ex>[r]^{\phi_{21}} &\diamondsuit_2 \ar@<0.4ex>[r]^{\phi_{32}}\ar@<0.4ex>[l]^{\phi_{12}} &\diamondsuit_3\ar@<0.4ex>[u]^{\psi_2} \ar@<0.4ex>[r]^{\phi_{43}}\ar@<0.4ex>[l]^{\phi_{23}} &\diamondsuit_4 \ar@<0.4ex>[r]^{\phi_{54}}\ar@<0.4ex>[l]^{\phi_{34}} &\diamondsuit_5 \ar@<0.4ex>[r]^{\phi_{65}}\ar@<0.4ex>[l]^{\phi_{45}} &\diamondsuit_6 \ar@<0.4ex>[r]^{\phi_{76}}\ar@<0.4ex>[l]^{\phi_{56}} &\diamondsuit_7 \ar@<0.4ex>[l]^{\phi_{67}}}
\end{aligned}
\end{equation}
\end{enumerate}
Explicitly, given a light stable representation $X$ of the quiver  $Q(G,\{\emptyset,\cdots,\emptyset,\{2\}, \emptyset,\cdots, \emptyset\})$, the representation $Y$ of the double quiver $\overline{G[i]}$ is  given by
\begin{equation}
 \begin{aligned}
  &Y_{\diamondsuit_j}=X_{\diamondsuit_j} &&Y_\bullet= X_\bullet, \\ & Y_{\psi_a}=X_{\psi_a}, && Y_{\phi_{ij}}= X_{\phi_{ij}}.
 \end{aligned}
\end{equation}
We claim that the representation $Y$ automatically satisfies the relations of the preprojective algebra $\cp(G[i])$.
One needs to check only the relations at the two nodes $\bullet$ and $\diamondsuit_{i}$ since nothing changes at the other nodes with respect to the pure SYM case. Then we must verify
\begin{equation}
 \begin{aligned}
  &\text{node }\diamondsuit_{i}\colon &&\sum_j\,\epsilon_j \phi_{ij}\phi_{ji}-\psi_1\,\psi_2=0\\
 &\text{node }\bullet\colon && \psi_2\,\psi_1=0.
 \end{aligned}
\end{equation}
The first relation is just $\partial_{B_{i}}\cw=0$ evaluated at $A_j=\mathrm{Id}$; indeed, 
\begin{equation}\cw=\sum_{k,j}\epsilon_k\, A_k\, \phi_{kj}\,B_j\,\phi_{kj}-B_{i}\,\psi_1\psi_2.\end{equation}
 The second relation is trivially satisfied since either $\psi_1$ or $\psi_2$ vanishes.

Therefore
\begin{equation}\label{cconclusion}
 X\ \text{a stable brick of }\mathscr{L}_0\quad\Longrightarrow\quad Y\ \text{a stable brick of }\cp(G[i]).
\end{equation}
Note, however, that $Y$ is a representation of $\cp(G[i])$ of a particular kind since we have the stronger condition that one of the two arrows connecting $\bullet$ and $\diamondsuit_{i}$ must vanish.
\medskip

A useful observation is that studying the light stable representations of the Coxeter--Dynkin quiver in the \textsc{rhs} of eqn.\eqref{cosseter} one also ends up with stable bricks of $\cp(G[i])$ of the special kind above under the mutated identification (cfr.\! eqn.\eqref{cosseter})
\begin{equation}
 Y_{\psi_1}=X_{\phi_2},\qquad Y_{\psi_2}=X_{\phi_2}.
\end{equation}

\subsection{Relations to Kac--Moody algebras}\label{sec:kac}

The augmented graph $G[i]$ is a tree. We know from \S.\,\ref{zzzzzz1} that for all representations $Y$ of $\cp(G[i])$
\begin{equation}
2\, \dim\,\mathrm{End}(Y) = (\dim Y,\dim Y)_{\widehat{C}} +\dim\,\mathrm{Ext}^1(Y,Y),
\end{equation}
where $(\cdot,\cdot)_{\widehat{C}}$ is the symmetric even integral quadratic form defined by the Cartan matrix $\widehat{C}$ of the Kac--Moody algebra with Dynkin diagram $G[i]$. Moreover, we also know that the spin\footnote{\ By the spin of a representation $X$ we mean the spin of the $SU(2)$ representation which tensored with a hypermultiplet gives the spin content of the corresponding BPS supermultiplet. Thus $s=0$ for a hypermultiplet, $s=1/2$ for a vector multiplet, and $s\geq 1$ for higher--spin supermultiplets.} $s$ of a light brick $X\in \mathscr{L}_0$ --- which is not in the closure of a family $\{ X_\zeta \:|\: X_\zeta\in \mathscr{L}_\zeta,\ \zeta\in\mathbb{P}^1\}$ --- is given in terms of the corresponding representation $Y$ of $\cp(G[i])$ by the Lusztig formula
\begin{equation}
 2\,s= \frac{1}{2}\,\dim\,\mathrm{Ext}^1(Y,Y)
\end{equation}
(note that the \textsc{rhs} is always an integer).
Therefore: \textit{If $Y\in \mathsf{mod}\,\cp(G[i])$ with $Y_\bullet\neq 0$ corresponds to a light (stable) supermultiplet of spin $s$, $\dim Y$ is a positive root of the Kac--Moody algebra $G[i]$ and
\begin{equation}
 (\dim Y,\dim Y)_{\widehat{C}} =2-4s.
\end{equation}
In particular, if $Y$ corresponds to a stable \emph{hypermultiplet}, $\dim Y$ is a \emph{real} root of $G[i]$. }
\vglue 12pt

For the implication in the other direction,\vglue 9pt

\textbf{Proposition.} \textit{Assume that $\dim Y$ is a root of the Kac--Moody algebra $G[i]$ 
which is a Schur root\footnote{\ Recall that a root is a \emph{Schur root} if it is the dimension vector of a brick.} for some orientation $\Omega$ of the Dynkin graph $G[i]$. Then in all weakly--coupled chambers there is a a light stable }
BPS state of charge vector
\begin{equation}\label{dimveeector}
(\dim Y_\bullet)\, s_\bullet+ \sum_j \dim Y_{\diamondsuit_j}\big(s_{\spadesuit_j}+s_{\diamondsuit_j}\big).
\end{equation}

\textit{
Moreover, if there is a light BPS particle of spin $\geq 3/2$, there are light BPS particles of arbitrarily high spin. `Light' higher spin BPS states organize themselves in Regge trajectories with spin $s\sim M^2$, where the mass $M=\mathrm{const.}\, n$, $n\in \mathbb{N}$.}
\vglue 12pt

\textsc{Proof.} A representation $Y$ of the (tree) quiver $G[i]_\Omega$ may be seen as a representation of $\cp(G[i])$. The corresponding light representation $X$ of the $G$--canonical algebra is stable (at weak coupling) if and only if $Y$ is stable as a representation of $G[i]_\Omega$ with respect to the induced central charge $Z(\cdot)_\text{ind}$. By the KS wall--crossing formula and the argument in \S.\,\ref{chambindependence}, the set of dimension vectors of stable light representations is independent of the chamber. Hence, if the representation $Y$ is stable for a \textit{particular} choice of a stability function $\tilde Z(\cdot)$ on the quiver $G[i]_\Omega$, there is a stable light representation $X$ of the original $G$--canonical algebra with dimension \eqref{dimveeector} in \textit{all} weakly--coupled chambers. 

We have only to check that given a Schur root $\alpha$ of $G[i]_\Omega$ we can find a \textit{particular} chamber in which a representation $Y$ of dimension $\dim Y=\alpha$ is stable. We choose our particular BPS chamber in the form
\begin{equation}\label{ppbbnnmp}
\tilde Z(\cdot)= Z_0(\cdot) + t\, \Big(\langle \cdot, \alpha\rangle_E-\langle \alpha,\cdot\rangle_E\Big),\qquad t\rightarrow +\infty.
\end{equation}
 where $\langle\cdot,\cdot\rangle_E$ is the Euler form of the quiver $G[i]_\Omega$. Let $Y$ be the generic representation of $G[i]_\Omega$ of dimension $\alpha$. The condition of stability of $Y$ with respect to $\tilde Z(\cdot)$ is then equivalent to the characterization of the Schur roots, see \cite{schofield} \textbf{Theorem 6.1}. 

The last statement follows from \textbf{Theorem 3.7} of \cite{schofield}: \textit{if $\alpha$ is an imaginary Schur root with $(\alpha,\alpha)_{\widetilde{C}}<0$, then $n\,\alpha$ is also an imaginary Schur root for all $n\in\mathbb{N}$.}  

Note that in the chamber \eqref{ppbbnnmp} all BPS states in the Regge trajectory of $\alpha$ will be \emph{simultaneously stable.} \hfill $\square$
\vglue 12pt

Let $G[i]$ be a Dynkin graph of \textit{negative type} (that is, neither an $ADE$ Dynkin graph nor an affine one) and let $M\subset \Gamma_\mathrm{root}$ be its fundamental set \cite{kac1,kac2}; the set of the imaginary roots is given by \cite{kac1,kac2}
$$\Delta^\mathrm{im}(G[i])=\bigcup_{w\in \mathrm{Weyl}(G[i])}w(M).$$ By \textbf{Lemma 1} of \cite{kac2} an imaginary root $\alpha\in M$ with $(\alpha,\alpha)_{\widetilde{C}}<0$ is always a Schur root. Therefore, if $G[i]$ is of negative type the light spectrum contains some (and hence infinitely many) higher spin states.
In conclusion, for `Lagrangian' $G$--canonical algebras we have three possibilities, corresponding to the three classes of Kac--Moody algebras:
\begin{description}
 \item[(P)] the light spectrum consists of $W$--bosons making one copy of the adjoint rep.\! of $G$ and (finitely many) hypermultiplets (`quarks') in definite representations of $G$ $\Longleftrightarrow$ the augmented Dynkin graph $G[i]$ is a Dynkin graph of the finite $ADE$ type;
\item[(Z)] the light BPS spectrum contains, besides the $W$--bosons of $G$, one additional vector multiplet, as well as infinitely many hypermultiplets, but \emph{no} higher spin BPS state
$\Longleftrightarrow$ the augmented Dynkin graph $G[i]$ is a an \emph{affine} $\widehat{A}\widehat{D}\widehat{E}$ Dynkin graph;
\item[(N)] the `light' (!!) BPS spectrum contains infinitely many states of arbitrarily large mass and spin organized into Regge trajectories $s\sim \mathrm{const.}\, m^2$.  
 \end{description}

Clearly, the case \textbf{(N)} cannot correspond to a UV complete QFT. Indeed, \textbf{(N)} is the same wild behavior which in the canonical case characterizes the models having a Landau pole in the UV.

\subsection{The Dynkin case \textbf{(P)}}

\subsubsection{Graphical criterion for aymptotic freedom}

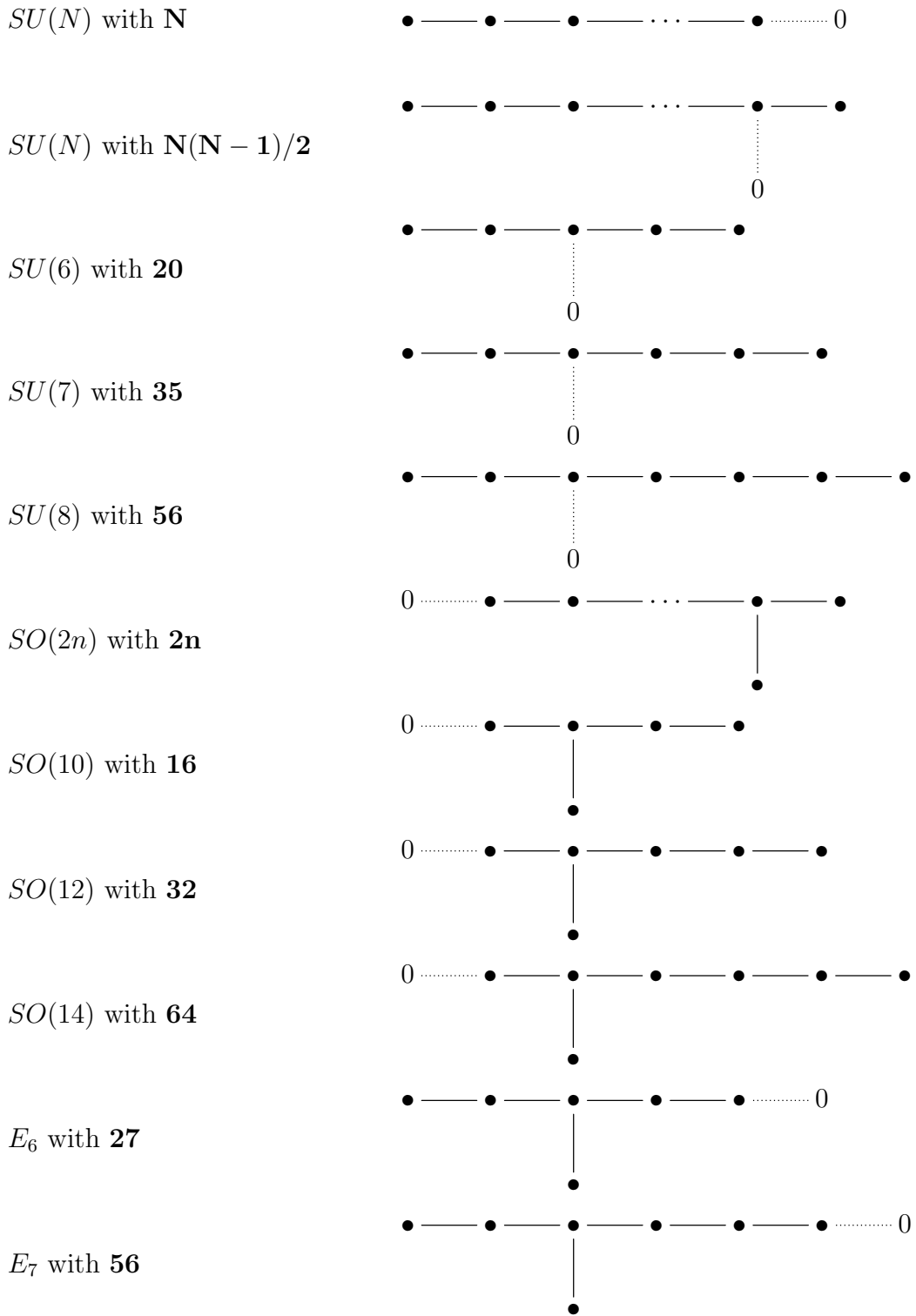
\begin{figure}
\begin{align*}
 &SU(N)\ \text{with }\mathbf{N}  &&\begin{gathered}\xymatrix{\bullet \ar@{-}[r] & \bullet\ar@{-}[r] & \bullet \ar@{-}[r] &\cdots \ar@{-}[r] & \bullet \ar@{..}[r] & 0}\end{gathered}\\
\\
&SU(N)\ \text{with }\mathbf{N(N-1)/2}  &&\begin{gathered} \xymatrix{\bullet \ar@{-}[r] & \bullet\ar@{-}[r] & \bullet \ar@{-}[r] &\cdots \ar@{-}[r] & \bullet \ar@{-}[r] & \bullet\\
& & & & 0\ar@{..}[u]}\end{gathered}\\
&SU(6)\ \text{with }\mathbf{20}  &&\begin{gathered} \xymatrix{\bullet \ar@{-}[r] & \bullet\ar@{-}[r] & \bullet \ar@{-}[r]  & \bullet \ar@{-}[r] & \bullet\\
& &  0\ar@{..}[u]}\end{gathered}\\
&SU(7)\ \text{with }\mathbf{35}  &&\begin{gathered} \xymatrix{\bullet \ar@{-}[r] & \bullet\ar@{-}[r] & \bullet \ar@{-}[r]  & \bullet \ar@{-}[r] & \bullet\ar@{-}[r]& \bullet\\
& &  0\ar@{..}[u]}\end{gathered}\\
&SU(8)\ \text{with }\mathbf{56}  &&\begin{gathered} \xymatrix{\bullet \ar@{-}[r] & \bullet\ar@{-}[r] & \bullet \ar@{-}[r]  & \bullet \ar@{-}[r] & \bullet\ar@{-}[r]& \bullet\ar@{-}[r] &\bullet\\
& &  0\ar@{..}[u]}\end{gathered}\\
&SO(2n)\ \text{with }\mathbf{2n}  &&\begin{gathered}\xymatrix{0 \ar@{..}[r] & \bullet\ar@{-}[r] & \bullet \ar@{-}[r] &\cdots \ar@{-}[r] & \bullet \ar@{-}[r] & \bullet\\
& & & & \bullet\ar@{-}[u]}\end{gathered}\\
&SO(10)\ \text{with }\mathbf{16}  &&\begin{gathered} \xymatrix{0\ar@{..}[r] & \bullet\ar@{-}[r] & \bullet \ar@{-}[r]  & \bullet \ar@{-}[r] & \bullet\\
& &  \bullet\ar@{-}[u]}\end{gathered}\\
&SO(12)\ \text{with }\mathbf{32}  &&\begin{gathered} \xymatrix{0 \ar@{..}[r] & \bullet\ar@{-}[r] & \bullet \ar@{-}[r]  & \bullet \ar@{-}[r] & \bullet\ar@{-}[r]& \bullet\\
& &  \bullet\ar@{-}[u]}\end{gathered}\\
&SO(14)\ \text{with }\mathbf{64}  &&\begin{gathered} \xymatrix{0 \ar@{..}[r] & \bullet\ar@{-}[r] & \bullet \ar@{-}[r]  & \bullet \ar@{-}[r] & \bullet\ar@{-}[r]& \bullet\ar@{-}[r] &\bullet\\
& &  \bullet\ar@{-}[u]}\end{gathered}\\
&E_6\ \text{with }\mathbf{27}  &&\begin{gathered} \xymatrix{\bullet \ar@{-}[r] & \bullet\ar@{-}[r] & \bullet \ar@{-}[r]  & \bullet \ar@{-}[r] & \bullet\ar@{..}[r]& 0\\
& &  \bullet\ar@{-}[u]}\end{gathered}\\
&E_7\ \text{with }\mathbf{56}  &&\begin{gathered} \xymatrix{\bullet \ar@{-}[r] & \bullet\ar@{-}[r] & \bullet \ar@{-}[r]  & \bullet \ar@{-}[r] &\bullet \ar@{-}[r] &\bullet\ar@{..}[r]& 0\\
& &  \bullet\ar@{-}[u]}\end{gathered}\\
\end{align*}
\caption{\label{graphicalrule2} The augmented Dynkin graphs which are themselves $ADE$ Dynkin graphs. }
\end{figure}

In figure \ref{graphicalrule2} (page \pageref{graphicalrule2}) we present all augmented graphs $G[i]$ which are finite--type
Dynkin diagrams, where the node $i\in G$ is identified with the fundamental representation $R_{i}$ of $G$ with highest weight the fundamental weight
$$\omega_{i}\equiv [0,\dots, 0,\overbrace{1}^{i}, 0,\cdots 0].$$ Comparing the figure \ref{graphicalrule2} and table \ref{niceres} we observe the remarkable\vglue 9pt

\textbf{Fact.} \textit{$\cn=2$ SYM with gauge group $G=ADE$ coupled to a quark in the fundamental representation $R_{i}$ is \emph{asymptotically free} if and only if the graph $G[i]$ is a Dynkin graph of finite--type. }
\smallskip

For $G=ADE$ there is just another asymptotically--free representation which is not fundamental, namely the two--index symmetric of $SU(n)$. At least at the level of state counting, this non--fundamental case looks quite similar to the fundamental ones with the $SU(n)$ Dynkin graph augmented to the $Sp(2n)$ one. However, following this suggestion would lead to a discussion of the non--simply laced case which is beyond my present understanding.

One also observes that, whenever a $\cn=2$ SYM coupled to a single quark in the representation $R_{i}$ of $G$ is a UV--superconformal QFT, the graph $G[i]$ is affine and $i$ is an extension node, that is, the `naive' heavy quark quiver would lead to the case \textbf{(Z)}. But in the conformal case that `naive' quiver is really \emph{too} naive to be trusted.

\subsubsection{Light BPS spectrum}

From \S.\,\ref{sec:kac} we know that the light BPS representations in the non--homogeneous $G$--tube of canonical type \eqref{specialcasse} have dimension vectors equal to eqn.\eqref{dimveeector} where $\dim Y$ is a positive root of $G[i]$. 
The map $\dim X\rightarrow \dim Y$ identifies the zero--magnetic charge sublattice of the original theory with the root lattice $\Gamma_{G[i]}$ of the enhanced gauge group $G[i]$. According to the discussion leading to eqn.\eqref{rwquiv}, we have two alternative natural identifications between the  vectors in
$\Gamma_{G[i]}$ and the physical quantum numbers, depending if $Y$ arises from a representation $X$ of the canonical or the Coxeter--Dynkin form of the quiver. From eqn.\eqref{uuuwuwu} one sees that the two identifications are related by the action of the simple reflection $s_\bullet\in \mathrm{Weyl}(G[i])$, which leaves invariant $\Gamma_{G[i]}$, so that, as expected, the physical spectrum is independent of the choice.

We use the simpler Coxeter--Dynkin identification of the physical quantum numbers. Let $f$ be the zero--eigenvector of the exchange matrix $B$ of the Coxeter--Dynkin $G$--quiver with $\{\kappa_j\}$ as in eqn.\eqref{specialcasse}. The vector $f$ is a purely flavor charge corresponding, up to normalization, to the baryon number $B$.
Let $\tilde C_{ab}$ be the Cartan matrix of $G[i]$ (with the convention that the index value $a=1$ corresponds to the enhancement node $\bullet$). One has the identity
\begin{equation}
(f_\bullet, f_{\diamond_1}, \cdots, f_{\diamond_{r(G)}})\,\tilde C= (1, 0,\cdots, 0),
\end{equation}  
where we chose the canonical normalization. Hence we see that under the identification of the zero--magnetic charge sublattice with $\Gamma_{G[i]}$ the baryon number is given by the Cartan generator of $G[i]$
\begin{equation}
B\equiv f_\bullet\, \alpha_\bullet^\vee+ \sum_i f_{\diamondsuit_i}\,\alpha^\vee_{\diamondsuit_i}\in \mathfrak{h}(G[i]),
\end{equation}
such that
\begin{equation}
\alpha_\bullet(B)=1,\qquad \alpha_{\diamondsuit_i}(B)=0.
\end{equation}

Then the matter light states in the $G$--tube of type \eqref{specialcasse} may be organized according to a fictitious Higgs symmetry breaking as follows.
One considers the Lie group $G[i]$ broken down to its subgroup $G\times U(1)_B$ by an adjoint representation Higgs field $\Phi$, of the form
\begin{equation}
\Phi = v\, B,\qquad v\gg 1.
\end{equation}
We decompose the adjoint representation of $G[i]$ into irreducible representations of $G\times U(1)_B$
\begin{equation}
 \big(\mathrm{adjoint}\: G[i]\big) = \big(\mathrm{adjoint}\: G\big) \oplus\left(\bigoplus_{B\geq 1} \big(R_B\oplus \overline{R}_{-B}\big)\right)  
\end{equation}

Then:
\textit{The light matter contained in a canonical $G$--tube of type \eqref{specialcasse} consists of hypermultiplets of baryon number $B=1,2,\cdots$ in the representations $R_B$ of the gauge group $G$. In particular, the states of baryon number $1$ correspond to a single quark in the representation $R_{i}$ as expected from the `naive' physical picture.}

Note that the matter content is always organized in complete representations of the gauge group $G$, as it should be.\medskip

In table \ref{ppppprrr} we write down the hypermultiplet spectra for all pairs $(G, R_{i})$ such that $G[i]$ is a Dynkin graph. In the last column we have put a check mark, \checkmark, whenever the spectrum coincides (up, possibly, to IR decoupled free fields) with the one expected on (naive) physical grounds. Comparing with table \ref{niceres} we see that this happens precisely when $R_{i_0}$ is a `nice' representation of $G$ in the sense of ref.\!\cite{Tack}. 

This makes sense: if $R_{i}$ is nice, the theory with one quark in the representation $R_{i}$ of the gauge group $G$ may be geometrically engineered in Type IIB, and hence has a quiver $(Q,\cw)$ by the arguments of \cite{CNV}. The physical properties of this quiver $(Q,\cw)$ forces us to identify it with the $G$--canonical quiver of type $(i_0,\{2\})$, see ref.\!\cite{ACCERV2}. But there is no reason to expect that SYM coupled to a generic representation $R$ will produce a \emph{quiver} $\cn=2$ theory: in general, we expect that we have to `complete' the theory with additional degrees of freedom to get a quiver $\cn=2$ theory. Then, in the non--`nice' cases, the above spectra may be thought of as a sort of `minimal' completion to such a quiver theory. Notice, however, that the `completion' in the last row of the table, marked with a double exclamation mark !! is not compatible with asymptotic freedom.  
 
\begin{table}
\begin{center}
 \begin{tabular}{|c|c|c|c|}\hline
$G$ & $R_{i_0}$ & $\bigoplus_{B\geq 1}R_B$ & \\\hline
$SU(N)\phantom{\Big|}$ & $\mathbf{N}$ & $\mathbf{N}_1$ & \checkmark\\\hline 
$SU(N)\phantom{\Big|}$ & $\mathbf{\tfrac{1}{2}N(N-1)}$ & $\mathbf{\tfrac{1}{2}N(N-1)}_1$ & \checkmark\\\hline 
$SU(6)\phantom{\Big|}$ & $\mathbf{20}$ & $\mathbf{20}_1\oplus \mathbf{1}_2$ & \checkmark\\\hline
$SO(2N)\phantom{\Big|}$ & $\mathbf{2N}$ & $\mathbf{2N}_1$ & \checkmark\\\hline
$S0(10)\phantom{\Big|}$ & $\mathbf{16}$ & $\mathbf{16}_1$ & \checkmark\\\hline
$SO(12)\phantom{\Big|}$ & $\mathbf{32}$ & $\mathbf{32}_1\oplus \mathbf{1}_2$ & \checkmark\\\hline
$E_6\phantom{\Big|}$ & $\mathbf{27}$ & $\mathbf{27}_1$ & \checkmark\\\hline
$E_7\phantom{\Big|}$ & $\mathbf{56}$ & $\mathbf{56}_1\oplus \mathbf{1}_2$ & \checkmark\\\hline\hline
$SO(14)\phantom{\Big|}$ & $\mathbf{64}$ & $\mathbf{64}_1\oplus \mathbf{14}_2$ & \\\hline
$SU(7)\phantom{\Big|}$ & $\mathbf{35}$ & $\mathbf{35}_1\oplus \mathbf{7}_2$ & \\\hline
$SU(8)\phantom{\Big|}$ & $\mathbf{56}$ & $\mathbf{56}_1\oplus \mathbf{28}_2\oplus \mathbf{8}_3$ & !!\\\hline
 \end{tabular} 
\end{center}
\caption{\label{ppppprrr} The BPS hypermultiplets of the $G$--canonical algebra of type $(i_0,\{2\})$.}
\end{table} 

\subsubsection{Independence from $\cw$}

There is a possible objection to the above conclusions which we wish to rule out. In many instances, when matching the known physical spectrum of a $\cn=2$ theory with $\mathsf{rep}(Q,\cw)$ for some  candidate pair $(Q,\cw)$, one finds all the physical BPS states of the given theory but also, in addition, some spurious states with higher charge (\textit{i.e.}\! dimension). In such cases a common remedy is to add higher order terms to $\cw$ which have no effect on the small dimension representations we wish to keep, but kill the unwanted higher dimensional ones.

May this remedy be used here to fix the hypermultiplet spectrum of naive `heavy quark' quiver $Q$ for the asymptotically--free non--`nice' representations? 

The answer is no. The point is that, for convenient choices of the weakly--coupled chamber, the offending stable representations are in fact representations of a Dynkin quiver $G[i]_\Omega$, that is, for each pair of opposite arrows in the double quiver $\cp(G[i])$ one arrow must vanish. In particular, the extra condition stated after eqn.\eqref{cconclusion} is automatically satisfied. This means than any cycle in $\cp(G[i])$ of length $>2$ has at least two arrows which vanish in such a representation. Going back to the original canonical quiver $Q(\{\kappa_i=2\,\delta_{i,i_0}\})$ we see that all cycles longer than those already present in $\cw$ have necessarily at least two arrows set to zero in any such representation. The gradient of all higher order terms then identically vanishes on the troublesome brick representations, which are not lifted away. Since the spectrum is weak--coupling chamber independent, considering more general chambers will not help.

\subsection{Fancier possibilities}
Up to now we have always chosen the special points $\zeta_{a,i}\in\mathbb{P}^1$ to be pairwise distinct. We may wonder what we would get but choosing some of them to be coincident, say
$$\zeta_{a_1,i_1}=\zeta_{a_2,i_2}=\cdots \zeta_{a_t,i_t}\equiv\zeta_0.$$  Clearly the light stable states contained in $\mathscr{L}_{\zeta_0}$ will be associated to (special) bricks of the preprojective algebra $\cp(G[i_1,\dots, i_t])$, where $G[i_1,\dots, i_t]$ is the unoriented graph obtained by attaching $t$ extra nodes to the Dynkin quiver of  $G$, the $\ell$--th such new node being attached by a single link to the $i_\ell$-th node of $G$. If $G[i_1,\dots, i_t]$ is also a finite--type Dynkin graph we are guaranteed that only finite many light hypers are present in the BPS spectrum. Their quantum numbers may be obtained following the branching of the adjoint representation of $G[i_1,\dots, i_t]$ into representations of the subgroup $G\times U(1)^t$. \textit{E.g.}\! for $t=2$ and $G=SU(N)$ we get (a part for a few exceptional cases for low $N$) the two matter contents $\mathbf{N}\oplus \mathbf{N}\oplus \mathbf{1}$ for $G[i_1,i_2]=A_{N+1}$, and $\mathbf{N}\oplus \mathbf{N}\oplus \mathbf{N(N-1)/2}$ for $G[i_1,i_2]=D_{N+1}$, both of which satisfy the `nice' bound $b\leq h(G)$.

\section{Non--Lagrangian canonical matter?}

The `Lagrangian' $G$--canonical algebras of previous section correspond (roughly) to the obvious $\cn=2$ models obtained by coupling colored quarks to SYM \cite{ACCERV2} with some subtlety when the matter is in large representations of the gauge group. Those models are not very interesting: The real purpose of classifying $\cn=2$ theories is to find the \emph{non}--obvious models which cannot be obtained by elementary constructions.

In the limited context of the $G$--canonical algebras we may ask which ones (if any) of the non--`Lagrangian' algebras correspond to meaningful QFTs. For $G=SU(2)$ we know the answer: a canonical algebra corresponds to a QFT if and only if its Euler characteristic $\chi\geq 0$. For higher rank gauge groups $G$ a similar statement is not known.
Certainly the $G$--canonical algebras have many properties typical of the algebra of a quiver $\cn=2$ QFT: they enjoy the Ringel property and, as we show in section \ref{Proofconjectr}, they are consistent with (the necessary part of the) $2d/4d$ correspondence. 

One may think that a good criterion for a $G$--canonical algebra to correspond to a QFT is that its light spectrum is physically \emph{reasonable}.
However here we find a conceptual difficulty. If $G\neq SU(2)$ the $G$--canonical algebras are not complete in the sense of \cite{CV11}, meaning that not all its chambers correspond to physically realizable situations. On the other hand, the light weak--coupling spectrum of a  non--`Lagrangian' $G$--canonical algebra is also chamber--dependent. A non--`Lagrangian' $G$--canonical algebra may have a spectrum which is physically reasonable in some (weak coupling) chambers and not reasonable in others. This would not be a problem as long as the chambers with a non--reasonable spectrum are not physically realizable: Indeed, \textit{all} $\cn=2$ quiver gauge theories with $G\neq SU(2)^k$ have some non--physical chambers with crazy (formal) BPS spectra\,! Thus the `meaningful spectrum criterion' is useless, unless one knows which chambers are physical. The physical chambers are known (at weak coupling) for a Lagrangian theory, where this information is not needed, but not in general.

It is easy to see that a chamber with a sensible spectrum exists for all $\mathsf{C}(G,\{\kappa_i\})$ with
\begin{equation}\kappa_{j}= \begin{cases}\{n_1,n_2,\cdots,n_k\} & j=i\\
\emptyset & j\neq i,\end{cases}
\end{equation}
 such that the fundamental representation $R_{i}$ of $G$ is `nice'. Again, localizing at a special point in $\mathbb{P}^1$, we are reduced to the case $\kappa_{i}= \{n\}$, $n\geq 3$. Then, going to the Coxeter--Dynkin version of the quiver
\begin{equation}\label{ccooddy}
 \begin{gathered}
  \xymatrix{  \cdots\ar[r] &\diamondsuit_{i} \ar[dr] & \cdots\ar[l]\\
&& \bullet_1\ar[ld] & \bullet_2 \ar[l] & \bullet_3 \ar[l] & \cdots \ar[l] & \bullet_{n-1} \ar[l] \\
\cdots \ar[r] & \spadesuit_{i} \ar@<1.1ex>[uu]\ar@<0.1ex>[uu] & \cdots\ar[l]}
 \end{gathered}
\end{equation}  
we see that we can construct a weak--coupling central charge $Z(\cdot)$ such that the light stable representations are those of the `Lagrangian' $n=2$ subquiver together with $n-2$ neutral hypermultiplets corresponding to the simple representations
$S_{\bullet_k}$, for  $k=2,3,\dots, n-1$ (cfr.\! figure \eqref{ccooddy}). These last hypermultiplets have no $G$--gauge interaction, but they do not decouple in the IR since, in the non--`Lagrangian' case, the matter sector is an interacting system in its own right. 

More generally, the light stable representations of the Coxeter--Dynkin quiver would consist in finitely many hypers in \textit{all} weakly--coupled chambers (covered by this quiver) under the assumption that the augmented Dynkin graph $G[i,\{n\}]$, obtained by attaching to the node $i$ of $G$ a branch of length $n$ (counting also the branching node), is still an $ADE$ Dynkin diagram, see table \ref{uuururur}.

\begin{table}
 \begin{center}
  \begin{tabular}{|c|c|c|}\hline
$G$ & $R_{i}$ & allowed $n\geq 3$\\\hline
$SU(N)$ & $\mathbf{N}$ & $n\geq 3$\\\hline
$SU(5)$ & $\mathbf{10}$ & $n=3,4,5$\\\hline
$SU(6)$ & $\mathbf{20}$ & $n= 3$\\\hline
$SU(7)$ & $\mathbf{35}$ & $n=3$\\\hline
$SO(2n)$ & $\mathbf{2n}$ & $n\geq 3$\\\hline  
$E_6$ & $\mathbf{27}$ & $n=3$\\\hline 
  \end{tabular} 
 \end{center}
\caption{\label{uuururur} Dynkin graphs of the form $G[i,\{n\}]$ with $n\geq 3$.}
\end{table}

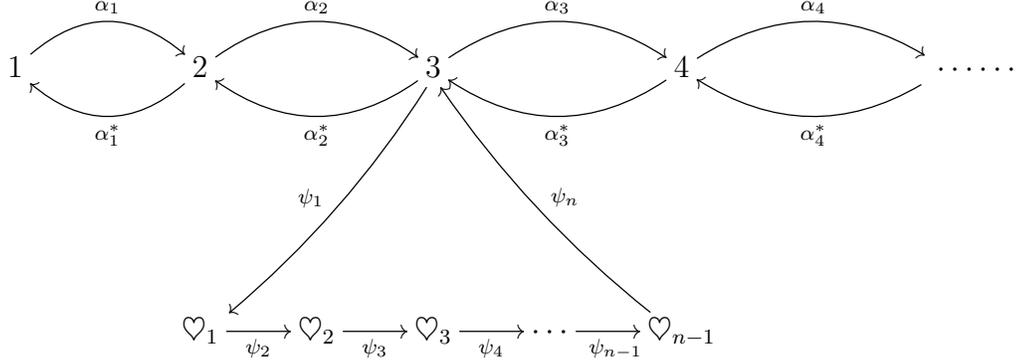
\begin{figure}
 \begin{equation*}
\begin{gathered}
\xymatrix{{1} \ar@/^1.5pc/[rr]^{{\alpha_1}} && {2}
\ar@/^1.5pc/[ll]^{{\alpha_1^*}}\ar@/^1.5pc/[rr]^{{\alpha_2}}
&& {3}\ar@<0.3ex>@/^0.5pc/[dddll]_{\psi_1}
\ar@/^1.5pc/[ll]^{{\alpha_2^*}}\ar@/^1.5pc/[rr]^{{\alpha_3}}
&& {4} 
\ar@/^1.5pc/[ll]^{{\alpha_3^*}}\ar@/^1.5pc/[rr]^{{\alpha_4}}
&& \ar@/^1.5pc/[ll]^{{\alpha^*_4}}\qquad\cdots\cdots\\
\\
\\
&& \heartsuit_1 \ar[r]_{\psi_2} & \heartsuit_2 \ar[r]_{\psi_3} &\heartsuit_3 \ar[r]_{\psi_4}& \cdots\ar[r]_{\psi_{n-1}} &\heartsuit_{n-1}\ar@<0.3ex>@/^0.5pc/[uuull]_{\psi_n}
}\end{gathered}
\end{equation*}
\caption{\label{luianexppl} $\cp(G[3])$ with the node $\bullet$ blown up in the $n-1$ nodes $\heartsuit_\alpha$.}
\end{figure}

However, this is by no means the end of the story. In the non--Lagrangian case the argument that the light stable representations of the Coxeter--Dynkin quiver are equivalent to those of the $G$--canonical quiver breaks down. In fact, it is easy to see that almost all the $G$--canonical algebras of table \ref{uuururur} do have (unphysical?) weak--coupling chambers with infinitely many light stable particles. Indeed, a light stable representation $X$ of the $G$--canonical quiver of type
$\kappa_i=\{n\}$, may be seen as a representation of the quiver obtained by `blowing up' the node $\bullet$ of $\cp(G[i])$ into $n-1$ nodes $\heartsuit_\alpha$, see figure \ref{luianexppl}.

If $X$ is a brick of the non--homogeneous $G$--tube, $\psi_n\psi_{n-1}\cdots \psi_1=0$. Consider, say, the class of representations $Y$ of the `blown--up' quiver with $\psi_k=0$. A brick in this class may be seen as a brick $Y$ in the preprojective algebra
of the graph $G[i,\{n,k\}]$ obtained by attaching to the node $i_0$ of $G$ \emph{two} branches of length $k-1$ and $n-k$, respectively. ($Y$ is a special brick of $\cp(G[i,\{n,k\}])$ satisfying the extra condition $Y_{\psi^*_j}=0$ for all $j$). If $G[i,\{n,k\}]$ is not a Dynkin graph for all $k=1,2,\dots,n$, there are necessarily some bricks $Y$ with $Y_{\psi^*_j}=0$ and $\dim Y$ an imaginary root of the Kac--Moody algebra $G[i,\{n,k\}]$.  By the same argument as in eqn.\eqref{ppbbnnmp} we may cook up a weakly coupled chamber where any such brick is stable; if $\dim Y\in M$ and $(\dim Y,\dim Y)_{\widetilde{C}}<0$ all particles in the infinite Regge trajectory
$n\,\dim Y$ would also be stable in the \emph{same} chamber. 

In conclusion, the only non--Lagrangian case in which we get --- in all weakly--coupled BPS chambers covered by the $G$--canonical form of the quiver --- a finite matter spectrum consisting only of hypermultiplets is $G=SU(N)$, $R_{i}=\mathbf{N}$ and $n=3$.
 This fact, by itself, does not guarantee that a corresponding QFT exists, nor the failure of this property would make the theory \emph{a priori} inconsistent, as long as we have no control on which BPS chambers may or may not be physically realized (the relevant BPS chambers have weak Yang--Mills coupling, but the theory being non--Lagrangian, the model is intrinsically strongly coupled in the matter sector).

\section{$2d/4d$ correspondence and the Coxeter element}\label{Proofconjectr}

The $2d/4d$ correspondence \cite{CNV} states that the quiver $Q$ of a four--dimensional $\cn=2$ theory is the BPS quiver of a two--dimensional $(2,2)$ theory \cite{CV92} with $\hat c_\mathrm{uv}\leq 2$.

In concrete terms \cite{CV92}, a 2d $(2,2)$ model is encoded in a Picard--Lefshetz mutation--class of {integral unipotent} matrices\footnote{\ $S$ is called the \emph{Stokes matrix}, since it is the Stokes matrix of the Riemann--Hilbert problem \cite{CV92} solving the $tt^*$ equations \cite{tt*} of the 2d $(2,2)$ model.} $S\in SL(n,\Z)$ such that
 the 2d quantum monodromy $M=(S^{-1})^tS$ has eigenvalues of absolute value $1$, or equivalently\footnote{\ Traditionally both the cyclotomic polynomials and the Coxeter element are denoted by the symbol $\Phi$. Hoping to avoid confusion, we use the boldface symbol $\boldsymbol{\Phi}$ for the cyclotomic polynomials, reserving $\Phi$ for the Coxeter element.},
\begin{equation}\label{cycloo}
\det\!\big[x-(S^{-1})^tS\big]=\prod_{m\geq 1} \boldsymbol{\Phi}_m(x)^{k_m},\qquad \Big[\boldsymbol{\Phi}_m(x)\ \text{the $m$--th cyclotomic polynomial}\Big].
\end{equation}
In the Kontsevich--Soibelman language \cite{KS} $S$ is group element associated to the arc $0\leq \theta <\pi$ in a stability condition on the Lie algebra $\mathfrak{sl}(n)$.

The implications of the restriction to $\hat c_\mathrm{uv}\leq 2$ are less obvious, but, seeing the $tt^*$ geometry as a variation of mixed Hodge structures and comparing weight filtrations, one easily convinces himself that the following condition is true for all 
 $\hat c_\mathrm{uv}\leq 2$ $(2,2)$ models
 \begin{equation}\label{cvvv}
 \text{\textbf{the minimal polynomial of} }(S^{-1})^tS\ \text{\textbf{divides} } (x-1)^3\left(\frac{x^\ell-1}{x-1}\right)^{\!\!2}\quad \text{for some }\ell\in \mathbb{N}.
 \end{equation}
 
 Then the $2d/4d$ correspondence just states that the exchange matrix of the quiver $Q$ of the 4d theory is 
 \begin{equation}
 B(Q)=S^t-S
 \end{equation}
with $S$ the Stokes matrix of the corresponding 2d one. The correspondence is consistent with mutations, since mutations in 2d and 4d agree (\textit{i.e.}\! tilting is an isometry of the Euler form \cite{ASS1}).\smallskip

In the context of the representation theory of a (basic) associative algebra $\mathcal{A}$ of finite global dimension there is an obvious realization of the above 2d structure: one defines the inverse Stokes matrix as
\begin{equation}
(S^{-1})_{ij}= \dim(e_i\,\mathcal{A}\, e_j),
\end{equation}
where $\{e_i\}_{i\in I}$ is a complete set of primitive orthogonal  idempotents of $\ca$. Then $S$ is the matrix of the Euler form on $\mathsf{mod}\,\ca$ (see appendix \ref{appeuler} for details)
\begin{equation}
(\dim X)_i\, S_{ij}\,(\dim Y)_j = \langle X, Y\rangle_E \equiv \sum_{k=0}^\infty (-1)^k \, \dim \mathrm{Ext}^k(X,Y).
\end{equation} 
With these identifications, the quantum monodromy $(S^{-1})^tS$ is equal to \emph{minus} the Coxeter element $\Phi$ of the algebra $\ca$, and (by Kronecker's theorem) the condition \eqref{cycloo} just says that $\Phi$ has spectral radius $1$. Eqn.\eqref{cvvv} is then a restriction on the size of the Jordan blocks of $\Phi$: they are at most $2\times 2$, except that the eigenvalue $-1$ may possibly have a $3\times 3$ Jordan block. 
An elementary but useful property is
\begin{equation} \label{tuqpoe}\boldsymbol{(\star)}\qquad\begin{aligned}&\text{the number of linear independent $-1$ eigenvectors of  $\Phi$} =\\
&=\text{the corank of }B(Q)=\\
&= \text{the rank of the flavor symmetry group} \end{aligned}
\end{equation}

Using the above identifications, the $2d/4d$ correspondence \cite{CNV} becomes a statement about the spectral properties of the algebra $\ca$: 
\smallskip

\textbf{Statement.}\label{statata} ($2d/4d$ correspondence) \textit{Let $(Q,\cw)$ be the quiver of a 4d $\cn=2$ gauge theory and $\ca_J$ its Jacobian algebra. Then $\ca_J$ is the $3$--CY completion of an algebra $\ca^\prime$ whose Coxeter element has spectral radius $1$ and Jordan blocks of size no more than $3\times 3$, blocks of size $3\times 3$ being associated to the eigenvalue $-1$. For \emph{complete} $\cn=2$ theories, the Jordan blocks are at most $2\times 2$ and a Jordan block of maximal size is associated to the eigenvalue $+1$.}
\smallskip

The last sentence follows from the fact  \cite{CV11} that a complete 4d $\cn=2$ theory corresponds to a 2d (2,2) model with $\hat c_\mathrm{uv}\leq 1$.
\vglue 6pt

\textbf{Remark.} In general, a Jacobian algebra $\ca_J$ may be seen as a $3$-CY completion of some basic algebra $\ca^\prime$ with $\mathrm{gl.dim.}\,\ca^\prime\leq 2$ in more than one way. However, the exchange matrix $B(Q)\equiv S^t-S$ will not depend on the choice of $\ca^\prime$, as it is equal to the exchange matrix of the quiver of $\ca_J$. The $2d/4d$ correspondence just requires the existence of one unipotent matrix $S$ such that $B(Q)\equiv S^t-S$ and $\mathrm{spec.radius}((S^{-1})^tS)=1$, and hence it is satisfied if we can find \textit{one} algebra $\ca^\prime$ with a Coxeter element with the stated properties.

\subsection{Proof of $2d/4d$ for the $G$--canonical theories}

Now we show that the $G$--canonical theories satisfy (the necessary part of) the $2d/4d$ correspondence as stated above. This is well known for the $SU(2)$--canonical models \cite{COXX,kell33}. Here we extend the argument to general $G$.

Let us consider first the special `Lagrangian' case of section \S.\,\ref{iiiiiaaa}. We begin by revisiting pure SYM.

\subsubsection{Triangle form of pure SYM}

It is convenient to describe the pure $G$--SYM in terms of different quivers in the same mutation--class as the square--tensor product $\mathbb{G}$, namely in terms of \emph{triangle} tensor products \cite{kellerP}\!\!\cite{CNV}.
Let $\overrightarrow{G}$ a quiver whose underlying graph is the Dynkin diagram $G$, and let $\C\overrightarrow{G}$ be the corresponding path algebra. Following Keller \cite{kellerP,kellerR}, we consider the tensor product algebra $\mathcal{A}_G= \C\overrightarrow{G}\otimes\C \widehat{A}(1,1)$, and let $\mathcal{A}_\mathrm{CY}=\Pi(\mathcal{A}_G)$ be its $3$--CY completion with respect to the obvious commutation relations\footnote{\ Notation: $e_i$ denotes the lazy path (idempotent) at the $i$--th node; $s,t\colon Q_1\rightarrow Q_0$ are the source and target maps, respectively.}
\begin{equation}
\big(\phi \otimes e_{t(\psi)}\big)\big(e_{s(\phi)}\otimes \psi\big)=\big(e_{t(\phi)}\otimes \psi\big)\big(\phi\otimes e_{s(\psi)}\big).
\end{equation} 

The completed pair $(Q,\cw)$ so obtained --- called the \textit{triangle tensor product} $\overrightarrow{G}\,\boxtimes\, \widehat{A}(1,1)$ of the two quivers $\overrightarrow{G}$ and $\widehat{A}(1,1)$  --- is mutation--equivalent to the standard quiver $\mathbb{G}$ of pure $G$ SYM \cite{kellerP,kellerR}\!\cite{CNV}.

In the following we take $\overrightarrow{G}$ to be the Dynkin quiver with the alternating orientation so that the reference node $i_0\in \overrightarrow{G}$ is a \emph{sink}. \textit{E.g.}\! for $G=A_4$ with $i_0=2$ we focus on the following form of the quiver (we use the standard convention that relations are represented in the quiver by inverse dashed arrows)
\begin{equation}\label{edfig}
 \begin{gathered}
  \xymatrix{\spadesuit_1\ar@<0.3ex>[d]\ar@<-0.3ex>[d] \ar[r] & \spadesuit_2\ar@<0.3ex>[d]\ar@<-0.3ex>[d]&\spadesuit_3\ar[l]\ar[r]\ar@<0.3ex>[d]\ar@<-0.3ex>[d]&\spadesuit_4\ar@<0.3ex>[d]\ar@<-0.3ex>[d]\\
\diamondsuit_1\ar[r] & \diamondsuit_2\ar@{..>}@<0.3ex>[ul]\ar@{..>}@<-0.3ex>[ul]\ar@{..>}@<0.3ex>[ur]\ar@{..>}@<-0.3ex>[ur] & \diamondsuit_3\ar[r]\ar[l] & \diamondsuit_4\ar@{..>}@<0.3ex>[ul]\ar@{..>}@<-0.3ex>[ul]}
 \end{gathered}
\end{equation}

A very convenient property of the triangle quiver is the factorization of the Euler forms
\begin{equation}\label{factorization}
 \big\langle \alpha\otimes x,\beta\otimes y\big\rangle_{\C\overrightarrow{G}\otimes\C \widehat{A}(1,1)}=
\big\langle \alpha, \beta\big\rangle_{\C\overrightarrow{G}}\:\cdot\: \big\langle x, y\big\rangle_{\C \widehat{A}(1,1)}
\end{equation}
As a consequence, one has\footnote{\ By abuse of language, we call the Stokes matrix (resp.\! Coxeter matrix) of the algebra $\ca_G\equiv \C\overrightarrow{G}\otimes\C \widehat{A}(1,1)$ the Stokes matrix (resp.\! the Coxeter matrix) of \emph{pure $G$ SYM}. }
\begin{equation}
S_\mathrm{SYM}=S_G\otimes S_{SU(2)}
\end{equation}
 where $S_G$ is the Stokes matrix for the Argyres--Douglas model of type $G=ADE$ and $S_{SU(2)}$ is the one for SYM with gauge group $SU(2)$. Consequently, the Coxeter element of pure SYM is
 \begin{equation}
 \Phi_\mathrm{SYM}=-\Phi_G\otimes \Phi_{\widehat{A}_1}
 \end{equation}
 where $\Phi_G$, $\Phi_{\widehat{A}_1}$ are the usual Coxeter elements of the Lie algebras $G$ and $\widehat{A}_1$, respectively.
\smallskip

Then\smallskip

\textbf{Corollary.} \cite{CNV} \textit{Pure SYM satisfies the $2d/4d$ correspondence. One has
\begin{equation}\label{purebeta}
(-1)^h\,\Phi_{SYM}^h(\alpha\otimes \varrho)=\alpha\otimes \varrho+ 2h\,m(\varrho)\,\alpha\otimes\delta,
\end{equation}where $h$ is the Coxeter number of $G$. The characteristic polynomial of $\Phi_{SYM}$ is $\chi_G(-x)^2$ where $\chi_G(x)$ is the characteristic polynomial for a Coxeter element $\Phi_G\in \mathrm{Weyl}(G)$ (see table \ref{lekele} for the factorization of $\chi_G(x)$ into cyclotomic polynomials $\boldsymbol{\Phi}_m(x)$). From eqn.\eqref{purebeta}
\begin{equation}
 \Big((-1)^h\,\Phi_{SYM}^h-1\Big)^2=0,
\end{equation}
and \underline{all} Jordan blocks have size $2$.} Comparing with the \textbf{Statement} on page \pageref{statata}, we see that SYM is a complete theory iff $h=2$ (of course, this also follows from the classification of \cite{CV11}).

\begin{table}
\begin{center}
 \begin{tabular}{|c|c|c|}\hline
Dynkin type & $\begin{array}{l}\text{cyclotomic}\\ \text{factorization}\end{array}$ & $h(G)$\\\hline
$A_n$ & $\prod\limits_{d|(n+1),\ d>1} \boldsymbol{\Phi}_d(x)$ & $n+1$\\\hline
$D_n$ & $\boldsymbol{\Phi}_2(x) \prod\limits_{d|2(n-1)\atop d\neq 1,\ d\neq n-1}\boldsymbol{\Phi}_d(x)$ & $2(n-1)$\\\hline
$E_6$ & $\boldsymbol{\Phi}_3(x)\,\boldsymbol{\Phi}_{12}(x)$ & $12$\\\hline
$E_7$ & $\boldsymbol{\Phi}_2(x)\,\boldsymbol{\Phi}_{18}(x)$ & $18$\\\hline
$E_8$ & $\boldsymbol{\Phi}_{30}(x)$ & $30$\\\hline  
 \end{tabular} 
\end{center}
\caption{\label{lekele} Characteristic polynomials $\chi_G(x)$ for the $ADE$ Coxeter elements.}
\end{table}

\subsubsection{`Lagrangian' canonical models as one--point (co)extension of algebras}

In the framework of the preceding section, coupling to SYM a massive quark in the representation $R_{i_0}$ just amounts to the $3$--CY completion of the one--point extension\footnote{\ For the benefit of the reader, the one--point (co)extensions are review in appendix \ref{oneppoiiint}.} algebra $\ca_G[M_{i_0}]$ at an indecomposable module $M_{i_0}$ of $\ca_G$  having dimension vector equal to the imaginary root $\delta_{i_0}$ of the $i_0$--th Kronecker subquiver, say, in the example of 
figure \eqref{edfig} with $i_0=2$
\begin{equation}\label{edfig2}
M_2\colon\quad \begin{gathered}
  \xymatrix{0\ar@<0.3ex>[d]\ar@<-0.3ex>[d] \ar[r] & \C\ar@<0.3ex>[d]^\lambda\ar@<-0.3ex>[d]_1&0\ar[l]\ar[r]\ar@<0.3ex>[d]\ar@<-0.3ex>[d]&0\ar@<0.3ex>[d]\ar@<-0.3ex>[d]\\
0\ar[r] & \C\ar@{..>}@<0.3ex>[ul]\ar@{..>}@<-0.3ex>[ul]\ar@{..>}@<0.3ex>[ur]\ar@{..>}@<-0.3ex>[ur] & 0\ar[r]\ar[l] & 0\ar@{..>}@<0.3ex>[ul]\ar@{..>}@<-0.3ex>[ul]}
 \end{gathered}
\end{equation}
where the choice of $\lambda\in \mathbb{P}^1$ is irrelevant (one gets isomorphic algebras). The quiver of
$\ca_G[M_{i_0}]$ is obtained by adding to the quiver  $\overrightarrow{G}\boxtimes\widehat{A}(1,1)$ a new node $\bullet$ (corresponding to the extra $U(1)$ symmetry given by the baryon number) connected by a solid arrow $\psi$ to the node $\spadesuit_{i_0}$ while a dashed arrow $\eta$ (\textit{i.e.}\! a relation) connects $\diamondsuit_{i_0}$ to $\bullet$. Dually (and equivalently for our purposes) we can consider the one--point \emph{co}extension in which the two arrows incident to $\bullet$ flip their dashed/undashed character. The quiver of the Lagrangian canonical theory of type $(i_0,\{2\})$ (\textit{i.e.}\! $\kappa_i=\emptyset$ for $i\neq i_0$, $\kappa_{i_0}=\{2\}$) corresponds to the (completed) Jacobian algebra
\begin{equation}
 \Pi(\ca_G[M_{i_0}]).
\end{equation}
 For the example in figures \eqref{edfig}\eqref{edfig2} this is
\begin{equation}\label{edfig3}
 \begin{gathered}
  \xymatrix{\bullet \ar@/^1.6pc/[rrr]^\psi&&\spadesuit_1\ar@<0.3ex>[d]\ar@<-0.3ex>[d] \ar[r] & \spadesuit_2\ar@<0.3ex>[d]^B\ar@<-0.3ex>[d]_A&\spadesuit_3\ar[l]\ar[r]\ar@<0.3ex>[d]\ar@<-0.3ex>[d]&\spadesuit_4\ar@<0.3ex>[d]\ar@<-0.3ex>[d]\\
&&\diamondsuit_1\ar[r] & \diamondsuit_2\ar@{..>}@/^3pc/[lllu]^\eta\ar@{..>}@<0.3ex>[ul]\ar@{..>}@<-0.3ex>[ul]\ar@{..>}@<0.3ex>[ur]\ar@{..>}@<-0.3ex>[ur] & \diamondsuit_3\ar[r]\ar[l] & \diamondsuit_4\ar@{..>}@<0.3ex>[ul]\ar@{..>}@<-0.3ex>[ul]}
 \end{gathered}
\end{equation}
together with a new term in the superpotential \begin{equation}\Delta\cw=\mathrm{Tr}[\eta(A-\lambda B)\psi].\end{equation} To couple $N_f$ such quarks we just reiterate the procedure $N_f$ times
\begin{equation}
 \Pi\Big(\ca_G\overbrace{[M_{i_0}][M_{i_0}]\cdots [M_{i_0}]}^{N_f\ \text{times}}\Big)
\end{equation}
choosing \emph{distinct} points $\lambda_a\in \mathbb{P}^1$. More generally, we may take the $M_i$'s to correspond to different nodes of $\overrightarrow{G}$, provided all of them are sinks; this correspond to a bunch of quarks in the corresponding fundamental representations $\{R_i\}$.  It is elementary that the quiver $(Q,\cw)$ we get is mutation--equivalent to the $G$--canonical one.
\smallskip

\textbf{Remark.} We stress that the above algebraic description of the quiver holds assuming the node $i_0\in \overrightarrow{G}$ is a sink. Indeed, in general, a one--point extension $\ca[X]$ is extended at the module of $X$ of $\ca$ such that, in the extended algebra $X\simeq \mathrm{rad}\, P(\bullet)$, where $P(\bullet)$ is the projective cover of the simple representation at node $\bullet$. Were $i_0$ not a sink, the above quiver and superpotential would correspond to a  
 $\mathrm{rad}\, P(\bullet)$ which is more complicated than just our $M_{i_0}$. 
\medskip

To settle the `Lagrangian' case it remains to show\smallskip

\textbf{Proposition.} ($2d/4d$ for Lagrangian $\cn=2$ QFT) \textit{Let $\ca_G[M_{i_1}]\cdots [M_{i_s}]$ be the (uncompleted) algebra associated to $s$ quarks in the fundamental representations $R_{i_1}$, $R_{i_2}$, ..., $R_{i_s}$ of the gauge group $G$. Then its Coxeter element $\Phi$ has spectral radius $1$. In fact the spectrum of $\Phi$ is equal to that of $\Phi_\text{SYM}$ for the pure SYM case plus $s$ additional $-1$ eigenvalues corresponding to the extra $U(1)^s$ flavor charges
(compare with property $\boldsymbol{(\star)}$ in eqn.\eqref{tuqpoe}).  Moreover, the Jordan blocks of $\Phi$ have at most size $2\times 2$.}\smallskip

\textsc{Proof.} In view of the property $\boldsymbol{(\star)}$ and the fact that pure SYM has no flavor charge, the proposition is equivalent to the statement that the characteristic polynomial of the Coxeter element of $\ca_G[M_{i_1}]\cdots [M_{i_s}]$ is
\begin{equation}\label{chiNf}
 \chi_{G,N_f}(x)= (x+1)^{N_f}\, \chi_G(-x)^2.
\end{equation}
Let us proof eqn.\eqref{chiNf}. As reviewed in the appendix, there are formulae to compute recursively the Coxeter element of successive one--point (co)extension of algebras \cite{RI}. Let $X$ be a $\ca[M]$--module; we set $(\dim X)_\bullet=n_\bullet$ and $\dim M|_\ca=\boldsymbol{n}$ (where $\bullet$ denotes the `new' node of the quiver of the algebra $\ca[M]$). Then the extended Tits form is
\begin{equation}
 q(\dim X)\equiv \langle \dim X,\dim X\rangle_E^{\ca[M]}= n_\star^2+ q_\ca(\boldsymbol{n})- \langle \boldsymbol{n}, \dim M\rangle_{E}^\ca.
\end{equation}
The Stokes matrix of the one--point extended algebra is
\begin{equation}\label{EuExt1}
 S_{\ca[M]}= \left(\begin{array}{c|c} 
                  1\phantom{\Big|} & 0\\\hline
-S_\ca\,\dim M & \phantom{\Big|}S_\ca
                 \end{array}
\right),
\end{equation}
while the Coxeter element is
\begin{equation}\label{coxex1}
 \Phi_{\ca[M]}=\left(\begin{array}{c|c}
                                             \phantom{\Big|} q_\ca(\dim M)-1 & -(\dim M)^t S_\ca\\\hline
\phantom{\Big|}-\Phi_\ca\dim M & \Phi_\ca
                                             \end{array}
\right).
\end{equation}
where $\Phi_\ca$ is the Coxeter of $\ca$ and $S_\ca$ is its Stokes matrix.
\vglue 12pt

For the sequence of extensions of algebras 
$$\ca^\ell\equiv\ca_G[M_{i_1}][M_{i_2}]\cdots[M_{i_\ell}],\qquad\ell=0,1,2,\dots, k,$$ 
we define the sublattice $\Gamma_e\in K_0(\mathsf{mod}\text{--}\ca^\ell)$
generated by the dimension vectors $v$ such that $v_{\bullet_t}=0$ for all added nodes $\bullet_t$, $=1,2,\dots, \ell$ and $v|_{\ca} \in \Gamma_G\otimes \delta$ where $\Gamma_G$ is the root lattice of $G$ and $\delta$ is the minimal imaginary root of $\widehat{A}(1,1)$. $\Gamma_e$ is called the `purely electric sublattice'.

The proof of eqn.\eqref{chiNf} is based on the following\smallskip

\textbf{Lemma.} ($e$ is a local Nother charge)
\textit{Let $\langle\cdot, \cdot \rangle_\ell$ be the Euler form of the recursively extended algebra $\ca^\ell\equiv \ca_G[M_{i_1}][M_{i_2}]\cdots[M_{i_\ell}]$, for $\ell=0,1,2,\dots, k$, and assume that the modules $M_{i_\ell}$ are as in the} \textbf{Proposition}. \textit{Then 
\begin{enumerate}
\item \emph{(locality)} for all $\ell=0,1,\dots, k$ one has 
\begin{equation}\label{gallaezero}
 \langle \cdot, \cdot\rangle_\ell\Big|_{\Gamma_e}=0,
\end{equation}
\item \emph{(monodromy invariance of $\Gamma_e$)}
 for all $0\leq \ell\leq k$ the Coxeter element $\Phi_{\ca^\ell}$ maps $\Gamma_e$ into $\Gamma_e$.
\end{enumerate}
}

\smallskip

\textsc{Proof.}
Induction on $\ell$. $\ca^0=\ca_G$, and the Euler form factorizes, see eqn.\eqref{factorization}; the factor associated to the Kronecker algebra vanishes since the imaginary root $\delta\in \mathrm{rad}\,q(\cdot)_{\C \widehat{A}(1,1)}$, so statement 1. is true for $\ell=0$. But from eqn.\eqref{EuExt1} we have
\begin{equation}
 \langle \alpha, \beta\rangle_\ell\Big|_{\Gamma_e}= \langle \alpha, \beta\rangle_{\ell-1}\Big|_{\Gamma_e}
\end{equation}
and 1. follows from induction. On the other hand,
$\Phi_{\ca^0}=-\Phi_G\otimes \Phi_{\widehat{A}_1}$, while $\Phi_{\widehat{A}_1}\delta=\delta$. Thus 2. is true for $\ca^0$. Eqn.\eqref{coxex1} implies
\begin{equation}
\Phi_{A^\ell} \Gamma_e= \left(\begin{array}{c}
-(\dim M_{i_\ell})^t S_{\ell-1}\Gamma_e\\\hline
\Phi_{A^{\ell-1}}\Gamma_e\end{array}\right)=\left(\begin{array}{c}
0\\\hline
\Phi_{A^{\ell-1}}\Gamma_e\end{array}\right)\subseteq \Gamma_e
\end{equation}
by induction. \hfill $\square$
\medskip

\textsc{Proof of the proposition.}
By the \textbf{Lemma}, $\Phi_{\ca^\ell}\dim M_{i_{\ell+1}}\in \Gamma_e$.
Hence 
\begin{multline}\label{rrtue}
 (\dim M_{i_{\ell+1}})^t S_{\ca^\ell} \big(\lambda-\Phi_{\ca^\ell}\big)^{-1}\Phi_{\ca^\ell}\dim M_{i_{\ell+1}}=\\
=\sum_{k=0}^\infty \lambda^{-(k+1)}\,\langle \dim M_{i_{\ell+1}}, \Phi_{\ca^\ell}^{k+1}\, \dim M_{i_{\ell+1}}\rangle_\ell=0,
\end{multline}
where we used eqn.\eqref{gallaezero}. From eqn.\eqref{coxex1}, 
we have
\begin{equation}\begin{split}
\det[\lambda-\Phi_{\ca^{\ell+1}}]&=\\
&= \det[\lambda-\Phi_{\ca^\ell}]\cdot
\Big(\lambda+1- (\dim M_{i_{\ell+1}})^t S_{\ca^\ell}(\lambda-\Phi_{\ca^\ell})^{-1}\, \Phi_{\ca^\ell}\dim M_{i_{\ell+1}}\Big)=\\
&=(\lambda+1)\, \det[\lambda-\Phi_{\ca^\ell}]\end{split}
\end{equation}
and the result follows from induction on $\ell$. \hfill $\square$

\subsubsection{General $G$--canonical algebras}

Next consider the general case in which we insert over each node $i\in G$ a canonical subquiver $Q(\{p_{1,i},p_{2,i},\cdots, p_{s_i,i}\})$. As in the $p=2$ case considered in the previous subsection, the computation reduces recursively to the case of just one non--trivial integer, that is to $Q(\{p\})$ inserted over a given node $i\in G$.

We focus on the subquiver over the $i$--th node of $G$, which we take in the Coxeter--Dynkin form  ($\ell\equiv p-1$). Define
$Q_\ell$ to be the quiver with relations
\begin{equation}
 \begin{gathered}
  \xymatrix{  \cdots\ar[r] &\diamondsuit_i \ar@{..>}[dr] & \cdots\ar[l]\\
&& *++[o][F-]{1}\ar[ld] & *++[o][F-]{2} \ar[l] & *++[o][F-]{3} \ar[l] & \cdots \ar[l] & *++[o][F-]{\ell} \ar[l] \\
\cdots \ar[r] & \spadesuit_i \ar@<0.5ex>[uu]\ar@<-0.5ex>[uu] & \cdots\ar[l]}.
 \end{gathered}
\end{equation}
where $\cdots$ stands for the rest of the SYM quiver in the triangle form.
The relevant $G$--canonical quiver is then the $3$--CY completion of $Q_p$.

$Q_\ell$ may be obtained inductively by successive one--point extensions. $Q_\ell$ is the one point--extension of $Q_{\ell-1}$ at the projective cover $P(\ell-1)$ of the simple with support on the node $\ell-1$, the case $\ell=1$ being the `Lagrangian' extension already studied. Thus we have the sequence of algebras
\begin{equation}
 \ca_\ell= \ca_{\ell-1}[P(\ell-1)],\qquad \ca_1=\ca_G[M_{i}].
\end{equation}
We extend $\ell$ down to $0$ by setting $\ca_0= \C\overrightarrow{G}\otimes \C\widehat{A}_1$.
Let $\Phi_\ell$ the Coxeter element of $\ca_\ell$ and
\begin{equation}
 \chi_\ell(x)=\det[x-\Phi_\ell]
\end{equation}
its characteristic polynomial. Then
\smallskip

\textbf{Proposition.} (\textbf{Proposition 3.4} of \cite{LLLL}.) \textit{For $\ell\geq 2$ we have the recursion relation
\begin{equation}
 \chi_\ell(x)= (1+x)\,\chi_{\ell-1}(x)- x\, \chi_{\ell-2}(x).
\end{equation}
}

Then\smallskip

\textbf{Corollary.} \textit{All $G$--canonical algebras are consistent with the $2d/4d$ correspondence. In particular, for a $G$--canonical theory of type $\{\kappa_i\}_{i\in G}$ with $\kappa_i=\{p_{1,i}, p_{2,i},\cdots, p_{s_i,i}\}$ the characteristic polynomial of the Coxeter element (of the non--complete algebra) is
\begin{equation}
 \det[x-\Phi]=\chi_G(-x)^2\: \prod_{i\in G} \left(\prod_{a=1}^{s_i} \frac{x^{p_{a,i}}-1}{x-1}\right).
\end{equation}}

\section{Conclusions: $\cn=2$ categorical tinkertoys}

In this paper we have motivated the basic claims of section \ref{intt}. Given a 4d quiver $\cn=2$ QFT, for each corner in its parameter space where it looks like a weakly coupled gauge theory with gauge group $G$ (there are typically many such corners, related by $\cn=2$ dualities \cite{Gaiotto}), the non--perturbative BPS category $\mathsf{rep}(Q,\cw)$ satisfies 
the \emph{Ringel property of type $G$}: there is an exact Abelian subcategory $\mathscr{L}\subset \mathsf{rep}(Q,\cw)$ whose stable objects correspond to the BPS states which have bounded masses as we approach the zero YM coupling limit, that is, they correspond to the perturbative spectrum in the given weakly--coupled corner. Properly speaking, the statement holds only after replacing each Abelian category by the corresponding derived one, since only the derived category is intrinsically defined (that is, an invariant under 1d Seiberg duality); alternatively, we have to make sure that the particular representative $(Q,\cw)$ of the quiver mutation class we use `covers' the particular corner in parameter space we are scanning, \textit{i.e.}\! that BPS particles in that regime are in fact stable objects in $\mathsf{rep}(Q,\cw)$.

Correspondingly, the category $D^b(\mathsf{rep}(Q,\cw))$ should contain \emph{all} the light subcategories $D^b(\mathscr{L})$ associated to the several decoupling corners
\begin{equation}\label{ffffrr}
\bigvee_{\text{decoupling}\atop \text{corners}} D^b(\mathscr{L})\subseteq D^b(\mathsf{rep}(Q,\cw)),
\end{equation}  
as we have illustrated in the tubular models --- whose most prominent example is $SU(2)$ SQCD with $N_f=4$ ---
see \S.\ref{ttubular}.

This structure seems to remain there even if our $\cn=2$ theory is \emph{not} a quiver theory.  The `perturbative' category $\mathscr{L}$ still makes perfect sense and may even be constructed explicitly for, say, punctureless $A_1$ Gaiotto theories. What is not obvious in that case is which category $\mathscr{T}$ (if any) gives the non--perturbative completion of $\mathscr{L}$. The guess is that one has just to replace the \textsc{rhs} of eqn.\eqref{ffffrr} by an appropriate triangulated category $\mathscr{T}$ of a more general form.

The light category $\mathscr{L}$ corresponds to the perturbative theory to which the full theory is asymptotic in the given corner. Hence we know  quite a good deal about $\mathscr{L}$: the stable objects should consists of vectors making one copy of $\mathrm{ad}\,G$ together with finitely many hypers. This  gives an orthogonal decomposition
\begin{equation}
\mathscr{L}=\bigvee_{\lambda\in N}\mathscr{L}_\lambda
\end{equation}
where $N$ is an algebraic variety with one irreducible $\mathbb{P}^1$ component for each simple factor of $G=G_1\times G_2\times\cdots \times G_s$. The category $\mathscr{L}_\lambda$ over the generic point $\lambda$ of the $k$--th component is a homogeneous $G_k$--tube, which is universal for a given simple group $G_k$. Homogeneous $G$--tubes are described by the two Propositions on pages \pageref{genX1} and \pageref{genX2}.
At a finite number of (closed) points $\lambda_\alpha\in N$ the categories $\mathscr{L}_{\lambda_\alpha}$ are non--homogeneous. However, one has an inclusion--reversing correspondence of the form
\begin{equation}\label{UIUUU1}
\lambda_\alpha\in \big(\text{the $k$--th $\mathbb{P}^1$ component of $N$}\big)\quad \Rightarrow\quad \mathscr{L}_{\lambda_\alpha}\supseteq \big(\text{the homogeneous $G_k$--tube}.\big)
\end{equation}
A point $\lambda_\alpha$ may be contained in several $\mathbb{P}^1$ components; the corresponding category $\mathscr{L}_{\lambda_\alpha}$ will contain all the homogeneous $G_{k_j}$--tubes associated to the various irreducible components crossing at $\lambda_\alpha$. Such a non--homogeneous $G_{k_1}\times\cdots \times G_{k_r}$--tube describes matter charged under all gauge group factors $G_{k_1},\cdots, G_{k_r}$. More precisely the matter BPS states are the stable objects of the subcategory
\begin{equation}\label{UIUUU2}
\mathsf{matter}_{\lambda_\alpha}= \mathsf{add}\Big(\text{bricks in }\mathscr{L}_{\lambda_\alpha}\setminus \bigcup\nolimits_\ell \big(\text{homogeneous }G_{k_\ell}\text{--tube}\big)\Big).
\end{equation}
Note that $\mathsf{matter}_{\lambda_\alpha}$ may be trivial and yet the light category $\mathscr{L}_{\lambda_\alpha}$ is physically relevant: in particular, its still carries a $G_{k_1}\times\cdots\times G_{k_r}$ symmetry.

A matter category $\mathsf{matter}_{\lambda_\alpha}$ has a global symmetry which is gaugeable iff it can be embedded in an Abelian category $\mathscr{L}_{\lambda_\alpha}$ such that eqns.\eqref{UIUUU1}\eqref{UIUUU2} hold (with the usual \emph{caveat} that a proper invariant treatment requires to pass to the derived categories). It is easy to see that the states of a $G$--gaugeable matter category form representations of $G$. 

The problem which motivated us in the first place, that is, the classification of all (quiver) $\cn=2$ gauge theories, then splits in two steps: \textit{i)} to classify the $G$--tubes, and \textit{ii)} to give the rules that govern the appropriate gluing of the various $G$--tubes over a given reducible variety $N$.

The simplest examples of non--homogeneous $G$--tubes is obtained by mimicking in the general case what Ringel did for $SU(2)$; one gets in this way the class of canonical $G$--tubes described in some detail in this paper. They are enough to describe SQCD with any $N_c,N_f$.

However, the canonical ones do not exhaust the list  of $G$--tubes; neither they are the most interesting ones, physically speaking. The other known (at least in principle) class of non--homogeneous $G$--tubes are given by the `fixtures' \cite{CD1,CD2} of the Gaiotto theories \cite{Gaiotto}. Indeed, as described in the main body of the paper, in the Gaiotto case the reducible one--dimensional variety $N$ is precisely     
the maximal degeneration of the Gaiotto curve defining the weak coupling limit of interest with the small $3$--punctured spheres contracted to points.

Our philosophy regards to the Gaiotto theories is that fixtures are better understood as non--homogeneous light categories $\mathscr{L}_{\lambda_\alpha}$
rather than as $\cn=2$ matter systems. The fact that the matter content of a fixture may be trivial, yet the fixture `carries' non--trivial global symmetry \cite{CD2} is an indication that this is the correct philosophy.

The $A_1$ fixtures make complete theories \cite{CV11} and are well understood. Higher rank ones are only implicitly known (as categories). Some steps toward their explicit construction were done in ref.\!\cite{ACCERV2}.

As to gluing rules for $G$--tubes, we certainly know that the surgeries induced by geometric degenerations of a smooth Gaiotto curve lead to a consistent theory. However, even in the $A_1$ case we know that there are finitely many consistent gluings of a more general kind (based on $\mathbb{P}^1$ less \emph{three} points, rather than \emph{two}) \cite{CV11}.
What are the rules for the non--complete case we don't know, even in the canonical case (except for $SU(N_c)$ SQCD with $N_f$ quarks, where one can give a purely representation--theoretical reason why the no--Landau--pole criterion $N_f\leq 2N_c$ should hold). However, the feeling is that the right rule should be a rather slight generalization of the geometric gluing of tinkertoys in the Gaiotto setting. This justify the name of \emph{categorical tinkertoys} for the light subcategories $\mathscr{L}_\lambda$.

The $G$--tubes which are not in the above two broad class of examples (if any), and the corresponding gluing rules, are at the moment \emph{terra incognita}.

There is a lot of work to be done.

\section*{Acknowledgements}

I have greatly benefited of discussions with
Murad Alim, Clay C\'ordova, Michele Del Zotto, Sam Espahbodi,
Ashwin Rastogi and Cumrun Vafa. I became aware of many crucial results in representation theory thanks to the kind suggestions of William  Crawley--Boevey, Bernhard Keller, and Jan Sch\"oer. I thank them all.

I thank the Simons Center for Geometry an Physics for hospitality during the completion of the present work.

\bigskip

\appendix

\section{Math technicalities}

\subsection{Krull--Schmidt $\C$--categories: Notations \cite{RI}}\label{TTTTTTT}

A category $\ck$ is said to be $\C$--additive provided it has finite direct sums and all the sets $\mathrm{Hom}(X,Y)$, for $X,Y$ objects of $\ck$ are finite--dimensional $\C$--vectorspaces with $\C$ acting centrally on $\mathrm{Hom}(X,Y)$ and such that the compositions $\mathrm{Hom}(X,Y)\times \mathrm{Hom}(Y,Z)\rightarrow \mathrm{Hom}(X,Z)$ are bilinear. The direct sums are written as $X_1\oplus X_2$, and a non--zero object $X$ is said to be \emph{indecomposable} if $X=X_1\oplus X_2$ implies $X_1=0$ or $X_2=0$. An indecomposable $X$ with $\mathrm{End}(X)=\C$ is called a \emph{brick}. Two objects $X,Y$ with
$\mathrm{Hom}(X,Y)=\mathrm{Hom}(Y,X)=0$ are sais to be \emph{orthogonal}.

A $\C$--additive category $\ck$ will be said to be a \emph{Krull--Schmidt category} if the endomorphism ring $\mathrm{End}(X)$
of any indecomposable object $X$ of $\ck$ is a local ring. In a Krull--Schmidt  category the decomposition of an object into a direct sum of indecomposable objects is unique (up to permutation of the summands).
Therefore, in a Krull--Schmidt category the classification up to isomorphism of objects reduce to the classification of the indecomposable ones.

Given a Krull--Schmidt category $\ck$, a full subcategory $\cl\subset \ck$ closed under direct sums ansd direct summands (and isomorpshims) is called an \emph{object class in $\ck$}. $\cl$ is itself a Krull--Schmidt category, and it is uniquely determined by the indecomposable objects belonging to $\cl$. Given a set $\cn$ of objects in $\ck$, $\mathsf{add}(\cn)$ is the smallest object class containing $\cn$: it is given by all the direct sums of the direct summands of all objects in $\cn$. 

If $\cn_1,\cn_2$ are two object classes, we write $\cn_1\vee \cn_2$ for the object class $\mathsf{add}(\cn_1\cup \cn_2)$.

\subsection{Euler forms and Coxeter transformations \cite{RI,ASS1}}\label{appeuler}

Given a finite--dimensional basic algebra $A$ its \emph{Cartan matrix}
$C=C_A$ is defined as
\begin{equation}
C_{ij}= \dim \mathrm{Hom}(P(i),P(j)),
\end{equation}
where $P(i)$, ($i=1,2,\dots, n$) are the projective covers of the (non--isomorphic) simple modules $S(i)$.
\footnote{ \ \textsc{Beware\,!!} If $A$ is the path algebra of a Dynkin quiver, the above Cartan matrix does not coincide with the Cartan matrix, Car, in the usual sense of Lie algebras, that is in the Kac sense \cite{kac1,kac2}. Rather
\begin{equation*}
\mathrm{Car}= C^{-1}+(C^{-1})^t.
\end{equation*}}
Hence the $j$--th column of the Cartan matrix is the dimension vector of the  representation $P(j)$
\begin{equation}\label{dimPi}
(\dim P(j))_i \equiv \dim \mathrm{Hom}(P(i),P(j))\equiv C_{ij}.
\end{equation}
On the other hand, using the Nakayama functor $\nu^-$,
\begin{equation}\begin{split}
(\dim Q(i))_j&=\dim \mathrm{Hom}(Q(i),Q(j))=\dim \mathrm{Hom}(\nu^- Q(i), \nu^-Q(j))=\\
&=\dim \mathrm{Hom}(P(i),P(j))= C_{ij},
\end{split}\end{equation}
and thus the $i$--th row of $C$ is the dimension vector of the injective envelope $Q(i)$ of $S(i)$.

If $A$ has finite global dimension, $C$ is invertible over $\Z$ (i.e.\! a unit of $\mathbb{M}(n,\Z)$, that is
$\det C=\pm 1$).

We define the \emph{Euler bilinear form} as
\begin{equation}
\langle x, y\rangle = x^t\, C^{-t}y,
\end{equation}
and the \emph{Tits quadratic  form} as
\begin{equation}
q(x)=\langle x, x\rangle_E.
\end{equation}

By induction on the projective dimension of $A$, one shows that
when $\mathrm{gl.dim}\,A<\infty$ one has
\begin{equation}
\langle \dim X, \dim Y\rangle_E =\sum_{r\geq 0} (-1)^r\, \dim \mathrm{Ext}^r(X,Y).
\end{equation}

In particular, if $A=\C Q$ is \emph{hereditary}
\begin{equation}
C^{-1}_{ji}=\langle S(i), S(j)\rangle_E= \delta_{ij}-\dim \mathrm{Ext}^1(S(i), S(j))
\end{equation}
and $q(\cdot)$ corresponds to the Tits form of the graph underling the quiver $Q$ (\textit{i.e.}\! the quadratic form associated to the Cartan matrix of the corresponding Lie algebra, \textit{cfr.}\! \cite{kac1,kac2}.
\smallskip

\textsc{Definition.} $A$ a finite--dimensional algebra of finite global dimension. Its \emph{Coxeter matrix} is
\begin{equation}
\Phi=- C^{t}C^{-1}.
\end{equation}

From the above discussion one see that
\begin{equation}
\Phi\,\dim P(i)=-\dim Q(i), \qquad \forall\, i.
\end{equation} 
\begin{equation}
\langle \dim X, \dim Y\rangle_E= -\langle \dim Y, \Phi\,\dim X\rangle_E = \langle \Phi\,\dim X, \Phi\, \dim Y\rangle_E.
\end{equation}

We note that a vector $v\in \mathrm{rad}\,q$ if and only if
$\Phi\,v=v$.
\smallskip

If $v\in \mathrm{rad}\, q(\cdot)$. we set 
\begin{equation}
\iota_v(-)= \langle v, -\rangle_E.
\end{equation}
One has
\begin{equation}
\iota_v(x)=\iota_v(\Phi\,x).
\end{equation}

\subsection{One--point (co)extensions of algebras \cite{RI,ASS3}}\label{oneppoiiint}

Let $A$ be a $\C$--algebra and $X$ a right $A$--module.\smallskip

The \emph{one--point extension of $A$ by the module $X$}, written $A[X]$, we mean the $2\times 2$ matrix algebra
\begin{equation}
A[X]=\begin{bmatrix} A & 0\\
{}_\C X_A & \C
\end{bmatrix}
\end{equation}
with the ordinary addition of matrices, and the
multiplication induced from the $\C-A$--bimodule structure of $X$.\smallskip

The \emph{one--point coextension of $A$ by the module $X$}, written $[X]A$, we mean the $2\times 2$ matrix algebra
\begin{equation}
[X]A=\begin{bmatrix} A & 0\\
DX & \C
\end{bmatrix}
\end{equation}
with the ordinary addition of matrices, and the
multiplication induced from the $A-\C$--bimodule structure of $DX=\mathrm{Hom}_\C(X,\C)$.\smallskip
%\end{defn}

The quiver of $A[X]$, $Q_{A[X]}$ contains the quiver of the algebra $A$, $Q$, as a full convex\footnote{\ A subquiver is convex is for any two points in the subquiver, $i$ and $j$, it contains all the paths in the quiver connecting them.} subquiver and there is precisely one more node which is a source. The same is true for the quiver $Q_{[X]A}$ of $[X]A$, except that the extra node is a sink.
\medskip

We focus on the one--point extensions (coextensions are dual). 
We call $0$ the `new' node whereas the `old' nodes of $Q_A$ are numbered from $1$ to $n$. Then the new arrows from $0\xrightarrow{\ \phi_{i}\ }i$ ($i\geq 1$) as well as the new relations involving the new arrows $\phi_i$ may be determined by the identification
\begin{equation}\label{radical}
X \simeq \mathrm{rad}\,P(0),
\end{equation}
where $P(0)$ is the (indecomposable) projective cover of the simple module $S(0)$ (whose underlying vector space is spanned by all paths starting at the new node $0$)
\begin{equation}
0\rightarrow \mathrm{rad}\, P(0) \rightarrow P(0)\rightarrow S(0)\rightarrow 0.
\end{equation}

\subsubsection{Euler form and Coxeter element of a one--point extension algebra \cite{RI}}
Let $C_A$ be the Cartan matrix of the algebra $A$. Then the Cartan matrix of the one--point extension $A[X]$ is given by
\begin{equation}
 C_{A[X]}= \left(\begin{array}{c|c} 
                  1 & 0\phantom{\Big|}\\\hline
\dim X & \phantom{\Big|}C_A
                 \end{array}
\right)
\end{equation}
\begin{equation}\label{EuExt1}
 C_{A[X]}^{-1}= \left(\begin{array}{c|c} 
                  1\phantom{\Big|} & 0\\\hline
-C_A^{-1}\,\dim X & \phantom{\Big|}C_A^{-1}
                 \end{array}
\right)
\end{equation}
\begin{equation}
\mathbf{Car}_{A[X]}\equiv C_{A[X]}^{-t}+ C_{A[X]}^{-1}= \left(\begin{array}{c|c} 
                  2\phantom{\Big|} & -(\dim X)^t C_A^{-t}\\\hline
-C^{-1}_A\,\dim X & \phantom{\Big|}\mathbf{Car}_{A}
                 \end{array}
\right)
\end{equation}
Let $M$ be a $A[X]$ module and $(\dim M)_0=n_0$ and $\dim M|_A=\boldsymbol{n}$ (recall that $0$ denotes the `new' node of the quiver $Q_{A[X]}$). Then the Tits form is
\begin{equation}
 q(\dim M)= n_0^2+ q_A(\boldsymbol{n})- \langle \boldsymbol{n}, \dim X\rangle_{E, A}.
\end{equation}
Finally, the Coxeter element is
\begin{equation}\label{coxex1}
 \Phi_{A[X]}\equiv -C_{A[X]}^tC_{A[X]}^{-1}=\left(\begin{array}{c|c}
                                             \phantom{\Big|} q_A(\dim X)-1 & -(\dim X)^t C_A^{-1}\\\hline
\phantom{\Big|}-\Phi_A\dim X & \Phi_A
                                             \end{array}
\right).
\end{equation}
\vglue 9pt

\textbf{lem}\label{ringlemma}
 Let $v\in \Gamma_{A[X]}$ with $v_0=0$. Then
\begin{equation}
 v\in \mathrm{rad}\, q_{A[X]}\quad\Leftrightarrow\quad v\in \mathrm{rad}\, q_A\ \text{and } \langle v,\dim X\rangle_{E,A}=0.
\end{equation}
%\end{lem}
\smallskip

\subsection{Canonical algebras \cite{RI,LE1}}\label{appcanonical}

We start with the $\star$--algebra $A_\star$ corresponding to a star--shaped quiver of type $(w_1,w_2,\dots, w_s)$ with the subspace orientation, namely
\begin{equation}\label{starquiver}
\begin{gathered}
\xymatrix{ & a_1\ar[r]^{x_1} & a_2\ar[r]^{x_1} & \cdots \ar[r]^{x_1} & a_{w_1-1}\ar[ddr]^{x_1} &\\
& b_1\ar[r]^{x_2} & b_2\ar[r]^{x_2} & \cdots \ar[r]^{x_2} & b_{w_2-1}\ar[dr]_{x_2} &\\
& \vdots & \vdots & \ddots & \vdots & \omega\\
& u_1\ar[r]^{x_{s-1}} & u_2\ar[r]^{x_{s-1}} & \cdots \ar[r]^{x_{s-1}} & u_{w_{s-1}-1} \ar[ur]^{x_{s-1}}&\\
& v_1\ar[r]^{x_{s}} & v_2\ar[r]^{x_{s}} & \cdots \ar[r]^{x_{s}} & v_{w_s-1}\ar[uur]_{x_{s}} &}
\end{gathered}
\end{equation}
and consider its one--point extension at the module
\begin{equation}\label{starmodule}
M\colon \quad\begin{gathered}
\xymatrix{ & \C\ar[r]^{1} & \C\ar[r]^{1} & \cdots \ar[r]^{1} & \C\ar[ddr]^{x_1} &\\
& \C\ar[r]^{1} & \C\ar[r]^{1} & \cdots \ar[r]^{1} &\C\ar[dr]_{x_2} &\\
& \vdots & \vdots & \ddots & & \C^2\\
& \C\ar[r]^{1} & \C\ar[r]^{1} & \cdots \ar[r]^{1} & \C \ar[ur]^{x_{s-1}}&\\
& \C\ar[r]^{1} & \C\ar[r]^{1} & \cdots \ar[r]^{1} & \C\ar[uur]_{x_{s}} &}
\end{gathered}
\end{equation}
where $x_i(\C)\subset \C^2$ are pairwise distinct lines in $\C^2$, or equivalently, distinct points in $\mathbb{P}^1$.

Of course, the algebra $A_\star[M]$ is precisely the canonical algebra introduced on page \pageref{canquiver}. 

\textbf{fact} A canonical algebra is equivalent to an affine one of type $ADE$ if and only if the star--quiver is a finite Dynkin diagram of type $ADE$; it is equivalent to an elliptic one iff the star--quiver is an affine Dynkin diagram (and hence one type $\widehat{D}_4$ or $\widehat{E}_r$).
%\end{fact}
\medskip

As always, we denote by $0$ the source we add trough the one--point extension process. Then the $P(i)$ of $A_\star[M]$ coincide for $i\neq 0$ with the projective indecomposable of $A_\star$ (since $Q_{A_\star}$ is a \emph{convex} subquiver of $Q_{A_\star[M]}$), while $P(0)$ is given by the non--split extension
\begin{equation}
0\rightarrow M\rightarrow P(0)\rightarrow S(0)\rightarrow 0.
\end{equation}
We number the nodes different from $0$ and $\omega$ by a pair of indices $(\ell,a)$ where $\ell=1,2,\dots, s$ numbers the branches of the star--quiver \eqref{starquiver} and $a=1,2,\dots, w_\ell-1$ the nodes on the $\ell$--th branch in the arrow order.

Then the Cartan matrix of $A_\star[M]$ is
\begin{align}
&C_{\omega, \omega}=(\dim P(\omega))_\omega=1 
&& C_{(\ell,a),\omega}=0 && C_{0,\omega}=0\\
& C_{\omega,(\ell,a)}=(\dim P(\ell,a))_\omega= 1 && C_{(\ell^\prime,b)(\ell,a)}= \delta_{\ell,\ell^\prime}\, \theta(a,b)\\
& C_{0,(\ell,a)}= 0 && C_{\omega,0}=(\dim P(0))_\omega=2\\
&C_{(\ell,a),0}= 1 && C_{0,0}=1,
\end{align}
where
\begin{equation}
\theta(a,b)=\begin{cases} 1 & a\geq b\\
0 & \text{otherwise.}
\end{cases}
\end{equation}
In other terms,
\begin{equation}
C=\begin{pmatrix}
\begin{array}{c|c}
1 & 0\\\hline
v & \phantom{\Big|}C_\star
\end{array}
\end{pmatrix}
\end{equation}
where $C_\star$ is the Cartan matrix of $A_\star$ and
$v=(1,1,\dots, 1,2)^t$. Then
\begin{equation}
C^{-t}=\begin{pmatrix}
\begin{array}{c|c}
1 & -v^t(C_\star)^{-t}\\\hline
0 & \phantom{\Big|}C_\star^{-t}
\end{array}
\end{pmatrix}.
\end{equation}
But $(C_\star^{-1})_{ij}$ is simply $1$ minus the number of arrows from $i$ to $j$. Then
\begin{equation}
(C^{-1})_{ij}v_j\equiv \gamma_i= \begin{cases} 1 & \text{if }i=(\ell,1)\\
(2-s) & \text{if }i=\omega\\
0 & \text{otherwise}.\end{cases} 
\end{equation}

Thus $C^{-t}$ is just the matrix of the Brenner form
\begin{equation}
\langle x, y\rangle= \sum_{i\in Q_0} x_iy_i-\sum_{\alpha\in Q_1} x_{s(i)}y_{t(i)}+\sum_r x_{s(r)}y_{t(r)}
\end{equation}
where the last sum is over a minimal set of relations.
\medskip

Let us consider the matrix of the symmetrized form
\begin{equation}
\mathbf{Car}\equiv C^{-1}+C^{-t}=\begin{pmatrix}
\begin{array}{c|c}
2 & -\gamma^t\\\hline
-\gamma & \phantom{\Big|}\mathbf{Car}_\star
\end{array}
\end{pmatrix}
\end{equation}
\begin{equation}
\det(\mathbf{Car})= 2\, \det(\mathbf{Car}_*- \gamma\,\gamma^t/2)
\end{equation}

\section{The matter subcategory of $\tfrac{1}{2}(\mathbf{2},\mathbf{2},\mathbf{2})$}\label{bricksofffB}

In this appendix we give the details of the computation of the matter subcategory of the light category $\mathscr{L}_0$ over the crossing point of the three $\mathbb{P}^1$'s for the $SU(2)^3$ gauge theory with a half--hyper in the $(\mathbf{2},\mathbf{2},\mathbf{2})$ representation whose canonical quiver is \eqref{prisms}.

The matter subcategory is the object class of the rigid bricks (bricks without self--extensions) in $\mathscr{L}_0$. Since the full Jacobian algebra is gentle, bricks are, in particular, string modules.

Let $X\in \mathscr{L}_0$ be a brick.
The fact that $\mathscr{F}_j(X)\in \ct_0$ for all $j$ implies that the vertical arrows in \eqref{prisms} are isomorphisms.
In turn this implies:
\begin{itemize}
\item vertical and horizontal arrows (either direct or inverse) {alternate} in $C$;
\item the first and last arrows are vertical.
\end{itemize}

Fix the length $2n+1$ and the starting node of the string $C$. Given the particular shape of the quiver \eqref{prisms}, the module $M(C)$ is uniquely determined by the map
$$\mathsf{ar}\colon \{1,2,\dots, n\}\rightarrow \{+1,-1\}$$ which specifies whether the $k$--th horizontal arrow is direct ($+1$) or inverse ($-1$).

Let $n_+$ (resp.\! $n_-$) the number of horizontal arrows $H$ in $C$ with $\mathsf{ar}(H)=+1$ (resp.\! $\mathsf{ar}(H)=-1$). We claim that $X$ being a light rigid brick implies $n_+\leq 1$ and $n_-\leq 1$. The claim is obvious if $n_-=0$ (resp.\! $n_+=0$). In this case $X$ has the form
$\mathscr{S}_j(X)$ for some $j$, and we get the claim by comparison with the known rigid bricks of $\widehat{A}(3,1)$.

In the general case it is easy to see that $\mathsf{ar}(a+1)=\mathsf{ar}(a)$ implies the existence of a non--zero nilpotent element in $\mathrm{End}(X)$. Hence direct and inverse horizontal arrows should alternate in the string $C$ of a brick. Hence a sincere brick in $M(C)\in \mathscr{L}_{0,0,0}$ with $C$ starting (say) in the node $2,1$ must contains as initial part of the string $C$ one of the following two
\begin{equation}\label{prisms3}
 \begin{gathered}
\xymatrix{ & i_2 &\\
i_1\ar[ur]^{H_3} & & i_5\\
\\
& i_3\ar[uuu]^{V_2}\\
i_0\ar[uuu]^{V_1} && i_4\ar[uuu]_{V_3}\ar[lu]_{h_1}}
 \end{gathered}\qquad 
 \begin{gathered}
\xymatrix{ & i_5 &\\
i_1 & & i_2\ar[ll]_{\hskip 1.14cm H_2}\\
\\
& i_4\ar[uuu]^{V_2}\\
i_0\ar[uuu]^{V_1} && i_3\ar[uuu]_{V_3}\ar[lu]_{h_1}}
 \end{gathered}
\end{equation} 
If we continue, say, the first string we would get 
\begin{equation}\label{prisms4}
 \begin{gathered}
\xymatrix{ & i_2 &\\
i_1\oplus i_6\ar[ur] & & i_5\ar[ll]\\
\\
& i_3\ar[uuu]\\
i_0\oplus i_7\ar@<1.5ex>[uuu]\ar@<-1.5ex>[uuu] && i_4\ar[uuu]\ar[lu]}
 \end{gathered}
 \end{equation}
Now the matrix $$\begin{pmatrix} 0 & 1\\ 0 & 0\end{pmatrix}$$ acting on the two isomorphic two--dimensional spaces is a nilpotent endomorphism of the representation, and so the string module is not a brick. If we further continue the string, the next horizontal arrow would be inverse, and hence the above transformation on the first two vectors of the spaces over the nodes $1,1$ and $2,1$ will remain a non--trivial endomorphism. Thus in the first case the length of $C$ is (at most) 5. The same argument applies to the second case. The claim $n_+,n_-\leq 1$ follows. This result may be equivalently stated as

\textit{An indecomposable representation $X\in \mathscr{L}_0$ is a brick iff $\dim X\leq (1,1,1,1,1,1)$.}

\section{Stable light strings for $\cn=2^*$ in regime (II) }\label{detailsforcn2*}

We have seen in the main text that,
focusing on the regime in eqn.\eqref{starstar}, the string/band $C$ of a stable representation of the gentle algebra $\mathscr{B}$  is --- except for `boundary' terms at the two ends --- a sequence of bullet configurations of the form
$\spadesuit\rightarrow \bullet\leftarrow\spadesuit$, or
$\spadesuit\rightarrow\bullet\rightarrow\diamondsuit$ and its inverse (recall that $\bullet$'s and $\spadesuit/\diamondsuit$'s alternate along $C$, and the arrows are uniquely determined up to $\Z_2$ automorphism).

Requiring $m(M(C))=0$, the following strings $C$ are ruled out
\begin{gather}
 \bullet \leftarrow \spadesuit \rightarrow \bullet\leftarrow \spadesuit\: \cdots\cdots\\
\cdots\cdots\: \spadesuit \rightarrow \bullet\leftarrow \spadesuit \rightarrow \bullet\\
\cdots\cdots\: \spadesuit \rightarrow\bullet\leftarrow \spadesuit \rightarrow \bullet \:\leftarrow \spadesuit \cdots\cdots
\end{gather}
moreover, $\spadesuit\rightarrow\bullet\leftarrow\spadesuit$ may be present at most once. Hence the surviving strings have the forms
\begin{gather*}
 (\bullet)\,\overbrace{(\spadesuit\,\bullet\,\diamondsuit)\,\bullet\,(\spadesuit\,\bullet\,\diamondsuit)\,\bullet\,\cdots\,\bullet\,(\spadesuit\,\bullet\,\diamondsuit)}^{n\ \text{times}}\,(\bullet)\\
 (\bullet)\,\diamondsuit\,\bullet\,\overbrace{(\spadesuit\,\bullet\,\diamondsuit)\,\bullet\,(\spadesuit\,\bullet\,\diamondsuit)\,\bullet\,\cdots\,\bullet\,(\spadesuit\,\bullet\,\diamondsuit)}^{n_1\ \text{times}}\,\bullet\,\spadesuit\,\bullet\,
\overbrace{(\spadesuit\,\bullet\,\diamondsuit)\,\bullet\,(\spadesuit\,\bullet\,\diamondsuit)\,\bullet\,\cdots\,\bullet\,(\spadesuit\,\bullet\,\diamondsuit)}^{n_2\ \text{times}}\,(\bullet)
\end{gather*}
where the initial and final bullets $(\bullet)$ may or may not be present. However, in the regime \eqref{starstar}, the strings of the second kind are unstable if they have a bullet at the ends, since $M(C)$ would have a quotient with support the node $\bullet$. By the same argument, stable string of the first kind cannot have a bullet at the end. Then we remain with  
\begin{gather*}
 (\bullet)\,\overbrace{(\spadesuit\,\bullet\,\diamondsuit)\,\bullet\,(\spadesuit\,\bullet\,\diamondsuit)\,\bullet\,\cdots\,\bullet\,(\spadesuit\,\bullet\,\diamondsuit)}^{n\ \text{times}}\\
 \diamondsuit\,\bullet\,\overbrace{(\spadesuit\,\bullet\,\diamondsuit)\,\bullet\,(\spadesuit\,\bullet\,\diamondsuit)\,\bullet\,\cdots\,\bullet\,(\spadesuit\,\bullet\,\diamondsuit)}^{n_1\ \text{times}}\,\bullet\,\spadesuit\,\bullet\,
\overbrace{(\spadesuit\,\bullet\,\diamondsuit)\,\bullet\,(\spadesuit\,\bullet\,\diamondsuit)\,\bullet\,\cdots\,\bullet\,(\spadesuit\,\bullet\,\diamondsuit)}^{n_2\ \text{times}}
\end{gather*}

Let us call the first string with the initial $\bullet$ present $C_1(n)$. One has
$$Z(C_1(n))= n\big(2\, Z_\bullet+Z_\spadesuit+Z_\diamondsuit\big).$$ Since $C_1(p)$ with $p<n$ defines a subrepresentation, $C_1(n)$ may be stable only for $n=0,1$. 

Let us call the first string without the initial bullet $C_2(n)$ ($n\geq 1$); one has
\begin{equation}
 \arg Z(C_2(n))= \arg[Z_\spadesuit+Z_\diamondsuit+(2-1/n)Z_\bullet]
\end{equation}
which, in the regime \eqref{starstar}, is decreasing with $n$. Hence for $p<n$
\begin{equation}
 \arg Z(C_2(p)) > \arg Z(C_2(n)),
\end{equation}
but $C_2(p)$ is a subrepresentation, and hence this is consistent with stability only if $n=1$.

Let us call the second representation $C_3(n_1,n_2)$. Note that $C_3(n_1,0)=C_1(n_1+1)$, so we may assume $n_2\geq 1$. One has
\begin{equation}
\arg Z(C_3(n_1,n_2))= \arg\left[Z_\spadesuit+Z_\diamondsuit+2\left(1-\frac{1}{n_1+n_2+1}\right)Z_\bullet\right]
\end{equation}
which is again a decreasing function of $n_1$, $n_2$. Since for $1\leq p< n_2$ we have a subrepresentation, we get
$n_2\leq 1$. By a similar argument, we have $n_1\leq 1$ also.

Then we have
\begin{align}
&C_1(0) &&\bullet\\
& C_1(1) &&\bullet \spadesuit \bullet\diamondsuit\\
& C_2(1)\simeq C_3(0,0) &&\spadesuit\bullet\diamondsuit\\
& C_3(0,1)&&\diamondsuit\bullet\spadesuit\bullet\spadesuit \bullet\diamondsuit\\
&C_3(1,1) && \diamondsuit\bullet\spadesuit\bullet\diamondsuit\bullet\spadesuit\bullet\spadesuit\bullet\diamondsuit
\end{align}
$C_3(1,1)$ has $C_3(0,1)$ as a subrepresentation. Stability of $C_3(1,1)$ requires
\begin{equation*}
\arg\left[Z_\spadesuit+Z_\diamondsuit+\frac{3}{2}\,Z_\bullet\right]=\arg Z(C_3(0,1))<
\arg Z(C_3(1,1))=\arg\left[Z_\spadesuit+Z_\diamondsuit+\frac{5}{3}\,Z_\bullet\right]
\end{equation*}
which is false in regime \eqref{starstar}.

Finally, the list of stable light string representations are (up to $\Z_2$ automorphism)
\begin{align}
&\bullet\\
&\bullet \spadesuit \bullet\diamondsuit \quad \text{or}\quad \spadesuit\bullet\diamondsuit\bullet\\
&\spadesuit\bullet\diamondsuit\\
&\diamondsuit\bullet\spadesuit\bullet\spadesuit \bullet\diamondsuit\quad \text{or}\quad \spadesuit\bullet\diamondsuit\bullet\diamondsuit\bullet\spadesuit
\end{align} 
where which of the two strings in each line --- having identical dimension vectors ---  is stable depends on the particular weak coupling chamber (belonging to regime (II)), and precisely one in each line is stable in each such chamber.

\section{Proof of $X\in\mathscr{L}^\mathrm{YM}$ $\Rightarrow$ $X\big|_{\mathbf{Kr}_i}\in \ct$}\label{applemma}

\subsection{The proof}
In this appendix we prove the 
\vglue 9pt

\textbf{Lemma 1.} \textit{$X\in \mathscr{L}^\mathrm{YM}$. Then $X\big|_{\mathbf{Kr}_i}\in\ct$ for all $i
\in G$.}
\medskip

We write $A_i,B_i$ for the two arrows of the representation $X$ belonging to the $i$--th Kronecker subquiver. Then the statement is implied by the following one\vglue 9pt

\textbf{Lemma 2.} \textit{$X\in\mathsf{rep}(Q,\cw)$. Assume that for some $i\in G$ and some integer $\ell\geq 1$, there are $\ell$ vectors $v_a\in X_{\spadesuit_i}$, $a=1,2\,\dots, \ell$, not \emph{all} zero,  such that
\begin{equation}\label{eeeche}
\begin{split}
A_i v_1&=0\\
A_i v_2&=B_i v_1\\
A_i v_3&=B_i v_2\\
\cdots & \cdots\\
A_i v_{\ell}&=B_i v_{\ell-1}\\
0&= B_i v_\ell
\end{split}\end{equation} 
Then $X\not\in\mathscr{L}^\mathrm{YM}$.}
\vglue 9pt

The implication $2\Rightarrow 1$ is easy to see. Let $X\in\mathsf{rep}(\mathbb{G})$ be such that $m_j(X)=0$ for all $j\in G$. If for some $i$, $X\big|_{\mathbf{Kr}_i}\not\in\ct$, then $X\big|_{\mathbf{Kr}_i}$ should have an indecomposable preinjective summand $Z\in \cq$
of the form
\begin{equation}
\xymatrix{\C^\ell \ar@<0.6ex>[rrr]^{[0,\mathbf{1}]}\ar@<-0.6ex>[rrr]_{[\mathbf{1},0]}&&& \C^{\ell-1}}
\end{equation}
where $\mathbf{1}$ stands for the $(\ell-1)\times(\ell-1)$ unit matrix.
  Then as the $v_a$'s we take the vectors of the standard basis of $Z_{\spadesuit_i}\simeq \C^\ell$, which satisfy \eqref{eeeche}, getting $X\not\in\mathscr{L}^\mathrm{YM}$.
\medskip

An important preliminary remark is the following. The vectors $\{v_1,v_2,\cdots, v_\ell\}$, not all zero, are not assumed to be linearly independent. However, given $\{v_1,v_2,\cdots, v_\ell\}$, not al zero, satisfying  \eqref{eeeche} we may construct a new set of vectors,
$\{\tilde v_1,\tilde v_2, \cdots, \tilde v_{\ell^\prime}\}$, which satisfy \eqref{eeeche} with $\ell$ replaced by some smaller $\ell^\prime\geq 1$ and \emph{are} linearly independent. Indeed, replacing $\{v_1,v_2,\cdots, v_\ell\}$ by $\{v_2,v_3,\cdots, v_\ell\}$, if necessary, we may assume $v_1\neq 0$. Then assume that
$\{v_1,v_2,\cdots, v_k\}$ are linearly independent, but $v_{k+1}$ is a linear combination of the preceeding $k$ vectors,
\begin{equation}
v_{k+1}=\sum_{s=1}^k \lambda_s\, v_s,\qquad \lambda_s\in\C.
\end{equation} 
The set
\begin{equation*}
\{\tilde v_1,\tilde v_2, \cdots, \tilde v_{k}\}=\left\{v_1,v_2-\lambda_k\,v_1,\cdots, v_{s}-\sum_{r=0}^{k-1}\lambda_{k-r}\,v_{s-r-1},\cdots, v_k-\sum_{r=0}^{k-1}\lambda_{k-r}\,v_{k-r-1}\right\}
\end{equation*}
(where $v_s=0$ for $s\leq 0$) of linearly independent vectors satisfies \eqref{eeeche} with $\ell$ replaced by $k\geq 1$. 
The remaining $\ell-k-1$ vectors $\{v_{k+2}-\sum_{r=0}^{k-1}\lambda_{k-r}\,v_{k-r+1},\cdots,
v_{\ell}-\sum_{r=0}^{k-1}\lambda_{k-r}\,v_{\ell-r-1}\}$ also satisfy \eqref{eeeche}, and we may repeat the same procedure recursively.

 \medskip

The proof of \textbf{Lemma 2}  proceeds by induction on $\ell$. 

If $\ell=1$, one has $0\neq v_1\in \mathrm{ker}\,A_i \cap \mathrm{ker}\,B_i$, hence the representation $Y$ with
$Y_{\spadesuit_i}=\C v_1$ and zero elsewhere is a subrepresentation of $X$ with $m_i(Y)=1>0$, and hence $X\not\in\mathscr{L}^\mathrm{YM}$.

Now assume the Lemma is true up to $\ell-1$. Let $v_a\in X_{\spadesuit_i}$  ($a=1,\dots, \ell$) be a set of vectors, not all zero, satisfying  eqn.\eqref{eeeche}. By the previous remark, we may assume the $\{v_a\}$ to be linearly independent since, otherwise, we may reduce ourselves to the set $\{\tilde v_a\}$ containing $\leq \ell-1$ vectors for with the Lemma holds in virtue of the induction hypothesis.

We write
\begin{equation}\label{opppa}
X_{\diamondsuit_i}\ni w_a= B_i v_a= A_iv_{a+1}, \qquad a=1,2,\dots \ell-1. 
\end{equation}
If the vectors $\{w_1,w_2,\cdots, w_{\ell-1}\}$ are all zero, the representation $Y$ with $Y_{\spadesuit_i}=\oplus_a\C  v_a$ and zero elsewhere is a subrepresentation of $X$ with
$m_i(Y)= \ell >0$ and $X\not\in\mathscr{L}^\mathrm{YM}$. Hence
we may assume the $w_a$'s not to be all zero.

For all $j\in G$ such that $C_{ij}=-1$ consider the square subquiver formed by the corresponding Kronecker subquivers
\begin{equation}
\begin{gathered}\xymatrix{X_{\diamondsuit_i} \ar[rr]^{\phi_{ji}} && X_{\spadesuit_j}\ar@<0.4ex>[dd]^{B_j}\ar@<-0.4ex>[dd]_{A_j}
\\
\\
X_{\spadesuit_i}\ar@<0.4ex>[uu]^{A_i}\ar@<-0.4ex>[uu]_{B_i} && X_{\diamondsuit_j}\ar[ll]^{\phi_{ij}}
}\end{gathered}
\end{equation}
Associated to this square we have the relation
\begin{equation}\label{reREL}
0=\epsilon_j\,\partial_{\phi_{ij}}\cw= A_j\,\phi_{ji}\,B_i-B_j\,\phi_{ji}\,A_i.
\end{equation}

From \eqref{opppa}\eqref{reREL} we see that the $\ell-1$ vectors $w_a$ satisfy
\begin{equation}\label{reREL2}
\begin{split}
A_j\phi_{ji}w_1&=0\\
A_j\phi_{ji}w_2&=B_j\phi_{ji}w_1\\
A_j\phi_{ji}w_3&=B_j\phi_{ji}w_2\\
\cdots &\cdots\\
0&=B_j\phi_{ji}w_{\ell-1}.
\end{split}
\end{equation}

If, \emph{for all $j\in G$ such that $C_{ji}=-1$},  the $\ell-1$ vectors $\{\phi_{ji}w_1,\phi_{ji}w_2,\cdots \phi_{ji}w_{\ell-1}\}$ are all zero, then the representation $Y$ with 
\begin{align}
&Y_{\spadesuit_i}=\bigoplus_{a=1}^k\C  v_a, &&Y_{\diamondsuit_i}=\mathrm{Span}\{w_a\:|\:a=1,2\dots, k-1\}\\ 
&Y_{A_i}= A_i\big|_{Y_{\spadesuit_i}}, &&Y_{B_i}= B_i\big|_{Y_{\spadesuit_i}},\end{align}
 is a subrepresentation of $X$ with $m_i(Y)\geq 1>0$ and, again, $X\not\in\mathscr{L}^\mathrm{YM}$. 
 \smallskip

Otherwise, there is a node $j\in G$ with $C_{ij}=-1$ such that the $\ell-1$ vectors 
\begin{equation}
z^{(j)}_a=\phi_{ji}\, w_a\in X_{\spadesuit_j},\qquad a=1,2,\cdots, \ell-1,
\end{equation}
are not all zero. Since by \eqref{reREL2} they satisfy eqn.\eqref{eeeche} with $\ell$ replaced by $\ell-1$ (and node $i$ replaced by node $j$\, !), by the induction hypothesis we are done. \hfill $\square$

\subsection{A stronger statement}\label{strongerstatement}

\textit{Let $X\in\mathsf{rep}(\mathbb{G})$ be stable at weak coupling. Then one of the following holds
\begin{align}
&\mathrm{(a)}\qquad X\big|_{\mathbf{Kr}_i}\in \cp\quad \forall\: i\in G\\
&\mathrm{(b)}\qquad X\big|_{\mathbf{Kr}_i}\in \ct\quad \forall\: i\in G\\ 
&\mathrm{(c)}\qquad X\big|_{\mathbf{Kr}_i}\in \cq\quad \forall\: i\in G. 
\end{align}
Moreover, if in addition $m_i(X)=0$ for all $i$, then for all subobjects $Y$ one has $m_i(Y)\leq 0$, that is $X\in\mathscr{L}^\mathrm{YM}$.}
\medskip

Indeed, let $X_\cq$ be the representation of $\mathbb{G}$ with, for all $i\in G$, $$\xymatrix{(X_\cq)_{\spadesuit_i}\ar@<0.4ex>[rr]^{(X_\cq)_{A_i}}\ar@<-0.4ex>[rr]_{(X_\cq)_{B_i}}&&(X_\cq)_{\diamondsuit_i}}$$
equal to the preinjective summand of the restriction $X|_{\mathbf{Kr}_i}$ and $(X_\cq)_{\phi_{ji}}$ equal to the restriction of  $X_{\phi_{ji}}$ which is well defined since, by the computations of the previous subsection,
\begin{equation}
 X_{\phi_{ji}}\,(X_\cq)_{\diamondsuit_i}\subseteq (X_\cq)_{\diamondsuit_j}\subset X_{\diamondsuit_j}.
\end{equation}
$X_\cq$ is then a subrepresentation of $X$ with
\begin{equation}
 m_i(X_\cq)\geq 0\qquad\text{and}\qquad m_i(X/X_\cq)\leq 0
\end{equation}
 which is compatible with stability at weak coupling if either $X_\cq=0$ or $X/X_\cq=0$. The second case corresponds to possibility (c), while the first one gives $X|_{\mathbf{Kr}_i}\in \cp\vee \ct$. The dual argument, with preinjective components replaced by the preprojective ones gives that either $X|_{\mathbf{Kr}_i}\in \cp$ or $X|_{\mathbf{Kr}_i}\in \ct\vee \cq$, completes the proof of the first part.

Suppose now that $X$ is stable at weak coupling with $m_i(X)=0$ and there is a subrepresentation $Y$ with $m_{i_0}(Y)>0$ for a certain $i_0$. Then the subrepresentation $Y_\cq\neq 0$ is a proper non--zero subrepresentation of $X$ with $m_i(Y)\geq 0$ for all $i$ and $m_{i_0}(Y)>0$ contradicting the stability of $X$ at weak coupling. \hfill $\square$


\newpage

\begin{thebibliography}{50}

\bibitem{SW1}
N.~Seiberg and E.~Witten, ``{Electric - magnetic duality, monopole
  condensation, and confinement in N=2 supersymmetric Yang-Mills theory},''
  \href{http://dx.doi.org/10.1016/0550-3213(94)90124-4,
  10.1016/0550-3213(94)90124-4}{{\em Nucl.Phys.} {\bfseries B426} (1994)
  19--52},
\href{http://arxiv.org/abs/hep-th/9407087}{{\ttfamily arXiv:hep-th/9407087
  [hep-th]}}.
%%CITATION = HEP-TH/9407087;%%.

\bibitem{SW2}
N.~Seiberg and E.~Witten, ``{Monopoles, duality and chiral symmetry breaking in
  N=2 supersymmetric QCD},''
  \href{http://dx.doi.org/10.1016/0550-3213(94)90214-3}{{\em Nucl.Phys.}
  {\bfseries B431} (1994) 484--550},
\href{http://arxiv.org/abs/hep-th/9408099}{{\ttfamily arXiv:hep-th/9408099
  [hep-th]}}.
%%CITATION = HEP-TH/9408099;%%.

\bibitem{CV11}
S.~Cecotti and C.~Vafa, ``{Classification of complete N=2 supersymmetric
  theories in 4 dimensions},''
\href{http://arxiv.org/abs/1103.5832}{{\ttfamily arXiv:1103.5832 [hep-th]}}.
%%CITATION = 1103.5832;%%.


\bibitem{ACCERV1}
M.~{Alim}, S.~{Cecotti}, C.~{Cordova}, S.~{Espahbodi}, A.~{Rastogi}, and
  C.~{Vafa}, ``{BPS Quivers and Spectra of Complete N=2 Quantum Field
  Theories},''
\href{http://arxiv.org/abs/1109.4941}{{\ttfamily arXiv:1109.4941 [hep-th]}}.
%%CITATION = ARXIV:1109.4941;%%.

\bibitem{ACCERV2}
M.~{Alim}, S.~{Cecotti}, C.~{Cordova}, S.~{Espahbodi}, A.~{Rastogi}, and
  C.~{Vafa}, ``{$N=2$ Quantum Field Theories and their BPS QUivers},''
\href{http://arxiv.org/abs/1112.3984}{{\ttfamily arXiv:1112.3984 [hep-th]}}.


\bibitem{Denef00}
F.~Denef, ``{Supergravity flows and D-brane stability},'' {\em JHEP} {\bfseries
  0008} (2000) 050,
\href{http://arxiv.org/abs/hep-th/0005049}{{\ttfamily arXiv:hep-th/0005049
  [hep-th]}}.
%%CITATION = HEP-TH/0005049;%%.

\bibitem{DM07}
F.~Denef and G.~W. Moore, ``{Split states, entropy enigmas, holes and halos},''
\href{http://arxiv.org/abs/hep-th/0702146}{{\ttfamily arXiv:hep-th/0702146
  [hep-th]}}.
%%CITATION = HEP-TH/0702146;%%.

\bibitem{DM}
M.~R. Douglas and G.~W. Moore, ``{D-branes, Quivers, and ALE Instantons},''
\href{http://arxiv.org/abs/hep-th/9603167}{{\ttfamily arXiv:hep-th/9603167}}.
%%CITATION = HEP-TH/9603167;%%.

\bibitem{Dia99}
D.-E. Diaconescu and J.~Gomis, ``{Fractional branes and boundary states in
  orbifold theories},'' {\em JHEP} {\bfseries 0010} (2000) 001,
\href{http://arxiv.org/abs/hep-th/9906242}{{\ttfamily arXiv:hep-th/9906242
  [hep-th]}}.
%%CITATION = HEP-TH/9906242;%%.

\bibitem{DFR1}
M.~R. Douglas, B.~Fiol, and C.~R{\"o}melsberger, ``{Stability and BPS
  branes},'' \href{http://dx.doi.org/10.1088/1126-6708/2005/09/006}{{\em JHEP}
  {\bfseries 0509} (2005) 006},
\href{http://arxiv.org/abs/hep-th/0002037}{{\ttfamily arXiv:hep-th/0002037
  [hep-th]}}.
%%CITATION = HEP-TH/0002037;%%.

\bibitem{DFR2}
M.~R. Douglas, B.~Fiol, and C.~R{\"o}melsberger, ``{The Spectrum of BPS branes
  on a noncompact Calabi-Yau},''
  \href{http://dx.doi.org/10.1088/1126-6708/2005/09/057}{{\em JHEP} {\bfseries
  0509} (2005) 057},
\href{http://arxiv.org/abs/hep-th/0003263}{{\ttfamily arXiv:hep-th/0003263
  [hep-th]}}.
%%CITATION = HEP-TH/0003263;%%.

\bibitem{FM00}
B.~Fiol and M.~Marino, ``{BPS states and algebras from quivers},'' {\em JHEP}
  {\bfseries 0007} (2000) 031,
\href{http://arxiv.org/abs/hep-th/0006189}{{\ttfamily arXiv:hep-th/0006189
  [hep-th]}}.
%%CITATION = HEP-TH/0006189;%%.

\bibitem{Fiol}
B.~Fiol, ``{The BPS spectrum of N = 2 SU(N) SYM and parton branes},''
\href{http://arxiv.org/abs/hep-th/0012079}{{\ttfamily arXiv:hep-th/0012079}}.
%%CITATION = HEP-TH/0012079;%%.

\bibitem{Denef}
F.~Denef, ``{Quantum quivers and Hall / hole halos},'' {\em JHEP} {\bfseries
  0210} (2002) 023,
\href{http://arxiv.org/abs/hep-th/0206072}{{\ttfamily arXiv:hep-th/0206072
  [hep-th]}}.
%%CITATION = HEP-TH/0206072;%%.


\bibitem{FHHI02}
B.~Feng, A.~Hanany, Y.~H. He, and A.~Iqbal, ``{Quiver theories, soliton spectra
  and Picard-Lefschetz transformations},'' {\em JHEP} {\bfseries 0302} (2003)
  056,
\href{http://arxiv.org/abs/hep-th/0206152}{{\ttfamily arXiv:hep-th/0206152
  [hep-th]}}.
%%CITATION = HEP-TH/0206152;%%.

\bibitem{FHH00}
B.~Feng, A.~Hanany, and Y.-H. He, ``{D-brane gauge theories from toric
  singularities and toric duality},''
  \href{http://dx.doi.org/10.1016/S0550-3213(00)00699-4}{{\em Nucl.Phys.}
  {\bfseries B595} (2001) 165--200},
\href{http://arxiv.org/abs/hep-th/0003085}{{\ttfamily arXiv:hep-th/0003085
  [hep-th]}}.
%%CITATION = HEP-TH/0003085;%%.

\bibitem{Feng:2001xr}
B.~Feng, A.~Hanany, and Y.-H. He, ``{Phase structure of D-brane gauge theories
  and toric duality},'' {\em JHEP} {\bfseries 0108} (2001) 040,
\href{http://arxiv.org/abs/hep-th/0104259}{{\ttfamily arXiv:hep-th/0104259
  [hep-th]}}.
%%CITATION = HEP-TH/0104259;%%.

\bibitem{HK05}
A.~Hanany and K.~D. Kennaway, ``{Dimer models and toric diagrams},''
\href{http://arxiv.org/abs/hep-th/0503149}{{\ttfamily arXiv:hep-th/0503149
  [hep-th]}}.
%%CITATION = HEP-TH/0503149;%%.

\bibitem{FHKVW05}
S.~Franco, A.~Hanany, K.~D. Kennaway, D.~Vegh, and B.~Wecht, ``{Brane dimers
  and quiver gauge theories},''
  \href{http://dx.doi.org/10.1088/1126-6708/2006/01/096}{{\em JHEP} {\bfseries
  0601} (2006) 096},
\href{http://arxiv.org/abs/hep-th/0504110}{{\ttfamily arXiv:hep-th/0504110
  [hep-th]}}.
%%CITATION = HEP-TH/0504110;%%.

\bibitem{Feng:2005gw}
B.~Feng, Y.-H. He, K.~D. Kennaway, and C.~Vafa, ``{Dimer models from mirror
  symmetry and quivering amoebae},'' {\em Adv.Theor.Math.Phys.} {\bfseries 12}
  (2008) 3,
\href{http://arxiv.org/abs/hep-th/0511287}{{\ttfamily arXiv:hep-th/0511287
  [hep-th]}}.
%%CITATION = HEP-TH/0511287;%%.


\bibitem{Aspinwall09}
P.~S. Aspinwall, T.~Bridgeland, A.~Craw, M.~Douglas, M.~Gross, A.~Kapustin,
  G.~W. Moore, G.~Segal, B.~Szendr{\H o}i, and P.~Wilson, {\em {Dirichlet
  Branes and Mirror Symmetry, (Clay mathematics monographs. Volume 4)}}.
\newblock American Mathematical Society, Clay Mathematics Institute, 2009.


\bibitem{derksen1}
H.~Derksen, J.~Wyman, and A.~Zelevinsky,
``Quivers with potentials and their representations I: Mutations,''
Selecta Mathematica \textbf{14} (2008) 59--119.

\bibitem{kellerA}
B.~Keller and S.~Yang,
``Derived equivalences from mutations of quivers with potentials,''
Adv. Math. \textbf{226} (2011), no. 3, 2118--2168;
\href{http://arxiv.org/abs/0906.0761}{{\ttfamily arXiv:0906.0761 [math.RT]}}.


\bibitem{Seiberg}
N.~Seiberg, ``{Electric - magnetic duality in supersymmetric nonAbelian gauge
  theories},'' \href{http://dx.doi.org/10.1016/0550-3213(94)00023-8}{{\em
  Nucl.Phys.} {\bfseries B435} (1995) 129--146},
\href{http://arxiv.org/abs/hep-th/9411149}{{\ttfamily arXiv:hep-th/9411149
  [hep-th]}}.
%%CITATION = HEP-TH/9411149;%%.

\bibitem{CNV}
S.~Cecotti, A.~Neitzke, and C.~Vafa, ``{R-Twisting and 4d/2d
  Correspondences},''
\href{http://arxiv.org/abs/1006.3435}{{\ttfamily arXiv:1006.3435 [hep-th]}}.
%%CITATION = 1006.3435;%%.


\bibitem{CV92}
S.~Cecotti and C.~Vafa, ``{On classification of N=2 supersymmetric theories},''
  \href{http://dx.doi.org/10.1007/BF02096804}{{\em Commun. Math. Phys.}
  {\bfseries 158} (1993) 569--644},
\href{http://arxiv.org/abs/hep-th/9211097}{{\ttfamily arXiv:hep-th/9211097}}.
%%CITATION = HEP-TH/9211097;%%.

\bibitem{kellerB}
B.~Keller,
``Deformed Calabi--Yau completions,''
\href{http://arxiv.org/abs/0908.3499}{{\ttfamily arXiv:0908.3499 [math.RT]}}.

\bibitem{Ring1}
C.M.~Ringel, 
``The spectral radius of the Coxeter transformations for a generalized Cartan Matrix,'' 
Math. Ann. \textbf{300} 331--339 (1994).

\bibitem{AD}
P.~C. Argyres and M.~R. Douglas, ``{New phenomena in SU(3) supersymmetric gauge
  theory},'' \href{http://dx.doi.org/10.1016/0550-3213(95)00281-V}{{\em
  Nucl.Phys.} {\bfseries B448} (1995) 93--126},
\href{http://arxiv.org/abs/hep-th/9505062}{{\ttfamily arXiv:hep-th/9505062
  [hep-th]}}.



\bibitem{felikson}
A.~Felikson, M.~Shapiro, and P.~Tumarkin,
``Skew--symmetric cluster algebras of finite mutation type,''
\href{http://arxiv.org/abs/0811.1703}{{\ttfamily arXiv:0811.1703 [math.CO]}}.

\bibitem{Gaiotto}
D.~Gaiotto, ``{N=2 dualities},''
\href{http://arxiv.org/abs/0904.2715}{{\ttfamily arXiv:0904.2715 [hep-th]}}.
%%CITATION = ARXIV:0904.2715;%%.


\bibitem{GMN09}
D.~Gaiotto, G.~W. Moore, and A.~Neitzke, ``{Wall-crossing, Hitchin Systems, and
  the WKB Approximation},''
\href{http://arxiv.org/abs/0907.3987}{{\ttfamily arXiv:0907.3987 [hep-th]}}.
%%CITATION = ARXIV:0907.3987;%%.

\bibitem{WW}
S.~Weinberg and E.~Witten,
``Limits on massless particles,''
Phys. Lett. \textbf{B96} (1980) 59.

\bibitem{king}
A.D.~King,
``Moduli of representations of finite--dimensional algebras,''
Quart. J. Math. Oxford Ser.(2) \textbf{45} (1994), 515--530.

\bibitem{RI}
C.M.~Ringel,
\textit{Tame Algebras and Integral Quadratic Forms,}
\textsc{Lecture Notes in Mathematics} \textbf{1099}, Springer, Berlin, (1984).



\bibitem{CD1}
O.~Chacaltana and J.~Distler, ``{Tinkertoys for Gaiotto Duality},''
  \href{http://dx.doi.org/10.1007/JHEP11(2010)099}{{\em JHEP} {\bfseries 1011}
  (2010) 099},
\href{http://arxiv.org/abs/1008.5203}{{\ttfamily arXiv:1008.5203 [hep-th]}}.

\bibitem{CD2}
O.~Chacaltana and J.~Distler, ``{Tinkertoys for the $D_N$ series},''
\href{http://arxiv.org/abs/1106.5410}{{\ttfamily arXiv:1106.5410 [hep-th]}}.



\bibitem{ASS}
D.~Simson and A.~Skowronski,
\textit{Elements of the Representation Theory of Associative Algebras. 2: Tubes and Concealed Algebras of Euclidean type,}
\textsc{London Mathematical Society Student Texts} \textbf{71}, Cambridge Univerty Press (2007).


\bibitem{butring}
M.C.R.~Butler and C.M.~Ringel,
``Auslander--Reiten sequences with few middle terms and applications to string algebras,''
Comm. in Algebra \textbf{15} (1987) 145--179.

\bibitem{kellerP}
B.~Keller,
``The periodicity conjecture for pairs of Dynkin diagrams,''
\href{http://arxiv.org/abs/1001.1531}{{\ttfamily arXiv:1001.1531 [math.RT]}}.
 .
\bibitem{CB}
W.~Crawley--Boevey,
``Lectures on Representations of Quivers,''
available on line at \href{http://www1.maths.leeds.ac.uk/~pmtwc/quivlecs.pdf}{{\ttfamily http://www1.maths.leeds.ac.uk/~pmtwc/quivlecs.pdf}}.


\bibitem{pre1}
I.M.~Gelfand and V.A.~Ponomarev, 
``Model algebras and representations of graphs,''
Funktsional. Anal. i Prilozhen. \textbf{13} (1979) 1--12.


\bibitem{pre2}
C.M.~Ringel, 
``The preprojective algebra of a quiver,''
in \textit{Algebras and modules, II} (Geiranger, 1996), 467--480, CMS Conf.\! Proc.\! \textbf{24}, Amer.~Math.~Soc., Providence, RI, 1998.


\bibitem{Tack}
Y.~Tachikawa and S.~Terashima,
``Seiberg--Witten Geometries Revisited,''
\href{http://arxiv.org/abs/1108.2315}{{\ttfamily arXiv:1108.2315 [hep-th]}}.

\bibitem{LE1}
H.~Lenzing,
``Hereditary categories,''
 in \textit{Handbook of Tilting Theory,} Ed.~by A.~H\"ugel, D.~Happel, and H.~Krause,
\textsc{London Mathematical Society Lecture Notes Series} \textbf{332}
Cambridge University Press, Cambridge 2007, pages 105--145. 

\bibitem{LE2}
W.~Geigle and H.~Lenzing,
``A class of weighted projective lines arising in representation theory of finite dimensional algebras,''
\textsc{Lectures Notes in Mathematcs} \textbf{1273}, (1987) pp. 81--104. 

\bibitem{kac1}
V.G.~Kac,
``Infinite Root Systems, Representations of Graphs and Invariant Theory,''
Inv. Math. \textbf{56} (1980) 56--92.

\bibitem{kac2}
V.G.~Kac,
``Root systems, representations of quivers and invariant theory,''
in \textit{Invariant Theory,} Proceeding, Montecatini 1982, \textsc{Lectures Notes in Mathematics}
\textbf{996}, Springer (1983) pp.74--107.


\bibitem{LZ}
H.~Lenzing and J.-A.~de la Pe\~na, 
``Spectral analysis of finite dimensional algebras and singularities,''
\href{http://arxiv.org/abs/0805.1018}{{\ttfamily arXiv:0805.1018 [math.RT]}}.

\bibitem{ASS1}
I.~Assem, D.~Simson and A.~Skowronski,
\textit{Elements of the Representation Theory of Associative Algebras. 1: Techniques of Representation Theory,}
\textsc{London Mathematical Society Student Texts} \textbf{65}, Cambridge Univerty Press (2006).


\bibitem{saito}
K.~Saito,
``Extended affine root systems.I. Coxeter transformations,'' 
Publ. Res. Inst. Math. SCi. \textbf{21} (1985) 75--179.


\bibitem{moody}
R.V.~Moody, S.~Eswara Rao and T.~Yokonuma,
``Toroidal Lie algebras and Vertex representations,''
Geom. Dedicata \textbf{35} (1990) 29=83--307.


\bibitem{CBBB}
W.~Crawley--Boevey,
``Kac's theorem for weighted projective lines,''
 J. Eur. Math. Soc. (JEMS) \textbf{12} (2010), no. 6, 1331--1345;
 \href{http://arxiv.org/abs/math/051207}{{\ttfamily arXiv:math/051207 [math.AG]}}.


\bibitem{ringelderived}
D.~Happel and C.M. Ringel,
``The derived category of a tubular algebra,''
in \textit{Representation Theory I. Finite Dimensional Algebras,}
\textsc{Lecture Notes in Mathematics} \textbf{1177} (1986) Springer.

\bibitem{tuderived}
H.~Lenzing and H.~Meltzer, 
``The automorphism group of the derived category for a weighted projective line,''
Comm. Algebra \textbf{28} (2000) 1685--1700.

\bibitem{tubcluster}
M.~Barot, D.~Kussin, and H.~Lenzing,
``The cluster category of a canonical algebra,''
Trans. Amer. Math. Soc. \textbf{362} (2010), no. 8, 4313--4330;
\href{http://arxiv.org/abs/0801.4540}{{\ttfamily arXiv:0801.4540 [math.RT]}}.

\bibitem{LE7}
H.~Lenzing,
``Hereditary Categories, Lectures 1 and 2,''
Advanced School and Conference on Representation Theory and related Topics, ICTP, Trieste Jannuary 2006,
available on line at \\
\href{http://www.math.jussieu.fr/~keller/ictp2006/lecturenotes/lenzing1.pdf}{{\ttfamily http://www.math.jussieu.fr/~keller/ictp2006/lecturenotes/lenzing1.pdf}}.


\bibitem{llll}
H.~Lenzing,
``Hereditary Categories, Lectures 3, 4 and 5,''
Advanced School and Conference on Representation Theory and related Topics, ICTP, Trieste Jannuary 2006,
available on line at \\
\href{http://www.math.jussieu.fr/~keller/ictp2006/lecturenotes/lenzing2.pdf}{{\ttfamily http://www.math.jussieu.fr/~keller/ictp2006/lecturenotes/lenzing2.pdf}}.


\bibitem{triangulation1}
S.~{Fomin}, M.~{Shapiro}, and D.~{Thurston}, ``{Cluster algebras and
  triangulated surfaces. Part I: Cluster complexes},'' {\em Acta Mathematica}
  {\bfseries 201} (Aug., 2006) 83--146,
  \href{http://arxiv.org/abs/arXiv:math/0608367}{{\ttfamily
  arXiv:math/0608367}}.



\bibitem{traingulation2}
D.~Labardini--Fragoso,
``Quivers with potentials associated to triangulated surfaces,''
Proc. London Math. Soc. \textbf{98} (2009) 797--839.
 \href{http://arxiv.org/abs/arXiv:0803.1328}{{\ttfamily
  arXiv:0803.1328 [math.RT]}}.


\bibitem{derksen}
H.~Derksen and R.~Owen, ``New graphs of finite mutation type'',
\href{http://www.arXiv.org/abs/0804.0787}{{\tt arXiv:0804.0787 [math.CO]}}.


\bibitem{assemgentle}
I.~Assem, T.~Br\"ustle, G.~Charbonneau--Jodoin and P-G.~Plamondon,
``Gentle algebras arising from surface triangulations,''
\href{http://www.arXiv.org/abs/0805.1035}{{\tt 0805.1035 [math.RT]}}


\bibitem{LF2}
D.~{Labardini-Fragoso}, ``{Quivers with potentials associated to triangulated
  surfaces, Part II: Arc representations},''
  \href{http://arxiv.org/abs/0909.4100}{{\ttfamily arXiv:0909.4100 [math.RT]}}

\bibitem{kellerapp}
B.~Keller,
``Quiver mutation in Java'', available from the author's homepage, {\tt http://www.institut.math.jussieu.fr/$\widetilde{\phantom{-}}$keller/quivermutation}

\bibitem{hubery}
A.~Hubery,
``Quiver representations respecting a quiver automorphism: a generalization of a theorem of Kac,''
 \href{http://arxiv.org/abs/math/0203195}{{\ttfamily arXiv:math/0203195 [math.RT]}}

\bibitem{geissX}
C.~Geiss, B.~Leclerc, and J.~Schro\"oer,
``Semicanonical bases and preprojective algebras,''
Ann. Sci. \'Ecole Norm. Sup. (4) \textbf{38} (2005), no. 2, 193--253;
 \href{http://arxiv.org/abs/math/0402448}{{\ttfamily arXiv:math/0402448 [math.RT]}}


\bibitem{gukovrrr}
S.~Gukov and E.~Witten,
``Gauge theory, ramification, and the geometric Langalands program,''
\href{http://arxiv.org/abs/hept-th/0612073}{{\ttfamily arXiv:hept-th/0612073 [hep-th]}}

\bibitem{KS}
M.~{Kontsevich} and Y.~{Soibelman}, ``{Stability structures, motivic
  Donaldson-Thomas invariants and cluster transformations},'' {\em ArXiv
  e-prints} (Nov., 2008) , \href{http://arxiv.org/abs/0811.2435}{{\ttfamily
  arXiv:0811.2435 [math.AG]}}.

\bibitem{DG}
T.~Dimofte and S.~Gukov, ``{Refined, Motivic, and Quantum},''
  \href{http://dx.doi.org/10.1007/s11005-009-0357-9}{{ Lett. Math. Phys.}
  {\bfseries 91} (2010) 1},
\href{http://arxiv.org/abs/0904.1420}{{\ttfamily arXiv:0904.1420 [hep-th]}}.
%%CITATION = ARXIV:0904.1420;%%.


\bibitem{CV09}
S.~Cecotti and C.~Vafa, ``{BPS Wall Crossing and Topological Strings},''
\href{http://arxiv.org/abs/0910.2615}{{\ttfamily arXiv:0910.2615 [hep-th]}}.
%%CITATION = ARXIV:0910.2615;%%.

\bibitem{DGS}
T.~Dimofte, S.~Gukov, and Y.~Soibelman, ``{Quantum Wall Crossing in N=2 Gauge
  Theories},'' \href{http://dx.doi.org/10.1007/s11005-010-0437-x}{{\em
  Lett.Math.Phys.} {\bfseries 95} (2011) 1--25},
\href{http://arxiv.org/abs/0912.1346}{{\ttfamily arXiv:0912.1346 [hep-th]}}.
%%CITATION = ARXIV:0912.1346;%%.

\bibitem{GMN10}
D.~Gaiotto, G.~W. Moore, and A.~Neitzke, ``{Framed BPS States},''
\href{http://arxiv.org/abs/1006.0146}{{\ttfamily arXiv:1006.0146 [hep-th]}}.
%%CITATION = ARXIV:1006.0146;%%.

\bibitem{ADJM1}
E.~Andriyash, F.~Denef, D.~L. Jafferis, and G.~W. Moore, ``{Wall-crossing from
  supersymmetric galaxies},''
\href{http://arxiv.org/abs/1008.0030}{{\ttfamily arXiv:1008.0030 [hep-th]}}.
%%CITATION = ARXIV:1008.0030;%%.

\bibitem{ADJM2}
E.~Andriyash, F.~Denef, D.~L. Jafferis, and G.~W. Moore, ``{Bound state
  transformation walls},''
\href{http://arxiv.org/abs/1008.3555}{{\ttfamily arXiv:1008.3555 [hep-th]}}.
%%CITATION = ARXIV:1008.3555;%%.

\bibitem{MPS}
J.~Manschot, B.~Pioline, and A.~Sen, ``{Wall Crossing from Boltzmann Black Hole
  Halos},'' \href{http://dx.doi.org/10.1007/JHEP07(2011)059}{{\em JHEP}
  {\bfseries 1107} (2011) 059},
\href{http://arxiv.org/abs/1011.1258}{{\ttfamily arXiv:1011.1258 [hep-th]}}.
%%CITATION = ARXIV:1011.1258;%%.

\bibitem{GMN11}
D.~Gaiotto, G.~W. Moore, and A.~Neitzke, ``{Wall-Crossing in Coupled 2d-4d
  Systems},''
\href{http://arxiv.org/abs/1103.2598}{{\ttfamily arXiv:1103.2598 [hep-th]}}.
%%CITATION = ARXIV:1103.2598;%%.

\bibitem{CBlemma}
W.~Crawley--Boevey,
``On the exceptional fibres of Kleinian singularities,''
Amer. J. Math. \textbf{122} (2000) 1027--1037.

\bibitem{GLS}
C.~Geiss, B.~Leclerc, and J.~Sch\"oer,
``Kac--Moody groups and cluster algebras,''
Adv. Math. \textbf{228} (2011), no. 1. 329--433;
\href{http://arxiv.org/abs/1001.3545}{{\ttfamily arXiv:1001.3545 [math.RT]}}.

\bibitem{LU}
G.~Lusztig,
``Quivers, perverse sheaves, and quantized enveloping algebras,''
 J. Amer. Soc. \textbf{4} (1991) 313--324; 

\bibitem{GAB}
P.~{Gabriel}, ``{Unzerlegbare Darstellungen},''
  \href{http://dx.doi.org/10.1007/BF01298413}{{Manuscripta Mathematica}
  {\bfseries 6} (1972) 71--103}.
  
  
 \bibitem{Len34} 
 H.~Lenzing and J.A.~de la Pe\~na, 
 ``Extended canonical algebras and fuchsian singularities,'' 
 \href{http://arxiv.org/abs/math/0611532}{{\ttfamily arXiv:math/0611532 [math.RT]}}.

\bibitem{schofield}
A.~Schofield,
``General representations of quivers,''
Proc. London Math. Soc. (3) \textbf{65} (1992) 46--64.

\bibitem{tt*}
S.~Cecotti and C.~Vafa,
``Topological anti--topological fusion,''
Nucl. Phys. \textbf{B367} (1991) 359--461.

\bibitem{kellerR}
B.~Keller,
``Cluster algebras, quiver representations, and triangulated categories,''
\href{http://arxiv.org/abs/0807.1960}{{\ttfamily arXiv:0807.1960 [math.RT]}}.

\bibitem{COXX}
R.~Stekolshchik,
\textit{Notes on Coxeter Transformations and the McKay Correspondence,}
\textsc{Springer Monographs in Mathematics}, Springer, Berlin, (2008).


\bibitem{kell33}
B.~Keller and S.~Scherotzke,
``Linear reccurrence relations for cluster variables of affine quivers,''
Adv. Math. \textbf{228} (2011), no. 3, 1842--1862;
\href{http://arxiv.org/abs/1004.0613}{{\ttfamily arXiv:1004.0613 [math.RT]}}.

\bibitem{LLLL}
H.~Lenzing and J.A.~de la Pe\~na, 
``Spectral analysis of finite dimensional algebras and singularities,''
\href{http://arxiv.org/abs/0805.1018}{{\ttfamily arXiv:0805.1018 [math.RT]}}. 


\bibitem{ASS3}
D.~Simson and A.~Skowronski,
\textit{Elements of the representation theory of associative algebras: Vol. 3, Representation--infinite tilted algebras,}
\textsc{London Mathematical Society Student Texts} \textbf{72},
Cambridge University Press (2007).

\end{thebibliography}
\end{document}